\documentclass[aps,letterpaper,nofootinbib]{revtex4-1}
\pdfoutput=1
\usepackage{bm}
\usepackage{subfigure}
\usepackage{amssymb,amsmath,graphicx,color}

\usepackage{lineno}

\modulolinenumbers[5]
\newcommand{\gsim}{\raisebox{-0.7ex}{$\stackrel{\textstyle >}{\sim}$ }}

\def\si{^1 \hskip -0.03in S _0}
\def\siii{^3 \hskip -0.025in S _1}
\def\diii{^3 \hskip -0.03in D _1}

\def\tnubb{$2\nu\beta\beta$}
\def\znubb{$0\nu\beta\beta$}

\def\cQ{{\mathcal Q}}

\def\cO{{\mathcal O}}

\def\cQ{{\mathcal Q}}

\def\sigmaPHYS{332.4^{+(5.4)}_{-(4.7)}}
\def\sigmaEXPT{334.2(0.5)}

\def\mud{1.41^{+(28)(4)}_{-(25)(4)}}

\def\betann{0.296^{+(19)(15)}_{-(18)(15)}}

\def\betadpm{0.70^{+(24)(4)}_{-(23)(4)}}

\def\dbetannCORR{-0.070^{+(6)(4)}_{-(9)(4)}}

\begin{document}

\hbox{MIT-CTP/5229, UMD-PP-020-4, ICCUB-20-018, FERMILAB-PUB-20-445-T \\[20pt]}

\title{
Nuclear matrix elements from lattice QCD\\ for electroweak and beyond--Standard-Model processes}

\author{Zohreh Davoudi}
\address{Department of Physics, University of Maryland, College Park, MD 20742, USA}
\address{RIKEN Center for Accelerator-based Sciences, Wako 351-0198, Japan}

\author{William Detmold} 

\author{Phiala Shanahan}
\email{phiala@mit.edu}

\address{Center for Theoretical Physics, Massachusetts Institute of Technology, Cambridge, MA 02139, USA}

\author{Kostas Orginos}
\address{Department of Physics, College of William and Mary, Williamsburg, VA 23187-8795, USA}
\address{Thomas Jefferson National Accelerator Facility, 12000 Jefferson Avenue, Newport News, VA 23606, USA}

\author{Assumpta Parre\~no}
\address{Departament de F\'isica Qu\`antica i Astrof\'isica and Institut de Ci\`encies del Cosmos, }
\address{Universitat de Barcelona, Mart\'i i Franqu\`es, 1, E-08028 Barcelona, Spain}

\author{Martin J. Savage}
\address{Institute for Nuclear Theory, University of Washington, Seattle, WA 98195, USA}

\author{Michael L. Wagman}
\address{Fermi National Accelerator Laboratory, Batavia, IL 60510, USA}

\begin{abstract}

Over the last decade, numerical solutions of Quantum Chromodynamics (QCD) using the technique of lattice QCD have developed to a point where they are beginning to connect fundamental aspects of nuclear physics to the underlying degrees of freedom of the Standard Model. In this review, the progress of lattice QCD studies of nuclear matrix elements of electroweak currents and beyond-Standard--Model operators is summarized, and connections with effective field theories and nuclear models are outlined. 

Lattice QCD calculations of nuclear matrix elements can provide guidance for low-energy nuclear reactions in astrophysics, dark matter direct detection experiments, and experimental searches for violations of the symmetries of the Standard Model, including searches for additional CP violation in the hadronic and leptonic sectors, baryon-number violation, and lepton-number or flavor violation. Similarly, important inputs to neutrino experiments seeking to determine the neutrino-mass hierarchy and oscillation parameters, as well as other electroweak and beyond-Standard--Model processes can be determined. The phenomenological implications of existing studies of  electroweak and beyond-Standard--Model matrix elements in light nuclear systems are discussed, and future prospects for the field toward precision studies of these matrix elements are outlined.

\end{abstract}

\maketitle

\newpage

\tableofcontents

\newpage

\section{Introduction}
\label{sec:intro}

Establishing reliable predictive capabilities for the properties and reactions of nuclei from the Standard Model (SM)~\cite{Glashow:1961tr,Weinberg:1967tq,Salam:1968rm,Politzer:1973fx,Gross:1973id}, which describes the strong and electroweak interactions in nature, is a defining challenge that bridges nuclear and particle physics~\cite{NANP2013,Geesaman:2015fha}.  
Nuclear interactions play an essential role in  the evolution of the universe, and  strong and electroweak effects conspire in key nuclear processes, such as in those that govern the  nuclear reactions in the first minutes after the big bang \cite{Iocco:2008va,Cyburt:2015mya}, in reactions that power stars like the Sun \cite{Adelberger:2010qa}, and in supernova and other extreme astrophysical environments~\cite{Burrows:2012ew,Janka:2006fh,doi:10.1146/annurev.ns.38.120188.000245}.
Understanding these reactions has been a long-standing challenge for nuclear physics as they are often difficult to probe in the laboratory.
Quantitative control of nuclear structure and reactions based in the SM is also essential to constraining beyond-Standard--Model (BSM) physics scenarios from experimental searches. In particular, nuclear targets are ubiquitous in intensity-frontier experiments \cite{Hewett:2014qja} from laboratory searches for dark matter candidates such as weakly interacting massive particles~\cite{Bertone:2004pz,Feng:2010gw,Cushman:2013zza}, to searches for violations of  fundamental symmetries of the SM \cite{Cirigliano:2013lpa,Cirigliano:2019jig}, to long-baseline neutrino experiments aiming to constrain the neutrino-mass hierarchy and oscillation parameters~\cite{Alvarez-Ruso:2017oui,Kronfeld:2019nfb}. 
For such experiments, there are compelling reasons to determine the relevant nuclear matrix elements from the SM with complete uncertainty quantification; without controlled SM expectations, new physics cannot be effectively constrained. 
The key challenge in reliably determining electroweak matrix elements in nuclear systems is computing the effects of the strong interactions, described by Quantum Chromodynamics (QCD), that bind the fundamental constituents together, first into protons and neutrons, and then into nuclei. The only known systematically-improvable first-principles approach to this challenge is via the numerical technique of lattice QCD (LQCD). During the last decades, LQCD~\cite{Wilson:1974sk,Balian:1974ts,Creutz:1979kf} has become established as a reliable and controlled method of computing many hadronic quantities in the low-energy, low-density regime of QCD~\cite{Gattringer:2010zz,Rothe:1992nt,DeGrand:2006zz,Montvay:1994cy}. Progress in, and prospects for, the application of LQCD to electroweak and BSM processes in nuclei are the subjects of this review.

Since nuclei are intricate systems with multiple physically-important scales, constraining their strong-interaction dynamics is a significant theoretical challenge. Nevertheless, the exact and approximate symmetries of QCD constrain the form of nuclear interactions and instill a hierarchy in the size of their contributions.  Predictions for low-energy nuclear structure and processes can be made using
phenomenological nuclear forces and effective field theories (EFTs) based on this hierarchy. Within their regimes of validity, these approaches, and the extensive suite of nuclear many-body techniques that implement them, can be tuned to reproduce a subset of experimental constraints and  predict related quantities  (for recent reviews see, for example, Refs.~\cite{Carlson:2014vla,Meissner:2015wva,Hammer:2019poc,Epelbaum:2019kcf,Tews:2020hgp}).    
Building on many theoretical and computational advances over the last decade, a particularly sophisticated application of this approach recently demonstrated that calculations using higher-order chiral potentials including multi-nucleon correlations and currents 
can achieve an accurate description of Gamow-Teller decay matrix elements in medium-mass nuclei~\cite{Gysbers:2019uyb,King:2020wmp}.
To address processes for which experimental data is limited or absent, or to extend calculations beyond the regimes of validity of EFT approaches, where operator hierarchies are less clear and where many-body effects conspire, constraints directly from LQCD are expected to play an increasingly important role. In recent years, the first LQCD studies of nuclear structure have been performed~\cite{Beane:2014ora,Beane:2015yha,Savage:2016kon,Shanahan:2017bgi,Tiburzi:2017iux,Chang:2017eiq}, albeit with significant uncertainties. Achieving reliable LQCD calculations of nuclear matrix elements will enhance the connection between the SM and low-energy nuclear physics and promises to provide a unified SM foundation for computing nuclear structure and processes. This program, which is at the heart of the nuclear-physics mission~\cite{Beane:2008dv,GCNP_exascale,daviddeanCNP2014,GCNP_ResReqs,Carlson:2017ebk,Detmold:2019ghl}, is in synergy with the ongoing program of constraining and understanding nuclear physics through phenomenology, EFTs, and nuclear models.

In the last decade, the predictive capabilities of LQCD have been revolutionized through the development of new and improved algorithms  and the growth of computing resources \cite{Joo:2019byq}.
The mass of the proton has been recovered with better than percent-level precision, including both QCD and Quantum Electrodynamics (QED) interactions~\cite{Borsanyi:2014jba}. Moreover, aspects of nucleon structure from the decomposition of its spin, mass, and  momentum~\cite{Alexandrou:2017oeh,Yang:2018nqn,Lin:2018obj}, to its electromagnetic form factors~\cite{Rajan:2017lxk,Jang:2019jkn,Djukanovic:2019jtp,Alexandrou:2018sjm,Kallidonis:2018cas} and pressure distributions~\cite{Shanahan:2018nnv}, to its scalar, axial, and tensor charges~\cite{Bhattacharya:2016zcn,Gupta:2018qil,Chang:2018uxx,Bali:2019yiy,Capitani:2017qpc,Durr:2011mp,Durr:2015dna,Yang:2015uis,Bali:2012qs,Freeman:2012ry,Junnarkar:2013ac,Hasan:2019noy}, have been studied with quantified uncertainties. Indeed, the Flavor Lattice Averaging Group (FLAG) review~\cite{Aoki:2019cca}, historically dedicated to providing summaries of LQCD results relating to flavor physics in the meson sector, now includes select properties of the nucleon. At the same time, there has been significant progress in LQCD studies of thermodynamics~\cite{Bazavov:2014pvz,Borsanyi:2013bia}, of the physics of hadrons containing heavy quarks~\cite{Bazavov:2016nty,Lattice:2015tia,Lattice:2015rga,Detmold:2016pkz,Detmold:2015aaa,Bazavov:2019aom,Bazavov:2018omf}, and in constraining the SM contributions to the anomalous magnetic moment of the muon~\cite{Meyer:2018til,Blum:2019ugy,Blum:2018mom,Davies:2019efs,Borsanyi:2020mff,Gerardin:2019rua}.

Conceptually, the strong-interaction physics of nuclei is no more complicated to compute in the lattice field theory framework than that of the proton; protons and nuclei emerge in the same way from the dynamics of the quarks and gluons encoded by QCD. In practice, however, nuclear LQCD calculations suffer from  increased computational complexity compared to those for the proton, and also experience sampling noise that grows exponentially with the size of the nuclear system under study~\cite{Parisi:1983ae, Lepage:1989hd,Beane:2009py,Beane:2009gs,Beane:2010em,Beane:2014oea}. 
Furthermore, the QCD coupling and quark masses of the SM are such that there are a number of fine-tunings and emergent symmetries that manifest in the dynamics and structure of nuclei, such as Wigner's symmetry~\cite{Wigner:1937zz,Wigner:1939zz}, beyond those explicit in the SM Lagrangian. Reproducing these intricate features requires precision calculations.
For these reasons, despite more than a decade of progress and development, the era of fully-controlled LQCD calculations of the structure and interactions of nuclei is only just beginning.

The first LQCD studies of systems with baryon number greater than one were attempted more than 25 years ago~\cite{Fukugita:1994na,Pochinsky:1998zi,Wetzorke:1999rt,Wetzorke:2002mx}.
In the 2000s, refined techniques with which to study two-baryon systems were developed~\cite{Beane:2006gf,Ishii:2006ec,Yamazaki:2009ua,Beane:2006mx,Beane:2009py}, nuclei were studied in quenched QCD \cite{Yamazaki:2009ua}, and calculations of the $H$-dibaryon (a spin and isospin  singlet with strangeness $|S|=2$)~\cite{Beane:2009py,Beane:2010hg,Inoue:2010es,Beane:2011zpa} were the first to clearly identify QCD bound states in such systems. These calculations were all undertaken at unphysical values of the quark masses in order to reduce the computational resource requirements. These investigations were followed by further studies of states in the $\si$, $\siii$, and coupled $\siii$-$\diii$ two-nucleon channels~\cite{Beane:2011iw,Beane:2012vq,Yamazaki:2012hi,Beane:2013br,Yamazaki:2015asa,Francis:2018qch,Wagman:2017tmp}, and extended to states in higher partial waves~\cite{Berkowitz:2015eaa}. 
Simultaneously, methods based on the construction of non-relativistic Bethe-Salpeter wavefunctions and potentials were developed to access scattering information~\cite{Luscher:1986pf,Lin:2001ek,Aoki:2005uf,Ishii:2006ec,Murano:2011nz,Aoki:2011gt,HALQCD:2012aa,Sasaki:2019qnh}.
There have been extractions of three-hadron forces based on LQCD calculations in both the meson~\cite{Beane:2007es,Detmold:2008yn,Detmold:2011kw,Blanton:2019vdk} and baryon sectors~\cite{Doi:2011gq,Beane:2012vq,Barnea:2013uqa,Contessi:2017rww,Beane:2009gs}. Calculations of systems up to atomic number $A=5$  have been performed over the past decade with a range of unphysically-large values of the quark masses~\cite{Beane:2012vq,Yamazaki:2012hi,Yamazaki:2015asa}.
These LQCD studies of light nuclei have been used  to constrain nuclear EFTs, allowing constraints on larger nuclei and on the quark-mass dependence of nuclear forces and bindings~\cite{Beane:2008dv,Beane:2010em,Beane:2012ey,Epelbaum:2012iu,Barnea:2013uqa,Contessi:2017rww,Lahde:2019yvr,Eliyahu:2019nkz}.
While ongoing efforts aim to obtain results at the physical values of the quark masses~\cite{Doi:2017zov}, importantly, the ability to undertake LQCD calculations with unphysical quark masses may also provide phenomenologically-important results~\cite{Beane:2002vs,Beane:2002xf,Kneller:2003ka,Chen:2010yt,Meissner:2013lpp,Meissner:2014pma,Epelbaum:2013wla,Bansal:2017pwn}. For example, an essential ingredient to \emph{ab initio} nuclear many-body studies of the Hoyle state (the first $0^+$ excitation of $^{12}$C) is the rate of the change of the two-nucleon scattering lengths with respect to the quark masses near their physical values~\cite{Epelbaum:2012iu,Epelbaum:2013wla,Bansal:2017pwn,Lahde:2019yvr}. While this rate cannot be determined from experiment, it could be from LQCD, and sufficiently precise LQCD determinations would provide insight into expected fine-tunings in the  reactions that produce  carbon and oxygen in nature, and in the placement of the Hoyle state in the vicinity of the $^8{\rm Be}+\alpha$ resonance~\cite{Epelbaum:2012iu}.

With LQCD studies of nuclei progressing, the first attempts to investigate nuclear structure directly from the dynamics of quarks and gluons have also been made, complementing the existing body of experimental data, phenomenological modeling, and EFT analyses. The isovector magnetic moments~\cite{Beane:2014ora,Beane:2015yha,Detmold:2015daa} and magnetic  polarizabilities~\cite{Chang:2015qxa} of nuclei up to $A=4$ have been computed at larger-than-physical quark masses, and gluonic aspects of nuclear structure have been investigated~\cite{Winter:2017bfs}. Furthermore, the simplest nuclear reactions, such as slow neutron capture ($np\rightarrow d\gamma$)~\cite{Beane:2015yha} and $pp$ fusion ($pp\rightarrow d e^+ \nu$)~\cite{Savage:2016kon}, have been computed from a combination of LQCD and EFT. 
The Gamow-Teller contributions to triton $\beta$ decay~\cite{Savage:2016kon} and the couplings of $A\le3$ nuclei to scalar and tensor currents~\cite{Chang:2017eiq} have also been investigated. Finally, studies of the neutrinoful ($2\nu\beta\beta$) and neutrinoless ($0\nu\beta\beta$) double-$\beta$ decay processes have begun~\cite{Tiburzi:2017iux,Shanahan:2017bgi,Nicholson:2018mwc,Feng:2018pdq,Detmold:2018zan,Tuo:2019bue,Detmold:2020jqv}.
Since the focus of this review article is the current status of, and future prospects for, the determination of nuclear matrix elements of electroweak and BSM  currents using LQCD, and their connection to few- and many-body studies in nuclear physics, the impressive progress of the last decade in constraining single-hadron matrix elements will not be reviewed.\footnote{The 2019 FLAG report \cite{Aoki:2019cca} provides a recent compilation of many results in the single-hadron sector.}

As the field of nuclear LQCD continues to develop, the level of insight that it provides will grow. With calculations at the physical values of the quark masses, and with full control of lattice discretization and finite-volume effects, the next generations of LQCD studies of electroweak and BSM nuclear matrix elements will 
impact many  key areas of nuclear physics. Beyond electroweak and BSM matrix elements, LQCD is also expected to quantitatively elucidate the QCD origin of important aspects of nuclear structure such as the EMC effect \cite{Aubert:1983xm}, i.e., the difference between the parton distributions of a nucleus and those of the constituent nucleons.
While LQCD calculations of nuclei are now benefiting from petascale high-performance computing resources for the first time, sustained exascale computing and beyond will be required to achieve some of the goals of the field with the precision and accuracy required to maximally impact nuclear and high-energy physics~\cite{GCNP_exascale,Savage:2010hp,GCNP_ResReqs}.

\section{Lattice QCD for nuclear physics}
\label{sec:tools}

This section provides an overview of LQCD and the challenges associated with applying this approach to nuclear systems, along with a brief description of recent theoretical and numerical developments in studies of nuclei.
The impact of QCD-based constraints on nuclear matrix elements using LQCD can be expanded through a matching program in which phenomenological models or low-energy EFTs of nuclear interactions, and nuclear responses to SM and BSM probes, are constrained systematically. These provide the starting point for extensions to systems not directly accessible to LQCD. An outline of this matching program, and a status report on studies of nuclei with LQCD, are also provided in this section. 

\subsection{Lattice QCD }
QCD can be defined as the
continuum limit of a discretized lattice gauge theory. This formulation provides both an ultraviolet regulator of the continuum field theory that is valid non-perturbatively, and a numerical method for evaluating the functional integrals which define physical observables.\footnote{There are a number of excellent textbooks on lattice field theory~\cite{Rothe:1992nt,Montvay:1994cy,smit_2002,DeGrand:2006zz,Gattringer:2010zz}. Lecture notes on computational strategies for LQCD can be found in  Ref.~\cite{Luscher:2010ae},
	and previous reviews of LQCD techniques for nuclei can be found in Refs.~\cite{Beane:2008dv,Beane:2010em,Beane:2014oea,Lin:2015dga,Drischler:2019xuo}. } The Euclidean QCD partition function is
\begin{eqnarray}
{\cal Z}  =  \int {\cal D} A_\mu {\cal D} \bar  q \,{\cal D}   q \; 
e^{- S^{(E)}_{\rm QCD} }\,,
\label{eq:ZQCD}
\end{eqnarray}
where
\begin{eqnarray}
S_{\rm QCD}^{(E)} = \int d^4x\, {\cal L}_{\rm QCD}
\end{eqnarray}
is the QCD action, and 
\begin{align}  
{\cal L}_{\rm QCD} = 
\sum_{f\in\{u,d,s\ldots\}} \bar{q}_f \left[ D_\mu \gamma_\mu + m_f\right] q_f\ 
+ \ \frac{1}{2g_s^2} {\rm Tr}[G_{\mu\nu} G^{\mu\nu}]
\label{eq:Lqcd}
\end{align}
is the Euclidean QCD Lagrangian density. Here $g_s$ is the gauge coupling defining $\alpha_s = g_s^2/(4\pi)$, $q_f$ denotes the fermion field representing quarks of flavors $f$ with corresponding quark masses $m_f$, and $\gamma_\mu$ are the Dirac matrices. $D_\mu$ is the covariant derivative which  acts on the quark fields as
\begin{eqnarray}
D_\mu q_f(x) = \partial_\mu q_f(x) + i A_\mu(x) q_f(x),
\label{eq:covariant}
\end{eqnarray}
where $A_\mu (x)  =  T^a A_\mu^a(x)$ is the gauge field (encoding the gluon degrees of freedom), with $T^a=\lambda^a/2$, where the $\lambda^a$ are the 8 generators in the fundamental representation of SU$(3)$ (i.e., the Gell-Mann matrices acting in color space).
The gluon field-strength tensor is defined in terms of the gluon field as
\begin{eqnarray}
G_{\mu\nu}(x)= G_{\mu\nu}^a (x)\ T^a = \partial_\mu A_\nu (x) - \partial_\nu A_\mu (x)
+ i  \left[  A_\mu (x), A_\nu (x) \right].
\label{eq:Ddef}
\end{eqnarray}
In Eq.~\eqref{eq:ZQCD}, the fermionic integration measure implicitly includes a product of integrations over each fermion flavor, e.g., ${\cal D}q = \prod_f {\cal D}q_f$.  

The QCD Lagrangian possesses an important symmetry in the limit of massless quarks, $m_f\to0$, namely that it is invariant under independent rotations of the left- and right-handed components of the quark fields. 
Defining the multiplet $q=\{q_u,q_d,q_s\}^T$, and the left- and right-handed quark-field components $q_f^{L}=\frac{1}{2}(1-\gamma_5)q_f$ and $q_f^{R}=\frac{1}{2}(1+\gamma_5)q_f$, the QCD action is invariant under the global rotations $q^L\rightarrow U_L q^L$ and $q^R\rightarrow U_R q^R$, where $U_{L,R}$ are independent SU(3)$_f$ matrices acting in flavor space.
This chiral symmetry, however, is broken spontaneously, resulting in the emergence of massless Goldstone bosons. While quarks are not massless in nature, the up- and down-quark masses are small compared to the QCD scale. Consequently, the SU(2)$_f$ chiral symmetry remains an approximate symmetry in the light-quark sector, and the corresponding pseudo-Goldstone bosons, namely the pions, remain light compared to other hadrons. The mass of the strange quark, while less than the chiral-symmetry--breaking scale, is large enough that SU(3)$_f$ breaking effects in low-energy quantities are not negligible.

\begin{figure}[tb]
	\begin{center}
		\includegraphics[scale=0.45]{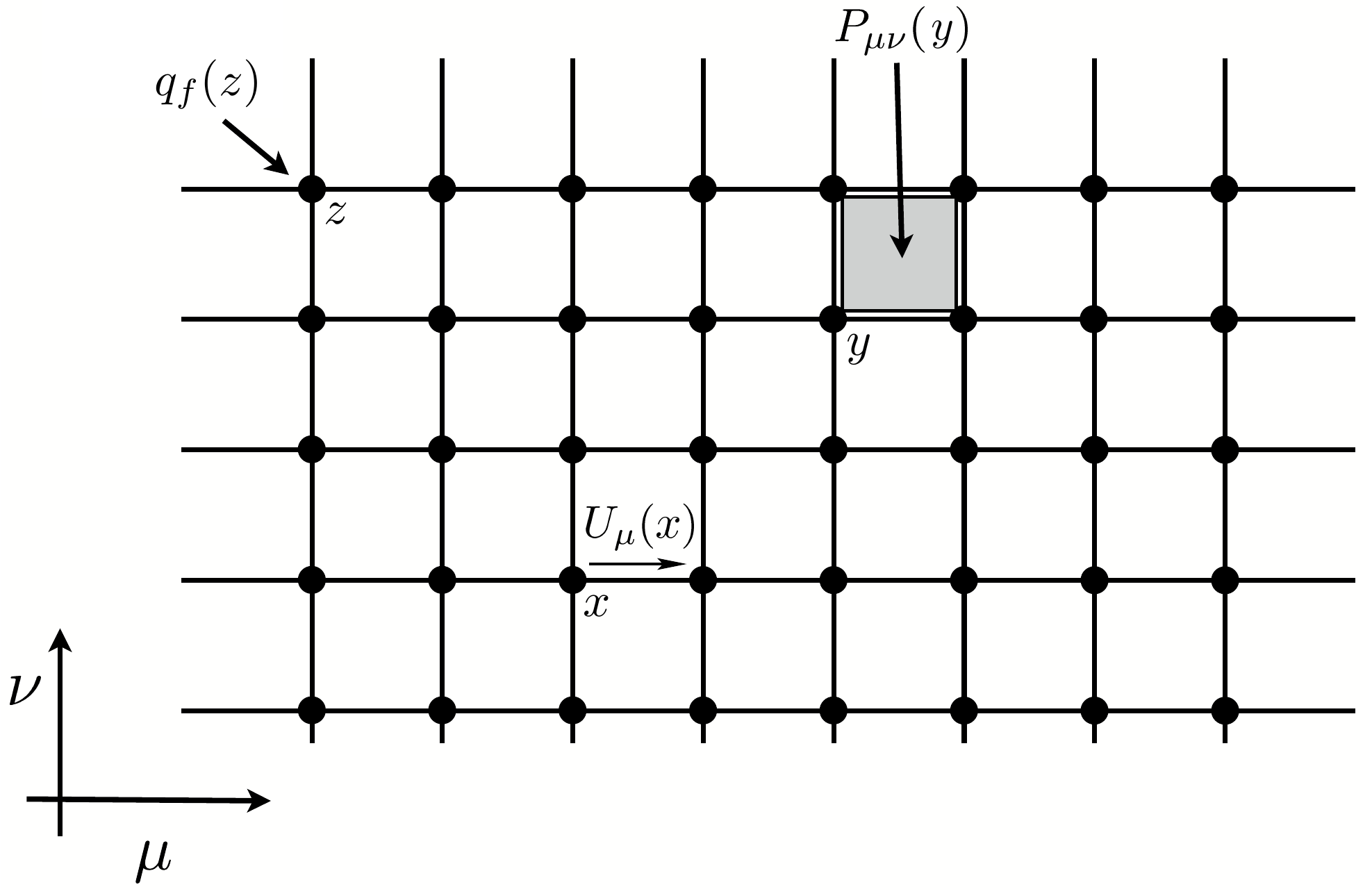}
	\end{center}
	\caption{
		A two-dimensional slice of the four-dimensional spacetime lattice. $U_\mu(x)$ denotes the gauge link from the lattice site $x$ to the site $(x+a\hat{\mu})$, where the subscript $\mu$ indexes the coordinate direction. $P_{\mu\nu}(y)$ denotes the $1\times 1$  plaquette beginning
		at $y$ and proceeding counter-clockwise around the $1\times1$ loop (see Eq.~\eqref{eq:plaquett}), and $q_f(z)$ denotes a quark field of flavor $f$ at the lattice site $z$.
		\label{fig:lattice}}
\end{figure}

Physical quantities in QCD can be calculated from expectation values of operators $\cal O$ that depend on the quark and gluon fields:
\begin{equation}
\label{eq:corrfn}
\langle {\cal O}\rangle = \frac{1}{\cal Z} \int {\cal D} A_\mu {\cal D} \bar
q \,{\cal D}   q \; {\cal O}[A_\mu,\bar{q},q]\; e^{- S^{(E)}_{\rm QCD} } .
\end{equation}
A rigorous definition of these correlation functions, and of the partition function in Eq.~\eqref{eq:ZQCD}, requires regularization and renormalization. A spacetime lattice $\Lambda_4=\{x_\mu=a n_\mu | n_\mu\in \mathbb{Z}\}$, discretized in units of the dimensionful lattice spacing $a$, provides a regulator which is valid even when the coupling is large. Physical results can be obtained in the limit when the discretization scale vanishes.
In this formulation, the gauge field is most naturally implemented  through SU$(3)$ group-valued  variables
\begin{equation}
U_\mu(x) = \exp\left( i \int_x^{x+a\hat\mu} dx^\prime
A_\mu (x^\prime)\right) 
\end{equation}
that are parallel transporters associated with the links between neighboring sites of the lattice (see
Fig.~\ref{fig:lattice}).

In the lattice formulation, multiple actions can be defined that lead to the same QCD action in the continuum limit. One example for the gauge-field degrees of freedom is the Wilson lattice action~\cite{Wilson:1974sk}, defined as 
\begin{equation}
S_g(U) ={\beta} \sum_{x,\mu<\nu}\left( 1 - \frac{1}{3}
{\rm Re }\left[ {\rm  Tr}\left[ {P_{\mu\nu}(x)} \right]\right]\right),
\label{eq:lattice_gauge_action}
\end{equation}
where the coupling $\beta$ is related to the bare gauge coupling as $\beta = 6 /g_s^2$, and the plaquettes $P_{\mu\nu}(x)$ are the
products of the links on the elementary $1\times1$ closed paths of the lattice, i.e., 
\begin{equation}
P_{\mu\nu}(x) = U_\mu(x)U_\nu(x+a\hat\mu)U^\dagger_\mu(x+a\hat\nu)U^\dagger_\nu(x),
\label{eq:plaquett}
\end{equation}
where $\hat{\mu}(\hat{\nu})$ is a unit vector in the $\mu(\nu)$ direction.
The naive continuum limit of this action, obtained by Taylor expanding the plaquettes around unity, is the continuum gauge action in Eq.~(\ref{eq:Lqcd}), with deviations at nonzero lattice spacing that are ${\cal O}(a^2)$. 
Alternative discretizations that contain larger loops, with coefficients
appropriately tuned,  can achieve  smaller discretization errors and  provide faster convergence to the continuum limit.
The systematic computation of these coefficients is known as the Symanzik improvement program~\cite{Symanzik:1983dc,Symanzik:1983gh,Luscher:1985zq}, which in the case of SU(3) gauge theory has been implemented both perturbatively~\cite{Luscher:1984xn,Alford:1995hw} and non-perturbatively \cite{Luscher:1996sc,Luscher:1996ug}.

Defining a lattice action for quarks (fermions) is a challenging problem due to the fermion doubling problem: in a  naive discretization of the fermion term in Eq.~\eqref{eq:Lqcd}, each fermion field exhibits  $2^d$ massless modes, where $d$ is the spacetime dimension~\cite{Wilson:1974sk}. 
The doubler modes, i.e. the additional light  fermion degrees of freedom,  can be removed  with different 
formulations of lattice fermions. Commonly used formulations are described below. 
\begin{itemize}
	\item The Wilson fermion formulation~\cite{Wilson:1974sk}
	adds an irrelevant dimension-five operator, $\bar{q}_f D^2q_f$,
	to the action, giving masses to the $2^d-1$  doubler modes that scale inversely with the lattice spacing, $a$. Consequently, as the continuum limit is approached, the doublers are removed from the low-energy spectrum,  leaving only one 
	light fermion.  However,  the additional dimension-five operator, known as the Wilson term, explicitly breaks chiral 
	symmetry and introduces lattice artifacts that scale linearly with the lattice spacing. 
	Following the Symanzik improvement program,  the Wilson action can be improved by adding an additional dimension-five
	quark bilinear operator, 
	${\cal O}_{\rm SW} = \frac{1}{2i}\bar{q}_f[\gamma^{\mu},\gamma^{\nu}] G_{\mu\nu} q_f$, 
	known as the Sheikholeslami-Wohlert (clover) term~\cite{Sheikholeslami:1985ij}, 
	with a coefficient, $C_{\rm SW}$.  As with improved gauge actions, this coefficient
	can be tuned so that the leading lattice artifacts, which scale as ${\cal O}\left(a\right)$,
	are removed~\cite{Luscher:1996sc}; 
	\item Twisted-mass fermions~\cite{Frezzotti:2000nk} are a variant of Wilson fermions in which lattice-spacing artifacts are reduced to  $O(a^2)$. In this formulation, the Wilson term and the physical quark-mass term are rotated by a relative twist angle in flavor chiral space. This rotation results in an isospin-breaking twisted mass term: $i\mu\bar{Q}\gamma_5\tau^3 Q$, where the field $Q$ describes a flavor doublet (e.g., combining $u$ and $d$, or $s$ and $c$ quark flavors),  $\tau^i$ are the Pauli matrices in flavor space, and $\mu$ is the twisted-mass parameter. A disadvantage of this approach is that it introduces isospin-breaking effects, e.g., a splitting between charged and neutral pions, even when the light quarks are degenerate;
	\item Kogut-Susskind (staggered) fermions~\cite{Kogut:1974ag} 
	constitute another way to remove some of the
	doublers and reinterpret the remaining degrees of freedom as four degenerate
	flavors.
	This approach is implemented by distributing the four components of each Dirac spinor to different lattice sites.
	This formulation preserves  a U(1) chiral symmetry, resulting in lattice artifacts  that scale as $O(a^2)$.  
	The remnant U(1) chiral symmetry is spontaneously broken, resulting in a single Goldstone boson that is massless in the chiral limit. 
	The remaining $2^d-1$ pions have masses that scale as $O(a^2)$ and become massless only in the continuum. 
	However, because QCD has two light flavors,  Kogut-Susskind fermions introduce complications in describing the low-energy spectrum of QCD, requiring the square root of the fermion determinant (see below) to be taken. Nevertheless, an extensive program of calculations based on rooted staggered fermions has been pursued, see for example Ref.~\cite{Bazavov:2009bb};  
	\item Finally,  domain-wall
	fermion~\cite{Kaplan:1992bt,Shamir:1993zy,Furman:1994ky} and overlap
	fermion~\cite{Narayanan:1994gw,Neuberger:1997fp} actions both preserve a lattice version of chiral symmetry that is valid at finite lattice spacing
	(they approximately or exactly satisfy the Ginsparg-Wilson relation~\cite{Ginsparg:1981bj}
	$\gamma_5 D + D\gamma_5 = a\,D\gamma_5 D$)
	and do not involve doubler modes.\footnote{Several other fermion actions, such as the
		fixed-point action~\cite{Hasenfratz:2006xi}, or the chirally-improved action~\cite{Gattringer:2000js,Gattringer:2003qx}, have been explored that 
		approximately satisfy the Ginsparg-Wilson relation.} Domain-wall fermions introduce a fictitious fifth dimension, $-L_5 < x_5 < L_5$, with physical degrees of freedom localized to $x_5=0$, and in their numerical implementation \cite{Shamir:1993zy,Brower:2012vk,Chiu:2002kj} induce chiral symmetry breaking that vanishes for infinite $L_5$. Such formulations are
	significantly more expensive computationally than the other fermion discretizations discussed above.
	
\end{itemize}
Regardless of the chosen fermion formulation, the lattice fermion action is of the form
\begin{equation}\label{eq:lattfermac}
S_f(\bar{q},q,U) = a^4\sum_{x,y} \bar{q}_f(x) D_f[U](x,y)  q_f(y),
\end{equation}
where the Dirac operator $D_f[U]$ acting on the fermion field is a sparse
matrix\footnote{In certain cases, such as with overlap fermions, the
	matrix is not sparse but has sparse-like properties,  i.e., the matrix-vector multiplication is a computationally-inexpensive operation.}
that depends on the specific action, on the gauge field $U$, and on the quark mass $m_f$.
As an explicit example, the Wilson fermion action is given by
\begin{align}
S_f^{\rm Wilson}(\bar{q},q,U)  =& 
-\frac{a^3}{2}\sum_{x,\mu} 
\bar{ q}_f(x) 
\left[ (1-\gamma_\mu)U_\mu(x) q_f(x+a\hat\mu)
+ (1+\gamma_\mu)U^\dagger_\mu(x-a\hat\mu) q_f(x-a\hat\mu) \right]
\nonumber \\
& \qquad + a^4\left(\frac{4}{a}+ m_f\right)\sum_x \bar{q}_f(x)  q_f(x) 
, 
\label{eq:WilsonS}
\end{align}
and the corresponding Wilson Dirac operator $D_f[U]$ can be read off by comparing Eqs.~\eqref{eq:lattfermac} and \eqref{eq:WilsonS}. 

The lattice partition function in the case of two degenerate light-quark flavors, $\ell=\{u,d\}$, and a strange quark, which is a good approximation to the low-energy physics of QCD, is
\begin{align}
{\cal Z} =& \int \prod_{\mu,x}dU_\mu(x)\prod_{x,f} 
d\bar{ q}_fd q_f
\; e^{-S_g(U)-\sum_f S_f(\bar{q}_f, q_f,U)} \nonumber\\
=&  \int \prod_{\mu,x}dU_\mu(x)\;
{\rm Det}\left(D_\ell[U]^\dagger D_\ell[U]\right) {\rm Det}\left(D_s[U]\right) \; e^{-S_g(U)}\,.
\end{align}
The integration over the Grassmann-valued quark fields has been performed analytically in the second equality.  Note that,  while the quark matrix
$D_\ell[U]$ represents one flavor, since the Wilson Dirac operator is $\gamma_5$-Hermitian,  ${\rm Det}\left(D_\ell[U]^\dagger
D_\ell[U]\right)= {\rm Det}\left(D_\ell[U]^\dagger\right) {\rm Det}\left(
D_\ell[U]\right)= {\rm Det}\left(D_\ell[U]\right) {\rm Det}\left(
D_\ell[U]\right)$ represents two mass-degenerate fermion flavors. Correlation functions, Eq.~\eqref{eq:corrfn}, after
integrating out
the quarks,  are similarly given by 
\begin{equation}
\langle{\cal O}\rangle = \frac{1}{\cal Z}
\int \prod_{\mu,x}dU_\mu(x)\; {\cal O}[D^{-1}_f[U],U]\;
{\rm Det}\left(D_\ell[U]^\dagger D_\ell[U]\right) {\rm Det}\left(D_s[U]\right)  e^{-S_g(U)}\, ,
\label{eq:CorFunc}
\end{equation}
where the field-dependent operators  ${\cal O}$  may depend on the inverse of the
Dirac operator for each flavor $f=\{u,d,s\}$, and on the gauge field. The strange quark can be numerically implemented as ${\rm
	Det}\left(D_s[U]^\dagger D_s[U]\right)^{1/2}$, although specific single-flavor algorithms exist \cite{Ogawa:2009ex,Chen:2014hyy,Jung:2017xef}.
The main numerical task
faced in LQCD calculations is the computation of Eq.~(\ref{eq:CorFunc}), which is challenging because the integral
over the gauge field is of extremely large
dimensionality. 
Given that QCD has a fundamental length 
scale of $\sim1~{\rm fm}$, 
calculations must be performed in lattice volumes that have a 
physical size much larger than this
in order to control finite-volume effects, and with 
lattice spacings much smaller than $\sim1~{\rm fm}$ in order to be close enough to the
continuum limit for a continuum extrapolation to be reliable.  To satisfy these constraints, state-of-the-art calculations target lattice volumes as large as $ (L/a)^3\times (T/a) \,\gsim 128^3\times 256$, where $L$ and $T$ are the spatial and temporal extents of the lattice, respectively, with lattice spacing $a\sim0.05$ fm. 
Accounting for the
color and spin degrees of freedom, such calculations involve
$\cO(10^{10})$ degrees of freedom  and challenge today's computational limits. 

Given the dimensionality, the only practical approach to the integration in Eq.~\eqref{eq:CorFunc} is using a Monte Carlo method. The combination of the quark determinant and the gauge action,
\begin{equation}
{\cal P}(U) =\frac{1}{\cal Z}\  {\rm Det}\left(D_\ell(U)^\dagger D_\ell(U)\right) {\rm Det} D_s(U) e^{-S_g(U)}\,,
\end{equation}
is a non-negative definite quantity, in the cases relevant to this review, that can be interpreted as a
probability measure, and hence importance sampling can be employed.  While there are many variants, the basic
algorithm is to produce $N_{\rm cfg}$ gauge-field configurations $\{U_i\}$ according to the
probability distribution ${\cal P}(U)$ using Markov Chain Monte Carlo algorithms such as hybrid Monte Carlo \cite{Duane:1987de}, and then to evaluate
\begin{equation}
\langle {\cal O} \rangle = \lim_{N_{\rm cfg}\rightarrow \infty} 
\frac{1}{N_{\rm cfg}}\sum_{i=1}^{N_{\rm cfg}} {\cal O}[{D^{-1}[U_i]},U_i] ,
\label{eq:average}
\end{equation}
where the evaluation of the right-hand side involves the computation of quark propagators $D^{-1}[U]$ on each of the configurations. 
At finite $N_{\rm cfg}$, the estimate of $\langle{\cal O}\rangle$ by Eq.~\eqref{eq:average} is approximate, with a statistical uncertainty
that decreases as ${\cal O}(1/\sqrt{N_{\rm cfg}})$ as $N_{\rm cfg}$ becomes large.  

Both for the generation of gauge-field configurations, and for the evaluation of the quark propagators needed in
Eq.~(\ref{eq:average}), the linear system of equations\
\begin{equation}
{D_f^\dagger(U)D_f(U)}\chi =  \phi
\label{eq:linearsyst}
\end{equation}
must  be solved. Historically, the vast majority of the resources used in LQCD calculations has been devoted to the solution of this linear system. Direct approaches are impractical for matrices of the  sizes relevant for LQCD calculations, however since the Dirac operator is a
sparse matrix, iterative solvers such as conjugate gradient can be used.  The efficiency of conjugate gradient and other Krylov-space based solvers is governed by the condition number of the Dirac matrix (the ratio of largest to smallest eigenvalues), which is inversely proportional to
the quark mass.  Since the physical quark masses for the up and down
quarks are quite small, the condition
number is large.
For quark-line disconnected diagrams, in which quarks propagate to and from the same point, the necessary ``all-to-all'' quark propagators from every lattice site to every other site are typically computed stochastically~\cite{Collins:2007mh,Alexandrou:2013wca,Bali:2009hu,Gambhir:2016jul,Bali:2015qya}.
The introduction of multi-grid methods in LQCD~\cite{Brannick:2007ue,Frommer:2013fsa} in the last decade significantly reduced  the computational cost of solving Eq.~\eqref{eq:linearsyst}, especially for light quark masses. In addition, specialized computing architectures, such as  graphics processing units~\cite{Clark:2009wm,Babich:2011np},
have greatly extended the range of computations that are currently possible.
For nuclear-physics calculations in particular, other  steps such as computing the required quark contractions, as discussed in the following sections, increase the computational resource requirements significantly.

\subsubsection{Correlation functions}
\label{subsec:baryonblocks}

A common example of the correlation function defined in Eq.~\eqref{eq:corrfn} is the two-point correlation function at zero spatial momentum, defined by\footnote{Throughout this review, $C_{\rm 2pt}^h(t)$ and $C_{\rm 2pt}^h(x,x_0)$ will be distinguished by their arguments. } 
\begin{equation}
\label{eq:2pt0mom}
C_{\rm 2pt}^h(t)= a^3\sum_{\vec{x}} C_{\rm 2pt}^h(x,x_0)\,,
\end{equation}
where $x=(\vec{x},t)$ and $x_0=(\vec{x}_0,0)$ (the correlation function is independent of $\vec{x}_0$ due to translational invariance), and
\begin{equation}
C_{\rm 2pt}^h(x,x_0) \equiv  \langle{\chi}_{h}(x)\bar{\chi}_{h}(x_0)\rangle = \frac{1}{\cal Z} \int  {\cal D}  q {\cal D} \bar  q {\cal D} { U}\;
{\chi}_{h}(x) \bar{\chi}_{h}(x_0) \; e^{-{ S}_{\rm QCD}[\bar{q},q,U]}.
\label{eq:2pt}
\end{equation}
Here, the interpolating operators, $\bar{\chi}_{h}(x_0)$ and
${\chi}_{h}(x)$, are composite operators constructed from quark and gluon fields that create and annihilate states with quantum numbers specified by $h$. The compound labels $h$ identify the states, including their momentum, angular momentum/irreducible cubic-group representation, isospin, and flavor; the label $h$ may also absorb spatial dependence of the interpolating operators if it is not specified explicitly. The interpolating operators are defined for the special case where they have fixed $x_4$ positions separated by Euclidean time $t$, and $S_\text{QCD}$ denotes the sum of lattice gauge and fermion actions. 
Calculations of two-point correlation functions enable the spectrum of states with the given quantum numbers to be extracted. Assuming an infinite temporal extent, insertion of a complete set of energy eigenstates leads to
\begin{align}
C_{\rm 2pt}^h(t)&= a^3\sum_{n} \sum_{\vec{x}}  \langle 0| {\chi}_{h}(x)|n\rangle \langle n|\bar{\chi}_{h}(x_0)| 0\rangle \nonumber \\
&= V\sum_n |Z^h_n|^2 e^{-E^h_n t}\,,
\label{eq:spectral_decomp}
\end{align}
where $V=L^3$ is the dimensionful spatial lattice volume, $Z^h_n = \langle n|\bar{\chi}_{h}(0)| 0\rangle$ is an ``overlap factor'' accounting for the overlap of the interpolating operator onto the specified energy eigenstate, and $E^h_n$ is the energy of the $n$th zero-momentum energy eigenstate with the quantum numbers $h$. Finite-volume states are normalized to unity throughout. To extract the spectral information, effective-mass functions that asymptote to the lowest energy eigenstate at large $t$ can be constructed as
\begin{eqnarray}
M^h_j(t) = \frac{1}{ j a}\,{\rm ln}\left[\frac{C_{\rm 2pt}^h(t)}{C_{\rm 2pt}^h(t+j a)}\right]
\stackrel{t\to \infty}{\longrightarrow} E^h_0\,,
\label{eq:effmassjuge}
\end{eqnarray}
where $j\in\mathbb{Z}\ne0$ is typically chosen to be in the range 1--3.

The computation of correlation functions relevant to the study of nuclei is a particularly challenging problem.
Monte Carlo evaluations of
correlation functions of multi-baryon systems converge slowly to the exact result,
requiring large statistics before useful
precision is obtained (see Sec.~\ref{sec:challenges} below).
In addition, systems with the quantum numbers of many nucleons and
hyperons are complex many-body systems with complicated spectra, and this
complexity manifests at the quark level.
In particular, after performing Grassmann integrals over the quark fields to express the correlation function in terms of quark propagators, $S_f(x,y)=\langle  q_f(x)\bar q_f(y)\rangle$ (suppressing spin and color indices, as well as dependence on the gauge field $U$), the number of quark contractions required to construct systems for large atomic numbers naively grows factorially, scaling as $n_u!n_d!n_s!$, where $n_{u,d,s}$ are the
numbers of up, down, and strange quarks required to construct the
quantum numbers of the state in question. In many cases, however, the large number of quark contractions can be significantly reduced by utilizing  symmetries. 

Quark-level nuclear interpolating operators can be constructed in a similar way  to  quark-model wavefunctions for
baryons~\cite{Feynman:1971wr}.  As shown in Ref.~\cite{Detmold:2012eu},
and first used in large-scale LQCD calculations earlier in Ref.~\cite{Beane:2012vq},  after performing symmetry reductions,
a  quark-level nuclear
interpolating operator with atomic number $A$ containing $n_q=3A$ quarks\footnote{Interpolating operators may also contain explicit gluon fields and additional quark-antiquark pairs.} has the form
\begin{equation}
\label{eq:gen_nuc_red}
\bar{\chi}_h(t)= \sum_{k=1}^{N_w} \tilde{w}^{(a_1,a_2,\ldots ,a_{n_q}),k}_h 
\sum_{\vec{i}} \epsilon^{i_1i_2\ldots i_{n_q}}
\bar{ q}(a_{i_1}) \bar  q(a_{i_2})\ldots \bar  q(a_{i_{n_q}}) \,,
\end{equation}
where $N_w$ is the total number of reduced weights $\tilde{w}$, $\vec{i}$
represents the $n_q$-plet $(i_1,i_2,\ldots ,i_{n_q})$, and
$\epsilon^{i_1i_2\ldots i_{n_q}}$ is a totally anti-symmetric tensor
of rank $n_q$ with
$\epsilon^{1234\cdots{n_q}} = 1$.
The $a_i$ are  compound
indices which combine the color, spinor, flavor, and spatial\footnote{Because calculations are
	performed on a lattice, the spatial degrees of freedom are
	finite and countable, and as a result an integer index can be used to
	describe them.  Here, the quark fields are assumed to be evaluated at the same time, $t$.} indices of
the quark fields, and the $n_q$-plet $\vec{a}=(a_1,a_2,\ldots,
a_{n_q})$ is an ordered list of indices that represents a class of
terms in Eq.~\eqref{eq:gen_nuc_red} that are permutations of each
other. 
The index $k$ on the weights $\tilde{w}^{(a_1,a_2,\ldots ,a_{n_q}),k}_h $ enumerates the number of
classes that $\tilde w_h$ decomposes into; for details see Ref.~\cite{Detmold:2012eu}.
The reduction of the number of non-trivial weights into classes  occurs primarily due to the Grassmannian nature of
the quark field, resulting in the explicit anti-symmetrization of the interpolating operator and the choice of simple spatial wavefunctions. 
For example, using a single-site spatial wavefunction,
the number of terms contained in the simplest interpolating operators
for the proton, deuteron, $^3$He, and $^4$He, are $N_w=9,\, 21,\, 9$,
and 1, respectively. 

In order to compute the reduced weights for a given set of 
quantum numbers, $h$,  an efficient approach is  to begin  by constructing baryon-level
interpolating operators from which the quark interpolating operators can then be derived. 
For a nucleus of atomic number $A$, an interpolating operator can be expressed as
\begin{equation}
\label{eq:gen_nuc_bar}
\bar{\chi}_h(t)= \sum_{k=1}^{M_w} \tilde{W}^{(b_1,b_2\cdots b_{A}),k}_h 
\sum_{\vec{i}} \epsilon^{i_1,i_2,\cdots,i_{A}}
\bar{B}(b_{i_1}) \bar B(b_{i_2})\cdots \bar B(b_{i_{A}}) \,,
\end{equation}
where $M_w$ is the number of hadronic reduced weights
$\tilde{W}^{(b_1,b_2\cdots b_{A}),k}_h$, $\bar{B}(b_i)$ are baryon
interpolating operators, and the $b_i$ are compound indices that include
parity, angular momentum, flavor, and spatial indices.
For simplicity, as well as efficiency of implementation of the resulting
nuclear interpolating operators, one can use a baryon interpolating operator selected so that it has large overlap with the single-baryon ground state, but is comprised of a small number of quark-level terms.
By substituting single-baryon interpolating operators in Eq.~(\ref{eq:gen_nuc_bar}), the reduced weights needed for Eq.~(\ref{eq:gen_nuc_red}) can be computed. 
Since nucleons are  effective degrees of freedom for low-energy nuclear physics, one may expect that interpolating operators that are
derived starting from Eq.~(\ref{eq:gen_nuc_bar}) will have large overlap
with nuclear ground states.

A standard choice for single-baryon interpolating operators is
\begin{equation}
\bar{B}(b)=  \sum_{k=1}^{N_{B(b)}} \tilde{w}^{(a_1,a_2,a_3),k}_b 
\sum_{\vec{i}} \epsilon^{i_1,i_2,i_3}
\bar{ q}(a_{i_1}) \bar  q(a_{i_2})\bar  q(a_{i_3})\,,
\end{equation}
where $N_{B(b)}$ is the number of terms in the
interpolating operator.  An example set of weights, $\tilde{w}^{(a_1,a_2,a_3),k}_b$, has been presented in Ref.~\cite{Basak:2005ir} (the color factors necessary for this formulation are not included in Ref.~\cite{Basak:2005ir} but can be added trivially).
Note that the weights also encode the spatial structure of the interpolating operators; a simple choice is to project the single-hadron interpolating operators onto a plane wave~\cite{Beane:2005rj,Beane:2006mx}.
However, this results in weights that are dense in the spatial indices and hence produces a large number of terms in Eq.~\eqref{eq:gen_nuc_red}. If such interpolating operators are used both as creation and annihilation operators, evaluation of Eq.~\eqref{eq:2pt} requires computation of a large number of terms, scaling as the spatial volume squared. 
Nevertheless, in meson-meson and multi-meson spectroscopy, such wavefunctions have been used~\cite{Beane:2011sc,Shi:2011mr,Detmold:2012wc,Dudek:2012gj}.
For multi-meson systems, special contraction methods were
required~\cite{Detmold:2010au,Shi:2011mr,Detmold:2012wc} to efficiently incorporate  the large number of terms. 
For multi-nucleon systems, one can simplify the problem by considering 
quark creation interpolating operators (sources) that have simple spatial
wavefunctions with degrees of freedom restricted to a few spatial points. In this way, using plane-wave projection for the hadronic interpolating operators used to construct the annihilation operator (sink), one can construct an efficient contraction algorithm for the nuclear correlation function that scales only linearly in the spatial lattice volume.  
This efficient algorithm proceeds via the construction of baryon building blocks. Using the quark propagator from a single source point, $x_0=({\vec x}_0,0)$, one can construct baryon blocks with quantum numbers
$b$ and momentum ${\vec p}$ as:
\begin{align}
\nonumber
{\cal B}_b^{a_1,a_2,a_3}({\vec p},t;x_0)= \sum_{\vec x}e^{i{\vec p}\cdot{\vec
		x}} 
\sum_{k=1}^{N_{B(b)}} \tilde{w}^{(c_1,c_2,c_3),k}_b &\sum_{\vec i} \epsilon^{i_1,i_2,i_3}\left[\right. \!
S(c_{i_1},x;a_1,x_0)\\  \label{eq:1}
&\times \left. S(c_{i_2},x;a_2,x_0)
S(c_{i_3},x;a_3,x_0)\right]
,
\end{align}
where $S(c,x;a,{x}_0)$ is the quark propagator from a source at $x_0=(\vec{x}_0,0)$ to a sink at $x=({\vec x},t)$ and $c_i$ and $a_i$ are the remaining combined
spin-color-flavor indices.\footnote{More complicated multi-hadron blocks have also been considered, for example in
	Ref.~\cite{Yamazaki:2009ua}, at the cost of increased storage requirements.}  The baryon block corresponds to the
propagation of a particular three-quark configuration from the source to the sink
where it is annihilated by the prescribed baryon interpolating operator.
The baryon is projected to definite  momentum  ${\vec p}$ allowing the total momentum of multi-hadron systems to be controlled by combining blocks of given momenta.

\begin{figure}[!t]
	\centering
	\subfigure[]{
		\includegraphics[width=0.45\textwidth]{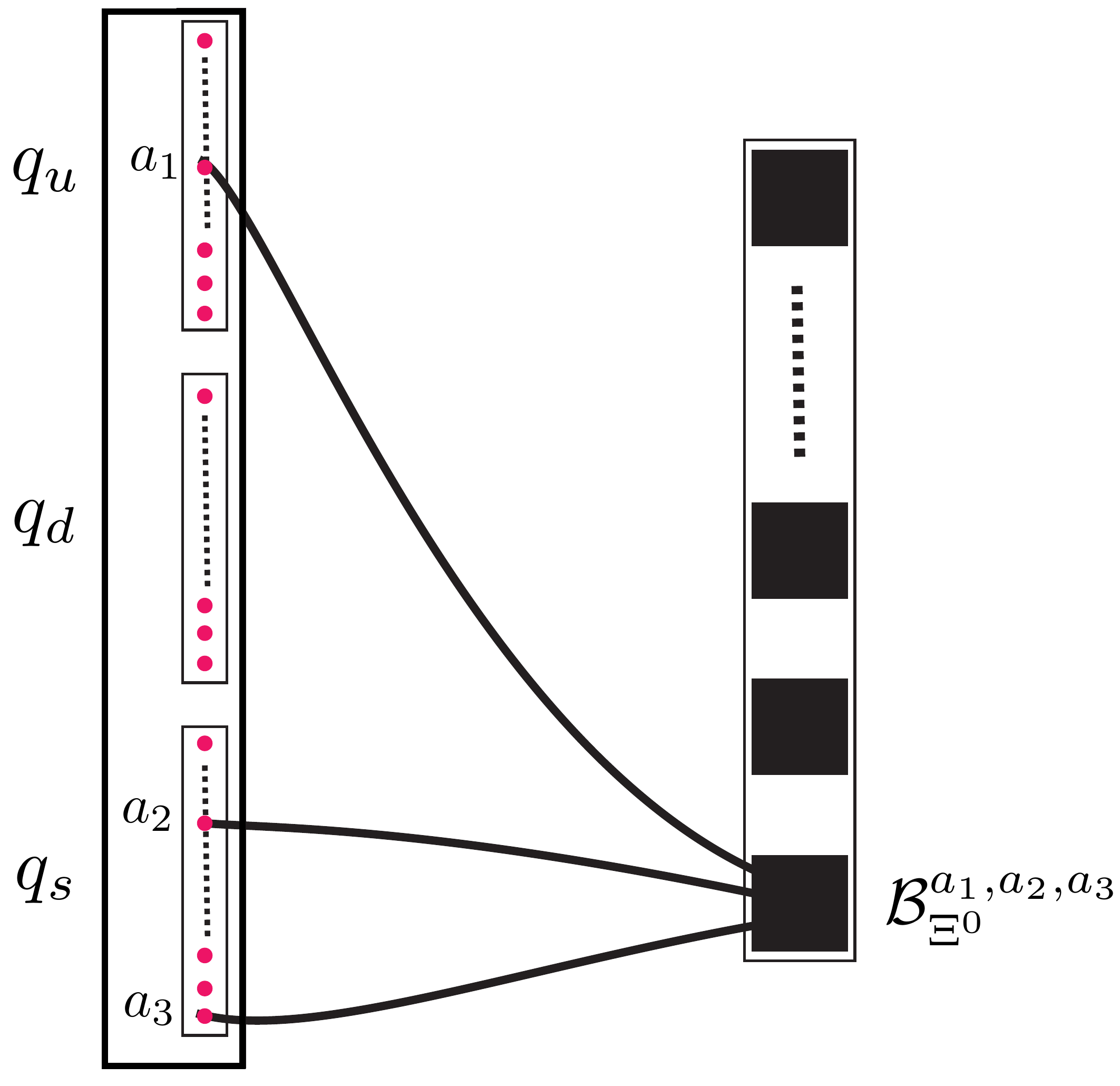}
	}
	\subfigure[]{
		\includegraphics[width=0.45\textwidth]{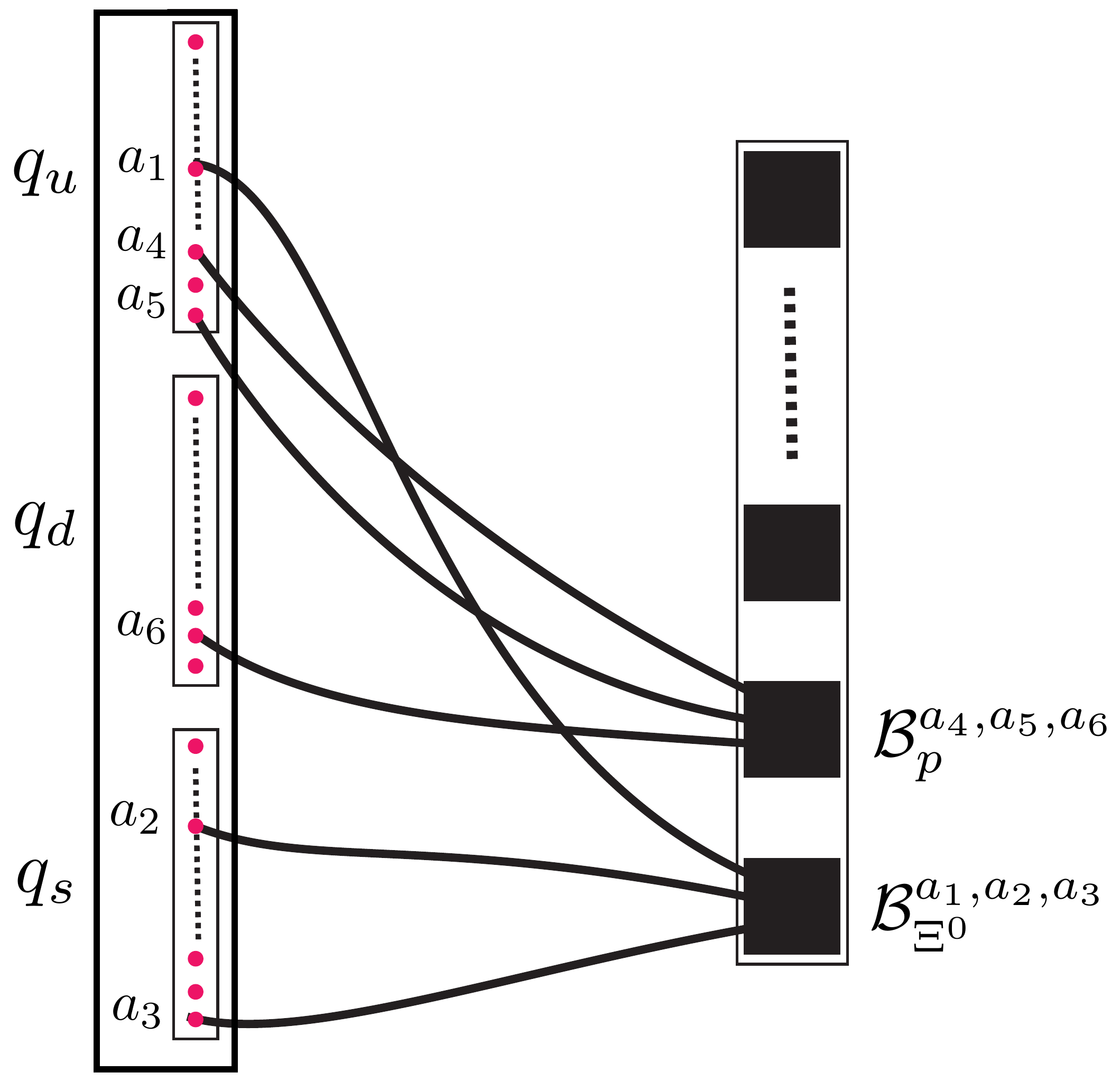}}
	\caption{Illustration of the first two steps  of the quark-hadron
		contraction method. The small circles in the left-hand section of each
		figure correspond to the quarks in the source interpolating operator,
		while the large squares in the right-hand sections, and lines extending from them, correspond
		to the hadronic blocks. In the first step (a), the indices of the first baryon block are contracted with the quark fields. In the second (b) and subsequent steps, additional baryon blocks are included and contracted with the remaining quark indices in the source wavefunction. [Figure modified from Ref.~\cite{Detmold:2012eu}.] }
	\label{fig:q-h}
\end{figure}

With the baryon blocks described above, correlation functions with interpolating operators describing products of momentum-projected baryons at the sink, and interpolating operators describing local products of $3A$ quark fields at the source, can be computed efficiently. Their evaluation is accomplished by iterating over all combinations of
source and sink interpolating operator terms and connecting the source
and sink with the appropriate sets of quark propagators. For each
pair of terms in the  source and sink interpolating operators, 
the product of all hadronic blocks and weights present at the sink is computed by selecting from each block the components dictated by the local source interpolating
operator. This selection must occur in all possible ways while keeping track of the sign changes arising from fermion exchanges. A detailed description of the process is given in Ref.~\cite{Detmold:2012eu}, and a diagrammatic illustration of the procedure is presented in Fig.~\ref{fig:q-h}.
This procedure has been used to perform the contractions needed for the large class of interpolating operators considered in the study of the spectrum of nuclei and hypernuclei up to $A=5$ in Ref.~\cite{Beane:2012vq}. In particular, a set of mutually orthonormal interpolating operators has been constructed~\cite{Detmold:2012eu} to generate the hadronic reduced weights and perform the single-baryon substitution in terms of quark fields. Similar approaches are discussed in Refs.~\cite{Doi:2012xd,Gunther:2013xj}.

\subsubsection{Nuclear matrix elements}
\label{subsec:backgroundfield}

Many physical quantities of interest in nuclear physics, such as charges, form factors, electroweak transition amplitudes, and moments of parton distributions, are defined in terms of matrix elements of local quark-bilinear operators in nuclear states. 
For initial and final states specified by the quantum numbers 
$h$ and $h^\prime$, and for a quark bilinear $\mathcal{O}_{f^\prime f}(x) = \bar{q}_{f^\prime}(x) \Gamma q_f(x)$, with Dirac structure $\Gamma$, a generic matrix element is given by
\begin{equation}
\label{eq:ME}
\mathcal{M}^{h^\prime h}_{\mathcal{O}_{f^\prime f}}=\langle h^\prime | {\cal O}_{f^\prime f} | h\rangle.
\end{equation}
For the purposes of this review, the primary focus is on forward matrix elements of flavor-conserving quark bilinears between identical states,
$\mathcal{M}^h_{\mathcal{O}_f}=\langle h| {\cal O}_f | h\rangle$.
A standard approach  to computing such matrix elements is via the construction of three-point correlation functions~\cite{Martinelli:1988rr,Martinelli:1987bh}, defined as 
\begin{equation}
\label{eq:3pt}
C^h_{{\rm 3pt},{\cal O}_f}(x;y;x_0)= \langle\chi_{h}(x)\mathcal{O}_f(y)  \bar{\chi}_{h}(x_0)\rangle.
\end{equation}
Ratios of three-point and two-point correlation functions, appropriately summed over the spatial positions $\vec{x}$ and $\vec{y}$, asymptote to the matrix elements $\mathcal{M}^h_{\mathcal{O}_f}$ in the limit of large temporal separations between the sources, operators, and sinks. For the case of an operator that does not change the quantum numbers of the state or inject momentum, an appropriate ratio is
\begin{equation}
\label{eq:3ptto2ptratio}
R_{3/2,{\cal O}_f}^{h}(t,\tau)= \frac{a^6\sum\limits_{\vec{x},\vec{y}} C^h_{{\rm 3pt},{\cal O}_f}(x;y;0)}{C_{\rm 2pt}^h(t)} \overset{t,\tau \rightarrow \infty}{\longrightarrow} c \, \mathcal{M}^h_{\mathcal{O}_f},
\end{equation}
where $t$ and $\tau$ correspond to the temporal coordinates of $x$ and $y$. In the $\{t,\tau,|t-\tau|\} \to \infty$ limit, contributions from excited states vanish exponentially rapidly, and the denominator cancels the exponential fall-off and overlap factors in the numerator, leaving the desired matrix element up to kinematic factors denoted by $c$ in Eq.~\eqref{eq:3ptto2ptratio}. More complicated ratios are required in the case of non-forward matrix elements.

Alternative approaches to computing matrix elements based on spectroscopy in a background field coupled to the desired operator~\cite{Fucito:1982ff,Bernard:1982yu,Martinelli:1982cb} are also possible. Two distinct background-field approaches have been considered in the context of nuclear matrix elements and are reviewed here.
The first approach is particularly suited to the determination of matrix elements of the electromagnetic current, although it can be generalized to other interactions \cite{Detmold:2004kw}. This method is implemented by modifying the SU(3) gauge links by multiplication by a particular set of external (classical) U(1) gauge links,
\begin{equation}
\label{eq:extfield}
U_\mu(x) \to U_\mu(x) U_\mu^{\rm ext}(x),
\end{equation}
where $\ U_\mu^{\rm ext}(x)\in {\rm U(1)}$ and the modified field is in the U(3) group rather than the SU(3) group.
The modifications of the links can be performed during the generation of the gauge-field configurations or it can be applied to existing SU(3) configurations. In the latter case, the coupling to sea-quark degrees of freedom through the quark determinant is not included.\footnote{This absence could be addressed using reweighting methods.} Nevertheless,
there are situations in which sea-quark contributions vanish and exact results can be obtained in this way. 
As a specific example that will be discussed at length in Sec.~\ref{sec:EM}, a magnetic field aligned along the $\hat z={x}_3$-direction can be implemented through the U(1) link field
\begin{equation}
\label{eq:Bfield}
U_\mu^{\rm ext}(x) = e^{ i Q e x_1 B \, \delta_{\mu, 2}} e^{- i Q e x_2 B L \, \delta_{\mu, 1} \, \delta_{x_1, L-a}}\,,
\end{equation}
where $Q$ is the quark charge in units of the electron charge $e$, and $L$ is the spatial extent of the gauge field configuration. The first term in this expression implements a constant magnetic field $\vec{B}=B\hat{z}$. However, on the discrete toroidal geometry of LQCD calculations, periodicity would lead to a non-constant $\vec{B}$ at the boundary. The second exponential factor in Eq.~\eqref{eq:Bfield} corrects for this artifact.  In order to ensure that the flux through each elementary plaquette of the lattice is uniform, the magnetic field in the expression above must  
satisfy the 't Hooft quantization condition~\cite{tHooft:1979rtg}, $Q e B = 2 \pi \, \tilde n / L^2$, where $\tilde n\in \mathbb{Z}$ is the flux quantum of the torus, for each fundamental charge $Q$.
There is flexibility in the choice of the external field;  for example, constant electric fields have also  been implemented~\cite{TIBURZI2009336}. More general choices of fields include spatially-varying electromagnetic fields, relevant for extracting matrix elements at non-zero momentum transfer and higher multipole properties \cite{Detmold:2004kw,Davoudi:2015zda,Davoudi:2015cba}. 

In this background-field approach, two-point correlation functions constructed from quark propagators determined from the modified link fields contain interactions with the external field to all orders in the field strength. Under appropriate conditions, the two-point correlation function has a spectral decomposition analogous to Eq.~\eqref{eq:spectral_decomp};
\begin{align}\nonumber
C_{\rm 2pt}^h(t,B) &\equiv 
a^3\sum_{\vec{x}}\langle  {\chi}_h(x) \bar\chi_{h}(x_0) 
\rangle_{B} 
\\\label{eq:BF2pt}
&= V\sum_n  |\langle 0 | \chi_h(0) | n\rangle_B|^2 e^{-E_n^h(B)t}\,,
\end{align}
where the subscript $B$ denotes a correlation function evaluated on gauge configurations modified by the U(1) field in Eq.~\eqref{eq:Bfield}. In the second line, a spectral decomposition over the energy eigenstates $n$ (which remains valid in the presence of this field) is employed, with the sum over states being over those with zero three-momentum. The eigenenergies and overlap factors depend on $B$ and their determination at multiple values of $B$ allows for information about the structure of the eigenstates to be extracted, as will be reviewed in Sec.~\ref{sec:EM}.
In the time regions in which the relevant correlation functions show ground-state dominance, the energy shift induced by the magnetic field, 
$\delta E^{h}({ B}) = E_0^{h}({ B}) - E^{h}_0({ 0})$,
can be obtained from ratios of the background-field correlation functions defined in Eq.~\eqref{eq:BF2pt} evaluated at zero and non-zero values of the field strength:
\begin{equation}
R^{h}(t,{ B})  =  
{ C_\text{2pt}^{h}(t, B) \over C_\text{2pt}^{h}(t, B=0) }
\,\stackrel{t\to\infty}{\longrightarrow} \,
Z^{h}({ B})\ e^{- \delta E^{h}({ B}) t },
\label{eq:ratcorr}
\end{equation}
where $Z^{h}({ B})$ is a time-independent, but $B$-dependent, quantity. 

An alternative background-field approach is based on constructing extended quark propagators. In this case, a generalized quark propagator is introduced that contains a single insertion of the operator proportional to a field-strength parameter $\lambda_f(z)$:
\begin{align}
\label{eq:extendedprop}
S^{\lambda_f}_{f,\mathcal{O}_f}(x;y)= S_f({ x};y) + \int d^4z\, \lambda_f(z)\, S_f(x;z)\Gamma S_f(z;y),
\end{align}
where again $\cO_f= \bar{q}_f\Gamma q_f$ \cite{Savage:2016kon,Bouchard:2016gmc}.
Here, $\lambda_f(z)$ can in general have spacetime dependence, but will henceforth be taken to be a constant for simplicity.\footnote{A useful non-constant choice is $\lambda_f(z)=\lambda_f \theta(z\in {\cal R})$, which is constant in a subregion, ${\cal R}$, of the lattice geometry and zero elsewhere. Recently, spatially-varying background fields have been used in this approach for single nucleons~\cite{Can:2020sxc}. Another generalization is to include flavor-changing background fields. }
This generalized propagator can now be used to build baryon blocks and two-point correlation functions for nuclei using the contraction strategies detailed above. Such two-point functions, 
\begin{equation}
C^{h,\mathcal{O}_f}_{\lambda_u,\lambda_d}(x,x_0)= C^h_{{\rm 2pt}}(x,x_0)[S_f\to S^{\lambda_f}_{f,{\cal O}_f}]\ ,
\label{eq:bfcorr}
\end{equation}
have polynomial dependence on $\lambda_f$. Here, the square brackets indicate the replacement of all propagators of flavor $f$ by generalized propagators with the specified values of the parameters $\lambda_f$. Different field-strength parameters can be used for different flavors of quarks, and, restricting to non-strange nuclei, the two-point functions built from these generalized propagators are a  polynomial in both of the variables $\lambda_u$ and $\lambda_d$:
\begin{equation}
\label{eq:Coflambda}    
C^{h,\mathcal{O}_f}_{\lambda_u,\lambda_d}(x,x_0)= \sum_{i_u=0}^{n_u}\sum_{i_d=0}^{n_d} \lambda_u^{i_u} \lambda_d^{i_d}\left. C^{h,\mathcal{O}_f}_{\lambda_u,\lambda_d}(x,x_0) \right|_{{\cal O}(\lambda_u^{i_u} \lambda_d^{i_d})}\,,
\end{equation}
where $n_f$ denotes the number of quarks of flavor $f$ in the interpolating operator and the $\left.\ldots\right|_{\cO(\lambda^n)}$ extracts the coefficient of $\lambda^n$. 
The coefficient proportional to $\lambda_f$ of the polynomial defined by the two-point correlation function of a particular nuclear state corresponds to the summation with respect to the operator insertion point, i.e.,
\begin{align}\label{eq:bf2pt}
\left. C^{h,\mathcal{O}_f}_{\lambda_u,\lambda_d}(x,x_0) \right|_{{\cal O}(\lambda_f)}
= a^4\sum_{{y}} \langle\chi_{h}(x)\mathcal{O}_f(y)  \bar{\chi}_{h}(x_0)\rangle = a^4\sum_{{y}}C^h_{{\rm 3pt},{\cal O}_f}(x;y;x_0).
\end{align}
Since the polynomial in Eq.~\eqref{eq:Coflambda} is of fixed order, this leading coefficient can be determined exactly given correlation functions computed with a sufficient number of different values of $\lambda_f$. It should be noted that the chosen values of $\lambda_f$ do not need to be real, or small in magnitude. In fact, it was shown in Ref.~\cite{Detmold:2012wc} in a similar context that it is advantageous to use complex field-strength values and perform a discrete Fourier transform to obtain all coefficients in the polynomial.

Given the background-field two-point correlation function of Eq.~\eqref{eq:bf2pt}, the ratio
\begin{align}
\label{eq:RO}
R^h_{3/2,{\mathcal{O}_f}}(t) = \frac{a^3\sum\limits_{\vec{x}}\left. C^{h,\mathcal{O}_f}_{\lambda_u,\lambda_d}(x,x_0) \right|_{{\cal O}(\lambda_f)}}{C^h_{\rm 2pt}(t)}\,
\end{align}
may be defined~\cite{Savage:2016kon,Bouchard:2016heu,Shanahan:2017bgi,Tiburzi:2017iux}, where, as above, $x_0=(\vec{x}_0,0)$, $x=({\vec x},t)$, and $C^h_{\rm 2pt}(t)$ is defined in Eq.~\eqref{eq:2pt0mom}. The sum over spatial sites in the numerator projects the background-field two-point correlation function to zero three-momentum.
An ``effective matrix element'' can be extracted from this ratio as 
\begin{align}
M^{h,\rm eff}_{\mathcal{O}_f}(t)= \frac{1}{ac}[ R^h_{\mathcal{O}_f}(t+a) - R^h_{\mathcal{O}_f}(t) ],
\end{align}
where $c$ denotes kinematic factors as in Eq.~\eqref{eq:3ptto2ptratio}. 
In the limit of large time $t$, this effective matrix element exponentially converges to the matrix element ${\cal M}_{{\cal O}_f}^h$  in Eq.~\eqref{eq:ME}.
At finite times, excited states contaminate the effective matrix element. These contributions can be parameterized as 
\begin{align}
M^{h,\rm eff}_{\mathcal{O}_f}(t) =  \mathcal{M}^h_{\mathcal{O}_f}\left[
1 + A e^{-\Delta t} + B t e^{-\Delta t} + \ldots\right],
\end{align}
where Eq.~\eqref{eq:RO} has been expanded in a Taylor series under the assumption that $e^{-\Delta t}\ll 1$, where $\Delta$ is the energy gap to the first excited state contributing to the correlation function. The ellipsis denotes additional contributions suppressed by larger energy gaps. Detailed descriptions of the behavior of excited-state contamination can be found in Ref.~\cite{Bouchard:2016heu}. Furthermore, a generalization of this background-field approach to a basis of interpolating fields, and to transition matrix elements, as well as a comprehensive discussion of excited-state contamination, can be found in Refs.~\cite{Bulava:2011yz,Tiburzi:2017iux}.

The particular computational approach described here is well-suited to the computation of matrix elements of nuclei and requires only two-point correlation function contraction codes which are relatively simple compared to those for nuclear three-point correlation functions. However, additional contractions must be performed for each value of $\lambda_f$. 
A further advantage of this approach is that it allows the extraction of matrix elements with different numbers of insertions of the operators into correlation functions within the same computation, as studied in Refs.~\cite{Shanahan:2017bgi,Tiburzi:2017iux} and discussed further in Sec.~\ref{sec:dbd}. As an example, the second-order response to an applied field of flavor $u$ projected to zero momentum is 
\begin{align}
C^{h,{\cal O}_f}_{\lambda_u,\lambda_d=0}(t) 
=&\,\,
a^3\sum_{\vec x}
\langle  \chi_{h}({\vec x},t) \bar\chi_{h}(0)  \rangle  
+ a^7\lambda_u
\sum_{{\vec x},{\vec y}}\sum_{t_1=0}^t
\langle  \chi_{h}({\vec x},t) {\cal O}_u ({\vec y},t_1)   \bar\chi_{h}(0) \rangle 
\nonumber \\ 
&+ a^{11}\frac{\lambda_u^2}{2}
\sum_{{\vec x},{\vec y},{\vec z}}\sum_{t_{1,2}=0}^t 
\langle  \chi_{h}({\vec x},t) {\cal O}_u ({\vec y},t_1)  
{\cal O}_u ({\vec z},t_2)   \bar\chi_{h}(0)  \rangle +  \cO(\lambda_u^3);
\label{eq:quad1}
\end{align}
this correlation function can be analyzed to extract the second-order matrix element $\langle h | \mathcal{O}_u\mathcal{O}_u | h \rangle$.

\subsection{Lattice QCD for few- and many-body systems
	\label{sec:EFT}}
A primary objective of LQCD studies of multi-nucleon systems is to constrain experimental observables of interest in nuclear physics. This is a challenging program for two primary reasons. First, it is likely that in the near future LQCD will not be able to directly access the properties of nuclei with $A>5$, as used in many experiments, given the computational cost of such calculations and other challenges arising from the unique features of nuclear systems that will be discussed in Sec.~\ref{sec:challenges}. Second, even for few-nucleon systems for which LQCD studies are viable, LQCD correlation functions, or combinations thereof, determine spectra and matrix elements that correspond to those of QCD in a finite Euclidean spacetime volume. The matching between LQCD results and physical observables that are defined in an infinite volume and Minkowski spacetime, such as scattering and transition amplitudes for electroweak and BSM processes, is generally non-trivial. This matching can proceed in at least three ways:
\begin{itemize}
	\item One approach applies L\"uscher's formalism~\cite{Luscher:1986pf,Luscher:1990ux},
	Lellouch-L\"uscher's formalism~\cite{Lellouch:2000pv}, and 
	generalizations~\cite{Rummukainen:1995vs,Beane:2003da, Kim:2005gf, He:2005ey,  Davoudi:2011md, Leskovec:2012gb, Hansen:2012tf, Briceno:2012yi, Gockeler:2012yj, Briceno:2013lba, Feng:2004ua, Lee:2017igf, Bedaque:2004kc, Luu:2011ep, Briceno:2013hya, Briceno:2013bda, Briceno:2014oea, Briceno:2017max, Polejaeva:2012ut, Briceno:2012rv,  Hansen:2014eka, Hansen:2015zga, Hammer:2017uqm, Hammer:2017kms, Guo:2017ism, Mai:2017bge, Briceno:2017tce,Doring:2018xxx, Briceno:2018aml, Romero-Lopez:2019qrt, Jackura:2019bmu, Hansen:2020zhy, Christ:2005gi, Meyer:2011um, Bernard:2012bi, Feng:2014gba, Briceno:2015tza, Briceno:2014uqa, Briceno:2015csa, Christ:2015pwa, Baroni:2018iau, Briceno:2019opb, Feng:2020nqj, Davoudi:2020xdv}, to provide a formal mapping between finite-volume energies and/or matrix elements and the infinite-volume scattering and transition amplitudes for few-hadron processes below certain inelastic thresholds.
	Beyond single channels, model dependence in such extractions enters in relating contributions evaluated at different kinematic points.
	These methods require that the interactions between fields are vanishing at the boundaries of the lattice volume, which requires the range of interaction $R<L/2$.  However, because off-axis distances between lattice images are larger than $L$, this formalism provides estimates with small systematic errors for 
	$R\gsim L/2$ for interactions whose strength decreases rapidly with the distance between the hadrons~\cite{Luscher:1986pf,Sato:2007ms}.
	\item 
	Another approach that has been successfully employed is to use an effective Hamiltonian derived from an EFT at a given order in the expansion and construct relevant finite-volume observables to match to those of the LQCD calculation(s), e.g. Ref.~\cite{Beane:2012ey}.
	In the case of scattering and reactions, the energy eigenvalues and matrix elements computed in the finite lattice volume(s) can be matched to those computed with the EFT to determine a finite number of LECs.  In contrast to L\"uscher's formalism discussed above, these methods apply even when the fields are interacting at the lattice boundaries~\cite{Sato:2007ms,Beane:2012ey,Lu:2018pjk}, i.e., beyond $R=L/2$.
	In principle, this approach should provide an equally 
	reliable method for providing QCD predictions; however, there is a limit to the achievable precision set by the order of the matching and the implementation of the EFT in many-body calculations.
	Pionless EFT~\cite{Kaplan:1996xu,Kaplan:1998tg,Kaplan:1998we,vanKolck:1998bw,Bedaque:1998kg,Chen:1999tn,Bedaque:2002yg,Bedaque:2002mn}, in which all mesons including the pion are integrated out, accurately describes few-baryon systems at low energies in nature and at somewhat larger energies in LQCD calculations with unphysically-large quark masses. In order to describe multi-baryon systems over the wider range of energies relevant for nuclear physics in nature, pions and sometimes $\Delta$ resonances must be included as explicit degrees of freedom in chiral EFT. There has been recent progress in constructing chiral potentials capable of reproducing experimentally-measured nucleon-nucleon phase shifts and nuclear-structure properties, as reviewed e.g., in Refs.~\cite{Epelbaum:2008ga,Machleidt:2011zz}. However, inconsistencies in the power counting of chiral potentials remain a limitation in making reliable predictions with a complete quantification of uncertainties using chiral EFT~\cite{Tews:2020hgp}.
	In particular, Weinberg's power counting prescription~\cite{Weinberg:1990rz,Weinberg:1991um} is not consistent at a fixed order in the chiral expansion (when iterated in the Lippmann-Schwinger or Schrodinger equation to compute observables)~\cite{Kaplan:1996xu}, but at sufficiently high orders the LECs required for renormalizability at a lower order are present, even in channels with a tensor force, including the $\siii$-$\diii$ coupled channels containing the deuteron.
	By promoting nucleon-nucleon contact interactions from the orders estimated using naive dimensional analysis (NDA) in sufficiently low partial waves~\cite{Kaplan:1996xu,Kaplan:1998tg,Kaplan:1998we}, and also nonperturbatively including iterated one-pion-exchange effects in low partial waves where singular potentials arise from tensor forces~\cite{Fleming:1999ee,Beane:2001bc,Nogga:2005hy,Birse:2005um,PavonValderrama:2016lqn,Wu:2018lai,Kaplan:2019znu}, it is expected that a chiral EFT power counting consistent with fixed-order renormalizability can be constructed for multi-nucleon systems that converges for momenta relevant to nuclear physics (see Ref.~\cite{Beane:2000fx} for further discussions and Ref.~\cite{vanKolck:2020llt} for a recent review).
	Kaplan-Savage-Wise (KSW)  power counting~\cite{Kaplan:1998tg,Kaplan:1998we}  is valid in spin-singlet channels such as $\si$,
	while for other channels a hybrid of Weinberg and KSW power-counting schemes was suggested in Ref.~\cite{Beane:2001bc} and extended to higher partial waves in Ref.~\cite{Nogga:2005hy}. In addition, a fractional-order power-counting scheme was suggested in Ref.~\cite{Birse:2005um}.
	There has also been recent progress in understanding renormalization constraints on nuclear matrix elements in chiral EFT, where NDA is similarly insufficient to guarantee renormalizability at fixed order in the EFT expansion~\cite{Valderrama:2014vra,Cirigliano:2018hja,deVries:2020loy}. More work is required to connect studies of the renormalization of few-nucleon systems with phenomenologically-successful descriptions of larger nuclei in chiral EFT, but progress in this direction is ongoing. 
	
	\item
	A third approach is to constrain models of nuclear forces and matrix elements by directly matching finite-volume energies and matrix elements to those of LQCD calculations.  This shares most of the features described in the previous point related to matching LECs in EFT, 
	but may utilize interactions that do not have systematic expansion parameters.  This is similar to constraining phenomenological nuclear forces from experimental data, but makes use of the additional ``parameter'' of the lattice volume.
	Such constrained interactions are expected to provide estimates of other quantities through interpolation, but it is not possible to provide a complete quantification of uncertainties using constrained models. Interactions of this type have been developed, constrained with precision by experiment, and successfully implemented in extensive many-body studies of nuclear systems.  LQCD calculations are expected to help refine components of these forces that are difficult or impossible to access experimentally. This may lead to significant near-term improvements in the accuracy of model predictions for matrix elements of electroweak and BSM currents in experimentally-relevant nuclei outside the current reach of systematically-controlled EFTs.
\end{itemize}

Each of these approaches has been successfully implemented in recent years, although only the second approach has been applied to nuclear matrix elements. The applicability and appropriateness of each method depends on the system under study and the ultimate goal of the analysis. L\"uscher's approach can be regarded as more general, as it is independent of any effective description of interactions, relying only on a parameterization of the scattering amplitude and other $n$-point functions. However, it has so far been limited to the two- and three-hadron sectors of QCD except in the perturbative regime of interactions. The model/EFT matching approach (directly or through the use of L\"uscher's method to constrain a model or EFT description of the scattering amplitude) combined with the use of many-body methods, in principle extends the reach of QCD-based predictions to the nuclear many-body sector. Specifying an EFT also allows for exponentially-small volume effects that are neglected in L\"uscher's approach to be incorporated. The limitation of this approach is its reliance on the validity of the model/EFT for the particular system considered~\cite{Bedaque:2006yi,Sato:2007ms,Beane:2000fx,Bansal:2017pwn,Drischler:2019xuo}. These approaches, along with select examples of their applications, are described in more detail in this section.

Another approach to two- (and three-) hadron interactions, which has been developed and applied by the HAL QCD Collaboration, takes advantage of Bethe-Salpeter wavefunctions obtained from LQCD computations of correlation functions of multiple baryons~\cite{Ishii:2006ec,Aoki:2011gt,HALQCD:2012aa,Gongyo:2017fjb,Miyamoto:2017tjs,Iritani:2018sra}. These wavefunctions are used to constrain non-local potentials in the form of a truncated velocity expansion. The potentials are then used to solve the Lippmann-Schwinger equation in infinite volume to obtain the physical scattering amplitudes. 
The HAL QCD potential method and its applications have been reviewed in Refs.~\cite{Aoki:2012tk,HALQCD:2012aa,Ikeda:2016zwx,Kawai:2017goq,Iritani:2018sra,Inoue:2014ipa,Sasaki:2013zwa,Inoue:2013nfe,Murano:2013xxa}, and the potential systematic uncertainties associated with the method are discussed in Refs.~\cite{Kurth:2013tua,Beane:2010em,Walker-Loud:2014iea,Yamazaki:2017gjl,Iritani:2018zbt,Davoudi:2017ddj} and briefly summarized in Sec.~\ref{subsec:fitting}. Since this approach does not currently extend to computations of matrix elements, it is not a focus of this review.

\subsubsection{Current status of studies of nuclei}
\label{sec:forces}
Over the last decade, the ground-state energies of light nuclei and hypernuclei have been computed by a number of groups~\cite{Beane:2010hg,Inoue:2010es,Beane:2011iw,Beane:2012vq,Yamazaki:2012hi,Yamazaki:2015asa,Berkowitz:2015eaa,Orginos:2015aya,Francis:2018qch,Inoue:2015dmi,Aoki:2012tk,HALQCD:2012aa,Inoue:2018axd,Ikeda:2016zwx,Kawai:2017goq,Iritani:2018sra,Inoue:2014ipa,Sasaki:2013zwa,Inoue:2013nfe,Murano:2013xxa} using the LQCD approach described in the previous sections. Because of the large computational resource requirements of such studies, all nuclear calculations to date have used larger-than-physical values of the quark masses corresponding to $300 \leq m_\pi \leq 800$~MeV (although HAL QCD have used close-to-physical quark-mass ensembles to constrain hyperon-hyperon and nucleon-hyperon potentials).
While infinite-volume extrapolations have been undertaken based on LQCD calculations at a fixed set of quark masses on a few lattice volumes, continuum extrapolations have not yet been performed. 
\begin{figure}[t!]
	\centering
	\includegraphics[width=0.98\textwidth]{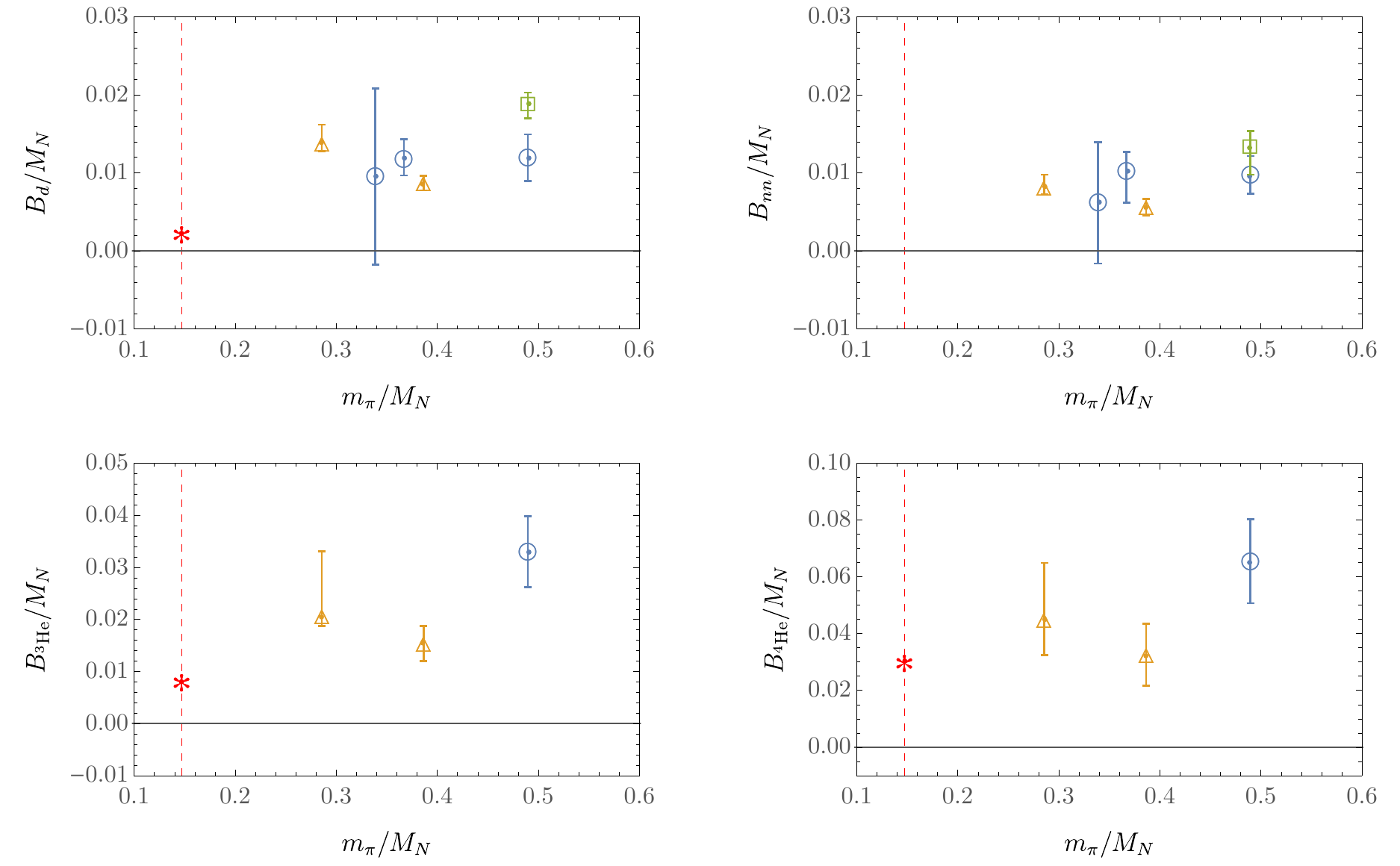}
	\caption[.]{
		The binding energies of light nuclei obtained with LQCD over a range of pion masses. Dimensionless quantities are used for each axis in order to minimize the effects of scale-setting choices when comparing results obtained in different studies. This summary figure shows results that have been extrapolated to infinite volume. Here, blue circles denote NPLQCD Collaboration results~\cite{Beane:2012vq,Orginos:2015aya,Beane:2011iw}, orange triangles show PACS-CS Collaboration results~\cite{Yamazaki:2012hi,Yamazaki:2015asa}, and green squares show CalLat Collaboration results~\cite{Berkowitz:2015eaa}. For the deuteron channel, the CalLat Collaboration \cite{Berkowitz:2015eaa} finds a second, shallow, state below the two nucleon threshold at $m_\pi=806$ MeV. Since this state is consistent with being a possible continuum state, it is not shown in the figure. The physical point is denoted by the dashed red line and experimental results are denoted by red stars. The HAL QCD Collaboration predicts that nuclei are unbound in the NN channels using the potential method at the unphysical quark masses they have studied.}
	\label{fig:nuclei}
\end{figure}

A summary of the state-of-the-art LQCD calculations of binding energies of light nuclei, $B_h = M_h - A M_N$, for nuclei $h$ with atomic number $A$, is shown in Fig.~\ref{fig:nuclei}. It is observed, in all studies that find bound states, 
that the binding energies of nuclei at larger-than-physical values of the quark masses are larger than those in nature. The HAL QCD Collaboration, using the potential approach, does not find bound $NN$ systems for any of the larger-than-physical quark masses that they have studied~\cite{Inoue:2015dmi,Aoki:2012tk,HALQCD:2012aa,Inoue:2018axd,Ikeda:2016zwx,Kawai:2017goq,Iritani:2018sra,Inoue:2014ipa,Sasaki:2013zwa,Inoue:2013nfe,Murano:2013xxa}. Besides this tension, discussed further below in Sec.~\ref{subsec:fitting}, studies performed by different collaborations using different lattice actions are broadly consistent, with an indication of a monotonic approach to the physical binding energies for each nucleus. 
Light hypernuclei for $A\leq4$ have also been studied. In particular, LQCD predictions have been made for the binding energy of the $H$-dibaryon (a six-quark state $uuddss$~\cite{Jaffe:1976yi}), whose existence may have interesting phenomenological consequences~\cite{Beane:2010hg,Inoue:2010es,Francis:2018qch,Beane:2011zpa,Shanahan:2011su,Haidenbauer:2011ah,Farrar:2002ic,Shanahan:2013yta}.

In addition to the nuclear spectrum and matrix elements, which are the primary focus of this review, LQCD calculations of nucleon-nucleon, as well as hyperon-nucleon, forces and scattering have developed rapidly, as detailed in the following subsection. Furthermore, LQCD studies of the gluon structure of light nuclei have been undertaken~\cite{Winter:2017bfs}, albeit as-yet at unphysical values of the quark masses. The goal of these studies is to provide reliable predictions for how the partonic structure of a nucleon is affected when bound in a nucleus, a subject that will be investigated with higher resolution than has been possible so far at the planned Electron-Ion Collider (EIC)~\cite{Accardi:2012qut}. Other interesting questions, such as the possibility of exotic states of matter in the form of quarkonium-nucleus bound states, have also been explored in first-principles studies of nuclei using LQCD~\cite{Beane:2014sda}.

\subsubsection{Scattering and transition amplitudes from finite-volume correlation functions
	\label{sec:directmapping}}
The quantum-mechanical approach of Lee, Huang, and Yang~\cite{Lee:1957zzb} in the 1950s established the connection between the elastic two-body scattering length and the energy eigenvalues in a finite volume. L\"uscher generalized this result to quantum field theory and to the full scattering amplitude below inelastic thresholds~\cite{Luscher:1986pf,Luscher:1990ux}. Further extensions of the formalism~\cite{Rummukainen:1995vs, Kim:2005gf, He:2005ey, Davoudi:2011md, Leskovec:2012gb, Hansen:2012tf, Briceno:2012yi, Gockeler:2012yj, Briceno:2013lba, Feng:2004ua, Lee:2017igf, Bedaque:2004kc, Luu:2011ep, Briceno:2013hya, Briceno:2013bda, Briceno:2014oea, Briceno:2017max} to boosted systems, coupled-channel processes, non-identical particles with arbitrary spin, as well as more general geometries and boundary conditions, have enabled LQCD determinations of phase shifts and inelasticities for a variety of two-hadron channels, see e.g., Refs.~\cite{Beane:2006gf, Beane:2009py, Beane:2010hg, Beane:2011iw, Beane:2012ey, Beane:2013br, Orginos:2015aya, Berkowitz:2015eaa, Wagman:2017tmp, Francis:2018qch, Junnarkar:2019equ, Wilson:2015dqa, Lang:2016hnn,Briceno:2016mjc,Wu:2017qve, Brett:2018jqw, Guo:2018zss, Skerbis:2018lew,Andersen:2018mau, Dudek:2016cru, Woss:2019hse}.

In a general form, L\"uscher's ``quantization condition'' for two-hadron states in a finite cubic volume with periodic boundary conditions (PBCs) can be written as
\begin{eqnarray}
{\rm Det}~[\mathcal{M}^{-1}+F]=0,
\label{eq:QC}
\end{eqnarray}
where the determinant is over all kinematically-allowed two-hadron channels, as well as over the total angular momentum $J$ and its azimuthal component $m_J$, the total partial wave $l$, and the spin $S$ of the system. $\mathcal{M}$ is the scattering amplitude and $F$ is a kinematic function:
\begin{eqnarray}
\left[F\right]_{Jm_J,lS;J'm_{J'},l'S';\rho,\rho'}=\frac{in_{\rho}k^*_{\rho}}{8\pi E^*}\delta_{S,S'}\delta_{\rho,\rho'}\left[\delta_{J,J'}\delta_{m_J,m_{J'}}\delta_{l,l'} +\frac{2i}{\pi\gamma}\sum_{l'',m''}(\tilde{k}_{\rho}^*)^{-l''-1}\mathcal{Z}^{\vec d}_{l''m''}[1;(\tilde{k}_{\rho}^*)^2]
\right.
\nonumber\\
\left . \qquad \times ~ \sum_{m_l,m_{l'},m_{S}}\langle lS,Jm_J|lm_l,Sm_{S}\rangle \langle l'm_{l'},Sm_{S}|l'S,J'm_{J'}\rangle \int d\Omega Y^*_{ l,m_l}Y^*_{l'',m''}Y_{l',m_{l'}}\right].
\label{eq:deltaG}
\end{eqnarray}
Here, $\tilde{k}_{\rho}^*=k_{\rho}^*L/2\pi$, where $k_{\rho}^*$ is the magnitude of the relative momentum of two hadrons in channel $\rho$ in the center of mass (CM) frame, $E^*$ and $E$ are the CM and laboratory-frame energies, respectively, $\gamma=E/E^*$ is the relativistic $\gamma$-factor, and $n_{\rho}=1/2~(1)$ if the particles in channel $\rho$ are identical (distinguishable). L\"uscher's $\mathcal{Z}$-function is defined as
\begin{equation}
\mathcal{Z}^{\vec d}_{lm}[s;x^2]= \sum_{\vec r} \frac{r^lY_{l,m}(\vec{r})}{(r^2-x^2)^s}\,,
\end{equation}
where $r=|\vec{r}\,|$. The sum is performed over ${\vec r}={\hat{\gamma}}^{-1}({\vec n}-\alpha_{\rho} {\vec d}\,)$, where $\vec{n}$ is a triplet of integers, $\vec d$ is the normalized boost vector ${\vec d}={\vec P}L/2\pi$, $\alpha_{\rho}=\frac{1}{2}\left[1+\frac{m_{\rho,1}^2-m_{\rho,2}^2}{E^{*2}}\right]$, and $\hat{\gamma}^{-1}{\vec x}\equiv{\gamma}^{-1}{\vec x}_{||}+{\vec x}_{\perp}$, with ${\vec x}_{||}~({\vec x}_{\perp})$ denoting the component of $\vec{x}$ that is parallel (perpendicular) to the total momentum, $\vec{P}$. $m_{\rho,1}$ and $m_{\rho,2}$ denote the masses of each hadron in channel $\rho$. When twisted boundary conditions are used, or the volume has asymmetric extents, a modified $\mathcal{Z}$ function is required~\cite{Bedaque:2004kc, Briceno:2013hya, Feng:2004ua, Detmold:2004qn,Lee:2017igf}.

In contrast to meson-meson scattering, nuclear-physics applications of L\"uscher's formalism took longer to develop, due to a delay in understanding that systems with large scattering lengths could also be reliably computed in a finite volume. As re-emphasized in Ref.~\cite{Beane:2003da}, L\"uscher's formula is a nonperturbative relation in the interaction strength, and is valid below inelastic thresholds as long as the range of interactions is smaller than $L/2$, as mentioned before. Explicitly, L\"uscher's relation in Eq.~(\ref{eq:QC}) is valid below the first inelastic threshold, e.g., the threshold for producing an on-shell pion in nucleon-nucleon scattering, with corrections suppressed by $\sim e^{-L/(2R)}$ where $R$ denotes the range of interactions, typically $R \sim m_{\pi}^{-1}$ in nuclear physics. Ref.~\cite{Beane:2003da} also provided a simple derivation of L\"uscher's formula that, although it takes advantage of an EFT description of the amplitudes, makes it clear that the details of the short-distance physics are irrelevant to the infrared physics associated with boundary effects. Employing L\"uscher's method, the first QCD determination of two-nucleon scattering amplitudes at low energies, albeit at unphysically large quark masses, appeared soon after~\cite{Beane:2006mx}. L\"uscher's method has continued to be used to study two-baryon scattering amplitudes is various channels and partial waves~\cite{Beane:2006gf, Beane:2009py, Beane:2010hg, Beane:2011iw, Beane:2012ey, Buchoff:2012ja, Beane:2013br, Orginos:2015aya, Berkowitz:2015eaa, Wagman:2017tmp, Francis:2018qch, Junnarkar:2019equ}.

Binding energies can be directly extracted from the spectral decomposition of LQCD two-point correlation functions as long as the state is deeply bound, in which case the binding energy, $B$, is exponentially close to the corresponding finite-volume energy. For a nonrelativistic two-body bound state, the binding momentum in a finite volume, $\kappa_L$, is 
\begin{equation}
\kappa_L=\kappa+\frac{Z^2}{L}\left[6\,e^{-\kappa L}+\frac{12}{\sqrt{2}}\,e^{-\sqrt{2}\kappa L}+\frac{8}{\sqrt{3}}\,e^{-\sqrt{3}\kappa L} \right]+\mathcal{O}\left(\frac{e^{-2\kappa L}}{L}\right),
\label{eq:extrapolation}
\end{equation}
where $\kappa$ is the infinite-volume binding momentum of the state ($\kappa=\sqrt{MB}$ for two identical hadrons with mass $M$) and $Z^2$ is the residue of the scattering amplitude at the bound-state pole. With slight modification, this relation can be extended to the case of relativistic bound states~\cite{Davoudi:2011md}. Relations for two-body systems with arbitrary masses moving in a finite volume can be found in Refs.~\cite{Bour:2011ef,Konig:2011nz,Davoudi:2011md,Briceno:2013bda}. This result, which can be intuitively understood from a phenomenological description of a bound-state wavefunction and the location of associated images in adjacent lattice volumes due to PBCs~\cite{Briceno:2013bda}, is a direct consequence of Eq.~(\ref{eq:QC}) when analytically continued as ${k^*}^2 \to -\kappa_L^2$. When a deeply-bound state is present in the spectrum, the volume dependence in Eq.~\eqref{eq:extrapolation} provides a check on the validity of the extracted energies~\cite{Wagman:2017tmp, Beane:2017edf, Davoudi:2017ddj}, see Sec.~\ref{sec:challenges}. Once the scattering amplitude is constrained in a given partial wave using L\"uscher's method, the binding energy can also be obtained from the location of the pole in the scattering amplitude in that partial wave. The comparison between the binding energy obtained from extrapolation of finite-volume energies using Eq.~(\ref{eq:extrapolation}) and that from the pole of the scattering amplitude provides a further consistency check on the calculations.

For quantitatively understanding matter in extreme environments, there is a need to improve constraints on three-, four-, and higher-body nuclear forces. Such constraints will improve knowledge of the equation of state in neutron stars and predictions for neutron-rich isotopes to be studied at the Facility for Rare Isotope Beams (FRIB). Generalizations of L\"uscher's formalism to three-hadron systems have been formulated using various approaches~\cite{Polejaeva:2012ut,Briceno:2012rv, Hansen:2014eka,Hansen:2015zga, Hammer:2017uqm,Hammer:2017kms, Guo:2017ism,Mai:2017bge,Briceno:2017tce,Doring:2018xxx,Briceno:2018aml,Romero-Lopez:2019qrt,Jackura:2019bmu,Hansen:2020zhy}. This program has developed significantly, and there is agreement in appropriate limits among different approaches, see Ref.~\cite{Hansen:2019nir} for a recent review. A number of these formalisms have been applied to constrain the parameterizations of three-pion interactions from LQCD~\cite{Mai:2018djl,Blanton:2019igq,Blanton:2019vdk,Mai:2019fba,Culver:2019vvu}. Additionally, relations for the three-hadron binding energies can be determined from the corresponding finite-volume formalisms, see e.g., Refs. \cite{Kreuzer:2010ti,Briceno:2012rv,Meissner:2014dea,Meng:2017jgx,Konig:2017krd}. 

Progress beyond three hadrons presents a challenge as  L\"uscher's formalism can not be straightforwardly extended to higher-body sectors. New ideas have appeared that do not rely on such an approach~\cite{Hansen:2017mnd,Bulava:2019kbi}, but instead are based on certain limits of a properly constructed smeared spectral function in a finite volume such that the Maiani-Testa no-go theorem~\cite{Maiani:1990ca} is circumvented in the infinite-volume limit. In the approach of Ref.~\cite{Bulava:2019kbi}, determination of scattering amplitudes for $n \to m$ processes requires LQCD calculations of $n+m$-point correlation functions that are numerically challenging, as well 
as an inverse transform of a discrete set of data, which necessarily involves model dependence. Nonetheless, this approach is formally straightforward to generalize to arbitrary elastic and inelastic processes. On the other hand, in the threshold region, i.e., when the interactions are weak such as in multi-pion systems in a maximal isospin state, quantum-mechanical perturbation theory can be used to relate the shift in the energy of $n$-boson systems in a finite volume to the two-boson scattering parameters defining the effective range expansion, the scattering length and the effective range parameter, and higher-body interactions~\cite{Beane:2007qr,Hansen:2016fzj,Beane:2020ycc}. With the use of this result, a three-hadron force parameter in the pionic system was constrained for the first time in Refs.~\cite{Beane:2007es,Detmold:2008fn}. This investigation has been extended in recent years with the use of nonperturbative $3\to 3$ quantization conditions in the three-pion sector~\cite{Mai:2018djl,Horz:2019rrn,Blanton:2019igq,Blanton:2019vdk,Mai:2019fba,Culver:2019vvu}.

The matching described above for scattering amplitudes in the two- and three-hadron sector is a necessary ingredient for the mapping between finite-volume nuclear matrix elements from LQCD and their infinite-volume counterparts. Lellouch and L\"uscher established that knowledge of finite-volume energies and matrix elements, as well as the energy dependence of the $2 \to 2$ scattering amplitude at those energies, are essential in connecting the $1 \to 2$ matrix elements of external operators in a finite Euclidean spacetime to the corresponding physical transition amplitude~\cite{Lellouch:2000pv}. This formalism was successfully applied to the weak process $K \to \pi \pi$~\cite{Bai:2015nea, Blum:2015ywa}, and its generalizations~\cite{Meyer:2011um, Briceno:2012yi, Bernard:2012bi, Briceno:2014uqa, Feng:2014gba, Briceno:2015csa, Briceno:2015tza} were applied to studies of the transition form factors of 
the $\rho$ resonance~\cite{Briceno:2016kkp}. Generalization to $\{0,1,2\} \to 2$ processes for relativistic systems with generic currents have since appeared~\cite{Briceno:2014uqa,Briceno:2015tza, Briceno:2015csa, Baroni:2018iau}. For two-nucleon transitions involving an electroweak current, the relevant mapping can be obtained from the general relation~\cite{Briceno:2015tza}
\begin{align}
\Big | \langle E_{n_f}, \vec{P}_f, L \vert  \mathcal J(0)  \vert E_{n_i}, \vec{P}_i, L \rangle \Big |^2 
=\frac{1}{L^6}~{\rm Tr}\left[
\mathcal R(E_{n_i}, \vec{P}_i)
{\mathcal{W}}_{L,{\rm df}}(P_i,P_f,L)
\mathcal R(E_{n_f}, \vec{P}_f)
{\mathcal{W}}_{L,{\rm df}}(P_f,P_i,L)
\right],
\label{eq:2to2}
\end{align}
for $2\overset{\mathcal{J}}{\longrightarrow}2$ processes.
Here, the left-hand side is the absolute value squared of the finite-volume matrix element of the Schr\"odinger-picture current $\mathcal{J}$ at the origin between initial and final two-hadron states with finite-volume energies $E_{n_i}$ and $E_{n_f}$ and total three-momenta $\vec{P}_i$ and $\vec{P}_f$, respectively. The function ${\mathcal{W}}_{L,{\rm df}}$ is defined as
\begin{align}
{\mathcal{W}}_{L,{\rm df}}(P_f,P_i,L)& \equiv {\mathcal{W}}_{\rm df}(P_f,P_i) +  \mathcal  M(P_f) \ [G(L) \cdot w](P_f,P_i) \ \mathcal M(P_i),
\label{eq:WLdf}
\end{align}
where $P_{i(f)}=(E_{n_{i(f)}},\vec{P}_{i(f)})$, and ${\mathcal{W}}_{\rm df}$ is a divergence-free infinite-volume transition amplitude in which the on-shell divergences associated with the $1\overset{\mathcal{J}}{\longrightarrow}1$ transitions on external-states hadrons are subtracted out. $\mathcal{M}$ is the $2\rightarrow 2$ elastic scattering amplitude of initial- and final-state hadrons, $G(L)$ is a new finite-volume function arising from the s-channel two-hadron loops with an insertion of the one-body current on the hadrons, defined in Ref.~\cite{Briceno:2015tza}, $w$ is the $1\overset{\mathcal{J}}{\longrightarrow}1$ transition amplitude, and $\mathcal R$ is defined as
\begin{equation}
\label{eq:Rdef}
\mathcal R(E_{n},\vec{P}) \equiv  \lim_{E \rightarrow E_{n}} \left[  (E_n - E) \frac{1}{F^{-1}(P,L) + \mathcal M(P)}\right]  ,
\end{equation}
where $P=(E,\vec{P})$, and $F$ is the finite-volume function defined in Eq.~(\ref{eq:deltaG}). All functions in Eq.~(\ref{eq:2to2}) are, therefore, evaluated at on-shell kinematics for the two-hadron system, giving access to the on-shell $2\overset{\mathcal{J}}{\longrightarrow}2$ transition amplitude. As will be reviewed in Sec.~\ref{sec:eft}, similar relations have been derived in the context of EFT expansion of two-nucleon electroweak transitions in Refs.~\cite{Detmold:2004qn,Meyer:2012wk,Briceno:2012yi}, such that the corresponding low-energy constants (LECs) can be constrained from LQCD matrix elements in two-nucleon systems.

For nuclear observables involving bi-local insertions of the electroweak current, such as the Compton scattering amplitudes of the nucleon and of nuclei, and $2\nu\beta\beta$ and $0\nu\beta\beta$ decays, the Lellouch-L\"uscher formalism must be generalized as potential intermediate multi-hadron states complicate the mapping between Minkowski and Euclidean spacetime quantities. Since the on-shell intermediate states are sensitive to the spacetime metric signature, their individual contributions must be reconstructed from LQCD determinations of two- and three-point functions separately and subtracted from the four-point function under study, as discussed in Ref.~\cite{Briceno:2019opb}. Under the assumptions that the kinematics of the process allow only zero, one-, and two-hadron intermediate states to go on-shell, the finite-volume effects have been identified, completing the connection to the physical bi-local transition amplitudes. This formalism was first developed and applied to studies of the $K_L-K_S$ mass difference and of rare kaon decays in Refs.~\cite{Christ:2014qaa,Christ:2015pwa,Christ:2016eae,Christ:2019dxu}, and was recently extended to \tnubb~\cite{Davoudi:2020xdv} and \znubb~\cite{Feng:2018pdq,Detmold:2020jqv} decays and other scenarios~\cite{Briceno:2019opb,Feng:2020nqj}.

\subsubsection{Matching to nuclear effective field theories and models}
\label{sec:eft}
A complementary approach to extracting nuclear spectra and properties from LQCD calculations is that of matching the numerical results to an appropriate description of the system under study using an EFT or a phenomenological model. This has the advantage that, in principle, more complicated systems can be addressed than through the direct approach discussed above. In part, this is possible because of the hierarchies that exist in nuclear physics and are encapsulated in nuclear models or nuclear EFTs. The forces that bind protons and neutrons together to form nuclei are dominated by two-body interactions, and three- and higher-body forces are subleading at normal nuclear densities. Similarly, where they are known, nuclear matrix elements are typically dominated by the coupling to single nucleons, with small but non-negligible contributions from the coherent coupling of two or more nucleons to the external probe. Since EFTs provide a systematic way to take advantage of such hierarchies, and since existing LQCD results have been primarily matched to EFTs in the few-nucleon sector, the focus of this section is on matching to EFTs rather than phenomenological models. Nonetheless, LQCD results for multi-nucleon correlation functions can also be used to constrain the parameters of phenomenological nuclear models, just as experimental data have been used to constrain and improve them.

While the precise power countings of nuclear forces and currents within EFTs that are consistent with the observed hierarchies in nature are still under development, existing EFTs have already provided the basis for a LQCD--EFT matching program. This approach has the distinct advantage that small nuclear systems that are computationally accessible in LQCD can be used in order to determine the unknown LECs, the parameters defining the forces and couplings to external currents in the EFT Lagrangian. Given LQCD results for spectra and interactions of nuclear systems, the LECs of EFT calculations can tuned to match the LQCD results. Such matching can be performed in the same finite volume as the LQCD calculations. Having determined the values of the LECs needed for a given process, the EFT interactions can be fed in to one of a range of many-body methods~\cite{Carlson:2014vla,Barrett:2013nh,Hagen:2013nca,Hagen:2015yea,Carbone:2013eqa,Soma:2013xha,Hergert:2015awm,Stro19ARNPS,Epelbaum:2011md,Quaglioni:2018odl} that enable predictions for nuclear systems considerably larger than those for which direct LQCD calculations are feasible.  Significant diagnostic efforts have been undertaken to characterize the differences between many-body methods based on different EFT Hamiltonians (with different dynamical degrees of freedom, cutoff scales, EFT order, parameter fitting and estimations, etc.), and these effects have been quantified for key observables, such as the lowest-lying spectra of light and medium-mass nuclei, their charge radii, and the nuclear saturation point and symmetry energy. In the remainder of this section, technical features of the EFT-based approaches in \emph{ab initio} calculations will not be reviewed; the reader is referred to a recent review~\cite{Tews:2020hgp} for a discussion of ongoing developments.

\begin{figure}[!t]
	\includegraphics[scale=0.575]{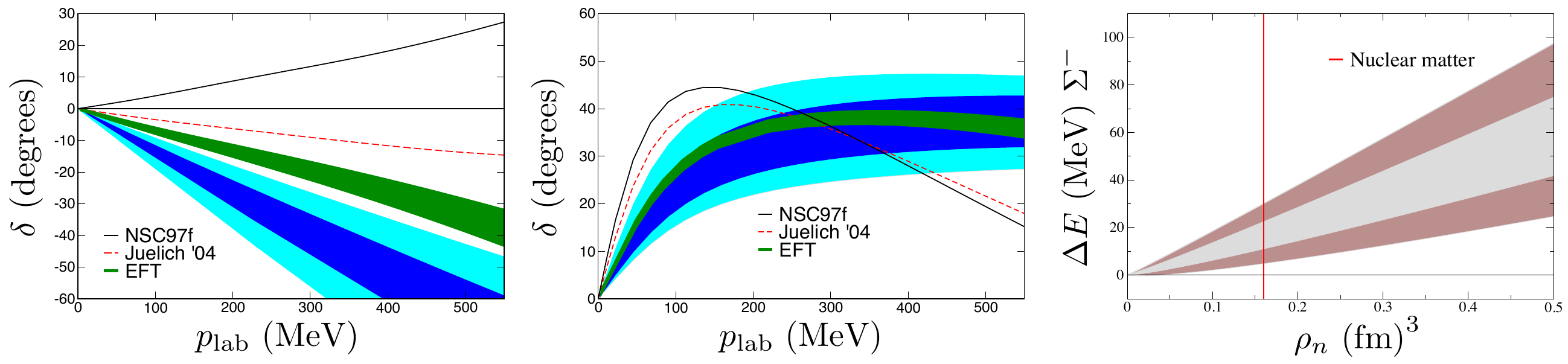}
	\caption[.]{
		The $\siii$ (left panel) and $\si$ (center panel) 
		$n\Sigma^-$ phase shifts versus laboratory momentum computed with LQCD and extrapolated to the physical pion mass using EFT (blue bands), along with other determinations (green bands are EFT fits to data, while the black and red curves are predictions from phenomenological models). The right panel shows the energy shift versus neutron density of a $\Sigma^-$ in a non-interacting Fermi gas of neutrons as determined from Fumi's theorem. The inner (outer) band encompasses statistical (systematic) uncertainties. [Figure from Ref.~\cite{Beane:2012ey}.]}
	\label{fig:YN}
\end{figure}

An early demonstration of the value of matching LQCD and nuclear EFTs was presented in
Ref.~\cite{Beane:2006mx}, where the calculated $NN$ phase shifts were matched to low-energy EFT in both spin channels. Another example of such matching was presented in Ref.~\cite{Beane:2012ey}, where the $N\Sigma$ scattering phase shifts and the role of the $\Sigma^-$-hyperons in the composition of dense nuclear matter were examined. In particular, the energies of $I=3/2$ $N\Sigma$ states were determined in two large volumes at a larger-than-physical quark mass. The spin-singlet $N\Sigma$ channel was found to be attractive, supporting a bound sate at the unphysically heavy quark mass considered. In contrast, a large positive energy shift was seen in the spin-triplet channel, indicating a highly repulsive interaction, potentially invalidating the condition on the range of interactions in L\"uscher's method. In Weinberg's power counting~\cite{Weinberg:1990rz,Weinberg:1991um}, the leading order (LO) EFT expansion of the hyperon-nucleon force comprises a contact interaction and a one-meson-exchange term~\cite{Savage:1995kv,Beane:2003yx}. For the $\siii$ channel, in addition to using L\"uscher's method, the three-dimensional Schr\"odinger equation was solved in the finite volume and the contact LECs were tuned so that the lowest-lying energy levels matched those from LQCD at a larger-than-physical quark mass. The constrained LEC was then used at the physical quark masses to obtain the phase shifts in infinite volume, assuming a negligible quark-mass dependence in the LEC of the momentum-independent interaction. A compilation of the results of this study is shown in Fig.~\ref{fig:YN}. Another example of an EFT matching was performed for two octet-baryon scattering at an unphysically large quark masses in Ref.~\cite{Wagman:2017tmp}, which led to interesting observations about the nature of nuclear and hypernuclear forces consistent with large-$N_c$ predictions~\cite{Kaplan:1995yg,Kaplan:1996rk}. With future advances in \emph{ab initio} methods using nuclear and hypernuclear Hamiltonians simultaneously \cite{Lonardoni:2014bwa}, these studies will help disentangle the nature of matter at the densities found in the interior of neutron stars. 

\begin{figure}[t!]
	\centering
	\includegraphics[scale=0.55]{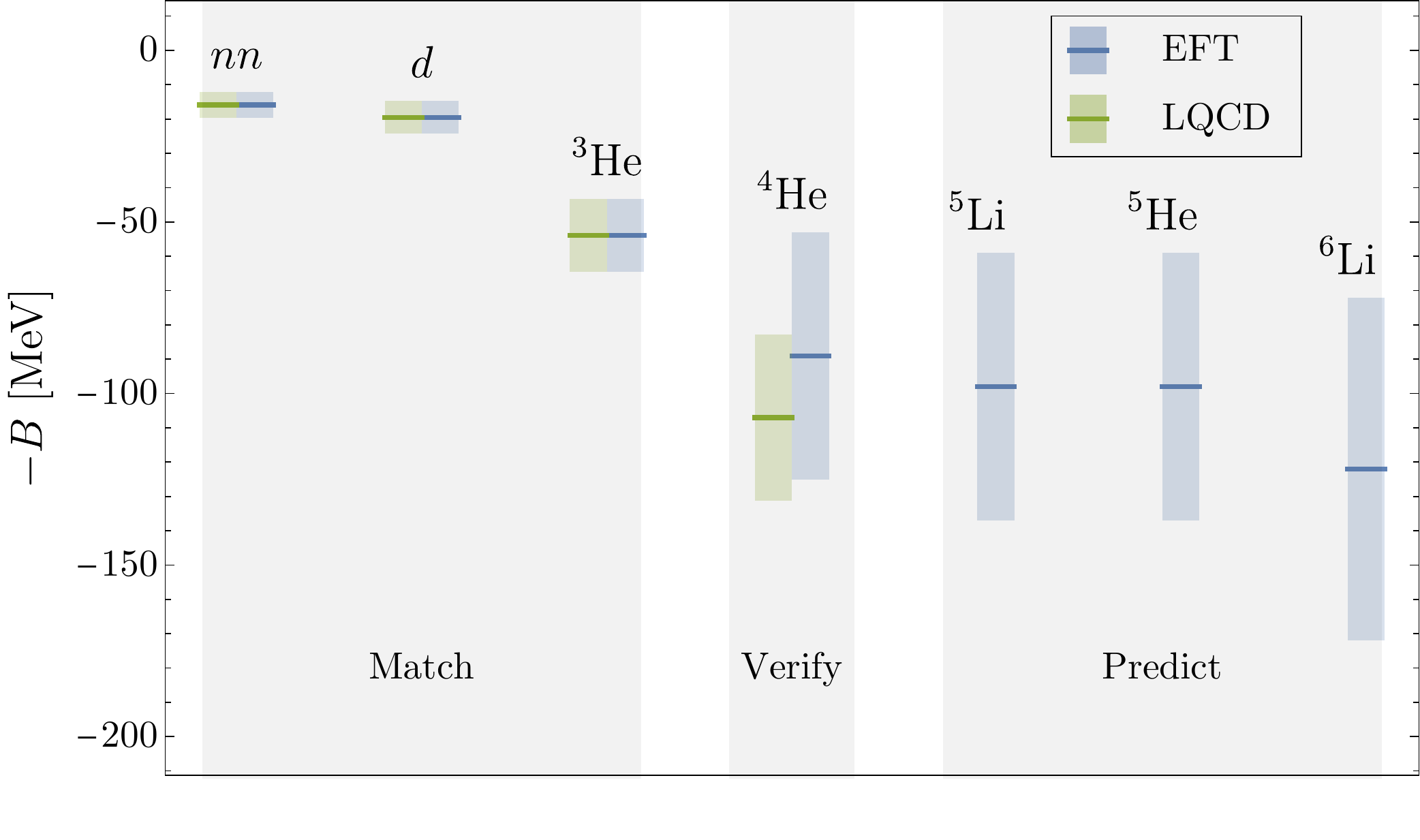}
	\caption{\label{fig:matching806mev}
		Binding energies of the deuteron, dineutron, $^3$He, and $^4$He at an SU(3) flavor-symmetric point obtained with LQCD at quark masses corresponding to a pion mass of $806$ MeV~\cite{Beane:2012vq} are used to obtain the binding energy of light nuclei up to $A=6$ using an AFDMC computation with the pionless EFT~\cite{Barnea:2013uqa}. The two- and three-nucleon LECs were constrained by the LQCD calculation of the $A\in\{2,3\}$ binding energies. 
		[Data from Refs.~\cite{Beane:2012vq,Barnea:2013uqa}.]}
\end{figure}

Another key example of the LQCD--EFT matching was presented in Refs.~\cite{Barnea:2013uqa,Kirscher:2015yda}, where the binding energies of a range of nuclei were predicted at unphysical values of the quark masses corresponding to $m_\pi= 806$ MeV using LQCD input. To achieve this, the LO two- and three-body interactions in pionless EFT~\cite{vanKolck:1998bw,Chen:1999tn,Bedaque:2002mn} were determined by matching to LQCD determinations of the binding energies of $A\in\{2,3\}$ systems~\cite{Beane:2012vq}. The ground-state energies of nuclei with atomic number $A\in\{4,5,6\}$ were then computed in Ref.~\cite{Barnea:2013uqa} using an auxiliary-field diffusion Monte Carlo (AFDMC) method, see Fig.~\ref{fig:matching806mev}. The binding energy of the $A=4$ system (${^4}$He) was found to be consistent, within uncertainties, with that obtained from LQCD, validating the approach. In extensions of this approach to larger nuclei, AFDMC~\cite{Contessi:2017rww} and a discrete-variable representation in the harmonic oscillator basis~\cite{Bansal:2017pwn} were used to compute the ground-state energy of a doubly-magic nucleus ${^{16}}$O (and ${^{40}}$Ca in the case of Ref.~\cite{Bansal:2017pwn}). The two approaches, performed at different orders in EFT, disagree on whether ${^{16}}$O remains bound at $m_{\pi} = 806$ MeV, indicating that a higher-order EFT calculation is needed when extrapolating LQCD results to larger nuclei. This progress, nonetheless, serves as a milestone in connecting LQCD and nuclear-structure studies. More precise LQCD input and increasingly reliable EFT Hamiltonians and many-body methods will result in refined predictive capabilities in upcoming years.

Matching LQCD to finite-volume few-body calculations to constrain nuclear models and/or EFTs appears a promising path, eliminating the step involving obtaining and matching the infinite-volume quantities. As an example, Ref.~\cite{Eliyahu:2019nkz} performs a direct matching of the pionless EFT in a finite volume for $A\in\{2,3\}$ nuclei to LQCD results at unphysical values of the quark masses, resulting in increased precision in the determination of the binding energies in the infinite-volume limit. Additionally, in Ref.~\cite{Klos:2016fdb} the ground- and first excited-state energies of the two-neutron system in a finite cubic volume with PBCs were computed. Once LQCD calculations of multi-neutron systems are available at the physical quark masses, such an approach can lead to constraints on multi-neutron forces from QCD. Furthermore, as argued in Ref.~\cite{Gandolfi:2017arm}, LQCD studies of the properties of neutron systems in small volumes could provide valuable input into the nature of the equation of state of cold neutron matter at high densities.

The LQCD--EFT matching approach can be extended to nuclear matrix elements. For example, provided a matrix element can be reliably described within an EFT framework, the procedure outlined above will provide constraints on the LECs associated with currents coupling to a single nucleon, or coherently to two or more nucleons. Such a mapping was first introduced in Ref.~\cite{Detmold:2004qn,Meyer:2012wk} for EM and weak processes in two-nucleon systems. In the presence of a background EM or weak field, the finite-volume energies of the nuclear systems are shifted. These shifts can be related to finite-volume matrix elements and thereby determine the LECs in an EFT expansion of the relevant transition amplitude. As will be discussed in Sec.~\ref{sec:EM}, this approach has been used~\cite{Savage:2016kon} to access the cross-section for the $np \to d \gamma$ process from LQCD. The extension of the formalism to a Lellouch-L\"uscher-type formula for two-nucleon electroweak transitions, such as $pp$ fusion, was presented in Ref.~\cite{Briceno:2012yi}. Future studies of the $np\rightarrow d\gamma$ process at lighter quark masses than those used in Ref.~\cite{Beane:2015yha} will require the application of this formalism. In Ref.~\cite{Kirscher:2017fqc}, this matching approach was further explored, and the two-nucleon EM-coupling LECs were fit to the isotriplet-isosinglet two-nucleon transition rate~\cite{Beane:2015yha} and the magnetic moment of the deuteron~\cite{Chang:2015qxa} obtained from LQCD at $m_{\pi} = 806$ MeV (along with the binding energies of $A\in\{2,3\}$ systems~\cite{Beane:2012vq} and the nucleon's isovector magnetic moment~\cite{Beane:2014ora}). Having constrained these LECs, the triton magnetic moment, as well as the magnetic polarizability of deuterium, were postdicted, and the polarizabilities of three-nucleon systems were predicted~\cite{Chang:2015qxa}.

\subsection{Technical challenges and developments in lattice QCD studies of nuclei}
\label{sec:challenges}

There are several technical challenges specific to the study of multi-nucleon systems in LQCD: such calculations suffer from i)  signal-to-noise (StN) at late Euclidean times that grows exponentially with the atomic number $A$~\cite{Beane:2009gs}, ii) dense excitation spectra arising from an accumulation of continuum scattering states
in the infinite-volume limit, and iii) a rapid growth in the number of contractions required to compute nuclear correlation functions, as discussed above in Sec.~\ref{subsec:baryonblocks}.
To address these issues requires methods to mitigate StN degradation, to analyze and control excited-state effects, and to reduce the cost of contractions in calculations relevant for nuclear physics.
These challenges, and strategies to mitigate them, are outlined in the following subsections.

\subsubsection{The signal-to-noise problem
	\label{sec:StN}} 

The statistical uncertainty of Monte Carlo calculations of a correlation function $\langle\mathcal{O}\rangle$ approaches $\text{Var}\langle\mathcal{O}\rangle/\sqrt{N_{\rm cfg}}$ as $N_{\rm cfg}\rightarrow \infty$, where $N_{\rm cfg}$ is the number of effectively decorrelated\footnote{See Refs.~\cite{Luscher:2010ae,Ramos:2018vgu,Wolff:2003sm} for discussions of autocorrelation in Markov Chain Monte Carlo samples, and techniques to account for them.} statistical samples.\footnote{As the spacetime volume of the lattice geometry $V$ increases, the number of source positions on a single gauge-field configuration that can be used to calculate approximately independent correlation functions increases.
	This volume averaging is particularly important for calculations of nuclear correlation functions;
	heavier systems tend to remain more localized in the vicinity of their sources, and as such experience a reduced sampling of the gauge configuration for any single source position. 
	This localization effect contributes to the relatively poor StN for baryonic quantities relative to mesonic quantities, but also means that a greater number of statistically-independent correlation functions can be extracted from a given configuration, as discussed for instance in Ref.~\cite{Beane:2009kya}.}
However, $\text{Var}\langle\mathcal{O}\rangle$ depends on the observable under study, and the StN ratios, $\langle \cO \rangle/\text{Var}\langle\mathcal{O}\rangle$, of nuclear correlation functions decrease exponentially at late  source/sink separation times, $t$~\cite{Parisi:1983ae,Lepage:noise}, and with baryon number, $A$~\cite{Beane:2009gs}.
This StN problem can be understood by analyzing the variance of (the real part of) nuclear two-point correlation functions:
\begin{equation}
\begin{split}\label{eq:var}
\text{Var}  \left< \text{Re}\left[ \chi_{A}(t) \bar{\chi}_{A}(0) \right] \right> & = \frac{1}{2} \left< \chi_{A}(t) \bar{\chi}_{A}(t)  \chi_{A}(0) \bar{\chi}_{A}(0)  \right>  \\
&\hspace{25pt} + \frac{1}{2} \left< \chi_{A}(t) \chi_{A}(t)  \bar{\chi}_{A}(0) \bar{\chi}_{A}(0)  \right> - \left< \chi_{A}(t) \bar{\chi}_{A}(0) \right>^2.
\end{split}
\end{equation}
Here, the nuclear interpolating operator $\chi_{A}$ has strangeness zero, and baryon number $A$ is explicitly specified. Additional quantum numbers such as spin and isospin that are irrelevant for this discussion are suppressed in this subsection. The spatial dependence of the interpolating operator has also implicitly been summed over to project to zero momentum.
As pointed out by Parisi~\cite{Parisi:1983ae} and Lepage~\cite{Lepage:noise} for the case of single-hadron states, and in Refs.~\cite{Beane:2009py,Beane:2009kya} specifically for the case of nuclei, the second term on the right-hand side of Eq.~\eqref{eq:var} involves operators at a single time with baryon number $2A$, but the first term involves operators with baryon number zero which decay exponentially more slowly with $t$ than $\left< \chi_{A}(t) \bar{\chi}_{A}(0) \right>^2$. Because the fermion integration in Eq.~\eqref{eq:CorFunc} is performed exactly, the lowest-energy state that contributes to $\left< \chi_{A}(t) \bar{\chi}_{A}(t)  \chi_{A}(0) \bar{\chi}_{A}(0)  \right>$ includes $3A$ pions.
The variance of a nuclear correlation function with baryon number $A$ is therefore proportional to $e^{-3A m_\pi t}$ in the large-$t$ limit, neglecting energy shifts arising from pion interactions.\footnote{By constructing nuclear correlation functions from nucleon blocks, at early times the variance of nuclear correlation functions scale as $\sim e^{-2 A M_N t}$ (neglecting nuclear binding energies).  See the subsequent discussions.}
It follows that StN ratios for Monte Carlo calculations of nuclear correlation functions are  proportional to $\sqrt{N_{\rm cfg}} e^{-A \left(M_N - \frac{3}{2}m_\pi\right) t}$ in the large $N_{\rm cfg}$ and $t$ limits, also neglecting further energy shifts arising from nuclear binding.
This naively suggests that LQCD calculations of nuclei require statistical ensembles whose size must grow exponentially with $A$ in order to maintain a fixed  StN ratio at a given time.

Although Parisi-Lepage scaling holds in the $t\rightarrow \infty$ and $N_{\rm cfg}\rightarrow \infty$ limits, excited-state contributions to the variance correlation function in Eq.~\eqref{eq:var} can significantly modify the finite-$t$ behavior of the variance.
High-statistics studies~\cite{Beane:2009kya,Beane:2009py,Beane:2009gs} showed that the exponential StN degradation at late times obtained in numerical LQCD calculations of two- and three-baryon systems is significantly slower than $e^{-A\left(M_N - \frac{3}{2}m_\pi\right)t} $ at intermediate times.
In particular, these studies exhibited the appearance of a ``golden window''~\cite{Beane:2009kya} 
where $t$ is both sufficiently large that excited-state effects in single-baryon correlation functions are sufficiently small that the variance decays much more rapidly with $t$ than $e^{-3Am_\pi t}$.
This behavior can be understood by considering the large-$t$ spectral representation of the variance correlation function.
The variance interpolating operator $\chi_{A} \bar{\chi}_{A}$ has overlap onto states of the form $AN + A\bar{N}$, $3A\pi$, and all intermediate combinations of the form $(A-k)N+(A-k)\bar{N}+3k\pi$, and therefore at large source/sink separation the variance of the real part of a nuclear correlation function is given by
\begin{align}
\nonumber
\text{Var}  \left< \text{Re}\left[ \chi_{A}(t) \bar{\chi}_{A}(0) \right] \right> & \rightarrow Z_{AN,A\bar{N}} \ e^{-2AM_N t} + Z_{(A-1)N,(A-1)\bar{N},3\pi}\ e^{-[2(A-1)M_N + 3m_\pi] t} + \ldots \\\label{eq:varZ}
& \hspace{20pt} + Z_{N,\bar{N},3(A-1)\pi} \ e^{-[2M_N + 3(A-1)m_\pi] t} + Z_{3A\pi} \ e^{-3Am_\pi t},
\end{align}
where the constants $Z_\alpha$ denote the overlap factors onto the state described by  $\alpha$, and interactions between hadrons and towers of states in which hadrons move with relative momenta are ignored.
The interpolating operator construction discussed in Sec.~\ref{subsec:baryonblocks} includes multi-baryon sinks built from products of baryon blocks in plane waves with definite relative momentum. 
This leads to a volume suppression of  $Z_{(A-k)N,(A-k)\bar{N},3k\pi} / Z_{AN,A\bar{N}} \sim (k!)^2 (m_\pi^3 V)^{-k}$, because a product of $N$ and $\bar{N}$ plane-wave interpolating operators only includes significant overlap with a $3\pi$ state from terms where two nucleons are localized in coordinate space within a hadronic volume $\sim m_\pi^3 V$, as discussed in Refs.~\cite{Beane:2009kya,Beane:2009py,Beane:2010em}. 
For finite $t$ and large $V$, contributions from terms in Eq.~\eqref{eq:varZ} besides those from the $3A\pi$ contribution can be numerically larger than the asymptotically dominant $3A\pi$ state.
If this is this case for a given interpolating operator and lattice volume, then there is a golden window where StN degradation is exponentially less rapid with $t$ than predicted by Parisi-Lepage scaling, facilitating LQCD calculations of nuclei with larger $A$.
Increasing $V$ decreases the ground-state overlap of nuclear variance correlation functions for this class of interpolating operators and enlarges the golden window.
However, this window ultimately shrinks with increasing $A$ due to the appearance of $(A!)^2$ in multi-pion to multi-nucleon overlap-factor ratios in the variance correlation function.
For large $A$ and large $t$, including overlap-factor scaling gives
\begin{equation}
\begin{split}
\text{StN}\left< \text{Re}\left[ \chi_{A}(t) \bar{\chi}_{A}(0) \right] \right> \rightarrow \frac{\sqrt{(V/a^3)^A N_{\rm cfg}}}{A!} e^{-A\left(M_N - \frac{3}{2}m_\pi\right)t},
\end{split}\label{eq:StNoverlap}
\end{equation}
indicating rapid StN degradation with $A$.

More recent theoretical analysis and numerical investigations have revealed that correlation-function noise has additional structure beyond Parisi-Lepage scaling of StN ratios.
Correlation-function probability distributions are defined as
\begin{equation}
\begin{split}
\mathcal{P}_{A}(c,t) &= \frac{1}{\mathcal{Z}} \int \mathcal{D}q\mathcal{D}\bar{q}\mathcal{D}U \ e^{-S_{\rm QCD}} \ \delta \left( \chi_{A}(t) \bar{\chi}_{A}(0) - c \right),
\end{split}\label{eq:probdef}
\end{equation}
and  encode the possible quantum fluctuations of a system and their likelihood.
The average correlation function scales as $\left< C_A(t) \right> = \int dc\ \mathcal{P}_A(c,t)\ c \rightarrow e^{-A M_N t}$, where $\rightarrow$ denotes proportionality at large $t$ neglecting multi-hadron interactions\footnote{Partial quenching effects arising from integrating out the quark fields before taking moments are also neglected in these scaling estimates.}, while Parisi-Lepage scaling predicts that the second moment scales as $\left< C_A(t)^2 \right> = \int dc\ \mathcal{P}_A(c,t)\ c^2 \rightarrow e^{-3 A m_\pi t}$.
This analysis has been generalized to higher moments of correlation functions~\cite{Savage:2010misc,Beane:2014oea}, with the result that even and odd moments scale differently as $\left< C_{A}(t)^{{2n}} \right>  \rightarrow e^{-3An m_\pi t}$ and $\left< C_{A}(t)^{2n+1} \right> \rightarrow e^{-A M_N t}e^{-3An m_\pi t}$ respectively, where $n\in \mathbb{N}$. 
This implies that the distributions of the real and imaginary parts of $C_{A}$, which is complex evaluated on a generic background gauge field (even though $\left< C_A \right>$ is real),  become increasingly broad and symmetric for large values of $A t$.
The distributions of the real parts of nucleon correlation functions are observed to be heavy-tailed and consistent with a Cauchy (Lorentzian) distribution at large source/sink separations~\cite{davidkaplanLuschertalk}.
Robust estimators may therefore prove useful for reliably determining nuclear correlation functions at large source/sink time separations, as discussed in Ref.~\cite{Beane:2014oea}.

The symmetric, heavy-tailed distributions of baryon correlation functions differ from the corresponding distributions of zero-momentum pion correlation functions, which are approximately log-normally distributed at large $t$~\cite{Hamber:1983vu,Guagnelli:1990jb}.
\begin{figure}[t!]
	\centering
	\includegraphics[width=0.32\textwidth]{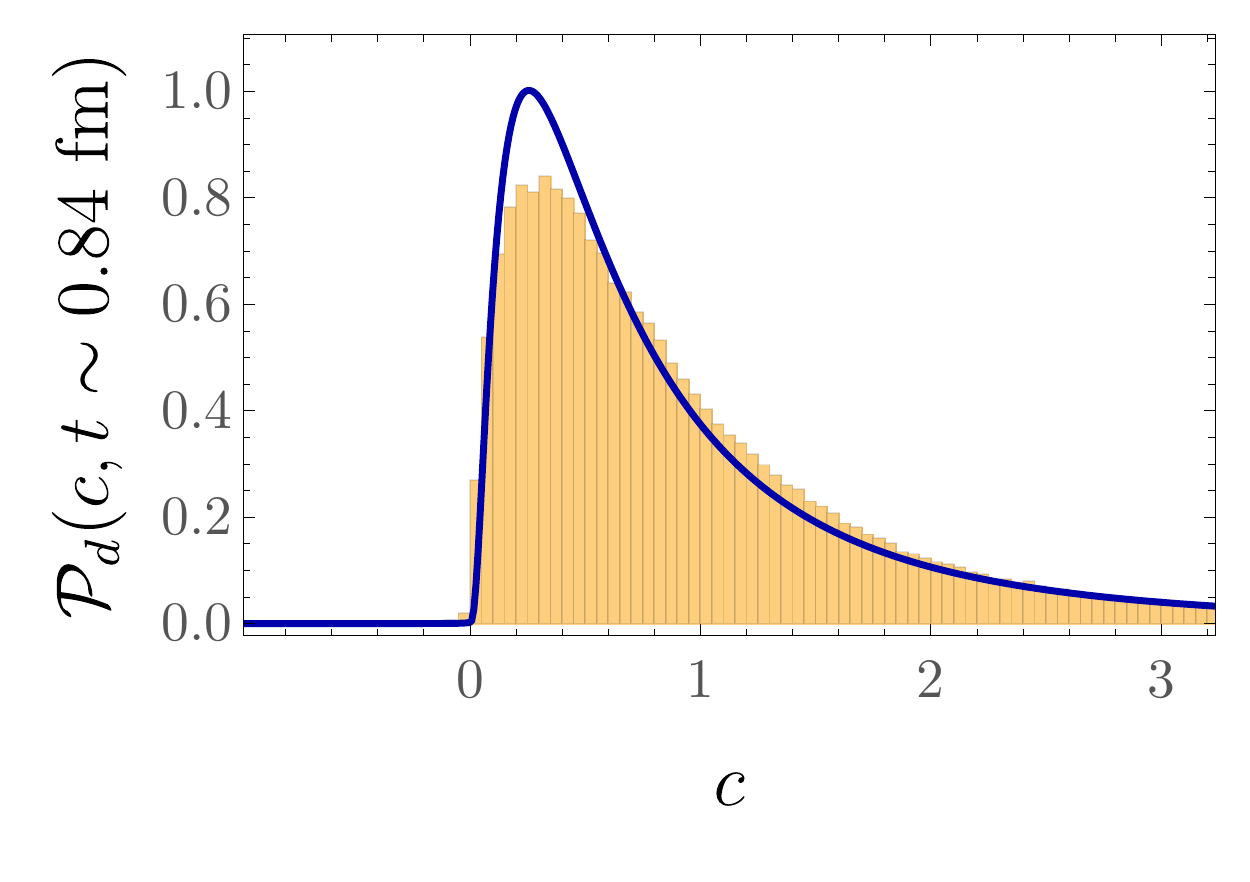}\hspace{5pt}
	\includegraphics[width=0.32\textwidth]{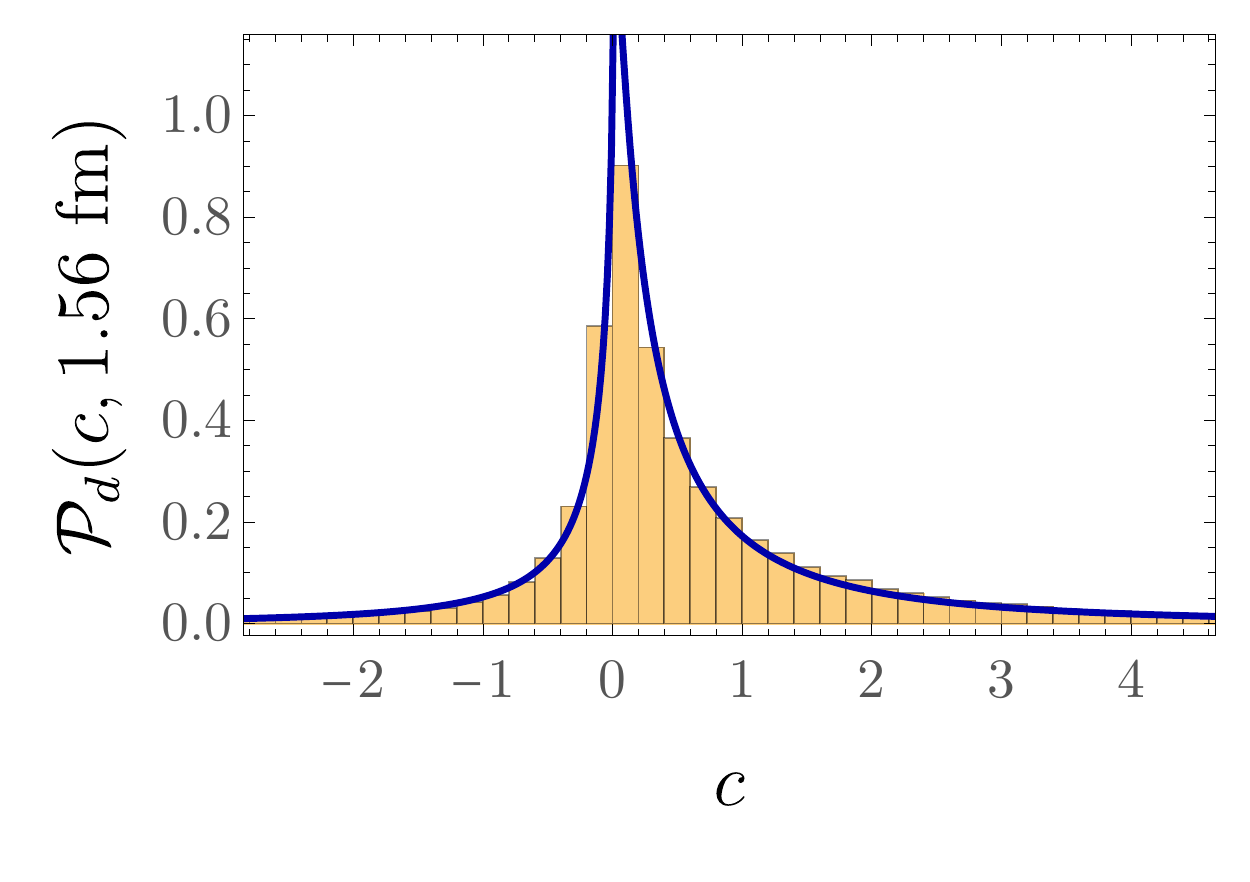}\hspace{5pt}
	\includegraphics[width=0.32\textwidth]{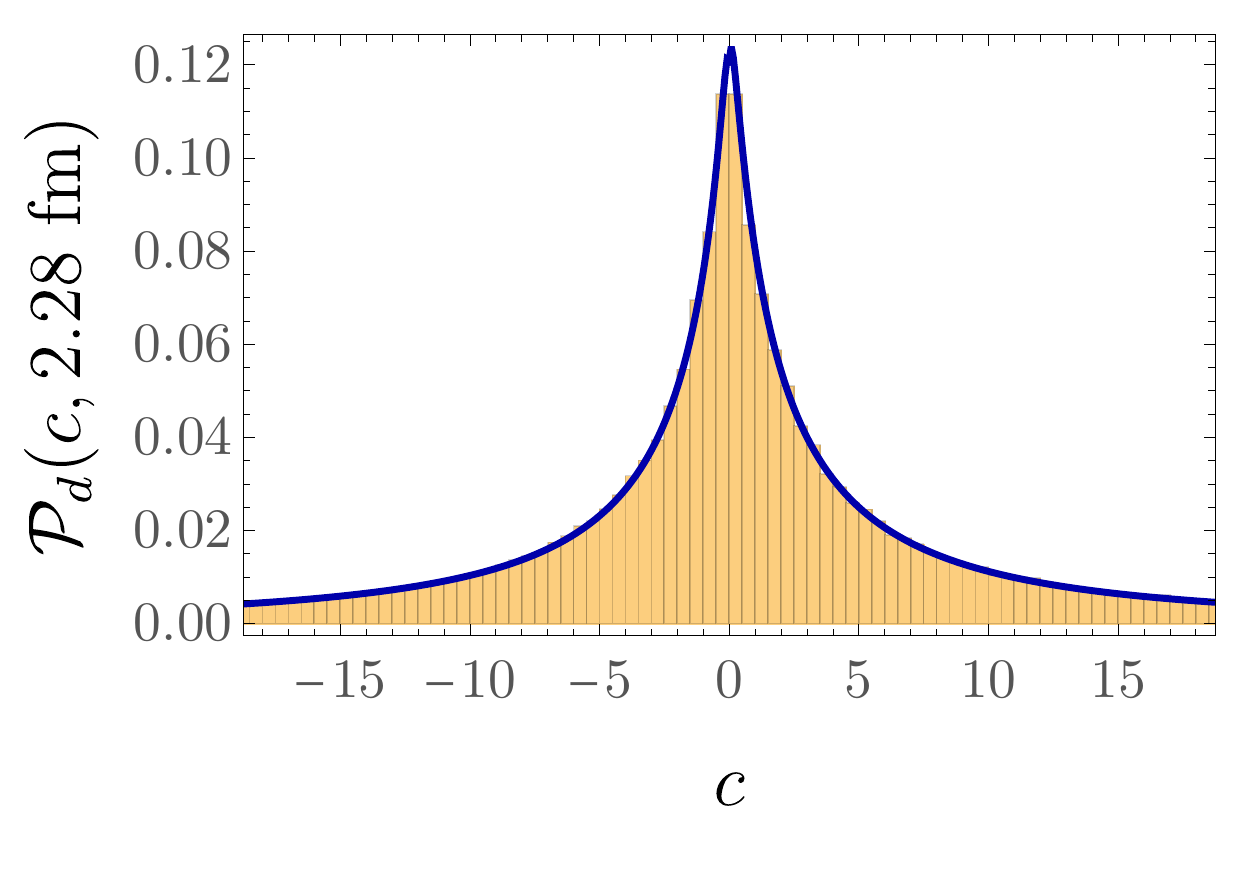}
	\caption{\label{fig:histograms}   Histograms of the real parts of deuteron correlation functions with $m_\pi = 450$ MeV from Ref.~\cite{Orginos:2015aya}, at three values of $t$  normalized independently by multiplying by $1/\left< \text{Re}\ C^d(t)\right>$. The blue curves show fits to complex log-normal distributions obtained as in Refs.~\cite{Wagman:2016bam,Wagman:2017jva} from a product of a log-normal distribution with mean $\mu_R = \left< r_d \right>$ and variance $\sigma_R^2 = \left< r_d^2 \right> - \mu_R^2$ times a wrapped-normal distribution with zero mean and variance $\sigma_\theta^2 = - \ln\left( \left< \cos\theta_d \right>^2 + \left< \sin\theta_d \right>^2 \right)$ obtained using sample mean estimators for $\mu_R$, $\sigma_R^2$, and $\sigma_\theta^2$ from LQCD results for $C_d = e^{r_d + i\theta_d}$. The distribution of the real part shown is obtained by marginalizing over the imaginary part of the resulting distribution for the complex variable $C_d$.}
\end{figure}
Log-normal distributions also describe correlation functions in theories of non-relativistic fermions~\cite{Endres:2011jm,Endres:2011er,Endres:2012cw,Drut:2015uua,Porter:2016vry} and the real parts of many-hadron correlation functions at small $t$~\cite{DeGrand:2012ik}.
For small $t$, both the log-magnitudes and phases of baryon correlation functions are approximately normally distributed; however, the width of the phase distribution grows with $t$, and for large $t$ the phase distribution approaches a uniform distribution on $(-\pi, \pi]$ and $\text{Re}[e^{i\theta_{h,A}}]$ is therefore approximately Cauchy distributed.
As shown in Fig.~\ref{fig:histograms}, a complex log-normal distribution obtained from the product of a normally distributed $R_{h,A}$ and a wrapped normal distribution (a sum over $2\pi$-periodic images of a normal distribution) for $\theta_{h,A}$  describes the real parts of multi-baryon correlation functions for a wide range of $t$~\cite{Wagman:2016bam}.

The role of complex-phase fluctuations in StN problems can be clearly seen from expectation values of the magnitudes and phase factors of nuclear correlation functions~\cite{Wagman:2016bam}.
Ensemble averages of magnitudes of nuclear correlation functions are observed to scale with $t$ as $\left< |C_A(t)| \right>\rightarrow e^{-\frac{3}{2}A m_\pi t}$ analogously to even moments of $C_{A}$, while ensemble-averaged nuclear correlation-function phase factors are observed to scale  with $t$ as $\left< e^{i \text{Arg}[C_A(t)]} \right> \rightarrow e^{-A (M_N - \frac{3}{2}m_\pi) t}$~\cite{Wagman:2016bam,Wagman:2017gqi}. 
Since $|e^{i \text{Arg}[C_{A}] }| = 1$ and $\text{Var}[e^{i \text{Arg}[C_{A}] }]$ is therefore ${\cal O}(1)$ for all $t$, this implies that the average phase factor has an exponential StN problem with the same severity as the full nuclear correlation function.
The magnitude and phase contributions to the effective mass are also seen to plateau much more slowly than the full correlation function.
The region of $t$ in which the correlation function is consistent with ground-state saturation, but the average phase factor has not yet reached the asymptotic value of $A(M_N - \frac{3}{2}m_\pi)$, corresponds to the golden window in which the StN degrades  slower than predicted by Parisi-Lepage scaling.

The existence of gauge-field--dependent phase fluctuations of $C_A(t)$ leads to a ``sign problem'' in the path integral in Eq.~\eqref{eq:2pt} defining $\left< C_{A}(t) \right>$: the full path integrand is not positive-definite and therefore cannot be interpreted as a probability distribution~\cite{Wagman:2016bam}.
Sign problems notoriously occur for partition functions of theories with complex actions, such as QCD with non-zero baryon chemical potential or with a CP-violating $\theta$ term and theories in Minkowski spacetime, and indicate that exponential increases in statistics are needed to achieve polynomial reduction in uncertainty (see Ref.~\cite{deForcrand:2010ys} for a review).
Although the occurrence of sign problems in path integrals defining observables does not obstruct standard Monte Carlo importance sampling strategies, the connection between phase fluctuations and StN problems suggests that improving one problem should improve the other.
Phase reweighting techniques~\cite{Wagman:2016bam,Wagman:2017gqi} similar to constrained path methods in nonrelativistic quantum Monte Carlo calculations~\cite{Zhang:1995zz,Wiringa:2000gb}, as well as phase unwrapping techniques combined with a cumulant expansion~\cite{Detmold:2018eqd} analogous to methods applied to QCD with non-zero baryon chemical potential~\cite{Ejiri:2007ga}, demonstrate that reducing correlation function phase fluctuations leads to exponential StN improvement but introduces additional systematic uncertainties.
An approach to reducing phase fluctuations without introducing additional systematic uncertainties has been introduced in Ref.~\cite{Detmold:2020ncp}, where integration-contour deformation techniques previously applied to improve sign problems in theories with complex actions~\cite{Cristoforetti:2012su,Alexandru:2018ddf,Alexandru:2018fqp} are used to construct ``deformed observables''. These observables have identical expectation values to their undeformed counterparts by Cauchy's theorem, but have deformation-dependent variance that can be exponentially reduced by optimizing the choice of deformation.
Future studies will explore whether contour deformations and other methods for improving sign problems can be used to improve the StN problems of nuclear correlation functions and matrix elements.

The spacetime structure of correlation functions has also been recently investigated and leveraged to propose new methods of improving StN problems.
Building off observations of the local coherence of the Dirac operator~\cite{Luscher:2007se}, hadron correlation functions have been shown to approximately factorize into products of correlation functions in which quark propagators only have support on a lattice subvolume~\cite{Ce:2016idq}.
The StN ratio of the correlation-function factor associated with each subvolume scales with the temporal extent of the subvolume rather than the full temporal extent and is, therefore, exponentially larger than the StN ratio of the full correlation function.
Multilevel integration algorithms have been developed, in which path integrals over subvolumes are performed and subsequently products of the subvolume results are averaged over the remaining degrees of freedom to construct correlation functions. This approach has been used to exponentially improve StN ratios for exactly factorizable observables in Yang-Mills theory~\cite{Luscher:2001up,Meyer:2002cd,DellaMorte:2007zz,DellaMorte:2008jd}.
Using this approximate factorization of quark propagators, and a similar approximate factorization of quark determinants, multilevel algorithms have been shown to exponentially improve nucleon correlation-function StN ratios~\cite{Ce:2016idq,Ce:2016ajy,Ce:2019yds}.
Applying multilevel integration to nuclear correlation functions is complicated by the presence of corrections to approximate quark-propagator factorization that must be accounted for in a suitably generalized nuclear contraction algorithm. If these challenges can be overcome, however, multilevel integration could lead to exponential improvement of the nuclear StN problem.

Even without modifying Parisi-Lepage scaling, it is possible to make significant practical improvements to the precision of nuclear correlation-function and matrix-element calculations.
Ref.~\cite{Detmold:2014hla} presented an approach to StN optimization based on construction of a variational basis of correlation functions similar to that discussed in Sec.~\ref{subsec:fitting} below.
This method can minimize the overlap onto the variance ground state, extend the golden window, and improve the precision of correlation functions at fixed $t$ that can be achieved with fixed computational resources.
Methods to reduce the computational cost of calculating correlation functions using machine learning are also being explored~\cite{Yoon:2018krb,Zhang:2019qiq}, although technical challenges remain.
The computational cost of calculating nuclear correlation functions could also be reduced by implementing more efficient linear-system solvers or accelerating the Hybrid Monte Carlo algorithm used to generate gauge-field configurations, for example, using Fourier acceleration~\cite{Cossu:2017eys,Christ:2018net} and machine learning methods~\cite{Boyda:2020hsi,Kanwar:2020xzo,Rezende:2020hrd,Bluecher:2020kxq,Shanahan:2018vcv,Albergo:2019eim,Kades:2019wtd,Urban:2018tqv,Zhou:2018ill,Yoon:2018krb,Tanaka:2017niz}.
Future application of these algorithms could provide significant practical improvements in the precision of nuclear correlation functions and matrix elements achievable with fixed computational resources. 
Hardware developments could also address these challenges. For example, a quantum computer of sufficient capability and capacity would enable real-time evolution of quantum systems, implemented either by the intrinsic dynamics of an analog simulator or by the gate-set of a digital quantum computer. This approach would not rely on Monte-Carlo importance sampling and therefore would not suffer from some of the bottlenecks encountered in present-day LQCD calculations. However, quantum simulation of QCD is not yet developed and may face other significant challenges. The first scientific and technological developments in this direction are now being made~\cite{Jordan:2011ne, Jordan:2011ci, Jordan:2014tma, Martinez:2016yna, Dumitrescu:2018njn, Klco:2018kyo, Lu:2018pjk, Bhattacharyya:2018bbv, Raychowdhury:2018osk, Klco:2019xro, Bauer:2019qxa, Klco:2019evd, Davoudi:2019bhy, Avkhadiev:2019niu, Klco:2019yrb, Lamm:2019uyc, Mueller:2019qqj, Lamm:2019bik, Kharzeev:2020kgc, Chakraborty:2020uhf, Ciavarella:2020vqm, Briceno:2020rar, Liu:2020eoa, Kreshchuk:2020dla, Klco:2020aud, Banuls:2019bmf}.

\subsubsection{Excited-state contamination} 
\label{subsec:fitting}
\begin{figure}
	\centering
	\includegraphics[width=0.7\columnwidth]{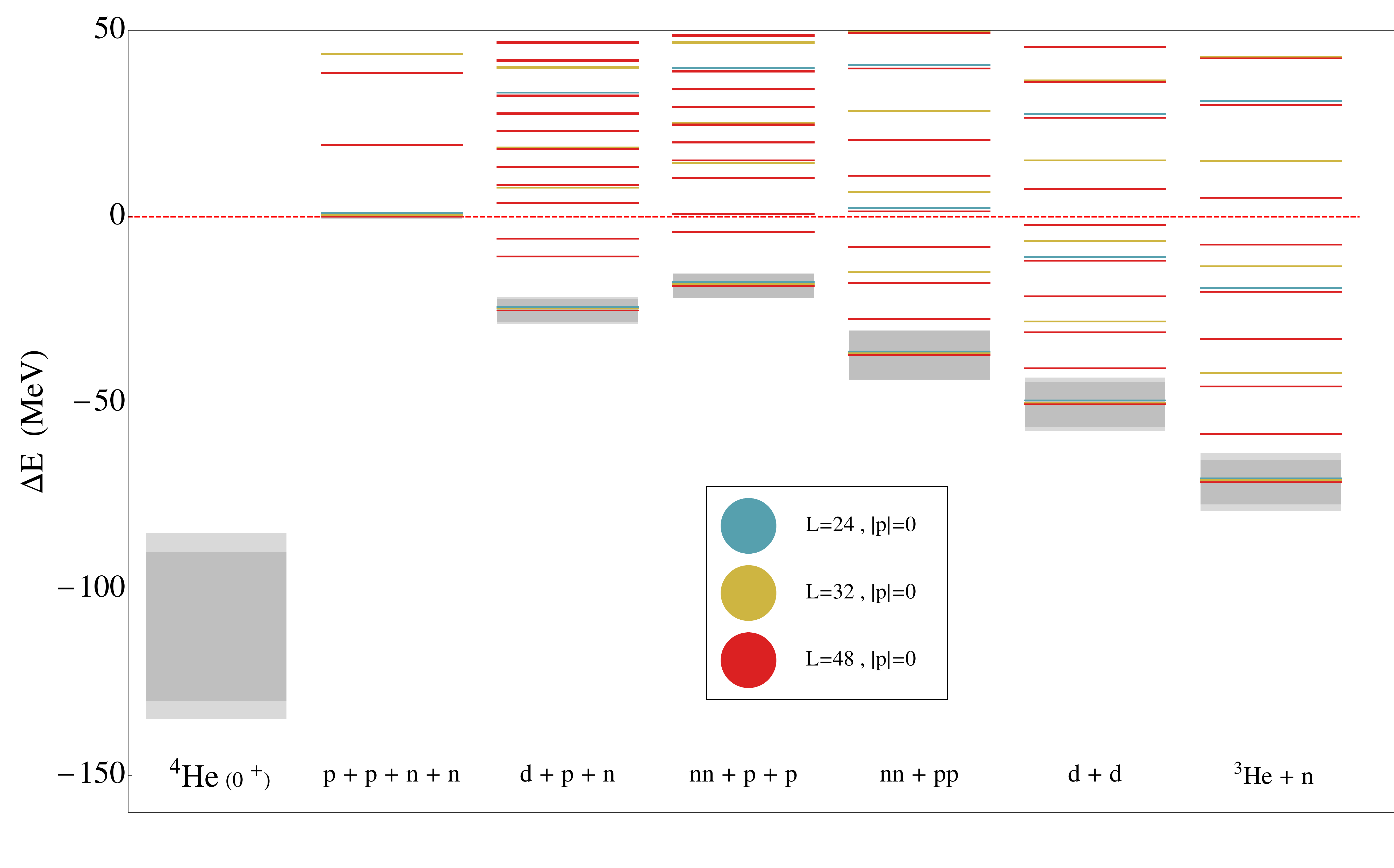}
	\caption{The expected excited-state level-spectra for correlation functions with the quantum numbers of $^4$He in three different volumes at $m_\pi=806$ MeV. The gray bands correspond to the lowest energy in each of the channels labeled at the bottom of the figure, computed as the sum of the energies of the individual components of the given channel. The colored lines correspond to the non-interacting energy levels that follow from these thresholds. [Figure from Ref.~\cite{Beane:2012vq}.]}
	\label{fig:dense}
\end{figure}

As the atomic number of a nucleus increases, its excitation spectra typically become more finely spaced. Furthermore, when scattering states are considered, which are present in many reaction processes, physical amplitudes can only be accessed in LQCD from a discrete finite-volume spectrum with L\"uscher's method and its generalizations. However, the density of states above the elastic threshold increases quickly as the lattice volume increases \cite{Beane:2012vq}. For example, the energy gap between the bound ground state (when present) and the first excited non-bound finite-volume state exponentially approaches the infinite-volume binding energy as the volume increases. On the other hand, the energy gap between the excited non-bound states approaches zero polynomially in inverse powers of the volume. An example of the expected level spectra of $^4$He at $m_\pi=806$~MeV is shown in Fig.~\ref{fig:dense}.
The closeness of states in the spectrum can potentially lead to the misidentification of states due to cancellations between exponential contributions from nearby states for non-symmetric correlation functions. For correlation functions with the quantum numbers of nuclei, exponential StN degradation limits the range of source/sink temporal separations that can be used in fits to the lowest-lying energies, hence making the issue of excited-state contamination particularly relevant. In fact, given the resource requirements of nuclear LQCD studies and the features outlined above, some LQCD results in two-nucleon systems studied to date remain inconsistent in their conclusions regarding the presence of bound states at large values of the quark masses~\cite{Beane:2012vq,Orginos:2015aya,Yamazaki:2012hi,Yamazaki:2015asa,Francis:2018qch,Aoki:2012tk,HALQCD:2012aa}. This discrepancy is seen primarily between two classes of studies employing L\"uscher's method~\cite{Luscher:1986pf} on one hand, and the potential method~\cite{Ishii:2006ec,Aoki:2008hh,Aoki:2011gt,HALQCD:2012aa,Gongyo:2017fjb,Miyamoto:2017tjs,Iritani:2018sra} on the other. Two origins have been proposed in the literature to account for the lack of agreement between these studies:
\begin{itemize}
	\item[]{i) L\"uscher's method, as discussed in Sec.~\ref{sec:EFT}, provides a model-independent mapping between finite-volume energy eigenvalues of two-hadron systems and the physical scattering amplitudes at those energies, as long as exponentially small corrections suppressed by $\sim e^{-m_{\pi}L}$ are negligible. Nonetheless, the inputs to L\"uscher's mapping are the energy eigenvalues extracted from a LQCD computation, which are subject to statistical and systematic uncertainties. As discussed above, a significant source of systematic uncertainty in nuclear correlation functions is the contribution from excited states at early Euclidean times when the signal is not yet overwhelmed by the noise at late times. In Refs.~\cite{Iritani:2017wvu,Iritani:2017rlk}, it is argued that all LQCD studies of two-nucleon systems at the time of those publications suffer from the false identification of plateaus in effective masses formed from two-nucleon two-point correlation functions. The argument is that the two-nucleon elastic excitation gaps at the physical values of the quark masses are of the order of a few MeV in currently-accessible lattice volumes, to be compared with a typical gap in nucleon inelastic excitations that is $\mathcal{O}(\Lambda_{\rm QCD})$. This means that the correlation functions will be dominated by the ground state only at very large Euclidean times beyond those accessible to current studies. The example of  two different interpolating operator structures, wall and smeared sources, was examined to provide evidence for source dependence of the plateaus observed, indicating a ``fake''/``mirage''-plateau problem. This criticism was rebutted in Refs.~\cite{Beane:2017edf,Wagman:2017tmp,Davoudi:2017ddj} where it was shown that optimized operators with large overlap onto the states studied can suppress excited-state contamination substantially, effectively providing a golden window at earlier times in which energies can be extracted before the statistical noise dominates (see also discussions in Sec.~\ref{sec:StN}). It was further shown in Ref.~\cite{Yamazaki:2017jfh} that results obtained with wall and smeared sources do, in fact, agree, in a high-precision example studied in that reference. Once the slow approach of the single-nucleon correlation function to its ground state is taken into account for the wall sources (when taking the ratio of interacting and non-interacting two-baryon correlation functions to extract ground-state energy shifts), consistent energies are obtained, albeit with much larger statistical uncertainties for the wall sources. Indeed, it was shown that the naive assignment of a plateau to the effective ratio in the case of wall sources gives rise to a volume dependence that follows that expected for scattering states, signaling significant excited-state contamination, while the energies obtained from the smeared sources follow an exponential volume dependence, signaling the presence of a bound state in the infinite-volume limit. Finally, in Refs.~\cite{Beane:2017edf,Wagman:2017tmp} the negligibly-small volume dependence of two-nucleon correlation functions at larger values of the quark masses was presented as crucial evidence for the existence of a bound state. It was argued that it is highly unlikely that intricate cancellations are in play between multiple exponential terms with nearly-equal energies, such that the net contribution conspires to create a single-exponential form, as such cancellation should work out identically for lattice volumes that are substantially different in size, and therefore have substantially different energies for scattering states. While the criticisms of Refs.~\cite{Iritani:2017wvu,Iritani:2017rlk} might not affect a number of LQCD studies for the reasons outlined, they strongly motivate the development of increasingly more reliable energy determinations from LQCD correlation functions in future studies, as will be reviewed below.
	}
	\item[]{ii) In the HAL QCD potential method~\cite{Ishii:2006ec,Aoki:2008hh,Aoki:2011gt,HALQCD:2012aa,Gongyo:2017fjb,Miyamoto:2017tjs,Iritani:2018sra}, energy-dependent but non-local two-baryon potentials are expressed in a derivative expansion. The first few terms in the expansion are used to form a truncated potential that is used in a Lippmann-Schwinger equation to solve for the scattering parameters in the infinite volume. The potential is derived from a Bethe-Salpeter wavefunction obtained from LQCD two-point functions. The result of this procedure provides a prediction of QCD only at the eigenenergies of the two-point functions, and in general may systematically differ from the physical scattering amplitudes at other energies. Despite recent studies attempting to quantify and control systematic uncertainties in the potential method~\cite{Iritani:2018zbt,Miyamoto:2019jjc}, major theoretical drawbacks~\cite{Beane:2010em,Walker-Loud:2014iea,Yamazaki:2017gjl} of the approach remain unsettled. In particular, as pointed out in Refs.~\cite{Beane:2010em,Walker-Loud:2014iea}, and more thoroughly argued in Refs.~\cite{Yamazaki:2017gjl,Yamazaki:2018qut}, the potential in quantum field theory is momentum dependent, and the physical scattering phase shifts only agree with the solution of the Lippmann-Schwinger equation at the corresponding energy of the potential. A derivative expansion of the potential, in particular, does not provide an expansion in momentum but in velocity, and the coefficients of this expansion are still momentum dependent. Solving for the phase shifts at all values of momenta from a potential that is only valid at the corresponding momenta of the finite-volume eigenstates could lead to uncontrolled systematic uncertainties. This also means that the derivative expansion does not necessarily provide a systematic expansion in a small parameter, and demonstrating the suppression of a higher-order contribution in the expansion is not sufficient to establish the suppression of other higher-order terms. The other closely-related issue is the dependence of the potential, particularly at short distances, on the interpolating operators used to extract the Bethe-Salpeter wavefunctions from LQCD two-point functions. Again this fundamental issue means that the operator independence of extracted amplitudes must be checked on a case-by-case basis and cannot be established {\it a priori}. Furthermore, such a potential may not be reliable for use in studies of dense nuclear systems with strong sensitivity to short-distance physics~\cite{Haidenbauer:2019utu}. Finally, the time-dependent potential method~\cite{HALQCD:2012aa} has been argued to be sensitive only to the nucleon's inelastic excitation gaps, hence only requiring correlation functions at much smaller Euclidean times than in other methods. However, as argued in Ref.~\cite{Yamazaki:2018qut}, a direct consequence of the momentum dependence of the true potential is to invalidate this statement. This puts the same requirement for ground-state saturation on the potential method that is demanded in the energy extractions required for approaches based on L\"uscher's method. Additionally, the time-dependent method requires the signal region to be free of contamination from states above the inelastic threshold, a criterion that cannot be mathematically demonstrated with a finite set of data at discrete times. 
	}
\end{itemize}
The arguments in favor of, and against, the points outlined continue~\cite{Aoki:2017yru,Drischler:2019xuo}, and no consensus has been reached to date.
Despite the current disagreements, it is important to note that as the computational resources dedicated to LQCD studies of nuclei increase toward the exascale computing era, there is in principle no impediment to calculating the spectrum and interactions of few-nucleon systems at lighter values of the quark masses in the upcoming years, using methods based solely in QCD. 

In the remainder of this section, the computational and analysis strategies developed in recent years to reliably extract the lowest-lying energy spectra from LQCD correlation functions are reviewed in more detail. As shown in Eq.~\eqref{eq:spectral_decomp}, Euclidean two-point correlation functions are guaranteed to have a spectral representation as a sum of exponentials. However, it is not possible to invert this relation exactly and obtain the full energy spectrum from finite-precision correlation functions determined over a finite range of $t$. In practice, it is necessary to fit correlation functions to a truncated spectral representation including the ground state and possibly a few excited states, and to select the  range of discrete $t$ to include in the fit. These, and other choices that must be made during fitting, lead to systematic uncertainties on the spectral results that are extracted. The concern regarding misidentification of the ground state can be ameliorated by the use of multiple different combinations of source and sink interpolating operators for a given set of quantum numbers, treated either independently, in a correlated manner, or used to build a variational basis as discussed below.

In order to precisely calculate nuclear binding energies and multi-hadron energy shifts, correlations between single- and multi-hadron  correlation functions can be exploited.
Fluctuations of gauge fields lead to correlated fluctuations of $\left< \chi_{h}(t) \bar{\chi}_{h}(0) \right>$ for different states $h$.
In particular, fluctuations of nuclear correlation functions are correlated with fluctuations of single-nucleon correlation functions, and correlated ratios:
\begin{equation}
\begin{split}
\mathcal{R}_{A}(t) = \frac{ \left< \chi_{A}(t) \bar{\chi}_{A}(0) \right> }{\left< \chi_{N}(t) \bar{\chi}_{N}(0) \right>^A }
\end{split}\label{eq:ratios}
\end{equation}
can often be determined significantly more precisely than $\left< \chi_{A}(t) \bar{\chi}_{A}(0) \right>$.
In the $t\rightarrow \infty$ limit when both the numerator and denominator are dominated by the ground-state contribution, $\mathcal{R}_{A} \sim e^{-\Delta t}$, where $\Delta =M_A-A M_N $ is the difference between the finite-volume energy of an $A$-nucleon system and $A$ times the nucleon mass. 
Fitting $\mathcal{R}_{A}(t)$ to a single-exponential form, or equivalently fitting $\frac{1}{a}\ln\left[ \mathcal{R}_A(t)/ \mathcal{R}_{A}(t+a)\right]$ to a constant, allows $\Delta$ to be determined more precisely than from an uncorrelated analysis of the correlation functions in the numerator and denominator of Eq.~\eqref{eq:ratios}. 
Both the numerator and denominator of Eq.~\eqref{eq:ratios} are contaminated by excited states for any finite $t$, and these excited-state contributions may partially cancel in ratios.
This means that $\mathcal{R}_{A}(t)$ can appear to be dominated by ground-state contributions even for small $t$ where excited-state effects on the individual correlation functions in Eq.~\eqref{eq:ratios} are significantly larger than statistical uncertainties on $\mathcal{R}_{A}$.\footnote{When the sources and sinks used in Eq.~(\ref{eq:ratios}) are symmetric, the numerator and denominator are separately convex, however their ratio does not need to be because of potential cancellations.}
Single-state fits to $\mathcal{R}_{A}(t)$ should only be performed at large enough $t$ that excited-state contamination is negligible in both the numerator and denominator of Eq.~\eqref{eq:ratios}.
Alternatively, single- or multi-state fits can be used to extract the ground-state energies of both correlation functions in the ratio separately, and a correlated difference between the resulting ground-state energies can be used to extract $\Delta$.
This strategy is advantageous because the individual fit functions can include all excited states that make resolvable contributions to either correlation function in the ratio, while statistical fluctuations are still analyzed in a correlated manner using jackknife or bootstrap resampling in order to improve the precision of  determinations of $\Delta$.

Rather than performing combined fits to correlation functions with multiple interpolating operator choices, it is also possible to build linear combinations of correlation functions with different interpolating operators that are optimized to maximize overlap onto a particular state of interest.
The Prony~\cite{prony,Fleming:2004hs,Fischer:2020bgv} and Matrix-Prony methods~\cite{Beane:2009kya,Fischer:2020bgv} construct optimized correlation functions from a vector of correlation functions with different sink operators.
With an $m\times m$ Hermitian matrix of different source and sink operators, it is possible to obtain up to $m$ QCD energy levels using the Generalized Eigenvalue Problem (GEVP) method~\cite{Michael:1982gb,Michael:1985ne,Luscher:1990ck,Blossier:2009kd,Schiel:2015kwa}.
Important steps in this direction have been made in Ref.~\cite{Francis:2018qch}, where correlation functions for the $H$-dibaryon were constructed from products of momentum-projected baryon operators at both the source and the sink, and GEVP methods were applied to the resulting matrices of correlation functions.
For the success of GEVP methods, it is essential that a set of interpolating operators with statistically-significant overlap onto all states in the spectrum, below a given energy, can be found. In future calculations, large operator sets that include operators overlapping strongly with both bound and scattering states will be needed in order to disentangle the dense spectra of low-lying states in analyses of nuclear correlation functions using GEVP methods. 

With a view to the future, possible applications of quantum computing to this challenge are also being investigated. For example, approaches to construct optimized interpolating operators using hybrid quantum-classical algorithms have been developed~\cite{Avkhadiev:2019niu}. Eventually, reliable large-scale quantum computers may provide an independent path to addressing the effects of excited states in lattice field theory calculations by providing direct access to S-matrix elements through real-time evolution, circumventing the challenges of identifying the ground and lowest-lying excited state contributions to an imaginary-time correlation function entirely~\cite{Jordan:2011ne, Jordan:2011ci, Jordan:2014tma}. Nevertheless, quantum-resource requirements for initial-state preparation and final-state spectroscopy have not yet been investigated for strongly-interacting quantum field theories such as QCD.

\subsubsection{Correlation-function complexity}

Over the last few years, significant algorithmic improvements have accelerated both gauge-field generation and the computation of quark propagators, which have historically dominated the computational resource requirements of LQCD calculations. In particular, the development of algebraic multigrid algorithms for LQCD \cite{Babich:2010qb,Osborn:2010mb,Boyle:2014rwa,PhysRevD.92.114516, Clark:2016rdz,Yamaguchi:2016kop,Bacchio:2017pcp,Brower:2018ymy,Richtmann:2019eyj}, which utilize an efficient approximation to the Dirac operator recursively defined on coarser levels and thereby exploit the finite correlation length of QCD, have enabled algorithms which are more efficient than traditional Krylov methods, such as conjugate gradient, by orders of magnitude, particularly for light quark masses. As a result of these improvements, for calculations of even modest-sized nuclei, contracting quark propagators to assemble correlation functions is now a significant cost. 

Ref.~\cite{Detmold:2019fbk} presents an algorithm to accelerate the contraction of quark propagators into correlation functions, motivated by the goal of reducing the numerical cost of computing multi-hadron correlation functions for LQCD calculations of nuclear physics. In that work, it was demonstrated that forming correlation functions from sparsened propagators defined on a coarsened lattice geometry enables significant speedups in the contraction stage of LQCD calculations for particular types of two- and higher-point correlation functions, reducing the cost of this task.  Specifically, a simple blocking prescription, where correlation functions are constructed from a sparse propagator defined from the full propagator on a coarse grid of sites, was shown to preserve the low-energy spectrum in hadronic and nuclear systems. 
Sparsened baryon blocks are defined by
\begin{equation}
\begin{split}
\label{eq:1sparse}
{\cal B}_{b,\rm{sparse}}^{a_1,a_2,a_3}({\vec p},t;s_1,s_2,s_3)&= \sum_{\vec x \in \widetilde{\Lambda}(N_{\rm sparse}) }\!\!\!e^{i{\vec p}\cdot{\vec
		x}} 
\sum_{k=1}^{N_{B(b)}} \tilde{w}^{(c_1,c_2,c_3),k}_b 
\sum_{\vec i} \epsilon^{i_1,i_2,i_3}
\\
&\hspace*{1cm}\times
S(c_{i_1},{ x};a_1,x_0^{(s_1)})
S(c_{i_2},{ x};a_2,x_0^{(s_2)})
S(c_{i_3},{ x};a_3,x_0^{(s_3)}),
\end{split}
\end{equation}
where the sparse spatial lattice $\widetilde{\Lambda}(N_{\rm sparse})$ is defined by
\begin{equation}
\widetilde{\Lambda}(N_{\rm sparse}) = \{ a(\widetilde{n}_1, \widetilde{n}_2, \widetilde{n}_3) \, | \,0 \leq \widetilde{n}_i < L/a,\ \widetilde{n}_i \equiv 0 (\text{mod}\ N_{\rm sparse}) \}.
\end{equation}
Sparsened baryon blocks can be computed using $N_{\rm sparse}^3$ fewer operations than needed for the standard baryon blocks defined by Eq.~\eqref{eq:1}.
Correlation functions produced from contractions of sparsened baryon blocks are identified as correlation functions using modified interpolating operators
\begin{equation}
\begin{split}
C_{\rm 2pt,\,sparse}^h(t,\vec{p}) \equiv a^3\sum_{\vec{x}\in\widetilde{\Lambda}} e^{i \vec{p}\cdot\vec{x}} C_{\rm 2pt}(x,x_0).
\end{split}
\end{equation}
These sparsened correlation functions are only approximately projected to the center-of-mass momentum $\vec p$, and include additional contributions from states with momentum $\vec{p} + \frac{2\pi N_{\rm sparse}}{L }(n_1,n_2,n_3)$ with $n_i \in \mathbb{Z}$.
These additional contributions alter the excited-state structure of correlation functions 
at small $t$, but do not modify the low-lying spectrum.
A simple numerically-inexpensive bias correction applied to the sparsened correlation functions can be applied to remove the modified excited-state effects~\cite{Detmold:2019fbk}.
The subvolumes over which sparsening can be effective depends on physical length scales, so as the continuum limit is approached, the improvement provided by this algorithm will increase.
Applying background-field methods in sparsened baryon blocks leads to background-field correlation functions that are only modified by the presence of additional excited-state effects and can be used to calculate nuclear matrix elements using the methods described in Sec.~\ref{subsec:backgroundfield}.

For larger numbers of baryons ($A>8$ protons and neutrons), it
is necessary to use multiple source locations because of the Pauli
exclusion principle. 
The block construction of Eq.~\eqref{eq:1} can be generalized to allow the quark propagators to originate from multiple different source locations, $\{x_0^{(1)},
x_0^{(2)},\ldots\}$, using
\begin{eqnarray}
\label{eq:6}
{\cal B}_b^{a_1,a_2,a_3}({\vec p},t;s_1,s_2,s_3)&=& \sum_{\vec x}e^{i{\vec p}\cdot{\vec
		x}} 
\sum_{k=1}^{N_{B(b)}} \tilde{w}^{(c_1,c_2,c_3),k}_b 
\sum_{\vec i} \epsilon^{i_1,i_2,i_3}
\\
&&\hspace*{1cm}\times
S(c_{i_1},{ x};a_1,x_0^{(s_1)})
S(c_{i_2},{ x};a_2,x_0^{(s_2)})
S(c_{i_3},{ x};a_3,x_0^{(s_3)}).
\nonumber
\end{eqnarray}
With these generalized blocks, the baryon-based algorithm discussed in Sec.~\ref{subsec:baryonblocks} allows for the construction of correlation functions for even very large nuclei, although the complicated spatial wavefunctions required because of the Pauli exclusion principle result in an exponential growth of complexity as $A$ increases.  

\begin{figure}[!t]
	\centering
	\includegraphics[width=6cm]{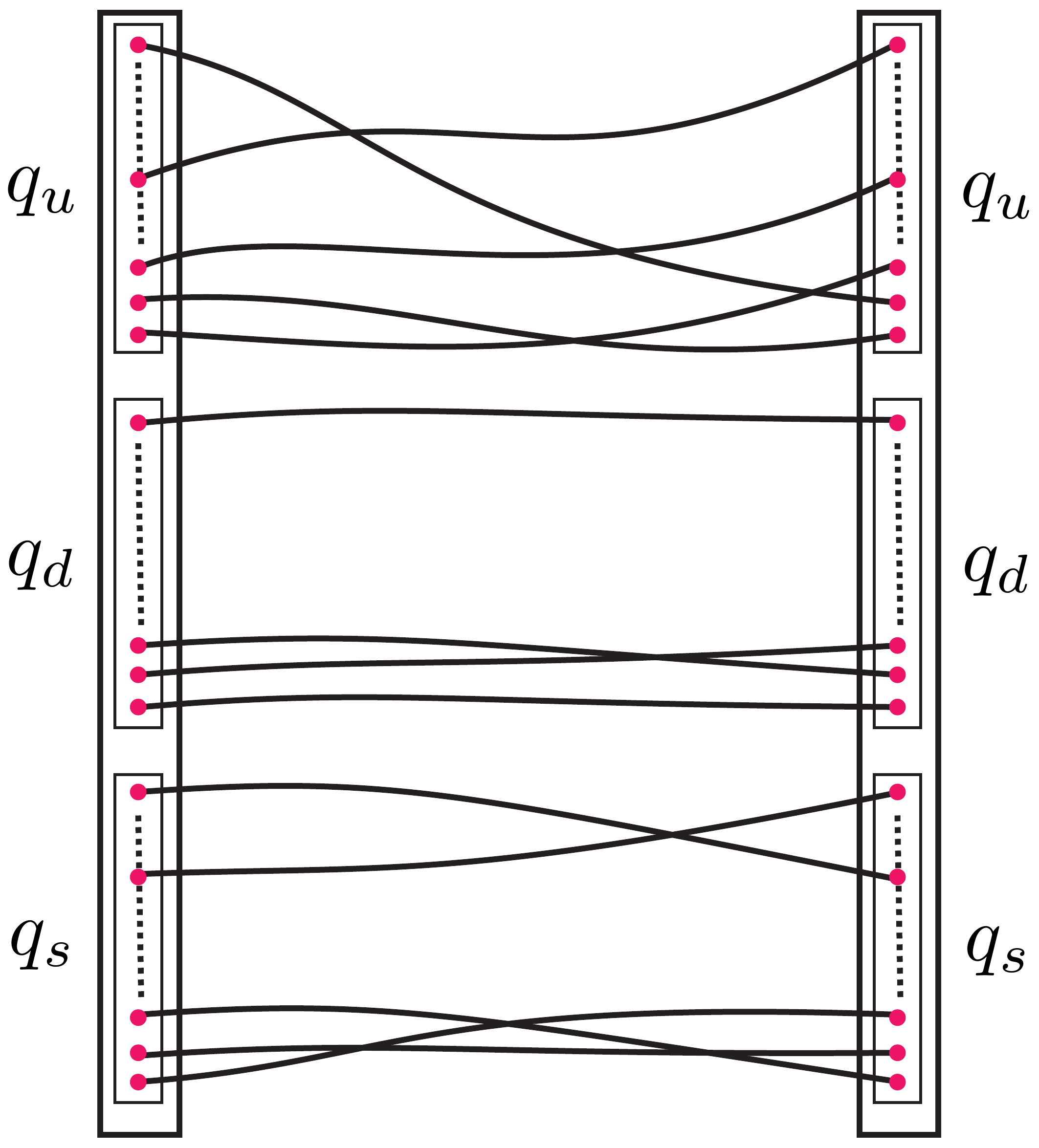}
	\caption{Illustration of the quark determinant-level contractions. The subblocks on each side (initial and final states), list the up, down, and strange
		quarks of a given pair of source and sink wavefunction terms in Eq.~\eqref{eq:det}. Within each block of a given flavor, all  permutations of contractions are performed by
		forming a determinant of the matrix of quark propagators as
		described in the text. [Figure modified from Ref.~\cite{Detmold:2012eu}.]}
	\label{fig:q-q}
\end{figure}

As shown in Ref.~\cite{Detmold:2012eu}, nuclear two-point correlation functions can be expressed as a determinant of a matrix $G$ whose matrix elements are constructed from the quark propagator $S$ as
\begin{equation}
G({\vec a}^\prime;{\vec a})_{j,i}= \left\{ \begin{array}{ll} S(a'_{j};a_{i}) & {\rm for } \;\; a'_j \in {\vec a}'\;\; {\rm and}\;\; a_i \in {\vec a} \\
\delta_{a'_j,a_i} & {\rm otherwise} \end{array}\right.,
\end{equation}
where, as before, ${\vec a}'=(a'_1,a'_2\ldots a'_{n_q})$ and ${\vec a}=(a_1,a_2\ldots a_{n_q})$. The non-trivial block of the matrix $ G({\vec a}^\prime;{\vec a})$ is of size $n_q\times n_q$, hence only this block is needed for computing its determinant.
Making use of this definition, the full nuclear correlation function can be written as
\begin{equation}
\langle{\chi}_{h_1}(t)\bar{\chi}_{h_2}(0)\rangle =  \int {\cal D} {U}\; {\cal P}( U)\;
\sum_{k'=1}^{N'_w}\sum_{k=1}^{N_w}
{\tilde{w}'_{h_1}}{}^{(a'_1,a'_2\cdots a'_{n_q}),k'}\;
\tilde{w}^{(a_1,a_2\cdots a_{n_q}),k}_{h_2} \times 
{\rm Det}{G({\vec a}';{\vec a})}\,.
\label{eq:det}
\end{equation}
Because of the flavor-blindness of the strong interaction, the matrix
$ G({\vec a}^\prime;{\vec a})$ is block diagonal in flavor space  resulting in  a product of smaller determinants, one for each flavor. This contraction approach is illustrated in Fig.~\ref{fig:q-q}.
Given the reduced weights determined above, and appropriate quark
propagators, the implementation of Eq.~(\ref{eq:det}) is very fast,
scaling polynomially with the number of terms in the source and sink
quark-level interpolating fields as well as the number of quarks per
flavor.
The total cost of this form of contractions naively scales as $n_u^3 n_d^3 n_s^3 \times N_w' N_w$, where $N_w'$, $N_w$, are the number of terms in the sink and source
quark interpolating fields respectively, and careful application of an algorithm such as LU decomposition is used to evaluate the determinant.\footnote{The expectation of polynomial scaling of 
	contractions was noted in Ref.~\cite{OFT}. However, the scaling of $N_w$ and $N_w'$  can grow exponentially with the atomic number $ A$.} Further improvements arise from using rank-1 updates (or higher-rank updates via the Woodbury formula) to relate determinants of similar matrices~\cite{Vachaspati:2014bda} and by caching; a large fraction of determinants can be evaluated with only 
${\cal O}(n_q)$ or ${\cal O}(n_q^2)$ cost in the number of quarks of a given flavor (although determining the optimal clustering that enables this is a challenging problem). As a result, if  interpolating fields with sufficiently small numbers of terms are found, correlation functions with a  large atomic number $A$ can be constructed; in Ref.~\cite{Detmold:2012eu}, explicit calculations are presented for systems as large as $A=28$.
It is possible that in the future the contraction problem for nuclei with larger $A$ could be addressed with novel algorithmic and hardware advances. For example, because of their ability to exploit superposition, in the future it is conceivable that the registers of quantum computers may be able to address the exponentially increasing number of nuclear contractions with only a polynomially increasing number of qubits.

\section{Electromagnetic interactions of light nuclei}
\label{sec:EM}

Some of the most fundamental properties of nuclei are encoded in electromagnetic (EM) matrix elements that describe the response of composite nuclear systems to external electromagnetic fields. 
These responses depend both on the currents induced by the fields and on the distribution of the quarks and gluons inside nuclei and thus provide important information about nuclear structure. 
For example, the  magnetic moments of $s$-shell nuclei are very well described by the phenomenological nuclear shell model~\cite{Caurier:2004gf}, which is based on the observation that, although nuclei are composed of quarks and gluons bound together by the strong force, many nuclear properties are qualitatively compatible with a simple model of nucleons interacting weakly in a mean-field potential. 
Despite shell-model descriptions being widely and successfully applied to a large variety of nuclei, this behavior is not yet understood at a fundamental level from QCD. LQCD calculations of electromagnetic properties can explain this phenomenon from QCD, reveal whether the same behavior holds at unphysical values of the quark masses, and investigate how these quantities evolve as the physical limit is approached. 

The electromagnetic structure of hadrons has been the target of LQCD investigations since the 1980s; the first LQCD computations of the nucleon's response to uniform magnetic fields were performed in the quenched approximation, starting with calculations of the nucleon magnetic moments~\cite{Bernard:1982yu,Martinelli:1982cb,Rubinstein:1995,Gadiyak:2001fe}, and more recently extending to other baryons in the baryon octet~\cite{Leinweber:1991} and decuplet~\cite{Lee:2005,PhysRevD.79.051502}. A series of subsequent calculations were able to extract not only magnetic moments, but also polarizabilities, for several members of the lowest-lying baryon and meson octets~\cite{FIEBIG198947,Christensen:2005,Lee:2005dq,Detmold:2009dx,Detmold:2010ts,Primer:2013pva,Lujan:2014kia,Freeman:2014kka,Parreno:2016fwu,Parreno:2017vbw,Bignell:2018,Bignell:2020,Engelhardt:2007ub,Luschevskaya:2016,Luschevskaya:2018chr}. Computations of EM properties involving more than one hadron have only been achieved recently, with the first calculation of the leading contribution to the magnetic-field response of $s$-shell nuclei being presented in Ref.~\cite{Beane:2014ora}, followed by a study of their magnetic polarizabilites in Ref.~\cite{Chang:2015qxa}. Extensions of these studies led to the first LQCD determination of the cross-section for the radiative radiative capture process $np \to d \gamma$~\cite{Beane:2015yha}, which enabled the isolation of subleading short-range modifications to the single-nucleon contributions to this process, and found consistency between LQCD and experimental measurements. 

This section will review the existing studies of nuclear responses to electromagnetic fields, all of which  were undertaken using two ensembles of gauge-field configurations generated using a L\"uscher-Weisz \cite{Luscher:1984xn} gauge action with clover fermions \cite{Sheikholeslami:1985ij}, the first at the SU(3)$_f$-symmetric point where $m_\pi = 806$ MeV~\cite{Beane:2012vq,Beane:2013br}, and the second with $N_f=2+1$ flavors corresponding to $m_\pi=450$ MeV~\cite{Orginos:2015aya}. For both ensembles, the gauge coupling is $\beta=6.1$, and the spacetime volume is $32^3\times 48$ for the SU(3)$_f$-symmetric ensemble and $32^3\times 96$ for the ensemble with $N_f=2+1$.
Despite the limited systematic control that can be achieved using calculations at a single  lattice spacing and lattice volume, and with larger-than-physical quark masses, phenomenologically-relevant results have already been obtained. Further impact can be expected given the controlled studies of the electromagnetic properties of nuclei which will be possible in the near future, as discussed in Sec.~\ref{subsec:EMfuture}. 

\subsection{Magnetic moments and polarizabilities}
\label{sec:fieldstrength}

The magnetic moments and polarizabilities of nuclei have been studied in LQCD using the background-field method described in Sec.~\ref{subsec:backgroundfield}. In this approach, spatially constant background magnetic fields in a given direction are constructed by multiplying the SU(3) gauge links of an ensemble by classical U(1) gauge links, i.e., $U_\mu(x) \to U_\mu(x) U_{\mu}^{\text{ext}}(x) \,$, where the form of $U_\mu^\text{ext}$ to create a magnetic field aligned along the $x_3$-direction is specified in Eq.~\eqref{eq:Bfield}.
Since this method incorporates the U(1) gauge links in the calculation after the gauge-field configurations have been generated,
the coupling to sea-quark degrees of freedom (and indirectly to gluons) through the fermionic determinant is missing.\footnote{Sea-quark contributions to the electric polarizability of hadrons were explored in Ref.~\cite{Freeman:2014kka} by means of a reweighting of sea-quark charges to allow them to couple to the background field, revealing important difficulties in the estimation of the reweighting factors due to the large stochastic noise.}
Nevertheless, there are situations in which sea-quark contributions exactly vanish. This is the case, for example, for SU(3)$_f$-symmetric calculations of the nuclear magnetic moments~\cite{Beane:2014ora}, and the $np \to d \gamma$ transition matrix element~\cite{Beane:2015yha}, where the sea-quark contributions, arising from expanding the fermionic determinant to linear order in the external field, are given by the product of a common mass factor (the quark mass is the same for all three quark flavors, $m_u = m_d = m_s$) and a charge factor, $\sum_f Q_f = Q_u + Q_d + Q_s$, which is exactly zero. With SU(3)$_f$-symmetry breaking, the sum of sea-quark current effects no longer vanishes because contributions from each flavor are no longer identical, and disconnected contributions generally appear. Nevertheless, in the isospin-symmetric case where $m_u=m_d$, the electromagnetic current can be decomposed into isoscalar and isovector contributions. Isovector quantities, such as the difference between proton and neutron magnetic moments and the $np \to d \gamma$ transition, are insensitive to disconnected contributions even away from the SU(3)$_f$-symmetric point.
Isoscalar quantities computed in this approach have a systematic bias from the missing sea-quark contributions,
although the omitted disconnected terms have been found to be small compared to the connected contribution in numerical studies of the EM structure of the nucleon~\cite{Abdel-Rehim:2015lha}.

In the absence of a background magnetic field, the energy eigenstates of a nuclear system are momentum 
eigenstates, and by choosing interpolating operators which project onto fixed three-momentum, one can extract  the ground-state energy of the system as discussed in Sec.~\ref{subsec:fitting}. To study the response of charged hadrons to a magnetic field, a natural projection is onto the lowest Landau level as explored in Refs.~\cite{Tiburzi:2012ks,Deshmukh:2018,Bignell:2018,Bignell:2020}. Such a projection enhances the overlap of the interpolating operators onto the state of interest whilst suppressing contributions from higher-energy states. This procedure has not yet been extended to nuclei; in this case, it is not clear how to combine Landau-projected proton blocks with momentum-projected neutron blocks to obtain operators with a better overlap onto nuclear Landau levels. A study of the quality of the overlap of nuclear interpolating operators onto Landau levels was undertaken in Ref.~\cite{Chang:2015qxa} (see Fig. 5 of that work). In that study, it was found that the ratio of overlap factors of different states at nonzero and zero background magnetic-field strengths is only weakly dependent on the field strength for neutral states, while for charged states it rapidly decreases with increasing magnetic-field strength, indicating that one must be cautious with the interpretation of extracted states. While it is clear that more effort needs to be invested in this direction to achieve complete systematic control of LQCD calculations of this type, the extractions of magnetic properties of light nuclei presented in Refs.~\cite{Beane:2014ora, Beane:2015yha, Chang:2015qxa} serve as benchmarks for future investigations. 

To be explicit, in a uniform background magnetic field in the $z$-direction ($x_3$-direction), i.e., $\vec{B}=B\hat{z}$, the energy eigenvalues of a hadron or a nucleus with spin $j\leq1$, polarized in the $z$-direction, and with magnetic quantum number $j_z$, can be expressed as
\begin{equation}
E_{h;j_z}(B) =
\sqrt{M_h^2 + P_\parallel^2 + (2 n_L +1) | Q_h e B|}
- \mu_h j_z B
- 2\pi \beta_h^{(M0)} | B|^2 
- 2\pi \beta^{(M2)}_h \langle j,j_z|\hat{T}_{33}|j,j_z \rangle B^2  + \ldots
\,. 
\label{eq:Eshift}
\end{equation}
Here, the ellipsis denotes terms that are higher order in the magnetic-field strength $B$, $M_h$ is the mass of the hadron or nucleus $h$, 
$P_\parallel$ is its momentum parallel to the magnetic field,
$Q_h$ is its charge in units of $e$,
and $n_L$ is the
quantum number of the Landau level that it occupies.  
When \mbox{$j\ge {1\over 2}$}, there is a contribution from the
magnetic moment, ${\vec \mu}_h = \mu_h \vec{j}$, that is linear in the magnetic field.
The scalar and tensor magnetic polarizabilities, $\beta_h\equiv\beta^{(M0)}_h$ and $\beta^{(M2)}_h$ respectively, contribute at 
${\cal O}({ B}^2)$, and $\hat T_{ij}  =  {1\over 2}\left[ \hat J_i \hat J_j +  \hat J_j \hat J_i - {2\over 3} \delta_{ij} \hat J^2 \right]
$ is a traceless symmetric tensor constructed from angular-momentum generators $\hat{J}_i$.

As can be inferred from the expansion in Eq.~(\ref{eq:Eshift}), at lowest order the difference in energy between the $j_z= \pm j$ states in a background magnetic field yields the magnetic moment. This quantity can be extracted  from ratios of correlation functions (given by Eq.~\eqref{eq:ratcorr}) with maximal spin projections $j_z= \pm j$~\cite{Primer:2013pva,Beane:2014ora}. 
The magnetic polarizabilities that govern the second-order response, on the other hand, can be obtained from spin-averaged
ratios where the leading magnetic moment contributions cancel~\cite{Chang:2015qxa}.

\begin{figure}[!t]
	\includegraphics[width=0.47\columnwidth]{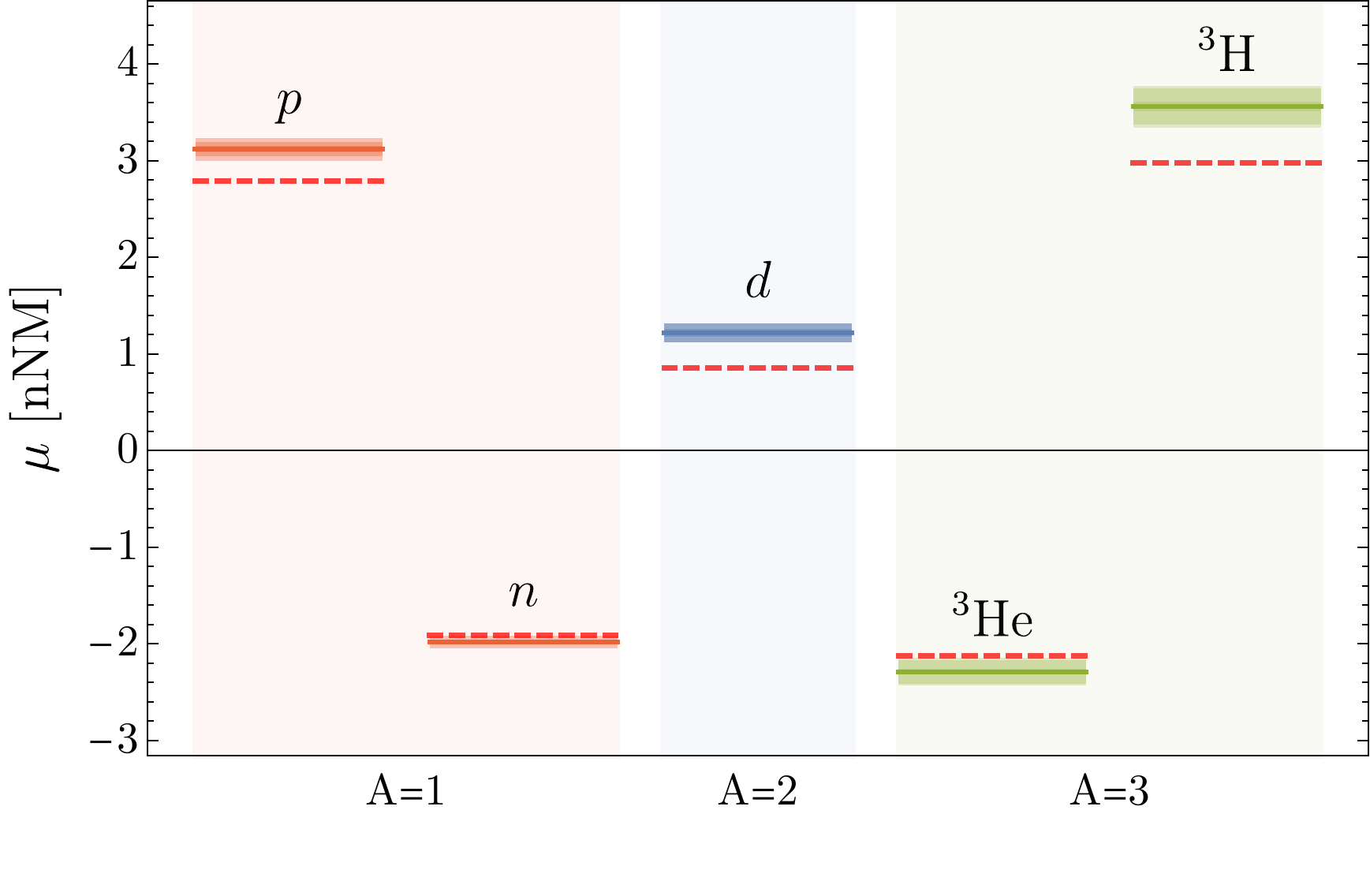}  \hspace*{0.5cm}
	\includegraphics[width=0.47\columnwidth]{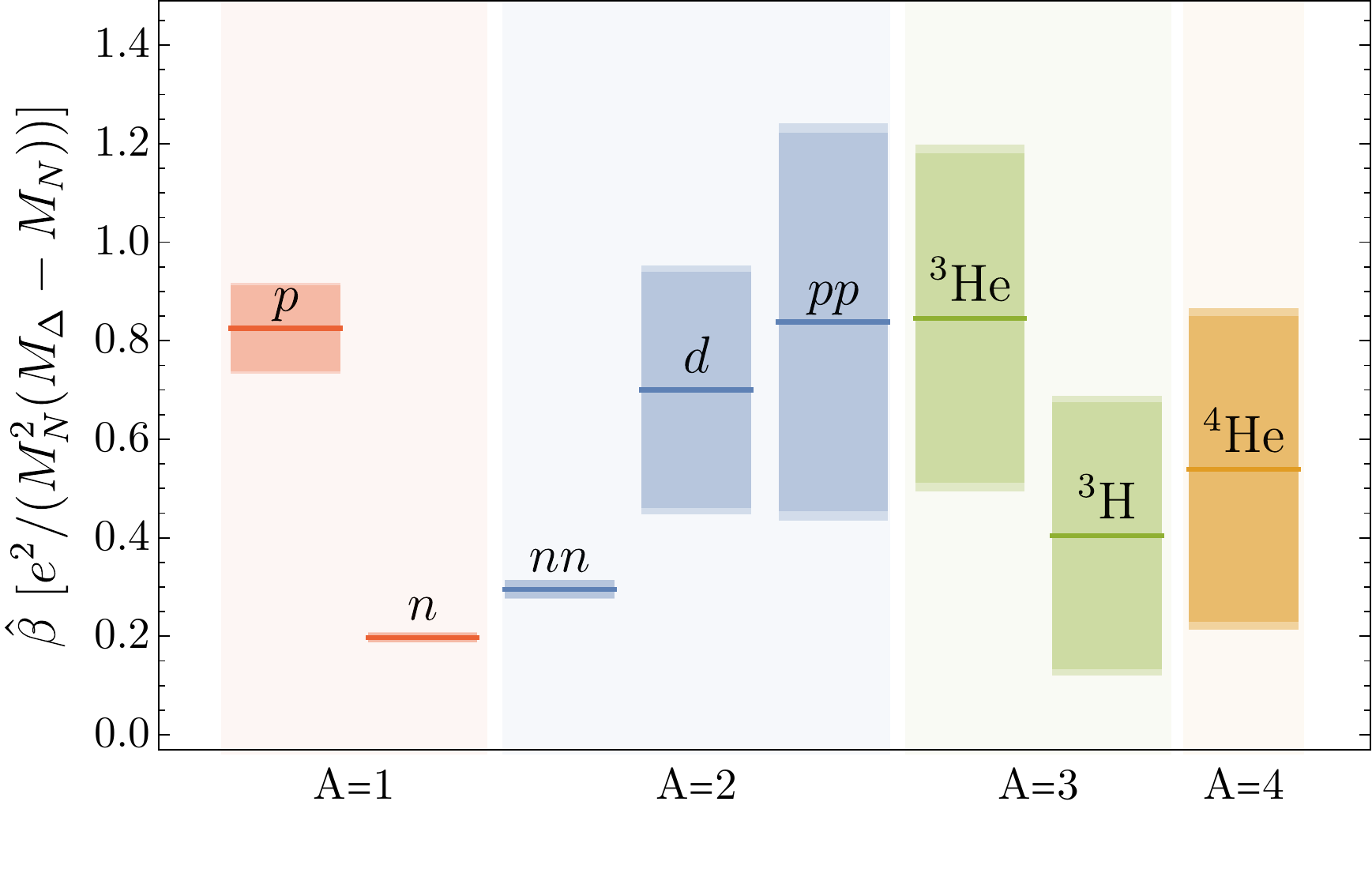}
	\caption{Magnetic moments (left) and magnetic polarizabilities (right) of nucleons and light nuclei calculated with LQCD at the SU$(3)_f$-symmetric point with $m_\pi= 806$~MeV. The results are given in natural units, as discussed in the text. The red dashed lines (left) correspond to the experimental magnetic moments. The darker shaded region represents the total uncertainty obtained by combining in quadrature the statistical and systematic uncertainties, and estimates of discretization and FV effects.
		[Data from Refs.~\cite{Beane:2014ora} and \cite{Chang:2015qxa}.]} 
	\label{fig:summaryMU}
\end{figure}

Figure~\ref{fig:summaryMU} shows  LQCD results for the magnetic moments and polarizabilities of light nuclei obtained in calculations performed at a single set of quark masses corresponding to a pion mass of $m_\pi= 806$ MeV (for nucleons~\cite{Lee:2005dq,Primer:2013pva,Bignell:2018,Bignell:2020,Engelhardt:2007ub} and other baryons in the $J^\pi=\frac{1}{2}^+$ octet~\cite{Parreno:2016fwu}, such studies have been undertaken for a range of quark masses). 
Note that the results are presented using natural units that for magnetic moments correspond to ``natural nuclear magnetons'' (nNM), $\hat \mu_h = \mu_h {M_N\over 2e}$, defined with respect to the nucleon mass at the quark masses used in the calculation. This choice avoids scale-setting uncertainties that arise when converting lattice units to nuclear magnetons using the physical nucleon mass.
For magnetic polarizabilitites, an appropriate dimensionless scale is given by the dominant $\Delta$-resonance pole contribution that is  ${\cal O}(e^2 /[ M_N^2 (M_\Delta -M_N)] )$, and dimensionless scalar and tensor magnetic polarizabilities are defined as $\hat\beta_h = \frac{M_N^2(M_\Delta-M_N)}{e^2} \beta_h^{(M0)} $ and $\hat\beta_h^{(2)} =  \frac{M_N^2(M_\Delta-M_N)}{e^2} \beta_h^{(M2)}$, respectively.
Since this quantity is only weakly dependent on the quark masses, one expects that it will provide appropriate units at any quark mass. Polarizabilities in physical units are presented in Fig. 23 of Ref.~\cite{Chang:2015qxa}.

Despite the use of a single, unphysical, set of quark masses, interesting features can be inferred from the existing LQCD investigations of the magnetic moments and polarizabilities of light nuclei. 
One feature of the results in Fig.~\ref{fig:summaryMU} is the approximate realization of naive shell-model expectations. In nature there is only a small difference between the experimental value and the shell-model prediction for the magnetic moment of $^3$He, which is very close to that of the neutron, with little contribution from the spin-paired protons. Similarly, the magnetic moment of the triton is very close to that of the proton, with little contribution from the spin-paired neutrons. From the left panel of Fig.~\ref{fig:summaryMU}, it is apparent that this similarity persists at the unphysical values of the quark masses used in the LQCD calculations, indicating that a shell-model-like description based on nucleon degrees of freedom remains appropriate. 
On the other hand, different trends were observed for the polarizabilities. For example, from Fig.~\ref{fig:summaryMU} it can be seen that the dineutron\footnote{The dineutron is bound at this set of quark masses, see Sec.~\ref{sec:forces}.} polarizability,  $\hat\beta_{nn}=\betann$, differs from twice that of the neutron by 
$\delta \hat\beta_{nn} \ \equiv\ \hat\beta_{nn} - 2\hat\beta_{n}  =   \dbetannCORR $. 
For the deuteron, which is described by the coupled \mbox{$\siii$-$\diii$} channels,  Eq.~(\ref{eq:Eshift}) shows that both the scalar and tensor polarizabilities contribute to the quadratic dependence of the deuteron energy on the magnetic-field strength. The magnetic moment and a combination of polarizabilities of the deuteron can be obtained from a coupled fit to the two $j_z=\pm 1$ states, giving the values
$\hat\mu_d = \mud$  and  $\hat\beta^{(M0)}_{d}  +  {1 \over 3} \hat\beta^{(M2)}_{d}  = \betadpm$. Comparing this result with the sum of the individual neutron and proton polarizabilities, 
$\hat \beta_p+\hat\beta_n\sim 1.02^{+(10)(5)}_{-(7)(5)}$, 
again illustrates the important role played by nuclear forces and electromagnetic interactions, which cause the bound $np$ system to be more magnetically rigid compared with the sum of its constituents.
The result obtained for the triton is significantly smaller than the sum of dineutron and proton polarizabilitites, allowing for a potential extraction of information on the two- and three-nucleon electromagnetic contributions. 
The polarizability computed for $^3$He, however, is compatible with the sum of the polarizabilitites of the diproton and neutron constituents within the uncertainties. That is, deviations from the simple one-body contributions to the magnetic polarizabilities, i.e., from short-range correlated two-nucleon responses to the field, could not be obtained given the large uncertainties involved in the determination of the $^3$He and $pp$ results.   
The $^4$He nucleus has no magnetic moment (since $J=0$ for this system), but its polarizability is determined to be between those of the $pp$ and $nn$ systems, with the same level of uncertainty as that characterizing the $A=3$ systems.
These LQCD determinations of the magnetic moments and polarizabilities, along with those known from experiment, were analyzed in pionless EFT in Ref.~\cite{Kirscher:2017fqc}.

\subsection{Two nucleons in strong magnetic fields}
\label{strongmf}

While the extraction of magnetic moments and polarizabilities is based on an expansion around zero field strength, LQCD calculations can also be performed for large values of the magnetic field. In fact, it is possible to utilze  strong magnetic fields with magnitudes comparable to the QCD scale, $|e { B}|\gtrsim\Lambda_{\rm QCD}^2$, in which the electrodynamics effects are comparable to the strong interaction effects.
These extreme fields may be encountered in natural astrophysical environments,  for example in magnetars~\cite{Duncan:1992hi}, which are rapidly rotating neutron stars with extremely large magnetic fields of up to ${\cal O}(10^{14})\ {\rm Gauss}$ at the surface that are conjectured to reach ${\cal O}(10^{19})\ {\rm Gauss}$  in the interior~\cite{Broderick:2000pe}. Very strong EM fields are also present in heavy-ion collisions~\cite{Kharzeev-NPA803-2008,McLerran:2013hla}, where the currents produced by relativistic nuclei, particularly during (ultra-)peripheral collisions, lead to fields within the projectiles that have also been estimated to be of the order of ${\cal O}(10^{19})\ {\rm Gauss}$.
From a phenomenological point of view, the asymptotic freedom of QCD \cite{Cohen:2008bk} dictates that nuclear systems under the effects of extremely large magnetic fields that are comparable to the QCD scale are expected to have eigenstates that are weakly-interacting up and down quarks in Landau levels. Hence, as the magnetic field tends to infinity, the ground-state energies of nuclei approach the sum of their constituents. The response of the spin-up and spin-down nucleon states to an external magnetic field was studied in Ref.~\cite{Chang:2015qxa}. 
Interestingly, it was observed that the ground-state energies of both $j_z=\pm \frac{1}{2}$ states for the proton exhibit significant nonlinearities for the whole range of magnetic field strengths that were explored, as expected due to the presence of Landau levels. The spin-up and spin-down neutron states behave very differently, however, with the spin-up state following a non-linear dependence, and the spin-down state a linear dependence, on the magnetic-field strength.

The effects of  strong magnetic fields were investigated for two-nucleon systems in LQCD in Ref.~\cite{Detmold:2015daa}.
The magnetic response of a two-nucleon system to an applied magnetic field can be obtained  by determining the energy shift 
\begin{equation}
\Delta_{NN}(B) \equiv \delta E_{{NN}}({ B}) - \sum\limits_{h\in{\{NN\}}} \delta E_{h}({ B}) \, {\rm ,}
\label{eq:delta-en}
\end{equation}
where the energy splittings are defined as $\delta E_h(B) = E_h(B) - E_h(0)$, with $E_h(B)$  given by Eq.~\eqref{eq:Eshift} (with $P_{||}=0$), 
and the sum ranges over the hadrons ($h$) contributing to the composite $NN$ system, e.g., for the deuteron, with $j_z=+1$, $d_{j_z=+1}$, the relevant hadrons are $p^\uparrow$ and $n^\uparrow$.
The shift defined in Eq.~\eqref{eq:delta-en} can be obtained from the large-time exponential decay of appropriate ratios of correlation functions.
\begin{figure}[!t]
	\centering
	\includegraphics[width=0.49\textwidth]{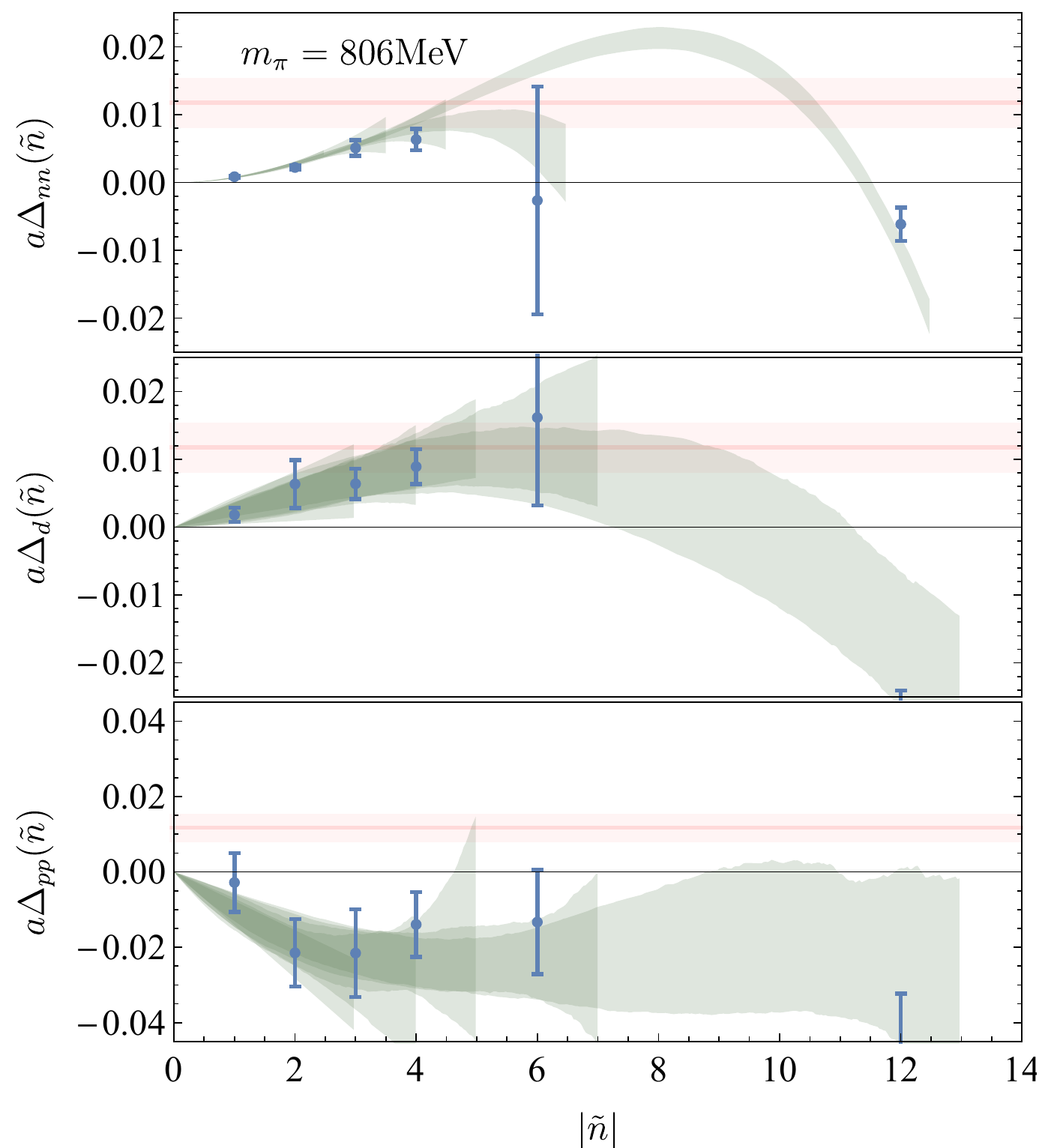}   
	\includegraphics[width=0.49\textwidth]{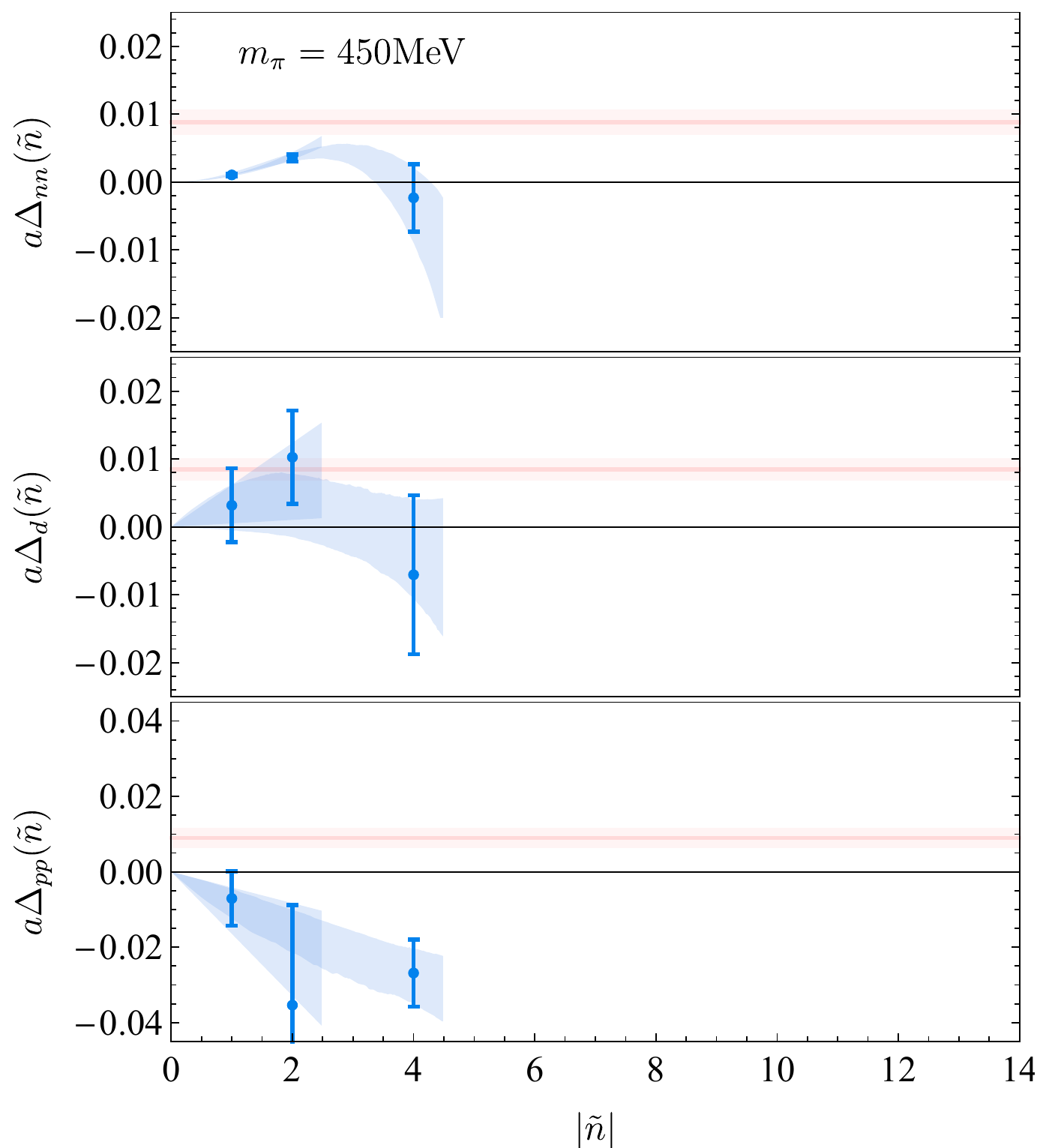}
	\caption{ 
		Response of the binding of the dineutron (top), of the $j_z=+1$ deuteron (center), and of the diproton (bottom) to applied magnetic fields, in lattice units. The figures in the left column show the results at $m_\pi=806$ MeV, while the ones on the right show results at $m_\pi=450$ MeV. The shaded regions correspond to envelopes of fits to the energy shifts using linear and quadratic polynomials in $\tilde n^2$ 	for the case of the dineutron, and in $\tilde n$ up to $4^{\text{th}}$ ($2^{\text{nd}}$) order for the $m_\pi=806$ (450) MeV ensemble, in the ranges indicated by the shaded regions. The horizontal pink bands indicate the binding thresholds. [Figure modified from Ref.~\cite{Detmold:2015daa}.]}
	\label{fig:Delta-NN-450-800}
\end{figure} 
Figure~\ref{fig:Delta-NN-450-800} shows the energy shifts determined using this approach for the dineutron, the $j_z=+1$ deuteron, and  the diproton for a magnetic field quantized as $|e{ B}|=6\pi |\tilde n|/(L^2)$, where $\tilde n=\{1,2,3,4,6,12\}$ for $m_\pi=806$ MeV and $\tilde n = \{1,2,4\}$ for $m_\pi=450$ MeV~\cite{Detmold:2015daa}. Since $B\sim{\cal O}(10^{19})$~Gauss for $\tilde n=1$ in this lattice volume, this corresponds to extremely large physical field strengths. For the dineutron and $j_z=+1$ deuteron, and for both values of the quark masses, as the strength of the applied magnetic field is increased, the ground state energies of the two-nucleon systems move closer to threshold, and in some cases may unbind once a critical field strength is reached. 
For both channels, the point of minimum binding decreases with the pion mass, suggesting that at the physical quark masses the (unbound) dineutron resonance is pushed even further into the continuum by an intense magnetic field, while the  deuteron becomes unbound due to the applied field.
The behavior in these channels is exemplary of the unitary regime in two-particle interactions, 
in which the binding energies decrease to zero and consequently the scattering lengths diverge. Near the values of the field strength at which 
the binding approaches zero, universal physics would emerge from the study of the low-energy dynamics of these systems. While this universality has been observed as {\it Feshbach resonances} in atomic physics~\cite{RevModPhys.82.1225}, an analogous observation has not been made in nuclear systems. For the diproton, as seen in the lowest two panels of Fig.~\ref{fig:Delta-NN-450-800}, an enhanced binding is observed at both quark masses as the field strength increases. This pattern would suggest that at the physical quark masses, the (unbound) diproton could overcome the Coulomb repulsion and form a bound state in sufficiently large magnetic fields, in agreement with the expectations of Ref.~\cite{Allor:2006zy}.

In Ref.~\cite{Detmold:2015daa}, magnetic-field effects on the channel with the quantum numbers of two $\Lambda$-baryons were also studied (see Fig.~5 in that reference). This doubly-strange channel contains a deeply bound state at the heavy quark masses used in the calculations~\cite{Beane:2010hg,Inoue:2010es,Beane:2012vq}. However, while it exhibits a slight reduction of the binding energy for intermediate field strengths, comparable in size to that of the dineutron system, it does not show resonant behavior in the range of field strengths that are probed as the binding energy is significantly larger than that in the two-nucleon case.

\subsection{The $np\rightarrow d\gamma$ radiative capture process
	\label{subsec:npdgamma}} 

Determinations of the energy levels of the $I_z=j_z=0$ $np$ states in the $\si$ and $\siii$--$^3D_1$ channels in a background magnetic field have direct implications for our knowledge of nucleon-photon interactions and are relevant to the evolution of the early universe. Specifically, since a magnetic field couples these two states, by determining the corresponding energy difference one can extract the short-distance two-body electromagnetic contributions to the low-energy radiative capture process,  $np \to d\gamma$, and to the photo-disintegration process $\gamma d \to np$~\cite{Detmold:2004qn,Chang:2015qxa,Beane:2015yha}. 
The relevant energy difference can be computed in LQCD by constructing a matrix of correlation functions generated from source and sink operators associated with  $\siii$ and $\si$  $I_z=j_z=0$ interpolating operators:
\begin{align}
{\bf C}(t,{B})  =  
\left(
\begin{array}{cc}
C_\text{2pt}^{\siii,\siii}(t,{ B}) & C_\text{2pt}^{\siii,\si}(t,{ B}) \\
C_\text{2pt}^{\si,\siii}(t,{ B}) & C_\text{2pt}^{\si,\si}(t,{ B}) 
\end{array}
\right),
\label{eq:cormat}
\end{align}
where the background-field two-point correlation functions are defined as in Eq.~\eqref{eq:BF2pt} (but with the individual specification of the quantum numbers of the source and sink interpolating fields).
Principal correlation functions, $\lambda_{\pm}(t,{ B})$, which exponentially converge to the eigenstates of the coupled system at large times, can then be obtained by diagonalizing this matrix at each value of $t$ and $B$. In this large-$t$ limit, the energy shift between the two eigenstates can be computed from the ratio
\begin{equation}
\delta R_{\, ^{3}S_1,^{1}S_0}(t,{ B}) \equiv \frac{\lambda_+(t,{ B})}{\lambda_-(t,{ B})}
\frac{C_\text{2pt}^{n,\uparrow}(t,{ B})C_\text{2pt}^{p,\downarrow}(t,{ B})}{C_\text{2pt}^{n,\downarrow}(t,{ B})C_\text{2pt}^{p,\uparrow}(t,{ B})}  \overset{t\rightarrow\infty}{\longrightarrow}  Z\ e^{-\delta E_{\, ^{3}S_1,^{1}S_0}({ B}) t}
\,,
\end{equation}
where $C_\text{2pt}^{p/n,\uparrow/\downarrow}(t,{ B})$ are the background-field correlation functions corresponding to the different polarizations of the proton and neutron, and $Z$ is a field-dependent overlap factor. The energy shift in the exponential can be written in terms of the energies of the two eigenstates and those of the polarized nucleons:
\begin{equation}
\delta E_{\, ^{3}S_1,^{1}S_0}\equiv\Delta E_{\, ^{3}S_1,^{1}S_0}-[E_{p,\uparrow}-E_{p,\downarrow}]+[E_{n,\uparrow}-E_{n,\downarrow}] 
\, {\rm ,}
\label{eq:eshift-delta}
\end{equation} 
where $\Delta E_{\, ^{3}S_1,^{1}S_0}$ is the (positive) energy difference between the two eigenstates.
At the large values of the quark masses, corresponding to $m_\pi= 806$~MeV and $m_\pi = 450$ MeV, used in the only existing LQCD determinations of this energy~\cite{Beane:2015yha}, the appropriate framework to relate this energy shift to the short-range two-nucleon interaction coefficient is pionless EFT.

Employing dibaryon fields to resum effective-range contributions~\cite{Kaplan:1996nv,Beane:2000fi}, the Lagrange density describing the interactions of the nucleon and the dibaryons with an external magnetic field can be written as: 
\begin{align}
{\cal L} =
{e\over 2 M_N} N^\dagger \left[ \kappa_0 + \kappa_1 \tau^3 \right] \sigma_i B_i \, N
+
{e\over M_N} {l_1\over\sqrt{r_1 r_3}} \left[ t_j^\dagger s_3 B_j \ +\ {\rm h.c.}\ \right]
+ 
{e\over M_N} 
{l_2 \over r_3}
i \epsilon_{ijk} t_i^\dagger t_j B_k,
\label{eq:traniL}
\end{align}
where $t_i$ and $s_3$ are the $i$th spin component of the isosinglet $^3S_1$ and the third isospin component of the isotriplet $^1S_0$ dibaryons, respectively. $\vec{\sigma}$ is the spin operator, and $\kappa_0=\frac{1}{2} \left(\kappa_p+\kappa_n\right)/2$ and \mbox{$\kappa_1=\left(\kappa_p-\kappa_n\right)/2$} are the isoscalar and isovector nucleon magnetic moments, respectively, in nuclear magnetons with $\kappa_p = 2.79285$ and $\kappa_n = -1.91304$. $r_1$ and $r_3$ are the effective ranges in the singlet and triplet channels, respectively. The NLO coefficients, $l_1$ and $l_2$\footnote{Note that in Ref.~\cite{Chang:2015qxa} $l_2$ has been replaced by ${ \tilde l_2} - r_3 \kappa_0$ to make explicit the deviation of the deuteron magnetic moment from the single-nucleon contribution.}, describe the coupling of the dibaryons to the magnetic field. This Lagrangian can be used to obtain the LO and NLO 
contributions to the $M1$ amplitude~\cite{Beane:2000fi}, which dominates the EM multipole expansion of the low-energy $np \to d\gamma$ cross-section~\cite{Bethe1950,Noyes1965}:
\begin{equation}
\sigma(np\rightarrow d\gamma)  = 
{e^2(\gamma_t^2+|{\vec p}\,|^2)^3\over M_N^4 \gamma_t^3 |{\vec p}\,|}
\left[\ 
|\tilde X_{M1}|^2  +  \ldots
\ \right].
\label{eq:npdgsigma}
\end{equation}
In this expression, $\gamma_t$ is the binding momentum of the deuteron, ${\vec p}$ is the momentum of each incoming nucleon in the CM frame, and 
$\tilde X_{M1}$ is the $M1$ amplitude, which is given by:
\begin{equation}
\tilde X_{M1}  = 
{Z_d\over  -{1\over \mathfrak{a}_1} + {1\over 2} r_1 |{\vec p}\,|^2 - i |{\vec p}\,|}
\left[\ {\kappa_1 \gamma_t^2\over \gamma_t^2 +|{\vec p}\,|^2}\left( \gamma_t - {1\over \mathfrak{a}_1} + {1\over 2} r_1 |{\vec p}\,|^2 \right)
+ {\gamma_t^2\over 2} l_1
\right].
\label{eq:npdgMone}
\end{equation}
Here, $Z_d = 1/\sqrt{1-\gamma_t r_3}$ is the square-root of the residue of the deuteron propagator at its pole and $\mathfrak{a}_1$ is the scattering length 
in the $\si$ channel. Contributions from higher-order multipoles in Eq.~\eqref{eq:npdgMone}, denoted by the ellipses, are suppressed at low energies.
The quantity $l_1 = \tilde l_1 - \sqrt{r_1 r_3} \kappa_1$ encapsulates the 
short-distance two-nucleon interactions through the coefficient $\tilde l_1$. %
This is the only quantity which is not determined by kinematics, single-nucleon properties, or scattering parameters.
As is discussed in Ref.~\cite{Beane:2015yha}, Wigner SU(4) symmetry allows this coefficient  to be related to the energy difference between the two  eigenstates of the coupled $\si$--$\siii$ system: 
\begin{eqnarray}
\Delta E_{^{3}S_1,^{1}S_0}({ B})  = 
2 \left( \kappa_1 + \gamma_t Z_d^2 \tilde l_1 \right) {e \over M} |{ B}| +{\cal O}(|{ B}|^2)
\, {\rm ,}
\label{eq:Esplit}
\end{eqnarray}
with $ \gamma_t Z_d^2 \tilde l_1 \equiv \bar{L}_1$ characterizing the two-nucleon contributions. Note that magnetic-field couplings to sea quarks do not contribute to this energy shift or to $\bar{L}_1$ in isospin-symmetric LQCD calculations.

\begin{figure}[!t]
	\centering
	\includegraphics[width=0.47\columnwidth]{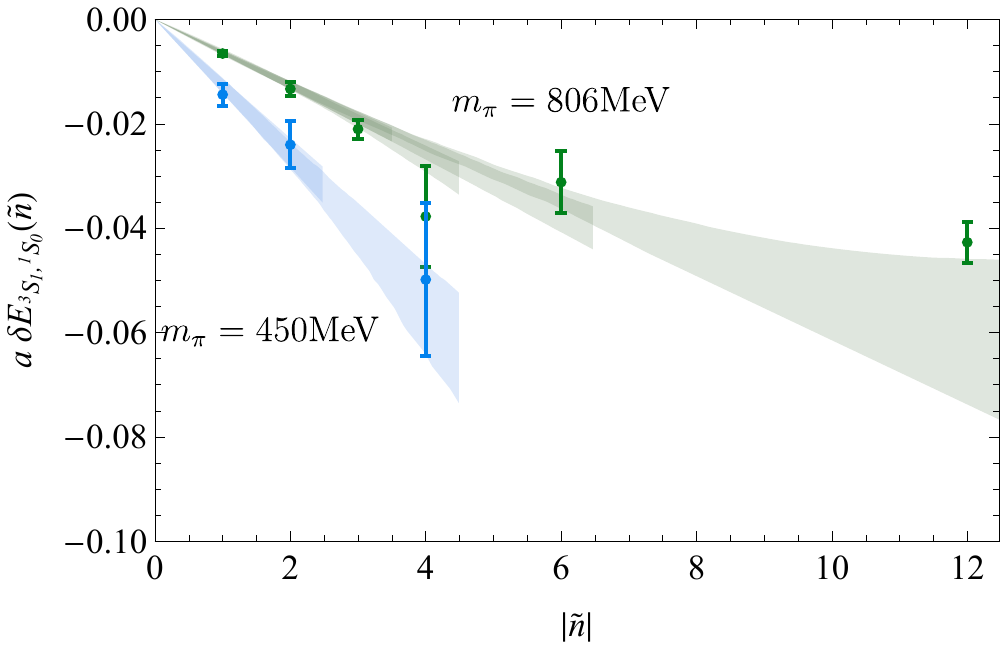} \hspace*{0.5cm}
	\includegraphics[width=0.45\textwidth]{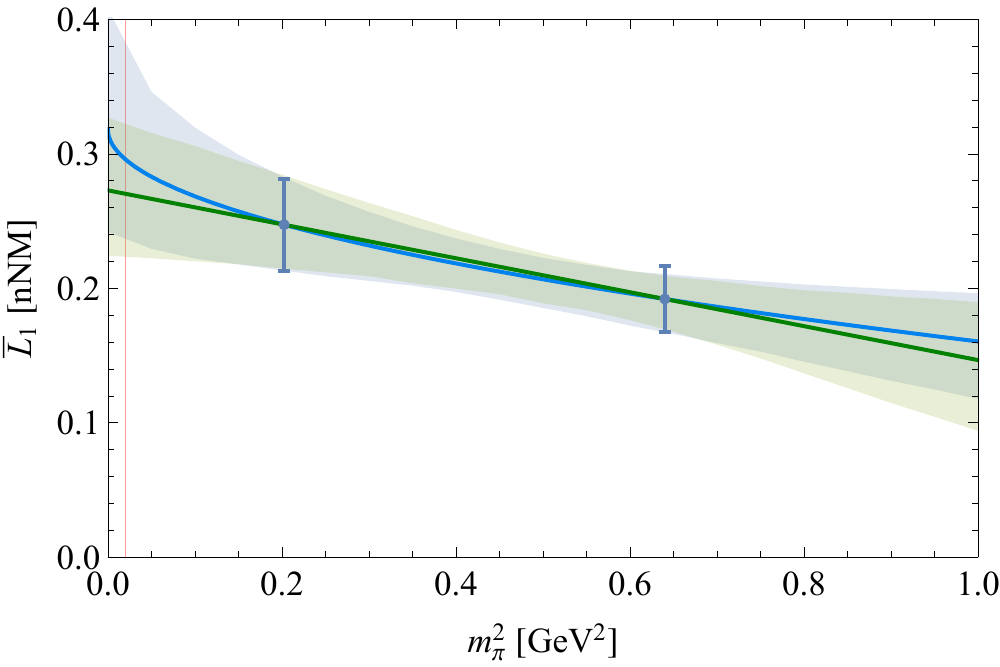} \ \
	\caption{
		LQCD results for the energy splittings between the two lowest-lying eigenstates of the $\siii$--$\si$ system (left), with the single-nucleon contributions removed	as a function of the magnetic field strength quantum $\tilde n$, along with the associated fits to extract the linear response.
		Also shown are the results of LQCD calculations of $\bar{L}_1$ (blue points) (right). The blue (green) shaded regions show extrapolations of $\bar{L}_1$ 
		to the physical pion mass (red vertical line) in natural nuclear magnetons (nNM) which are linear (quadratic) in $m_\pi$.
		[Figure modified from Ref.~\cite{Beane:2015yha}.]
	}
	\label{fig:L1parameter}
\end{figure}

Combining Eqs.~(\ref{eq:eshift-delta}) and ~(\ref{eq:Esplit}), the value of $\bar{L}_1$ can be extracted from the slope of the field-strength dependence of $\delta E_{ \, ^{3 \!}S_1,^{1\!}S_0}$, as shown in the left panel of Fig.~\ref{fig:L1parameter} for LQCD calculations performed with  $m_\pi=450$ and 806 MeV. To obtain a prediction for the $\bar{L}_1$ LEC from these results requires extrapolation to the physical values of the  light-quark masses. While the form of this extrapolation is not known {\it a priori}, the mild $m_\pi$ dependence that is observed in the right panel of 
Fig.~\ref{fig:L1parameter}, together with the mild variations shown by the magnetic moments (when expressed in units of natural nuclear magnetons), suggest that linear and quadratic forms could be reasonable choices to extrapolate to the physical point, as shown in the figure.
These two functional forms yield consistent values at the physical quark masses and allow an estimate of the extrapolation uncertainty. The extrapolated value is $\bar{L}_1^{\rm LQCD} = 0.285^{+(63)}_{-(60)}$~nNM, where the uncertainty incorporates the mass-extrapolation uncertainty in addition to statistical, correlation function fitting, and field-strength dependence fitting uncertainties. 
This value leads to 
$l_1^{\rm LQCD}=-4.48^{+(16)}_{-(15)}~{\rm fm}$, 
which  can be used in Eq.~(\ref{eq:npdgMone}) to extract the value of the $M1$ amplitude needed to obtain the cross-section for the $n p \to d \gamma$ process through Eq.~(\ref{eq:npdgsigma}), $\sigma^{\rm LQCD}  =  \sigmaPHYS~{\rm mb}\,$, which is in agreement with the experimental determination $\sigma^{\rm expt} = \sigmaEXPT~{\rm mb}$~\cite{Cox1965497}, both at an incident neutron speed of $v=2,200$~m/s.

\subsection{Future impact}
\label{subsec:EMfuture}

The response of a nucleus to an EM probe reveals a number of aspects of nuclear structure, including magnetic moments, polarizabilities, and the nuclear EM response functions and form factors. 
Experimental determinations of the magnetic moments of light nuclei led to insights into nuclear structure and provided early validation of nuclear shell-model frameworks.
By determining how these nuclear properties emerge from the SM, LQCD can provide further insights into the quark and gluon structure of nuclei and the  emergence of nucleons as effective degrees of freedom in nuclei.
The results reviewed above, in conjunction with future calculations at physical values of the light quark masses and including multiple physical volumes and lattice spacings, will provide decisive constraints on physical quantities that cannot be accurately measured experimentally and will increase our understanding of the behavior of 
nuclear systems and reactions in the presence of magnetic fields of different intensities.

For example, given the lack of free neutron targets, neutron polarizability determinations rely on experiments involving light nuclei. LQCD can provide insight into the electromagnetic structure of the deuteron and other light nuclei needed to understand and correct for nuclear effects in these experiments.
The LQCD calculations at unphysically large values of the light-quark masses described above show that the sum of the neutron and proton polarizabilities is larger than the deuteron polarizability, and therefore that the deuteron is more magnetically rigid than its components. 
Future LQCD calculations at the physical quark masses will determine whether this effect persists and predict the electromagnetic polarizabilities of the deuteron and other light nuclei in nature.
Such calculations will also determine the role of nuclear forces and gauge-invariant electromagnetic two-nucleon operators in the electromagnetic structure of light nuclei, and complement the constraints from upcoming experiments that will extract deuteron polarizabilities at the HI$\gamma$S facility~\cite{Weller:2009zza}, MAX-Lab at Lund~\cite{MAXLab}, and at MAMI in Mainz~\cite{Downie:2011mm}, as well as next-generation of Compton scattering experiments~\cite{Myers:2012xw,Myers:2014qhi,Myers:2014qaa} that will extend those studies to different nuclear targets.

The LQCD studies discussed above have explored the simplest slow-neutron capture process, $np \rightarrow d\gamma$, which is the first step of the big-bang nucleosynthesis (BBN) chain reaction that created the first elements in the seconds after the big bang.
Future calculations of nuclear reactions in light nuclei will provide QCD predictions for reaction rates that are less well known experimentally.
These first-principles theoretical constraints on BBN processes could shed light on the fine-tuning of the evolution of the universe and perhaps on the deficit in the measured abundance of ${}^7$Li in the universe compared with theory predictions~\cite{Steigman:2007xt,Pospelov:2010hj}.

LQCD determinations of the responses of nuclear systems to intense magnetic fields will provide valuable information for constraining phenomenological models of nuclei in extreme environments in nuclear astrophysics such as magnetars and in relativistic heavy ion collisions.
The results above show that there are significant changes in the binding of two-nucleon systems immersed in strong magnetic fields at two unphysical values of the quark masses, with the deuteron becoming unbound and diproton becoming bound at particular values of the magnetic-field strength.
Future LQCD calculations will determine whether this behavior and associated realizations of unitary physics persist at the physical light-quark masses and explore its consequences for nuclear astrophysics.

Precise measurements of the electromagnetic structure of light nuclei can also be used to search for new physics beyond the SM~\cite{Delaunay:2016brc,Frugiuele:2016rii,Berengut:2017zuo,Stadnik:2017yge,Delaunay:2017dku}.
The charge radii of light nuclei, in particular the deuteron, $^3$He, and $^4$He, have been extracted from laser spectroscopy measurements of the Lamb shifts of muonic atoms~\cite{Pohl:2016xsr,Pohl1:2016xoo,2018arXiv180807240S} as well as in electronic atom spectroscopy \cite{Blaum:2012ed} and electron-scattering experiments \cite{Nortershauser:2011zz,Sick:2014yha}. 
Tensions exist between muonic and electronic measurements of the charge radii of light nuclei with similar significance to tensions in proton-radius measurements, and comparisons between results for the proton and various nuclei can test possible explanations of these discrepancies.
Lepton universality requires the same charge radius to enter electronic and muonic observables. Therefore, while there are multiple possible sources of these discrepancies, new electron-muon universality-violating physics is an interesting scenario.
Nuclear-structure effects, including those present in two-photon exchange contributions to the Lamb shift, add significant theoretical uncertainties to extractions of nuclear charge radii using muonic-atom spectroscopy, and in some cases these nuclear-theory uncertainties dominate the total uncertainty of experimental extractions of nuclear radii~\cite{Diepold:2016cxv,Ji:2018ozm}.
Precise SM predictions for the EM radii of light nuclei and other EM structure properties from LQCD could provide key insights into these discrepancies and other fundamental-symmetry tests.

\section{Nuclear matrix elements of weak currents }
\label{sec:weak}
\noindent
Low-energy nuclear reactions induced by the weak interactions of the SM are at the core of accurate descriptions of Big Bang and supernova nucleosynthesis and of the burning mechanism and energy production in stars, and are key inputs in astrophysical and terrestrial neutrino-flux models.  LQCD offers the possibility to reliably constrain the cross-sections, or more directly the relevant nuclear matrix elements, for processes that cannot be measured experimentally given the extreme conditions under which they proceed, such as the $pp$-fusion process that occurs in stars like the Sun. Further, LQCD calculations can isolate and constrain the short-distance effects, such as two- and multi-nucleon currents, in reactions of light nuclei to provide constraints on the EFTs or phenomenological models that can be fed into modern \textit{ab initio} nuclear-reaction studies in larger nuclei~\cite{Navratil:2016ycn,Gandolfi:2020pbj,King:2020wmp}. The first LQCD results for the Gamow-Teller matrix element relevant to tritium $\beta$ decay and $pp$ fusion, as well as the flavor-separated axial charges for nuclei with $A<3$, are now available~\cite{Savage:1995kv}, albeit at unphysical values of the quark masses. These results will be reviewed in this section.

Given that the dominant contribution to weak matrix elements in nuclei arises from the coupling of a single nucleon to the weak current, characterized by the nucleon's axial charge $g_A$, progress in the determination of $g_A$ from LQCD is also summarized. 
LQCD determinations of $g_A$ rely on different methods to extract the nucleon matrix elements. Many studies, such as those of Refs.~\cite{Gupta:2018qil, Bhattacharya:2016zcn}, calculate two- and three-point correlation functions as in  Eqs.~\eqref{eq:2pt} and \eqref{eq:3pt}, and extract the axial charge from the time dependence of their ratio. Variants of the method based on modified propagators, defined in Eq.~\eqref{eq:ratcorr}, have been employed in Refs.~\cite{Chang:2018uxx, Berkowitz:2017gql,Savage:2016kon}. In all cases, a dominant source of systematic uncertainty arises from the contributions of excited states as discussed in Sec.~\ref{subsec:fitting} (see Ref.~\cite{Green:2018vxw} for a review of these effects in the context of the nucleon). 
The latest FLAG report~\cite{Aoki:2019cca} provides community-consensus values of recent LQCD determinations of $g_A$ using ensembles with $N_f\in\{2, 2+1,2+1+1\}$ quark flavors. Among the most precise determinations to date are those with $N_f=2+1+1$ by the CalLat collaboration~\cite{Chang:2018uxx, Berkowitz:2017gql} (one at the 1\% level) and the PNDME collaboration~\cite{Gupta:2018qil, Bhattacharya:2016zcn} (at the 3\% level) using the same ensembles of gauge-field configurations but with different valence-quark actions. The FLAG value, $g_A=1.251(33)$, is consistent with the considerably more precise experimental determinations, i.e., $g_A/g_V=1.2772(20)$ by the UCNA collaboration~\cite{Mendenhall:2012tz, Brown:2017mhw} and $g_A/g_V=1.2761^{+14}_{-17}$ by PERKEO II~\cite{Mund:2012fq}, where $g_V$ is the isovector vector charge and is equal to one up to very small corrections due to isospin breaking.
The LQCD determinations of $g_A$ have so far served as a testbed for the validation of LQCD methods and technologies in accessing nucleon matrix elements. Nonetheless, as future determinations are anticipated to reach sub-percent precision, comparison to high-precision neutron-decay measurements \cite{Mendenhall:2012tz,Brown:2017mhw} may allow for tests of the SM and constraints on BSM effects, such as right-handed currents \cite{Cirigliano:2019jig,Gonzalez-Alonso:2018omy}. New systematic uncertainties such as QED effects and isospin splittings will need to be fully addressed to reach sub-percent precision.

While precision determinations of the nucleon axial charge from LQCD in recent years are promising, in nuclear-physics contexts it is the multi-nucleon contributions to axial matrix elements that are often ill-constrained and this is where LQCD can play the largest role. Explicitly, LQCD will soon provide constraints that are more precise than those from experiment alone, hence enhancing the predictive capabilities of studies of weak processes in nuclei.

\subsection{Proton-proton fusion}
\label{subsec:ppfusion}
The production of deuterium in the $pp$-fusion process, $pp \to d e^+ \nu_e$, is the first step in the chain of reactions that produces energy in stars with masses similar to, or smaller than, that of the Sun~\cite{Adelberger:2010qa}. However, given the low incident velocities of protons in the stellar interior, the weak nature of the process, and the Coulomb barrier, the rate of this process is extremely low and has not been measured in the laboratory at the relevant energies.  As this rate is an important input into the Standard Solar Model, which predicts the Sun's neutrino flux for terrestrial neutrino-oscillation experiments, accurate theoretical determinations of the near-threshold cross-section are valuable (see e.g., Refs.~\cite{Adelberger:2010qa,Kubodera:2010qx} for recent reviews). It is known that the single-nucleon contribution, i.e., the conversion of the proton to a neutron in the process, dominates the cross-section, and the two-nucleon effects, described in phenomenological models by meson-exchange currents and in low-energy EFTs by such currents as well as by local two-nucleon operators, contribute only at the percent level~\cite{Adelberger:2010qa}. Nonetheless, the need for sub-percent precision on the cross-section has prompted investigations to constrain this two-body effect~\cite{Schiavilla:1998je,Carlson:1991ju,Butler:2001jj,Park:2002yp,Butler:2002cw,Ando:2008va,Marcucci:2013tda,Chen:2012hm,Acharya:2016kfl,De-Leon:2016wyu,Chen:2002pv, Acharya:2019fij}. Similar higher-body effects further contribute to an understanding of the problem of the phenomenological quenching of the axial charge in larger nuclei, see e.g., Ref.~\cite{Gysbers:2019uyb}, and are hence important to constrain at the microscopic level.

Within a chiral EFT approach, the leading two-nucleon operators in the axial current are related to operators appearing in two- and three-nucleon potentials. This leads to constraints on the two-body contributions to electroweak processes, albeit with large uncertainties~\cite{Park:2002yp,Acharya:2016kfl}. The low-energy process of $pp$ fusion is also suitable for analysis in pionless EFT \cite{Kaplan:1996nv,Kaplan:1998tg, Kaplan:1998we, vanKolck:1998bw, Chen:1999tn, Beane:2000fi}, in which the coupling of the axial-vector current to two nucleons at low energies is characterized by a single momentum-independent LEC, $L_{1,A}$. The first constraint on $L_{1,A}$ was obtained from analysis of  Sudbury Neutrino Observatory (SNO) and Super-Kamiokande data on charged-current and neutral-current neutrino-deuteron scattering reactions~\cite{Chen:2002pv}, since the same LEC contributes to these processes as well. Constraints on $L_{1,A}$ have been further improved using an approach based on consistent treatment of the tritium $\beta$-decay rate which is known precisely from experiment, and which shares the same LEC within the pionless EFT~\cite{De-Leon:2016wyu}. A new determination using an improved calculation of (anti)neutrino-deuteron inelastic scatterings has also appeared~\cite{Acharya:2019fij}. Despite these advances, the uncertainty on $L_{1,A}$ remains large and is comparable to its central value. Since muon capture on the deuteron  is also sensitive to  $L_{1,A}$, it is expected~\cite{Chen:2005ak} that a precise measurement that is underway in the MuSun experiment will provide a significant improvement in the precision of $L_{1,A}$~\cite{Kammel:2010zz,Andreev:2010wd,Rachelryan:2019}. A critical (and realistic) goal for LQCD in nuclear physics is to provide a QCD-based determination of $L_{1,A}$ that competes with, or improves upon, the best phenomenological values.

The matrix element of the axial-vector current $J_i^-=\bar{q} \gamma_i\gamma_5\tau_-q$ with $q=(q_u,q_d)^T$ and flavor matrix $\tau_-=\tau_1-i\tau_2$, between the $j$th spin component of the deuteron and the two-proton system, can be written as
\begin{eqnarray}
\left|\left\langle d;j\left| J_i^- \right|pp\right\rangle\right|
\equiv
g_A C_\eta \sqrt\frac{32 \pi}{\gamma_t^3} \,\Lambda(k) \,\delta_{ij},
\label{eq:ppMATdef}
\end{eqnarray}
where all the factors except for the quantity $\Lambda(k)$ are precisely known~\cite{Butler:2001jj}. In particular, $\gamma_t=\sqrt{M_NB_d}$ is the deuteron binding momentum ($B_d$ is the binding energy), $C_\eta$ is the QED Sommerfeld factor, and $k$ denotes the momentum of each proton in the CM frame. For $pp$ fusion at low incident velocities, $\Lambda(0)$ provides the dominant contribution~\cite{Adelberger:2010qa}. In pionless EFT, the momentum-independent single- and two-nucleon isovector axial-vector currents are
\begin{align}
J_{k}^{-(1)} &= \frac{g_{A}}{2} \, N^\dagger \tau ^{-}\sigma_{k}N,
\\
J_{k}^{-(2)} &= L_{1,A}\left( N^{T}P_{k}N\right) ^{\dagger }\left( N^{T}
\bar{P}^{-}N\right),
\label{eq:axial-currents}
\end{align}
respectively~\cite{Butler:1999sv}. Here,  $P_i \equiv \frac{1}{\sqrt{8}}\sigma^2\sigma^i\tau^2$ ($\bar{P}^- \equiv \frac{1}{\sqrt{8}}\sigma^2\tau^2\tau^-$) are projectors into the $\siii$ ($\si$) two-nucleon channel. With this characterization of momentum-independent currents in the EFT, the threshold amplitude in Eq.~(\ref{eq:ppMATdef}) at NLO can be written as~\cite{Kong:2000px,Butler:2001jj,Ando:2008va}\footnote{For improved convergence, effective-range contributions to the amplitude are resummed to all orders~\cite{Beane:2000fi,Phillips:1999hh,Ando:2008va,De-Leon:2016wyu}.}
\begin{eqnarray}
\Lambda(0) =
{1 \over\sqrt{1- \gamma_t \rho}}  
\{ e^\chi - \gamma_t \mathfrak{a}_{pp} [ 1-\chi e^\chi \Gamma(0,\chi) ]+
{1\over 2} \gamma_t^2 \mathfrak{a}_{pp} \sqrt{r_1 \rho} \}- {1\over 2g_A }\gamma_t \mathfrak{a}_{pp} \sqrt{1-\gamma_t \rho} \ L^{\rm sd-2b}_{1,A}.
\label{eq:Lambda-0}
\end{eqnarray}
Here, $\chi=\alpha M_p /\gamma_t$, where $\alpha$ is the QED fine-structure constant, $\Gamma(0,\chi)$ is the incomplete gamma function, $\mathfrak{a}_{pp}$ is the $pp$ scattering length, and $\rho$ is the effective range in the $\siii$ channel expanded around the deuteron pole. The solely two-nucleon short-distance axial coupling $L^{\rm sd-2b}_{1,A}$ is linearly dependent upon the $L_{1,A}$ coupling via a known relation, see Ref.~\cite{Savage:2016kon}, and its determination is the goal of the LQCD study discussed below.

A first LQCD determination of the $pp$-fusion process was presented in Ref.~\cite{Savage:2016kon}.
The transition between two-nucleon systems in the isosinglet and isotriplet channels have been studied from background-field correlation functions that are generalizations of those presented in Eq.~\eqref{eq:Coflambda} to transitions between different states. These correlation functions are constructed from extended propagators with an insertion of the axial current, i.e., $\Gamma=\gamma_3\gamma_5$ in Eq.~\eqref{eq:extendedprop}. In particular, transition correlation functions $C_{\lambda_u;\lambda_d=0}^{(\si,\siii)}(t)$ and $C_{\lambda_u=0;\lambda_d}^{(\si,\siii)}(t)$ can be shown to vanish for $\lambda_{u,d} = 0$, and to be third-order polynomials in $\lambda_{u}$ and $\lambda_{d}$, respectively. Calculations of the correlation function at at least three values of $\lambda_{u(d)}$, as well as the analogous correlation functions for the time-reversed transition, allow the extraction of the linear terms, $\left. C^{(\siii,\si)}_{\lambda_{u(d)};\lambda_{d(u)}=0}(t)\right|_{{\cal O}(\lambda_{u(d)})}$ and the time-reversed analog, from which the transition matrix element of the axial current  can be determined. Explicitly,  these correlation functions can be shown to have the form 
\begin{align}
\left.
C^{(\siii,\si)}_{\lambda_u;\lambda_d=0}(t)\right|_{{\cal O}(\lambda_u)} 
&=
Z_d Z_{np(\si)}^\dagger e^{-\bar{E}t}
\Bigg[
\sinh\left(\frac{t\Delta}{2}\right) \left\{ \frac{\langle d| \tilde{J}_3^{(u)} | np(\si) \rangle}{\Delta/2} + c_-\right\}
\nonumber \\
& \hspace{4.5cm}+ \cosh\left(\frac{t\Delta}{2}\right) c_+
+
{\cal O}(e^{-\tilde{\delta}\, t}) \Bigg],
\label{eq:C1s03s1}
\end{align}
and similarly for $\left. C^{(\siii,\si)}_{\lambda_u=0;\lambda_d}(t)\right|_{{\cal O}(\lambda_d)}$ under the replacement $\tilde{J}_3^{(u)} \to \tilde{J}_3^{(d)}$. Here, \begin{equation}\label{eq:Jtil}
\tilde{J}_3^{(f)} \equiv \int d^3x\, \bar{q}_f(\vec{x},t=0)\gamma_3\gamma_5 q_f(\vec{x},t=0),
\end{equation}
and $|np(\si) \rangle$ and $|d \rangle$ refer to the ground state of the isotriplet channel and to the $m=0$ component of the deuteron, respectively. $Z_d$ and $Z_{np (\si)}$ are the overlap factors of the source and sink interpolating operators onto the ground states of the $\siii$ and $\si$ channels, respectively. $\Delta=E_{np(^1S_0)}-E_d$, $\bar{E}=(E_{np(^1S_0)} +E_d)/2$, and $\tilde{\delta}$ denotes the generic gap between the ground states and first excitations of the two-nucleon systems. $c_{\pm}$ are $t$-independent factors involving energy gaps, ratios of overlap factors, and transition matrix elements between the ground and excited states. 
As is evident from Eq.~(\ref{eq:C1s03s1}), in the limit of exact SU(4) Wigner symmetry in which 
$\Delta\to 0$, the correlation function receives no contribution from $c_-$. 
Away from this limit, the $c_-$ term contaminates the extraction of the ground-state to ground-state matrix element, $\langle d| \tilde{J}_3^{(u)} | np(\si) \rangle$, and its effects must be estimated carefully.

Given the transition correlation functions, the isovector ratio
\begin{equation}
R^\pm_{\siii,\si}(t)=\frac{1}{2} \frac{\left . C^\pm_{\lambda_{u};\lambda_{d}=0}(t)\right|_{{\cal O}(\lambda_u)}
	-
	\left. C^\pm_{\lambda_{u}=0;\lambda_{d}}(t)\right|_{{\cal O}(\lambda_d)} }{\sqrt{C_{\rm 2pt}^{(\siii)}(t)C_{\rm 2pt}^{(\si)}(t)}}
\label{eq:Rpm}
\end{equation}
can be formed, where 
\begin{eqnarray}
\left.C^\pm_{\lambda_u;\lambda_d=0}(t)\right|_{{\cal O}(\lambda_u)} =
\frac{1}{2}\left[  
\left.C^{(\si,\siii)}_{\lambda_u;\lambda_d=0}(t)\right|_{{\cal O}(\lambda_u)}
\pm
\left.C^{(\siii,\si)}_{\lambda_u;\lambda_d=0}(t)\right|_{{\cal O}(\lambda_u)}
\right],
\label{eq:cpm}
\end{eqnarray}
and a similar expression defines $C^\pm_{\lambda_{u}=0;\lambda_{d}}(t)\big|_{{\cal O}(\lambda_d)}$.
The overall exponential behavior in 
Eq.~\eqref{eq:C1s03s1} cancels in this ratio, in analogy to the general procedure described in Sec.~\ref{subsec:backgroundfield}. 
The ground-state transition matrix element can be isolated as the coefficient of the term linear in $t$ in $R^+_{\siii,\si}(t)$ using
\begin{eqnarray}
\overline{R}^+_{\siii,\si}(t) \equiv \frac{1}{a}\left[R^+_{\siii,\si}(t+a)-R^+_{\siii,\si}(t)\right] &\stackrel{t \to \infty }{\longrightarrow}& \frac{1}{Z_A}
\langle d,3| \tilde{J}_3^{+}  | pp\rangle+\mathcal{O}\left( \frac{1}{N_c^4} \right),
\label{eq:pptodeff}
\end{eqnarray}
where $Z_A=0.867(47)$~\cite{Savage:2016kon} is the axial-current renormalization factor and where isospin symmetry has been used to relate the $\langle d,3| \tilde{J}_3^{+}  | pp\rangle$ and $\langle d| \tilde{J}_3^{(u)} | np(\si) \rangle$ matrix elements.
While Wigner symmetry is not exact, $\Delta\sim {\cal O}(1/N_c^2)$ in the large-$N_c$ limit of QCD, and the time-reversal even ($T$-even) combination $\left.C^+_{\lambda_u;\lambda_d}(t)\right|_{{\cal O}(\lambda_f)}$ can be argued to receive only $1/N_c^4\sim 1$\% corrections, see Ref.~\cite{Tiburzi:2017iux}. Additionally, the $T$-odd combination, $\overline{R}^-_{\siii,\si}(t)$, defined analogously using $R^-_{\siii,\si}(t)$, can provide a numerical estimate of the magnitude of the $\cO(1/N_c^4)$ contamination.

\begin{figure}[!t]
	\centering
	\includegraphics[scale=0.568]{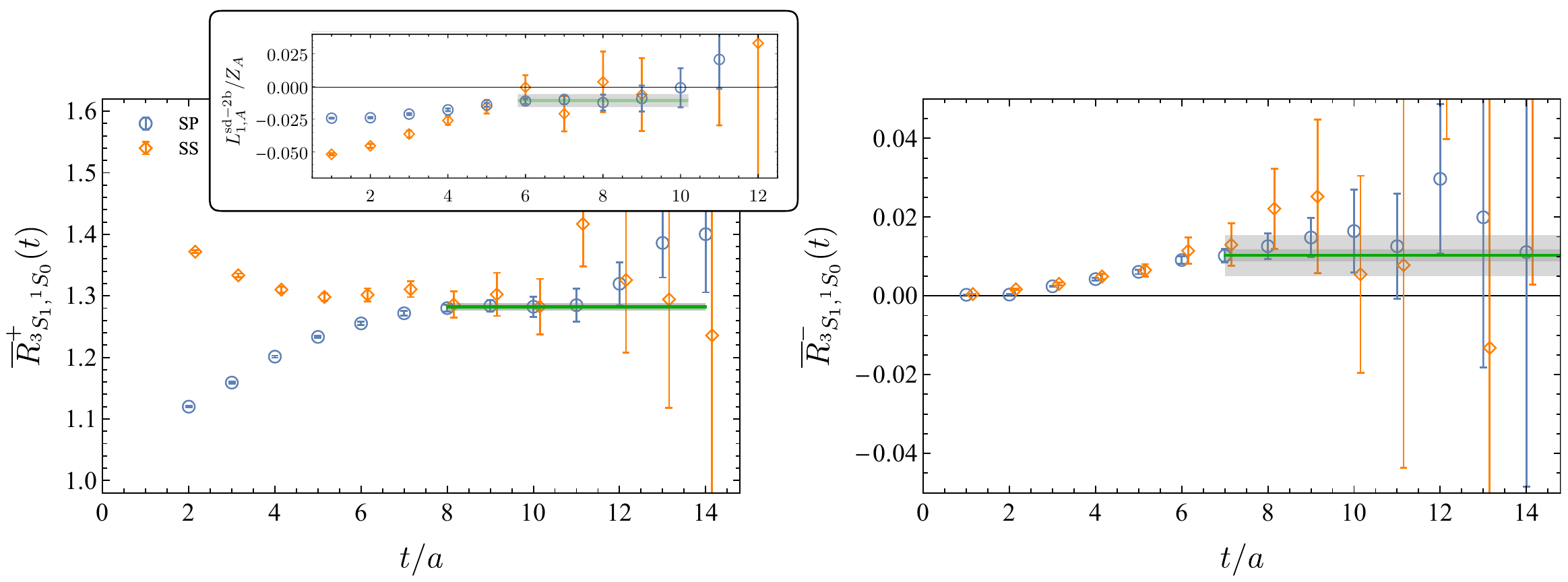}
	\caption{The quantities $\overline{R}^+_{\siii,\si}(t)$ (left) and $\overline{R}^-_{\siii,\si}(t)$ (right). The former asymptotes to the $pp\to d$ bare matrix element at large times, while the latter gives an estimate of dominant excited-state contaminations to the desired matrix element. Shown in the inset in the left is the  (unnormalized) difference between the full matrix element and the single-nucleon contribution, giving rise to quantity $L^{\rm sd-2b}_{1,A}/Z_A$ at large times. Different colors (blue and orange) represent two different choices for the sink interpolating operators. For further detail, see Refs.~\cite{Savage:2016kon,Shanahan:2017bgi}. 
		[Figure modified from Ref.~\cite{Savage:2016kon}.]}
	\label{fig:pptod}
\end{figure}

The main results of the LQCD study are reproduced in Fig.~\ref{fig:pptod}. This study made use of the same ensemble of isotropic clover gauge-field configurations that were discussed in the previous section, with SU(3)$_f$-symmetric quark masses corresponding to $m_\pi=806$ MeV, a lattice volume of $32^3\times48$, and a lattice spacing of $a\sim0.145$ fm. The quantity $\overline{R}^+_{\siii,\si}(t)$, which asymptotes to the $pp\rightarrow d$ axial transition matrix element at large times, is shown in the left panel. Shown in the inset is the (unrenormalized) matrix element when the single-body contribution is subtracted out, giving rise to the quantity $L^{sd-2b}_{1,A}/Z_A$.  In addition, the quantity $\overline{R}^-_{\siii,\si}(t)$ is shown in the right panel. This quantity asymptotes to a value that estimates the effects of excited states contaminating the extraction of the $pp\to d$ transition matrix element, and is seen to be small. This provides further support for the claim that the contribution of $c_-$ in Eq.~(\ref{eq:C1s03s1}) is $\cO(1/N_c^4) \sim \cO(1\%)$ of the dominant term. 

Without results at quark masses closer to the physical quark masses, a controlled determination of the LEC $L_{1,A}$ is not possible. Nonetheless, the mild quark-mass dependence of the similar LEC in the two-nucleon sector for the vector-current transitions, as discussed in Sec.~\ref{subsec:npdgamma}, suggests that the constraint obtained on $L_{1,A}$ at $m_{\pi} = 806$ MeV may be close to its value in nature. By assigning a conservative $50\%$ additive uncertainty for quark-mass extrapolation to the physical quark masses,  and combining the extracted matrix element with Eqs.~\eqref{eq:ppMATdef}--\eqref{eq:Lambda-0}, Ref.~\cite{Savage:2016kon} obtains a constraint on the $L_{1,A}$ coupling in nature evaluated at the scale $\mu = m_{\pi}^{\text{phys}}$:
\begin{eqnarray}
L_{1,A}= 3.9(0.2)(1.4)~{\rm fm}^3,
\label{eq:L1Anplqcd}
\end{eqnarray}
where the quoted uncertainties are statistical and systematic, respectively. This result can be compared with a recent phenomenological value obtained from $(\bar{\nu})\nu-d$ scattering: $L_{1,A}=4.9^{+(1.9)}_{-(1.5)}~{\rm fm}^3$ \cite{Acharya:2019fij}. Clearly the LQCD and phenomenological results are compatible and have similar overall uncertainties. Future LQCD studies of $pp$ fusion offer the prospect of significantly improving on the phenomenological result and thereby better constraining the solar burning process and concomitant neutrino emission. 

As discussed in Sec.~\ref{sec:directmapping}, as one approaches lighter quark masses, nuclear binding decreases towards the physical values~\cite{Beane:2012vq,Orginos:2015aya,Yamazaki:2015asa}, and in particular the two-nucleon system will be only slightly bound (in the isosinglet channel) or unbound (in the isotriplet channels). This difference makes determination of nuclear matrix elements at the physical point more involved, requiring the finite-volume technologies developed in e.g., Refs.~\cite{Briceno:2012yi,Briceno:2015tza} to relate the LQCD matrix element to its counterpart in an infinite Minkowski spacetime.

\subsection{Tritium $\beta$ decay}
Tritium $\beta$ decay, the $^3\text{H} \to {^3}\text{He}\,e^- \bar{\nu}$ process, is a super-allowed process in the SM. Increasingly precise measurements of its rate and of its final-state energy spectrum, such as at the Karlsruhe Tritium Neutrino experiment (KATRIN), will lead to precise constraints on the absolute mass of the electron neutrino~\cite{Aker:2019uuj,Drexlin:2013lha}. Such precise measurements will also provide a valuable means to search for signals of new physics~\cite{Stephenson:2000mw,Bonn:2007su,Ludl:2016ane,Arcadi:2018xdd} such as sterile neutrinos~\cite{Mertens:2014nha,Barry:2014ika}. Furthermore, in EFT, the Gamow-Teller matrix element contributing to this process shares common LECs  with the $pp$-fusion process discussed in the previous subsection, as well as with other weak processes in the two-nucleon system, providing a means to constrain unknown LECs such as $L_{1,A}$. A systematic treatment of the decay within EFT~\cite{Klos:2016omi,Baroni:2016xll,Baroni:2018fdn,De-Leon:2016wyu} will thus allow the two-nucleon short-distance contributions to the $\beta$ decay of larger nuclei to be quantified. Historically,  Gamow-Teller transitions measured in medium-mass nuclei have been challenging to reproduce from  theory \cite{Wildenthal:1983zz,Buck:1975ae,MartinezPinedo:1996vz}, initially requiring an \textit{ad hoc} modification of the axial charge of the nucleon in nuclei, known as quenching of the axial charge~\cite{Towner:1987zz}, to account for differences of tens of percent from single-nucleon estimates in which nuclear ground states with non-interacting nucleons occupy only the lowest shell-model states. Recently this problem has been resolved in some nuclei by a complete EFT treatment~\cite{King:2020wmp,Baroni:2018fdn,Gysbers:2019uyb} including both a two-nucleon axial coupling and correlations in the nuclear wavefunctions. LQCD studies can constrain the Gamow-Teller matrix element contributing to tritium $\beta$ decay directly, which will provide a valuable check on the phenomenological constraints. 
As is the case for the axial charge of the nucleon, a sufficiently precise calculation of the Gamow-Teller matrix element in the triton could be compared to phenomenological extractions from experiment and thereby serve as a test of the SM~\cite{Cirigliano:2019jig}, although the precision requirements are challenging.

The first LQCD determination of the Gamow-Teller matrix element relevant to the tritium $\beta$ decay, defined as $\langle {}^3{\rm He}| \bar{q}\gamma_k \gamma_5 \tau^+ q | {}^3{\rm H}\rangle \equiv \bar{u} \gamma_k \gamma_5 \tau^+ u \ g_A \langle{\bf GT}\rangle$ (where $u$ and $\bar{u}$ denote the $^3$He and $^3$H spinors) was performed in Ref.~\cite{Savage:2016kon}. The reduced matrix element, $\langle{\bf GT} \rangle$, along with the Fermi contribution,  $\langle{\bf F} \rangle$, from the vector-current matrix element, determines the half life of the decay, $t_{1/2}$, as~\cite{Schiavilla:1998je}
\begin{eqnarray}
\frac{(1+\delta_R) f_V }{ K/G_V^2} t_{1/2} = \frac{1}{\langle {\bf F} \rangle^2 +f_A/f_V\,g_A^2\langle {\bf GT}\rangle^2}.
\label{eq:halftime}
\end{eqnarray}
The Fermi contribution is constrained to be unity in the limit of exact isospin symmetry, and deviations from that limit are estimated to be at the sub-percent level by the Ademollo-Gatto theorem~\cite{Ademollo:1964sr}. From phenomenology, the most precise constraint on the Gamow-Teller matrix element is  $\langle{\bf GT}\rangle^{\text{phys}} = 0.9511(13)$~\cite{Baroni:2016xll}.  All other factors in Eq.~(\ref{eq:halftime}) are known precisely from experiment. 

In the isospin limit, the Gamow-Teller matrix element can be obtained from the axial charge of the triton in the same way that the neutron $\beta$ decay amplitude is related to the nucleon's vector axial charge. This quantity can be computed using the extended-propagator technique of Sec.~\ref{subsec:backgroundfield} with an insertion of the axial-vector current.
This determines the desired matrix element from the solution to a set of polynomial equations for different values of the background-field strengths $\lambda_{u(d)}$.  After forming appropriate ratios of three- and two-point functions, as outlined in Sec.~\ref{subsec:backgroundfield}, the (unrenormalized) axial charge of the triton is obtained from a constant fit to the linear time dependence of the isovector combination $\frac{1}{2}(R^{{^3}\text{H}}_{\cO_u}-R^{{^3}\text{H}}_{\cO_d})$ at large Euclidean time, where $R^{{^3}\text{H}}_{\cO_f}$ is defined in Eq.~(\ref{eq:RO}) and $\cO_f= \bar{q}_f\gamma_3\gamma_5 q_f$. Further, this quantity can be divided by the analogous quantity for the proton, which gives access to the reduced Gamow-Teller matrix element, $\langle{\bf GT}\rangle$, at large Euclidean times. This ratio is independent of the renormalization of the lattice axial-vector current.

\begin{figure}[!t]
	\includegraphics[scale=0.624]{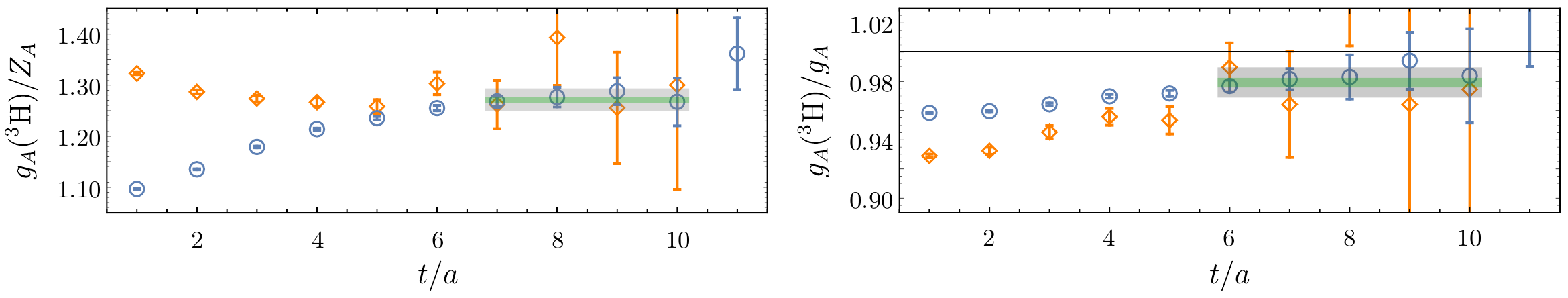}
	\caption{\label{fig:gAtrit} 
		Ratios of correlation functions that asymptote to the (unrenormalized) isovector axial matrix element in $^3$H at large times (left), and ratio of the isovector axial matrix element in $^{3}$H to that in the proton (right), with a horizontal line shown at 1 to guide the eye. Different colors (blue and orange) represent two different choices for the sink interpolating operators. [Figure modified from Ref.~\cite{Savage:2016kon}.]}		
\end{figure}

In Ref.~\cite{Savage:2016kon}, the first study of these quantities was reported using the same gauge-field ensembles as described above for the $pp$-fusion study, as shown in Fig.~\ref{fig:gAtrit}. That study obtains
\begin{align}
g_A({}^3{\rm H})/Z_A &= { 1.272(6)(22) }, \\
\langle{\bf GT}\rangle &=  { 0.979(3)(10) },
\label{eq:GTnplqcd}
\end{align}
at $m_{\pi} = 806$ MeV.
In both cases, the first uncertainty is statistical and the second is the systematic uncertainty from variations in the fit windows chosen and the analysis techniques applied. 
The value of the Gamow-Teller matrix element in tritium $\beta$ decay that is obtained is within three percent of the experimental value. This is  an interesting observation, considering that the values of the quark masses used in this study are far from those in nature. This points to the same mild quark-mass dependence as that observed for the Gamow-Teller matrix element of the proton, as well as for other structure properties of the nucleon and light nuclei that have been studied so far, such as the magnetic moments (when normalized in appropriate units) and the two-body contribution to the $np\to d\gamma$ transition rate, as discussed in Sec.~\ref{sec:EM}.

\subsection{Flavor-separated axial charges of light nuclei}

\begin{figure}[!t]
	\centering
	\includegraphics[scale=0.7]{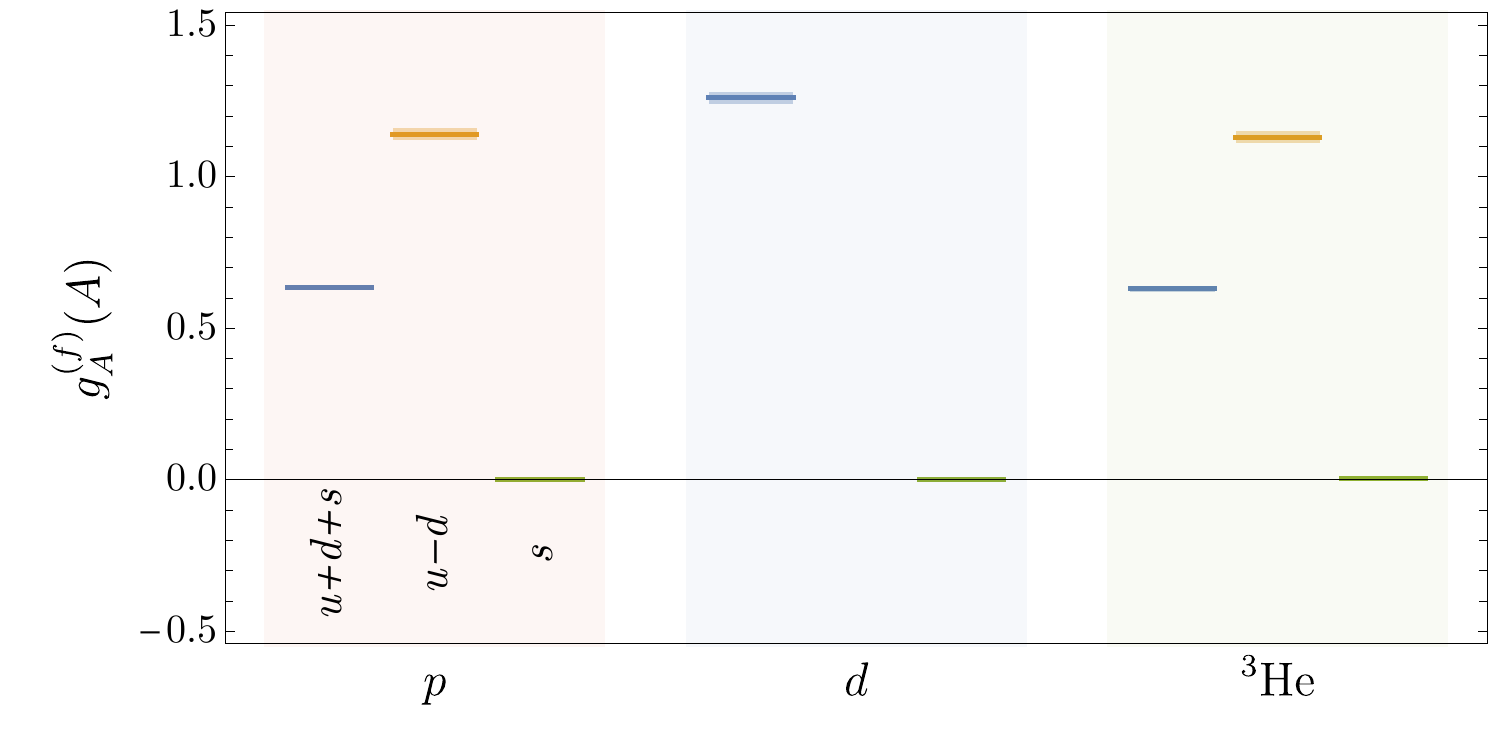}
	\caption{Flavor separation of the axial-current matrix elements in the proton, deuteron, and $^3$He, determined in Ref.~\cite{Chang:2017eiq} using calculations with $m_\pi =806$~MeV. Since the deuteron is isoscalar, it has no isovector charge. The strange quark charges are very small and consistent with zero in each case.
		[Data from Ref.~\cite{Chang:2017eiq}.]}
	\label{fig:flavour_axial}
\end{figure}

The calculations described above have also been extended from the isovector current to a full flavor decomposition of the axial current in the proton, deuteron, and $^3$He  systems~\cite{Chang:2017eiq}. The study of Ref.~\cite{Chang:2017eiq} was performed using the same background-field approach as described above for the connected quark contractions, and made use of hierarchical probing \cite{doi:10.1137/120881452,Gambhir:2016uwp} for calculation of the disconnected quark contractions. The resulting flavor-dependent charges are shown in Fig.~\ref{fig:flavour_axial}. As can be seen, the axial charges of the nuclei scale approximately with the total spin of the system, as would be the case for non-interacting nucleons in the lowest shell-model states. As for the isovector charge in tritium $\beta$ decay, deviations from this scaling are resolved from zero and shown in Fig.~\ref{fig:SATMEs} in Sec.~\ref{sec:BSM}.

\subsection{Hadronic parity violation}
\label{subsec:parityviolation}

Parity violation in the SM, while understood precisely at the quark and lepton level, is poorly constrained at the hadronic level due to the complicated non-perturbative nature of the strong interactions. Prior to the  UA1 and UA2 discovery of the neutral weak gauge boson~\cite{Arnison:1983mk,Bagnaia:1983zx}, hadronic parity violation, using the quantum numbers of nuclear levels as filters, provided  a prime laboratory to search for neutral weak currents~\cite{Walecka:1977us}.  While this motivation is no longer relevant, the emphasis of hadronic studies of parity violation has changed from driving a fundamental discovery to exploring their potential for high-sensitivity processes.
Neutral-current weak interactions are more challenging to investigate experimentally than the charged-current weak interactions. Given the $\cO(10^{-7})$ suppression of the weak-interaction effects compared with strong-interaction effects, most nuclear parity-violation experiments have been focused on larger nuclei since  enhancements of parity-violating observables are expected (and found) in a nuclear medium (see Ref.~\cite{Haxton:2013aca} for a review). However, the connection to the quark and gluon degrees of freedom of the SM is challenging to establish for hadronic parity-violating processes in  nuclei. 
Recently, two-nucleon systems have been the focus of experimental investigations, such as searches for the longitudinal asymmetry in polarized $\vec{p}p$ scattering~\cite{Eversheim:1991tg} and the photon asymmetry in $\vec{n}p \to d \gamma$~\cite{Blyth:2018aon}. These processes are more straightforward to connect to the underlying SM mechanism. In this context, EFTs have provided a systematic framework to identify leading parity-violating interactions at low energies~\cite{DANILOV196540,Desplanques:1976mt,Dai:1991bx, Kaplan:1992vj, Savage:1998rx, Savage:1999cm, Phillips:2008hn, Tiburzi:2012hx, deVries:2013fxa, Viviani:2014zha, Phillips:2014kna, Schindler:2015nga, Vanasse:2018buq, Guo:2018aiq}, improve upon conventional phenomenological approaches~\cite{DESPLANQUES1980449}, and pave the way for a systematic extension to larger nuclei. These EFTs, however, introduce new two-nucleon LECs that need to be constrained by experimental data where it exists, or by LQCD. A first LQCD attempt in constraining the LO parity-violating pion-nucleon coupling, needed for constructing the two-nucleon potential with $\Delta I =1$, was conducted in Ref.~\cite{Wasem:2011zz}, albeit subject to large statistical and systematic uncertainties.

Based on chiral and large-$N_c$ arguments, it has recently been realized that the dominant contributions to parity violation in two-nucleon systems arise from $\Delta I \in \{0,2\}$ operators, rather than the $\Delta I=1$ operator that was considered before~\cite{Phillips:2014kna, Schindler:2015nga}.  In fact, as was shown in Ref.~\cite{Schindler:2015nga}, the current experimental results on the longitudinal asymmetry in $\vec{p}p$ scattering already constrain a  linear combination of the LECs, effectively reducing the number of unknown LECs to one. This suggests that the $\Delta I =2$ channel is most relevant from the LQCD perspective, where the lack of quark-line disconnected diagrams provides a computational simplification. An exploratory LQCD study in this channel has been attempted, with preliminary results reported in Ref.~\cite{Kurth:2015cvl}, and the renormalization of relevant operators studied in Ref.~\cite{Tiburzi:2012xx}. Precise LQCD constraints on the low-energy EFTs or phenomenological models of hadronic parity violation in the single- and two-nucleon sectors will help explain and predict parity violation in large nuclei.

\subsection{Future impact}
The examples of determinations of weak-interaction nuclear matrix elements from LQCD discussed in this section demonstrate the powerful interplay between LQCD and EFT descriptions of the weak matrix elements.  Over the next decade, it is likely that the weak-interaction processes considered here, and others, will be studied directly from LQCD in nuclei with $A \leq 4$, at or near to the physical values of the quark masses, with fully controlled uncertainties. With sufficient precision, a meaningful connection to nuclear phenomenology, especially in the context of nuclear astrophysics, will be possible with the use of the  EFTs and phenomenological models. While for $L_{1,A}$ the necessary precision is known to be $\sim 10\%$, in other cases, the crucial EFT input entering different processes must be identified, and the target uncertainty of the corresponding LQCD determinations is yet to be defined. A notable aspect of LQCD is that it gives direct access to nuclear matrix elements at given kinematics and does not rely on model assumptions, such as the threshold expansion in the $pp$-fusion process. It can therefore provide a means to test such assumptions and to give access to higher-order corrections. LQCD can also calculate the rate of weak reactions of light hypernuclei that are relevant to experiments at J-PARC~\cite{Tamura:2012dka,Achenbach:2019shq} as well as for astrophysical investigations into the nature of dense matter~\cite{SchaffnerBielich:2008kb}.

Finally, it is notable that the progress in the nuclear-reaction program from LQCD has direct impact on the ongoing theoretical studies of neutrino-nucleus scattering cross-sections for e.g, long baseline underground neutrino experiments such as the Deep Underground Neutrino Experiment (DUNE)~\cite{Acciarri:2016crz,Abi:2020wmh} and Hyper-Kamiokande~\cite{Abe:2011ts,Yokoyama:2017mnt}, which aim to constrain the neutrino mass hierarchy and oscillation parameters with unprecedented precision by reconstructions of the neutrino energy from the final state. An essential theory input into this reconstruction is the axial charge, and more generally the axial radii and form factors, of the nucleus used in the detector. Constraining the axial form factors of the relevant nuclei, such as Argon, from the SM is thus a critical goal, as are studies of more complex subprocesses such as $N\rightarrow\Delta$ and $N\rightarrow N\pi$ axial transition matrix elements.
QCD input on weak matrix elements in light nuclei will be essential for constraining EFTs and phenomenological models capable of describing the experimentally-relevant medium-mass nuclei and grounding predictions for neutrino-nucleus scattering in the SM~\cite{Kronfeld:2019nfb}.

Future calculations of hadronic parity-violating transitions will also be useful in understanding broader aspects of the weak interactions in nuclei \cite{Schindler:2013yua,Haxton:2013aca}. 
Extensions of such studies to the {anapole moment} \cite{Haxton:2001ay} that corresponds to a $P$-odd, $T$-even, transverse spin-dependent interaction with an external EM field will be valuable in the context of understanding atomic and nuclear parity-violation experiments and constraining parity-violating pion-nucleon and nucleon-nucleon interactions.

\section{Neutrinoful and neutrinoless double-$\beta$ decays}
\label{sec:dbd}

\begin{figure}
	\centering
	\includegraphics[width=0.5\columnwidth]{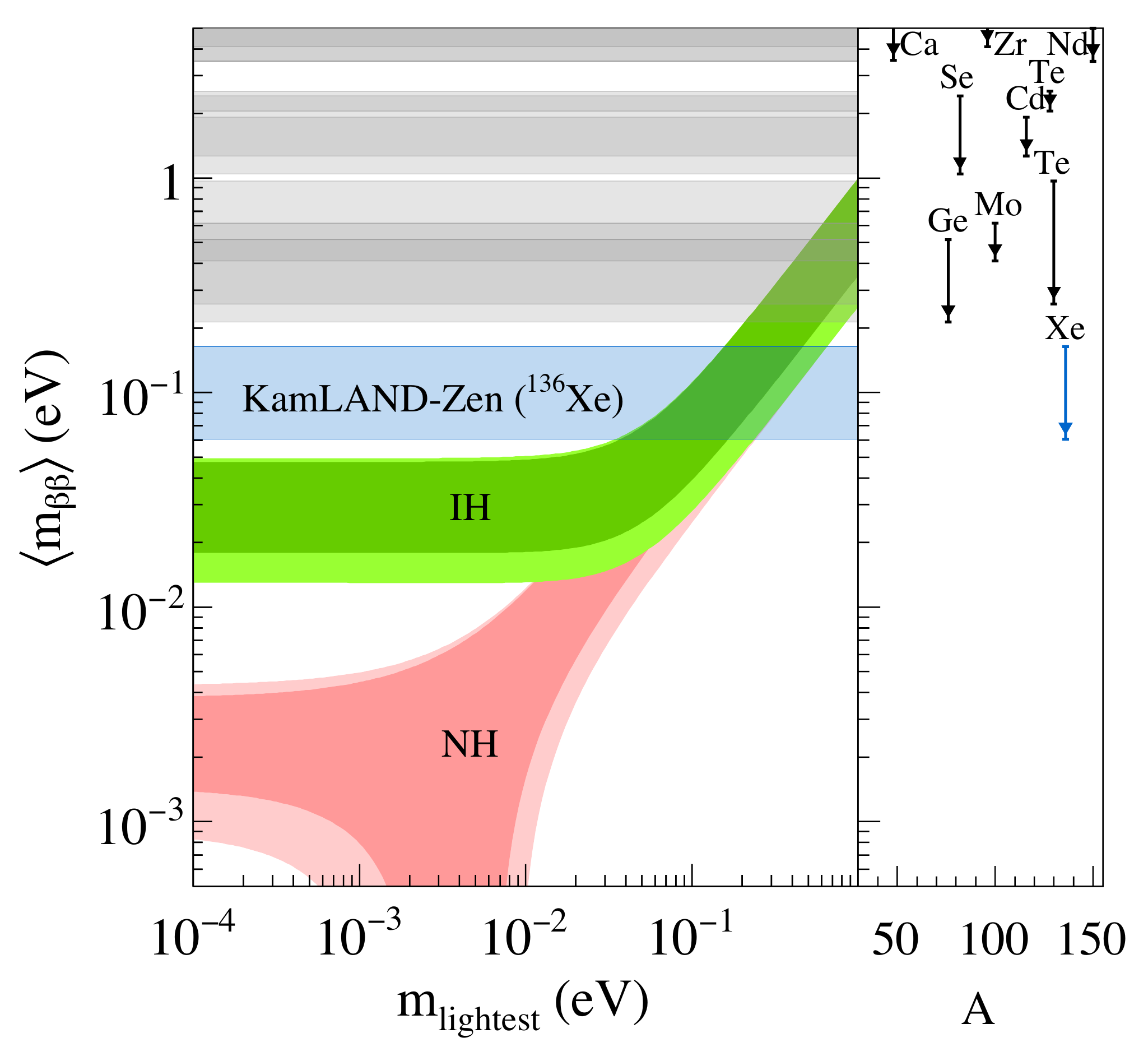}
	\caption{Constraints on, and expectations for, the quantity $m_{\beta\beta}$ (defined in the text) and the lightest neutrino mass, $m_{\rm lightest}$. 
		The pink and green regions show the allowed values consistent with neutrino-oscillation phenomenology for both the normal (NH) and inverted (IH) hierarchies of neutrino masses. In the normal hierarchy, the two neutrino masses with the smaller splitting are smaller
		than the third mass, while the inverted hierarchy refers to the opposite scenario.
		The blue and gray bands show the limits on $m_{\beta\beta}$ from searches using various nuclei. The width of these bands indicates estimated uncertainties in relating bounds on half-lives to $m_{\beta\beta}$, which are dominated by uncertainties in the requisite nuclear matrix elements. (Figure from Ref.~\cite{KamLAND-Zen:2016pfg}) 
		\label{fig:znubbconstraints}}
\end{figure}

Neutrinoful double-$\beta$ decays of nuclei of atomic number $A$ and proton number $Z$, $(A,Z) \to (A,Z+2)\, e^- e^- \bar\nu_e \bar\nu_e$,  are the rarest subatomic  processes observed experimentally. They serve as   intricate tests of our understanding of the physics of the weak interactions of nuclei and enable probes of deficiencies in that understanding. These decays   occur through  two SM weak transitions and are only observable in the handful of nuclei where single-$\beta$ decay is energetically forbidden. Using  sensitive experimental techniques, \tnubb\ has been observed for about a dozen nuclei, with half-lives $\tau \sim {\cal O}(10^{21})\ {\rm yr} $ \cite{Engel:2016xgb}.
The neutrinoless decay mode, $(A,Z) \to (A,Z+2)\, e^- e^- $, is also sought in experiment.
This mode requires lepton number violation (LNV) by two units; since the difference between baryon number $B$ and lepton number $L$ is conserved in the SM, observation of this $(B-L)$-violating process would imply new physical principles in nature and potentially elucidate a central aspect of the origin of the matter-antimatter asymmetry of the universe. Furthermore, an observation of this process would immediately imply that neutrinos are Majorana particles~\cite{Schechter:1981bd}, and could provide insight into the absolute scale of neutrino masses and the mechanism(s) for neutrino mass generation. Consequently, an extensive program of experiments
~\cite{Gando:2012zm,Agostini:2013mzu,Albert:2014awa,Andringa:2015tza,KamLAND-Zen:2016pfg,Elliott:2016ble,Agostini:2017iyd,Aalseth:2017btx, Albert:2017owj,Alduino:2017ehq,Agostini:2018tnm, Azzolini:2018dyb,Anton:2019wmi} have sought, and continue to seek, evidence for \znubb\  decays. At this time, the \znubb\ lifetime of ${}^{136}$Xe  is bounded by $t^{0\nu}_{1/2}>1.07\times 10^{26}$ yr
\cite{KamLAND-Zen:2016pfg}, and next-generation, ton-scale, experiments 
aim to increase sensitivity by up to two orders of magnitude. 
A recent summary of constraints on the combination $m_{\beta \beta} = \left|\sum_i  U_{ei}^2 m_i\right|$ as a function of $m_{\rm lightest}=\min_i m_i$ is reproduced in Fig.~\ref{fig:znubbconstraints}. Here, $m_i$ are the neutrino masses and $U$ is the Pontecorvo-Maki-Nakagawa-Sakata  (PMNS) matrix \cite{Pontecorvo:1967fh,Maki:1962mu}. Note that in obtaining these constraints, it is assumed that the primary mechanism for the \znubb\ decay is that involving the light neutrinos coupling to matter through the left-handed weak currents of the SM. Other mechanisms are possible, as discussed below, and each scenario requires certain nuclear matrix elements to be constrained from theory.

For both  \tnubb\ and \znubb\ decays,  critical components in determining the decay rate are the nuclear matrix elements of the interactions that give rise to the decay.   In recent years, LQCD studies of  \tnubb\ and \znubb\ decay matrix elements in various scenarios have begun and will be discussed below.  Combined with EFT methods and phenomenological models, these calculations promise to better constrain nuclear inputs for these processes. Refs.~\cite{Engel:2016xgb} and \cite{Cirigliano:2020yhp} provide a more comprehensive review of the connections of this LQCD effort to nuclear many-body approaches and phenomenology.

\subsection{Neutrinoful double-$\beta$ decay}

\label{sec:tnubb}

The \tnubb\ decay mode is a crucial test of our understanding of weak interactions in nuclei. In particular, deviations of \tnubb\ rates from the naive scaling from single-$\beta$ decay are observed and these differences are difficult to incorporate in phenomenological nuclear models \cite{Engel:2016xgb}.
Achieving controlled predictions of $2\nu\beta\beta$ decay rates from the SM is a challenging goal, as the nuclei which undergo this decay are too large for the direct application of LQCD in the foreseeable future. On the other hand, the more phenomenological many-body methods which have so far been applied to this problem have  significant model-dependent uncertainties that are difficult to quantify. To improve the reliability of these predictions, a promising path is to use LQCD to constrain and test EFTs and phenomenological models that are able to access experimentally-relevant nuclei.
Initial steps have recently been made towards this goal with  the first LQCD calculation of the second-order weak double-$\beta$ decay matrix element of the two-nucleon system, $nn\to pp e^-e^-\bar\nu_e \bar\nu_e$~\cite{Shanahan:2017bgi,Tiburzi:2017iux}. Future calculations of this and other few-body \tnubb\ decay matrix elements,  that are sufficiently precise and systematically controlled, will allow the free parameters of  nuclear EFTs and phenomenological models to be constrained from LQCD. Through this approach, the model dependence that exists beyond the $g_A^2$ contributions in current  many-body calculations of \tnubb\ decay rates will be reduced, and the associated uncertainties will be quantified more reliably.

The second-order weak interaction responsible for \tnubb\ is dominated by  the 
Gamow-Teller (axial-vector) piece of the weak current since the Fermi (vector) piece is suppressed by isospin symmetry.
Neglecting lepton-mass and isospin-breaking effects, the inverse half-life of the neutrinoful double-$\beta$ decay is expressed as~\cite{Engel:2016xgb}
\begin{eqnarray}
[t^{2\nu}_{1/2}]^{-1}=
G_{2\nu}(E_i-E_f,Z_i) | {\cal M}_{\rm GT}^{2\nu}|^2,
\label{eq:MGT1}
\end{eqnarray}
where $G_{2\nu}(E,Z)$ is a known phase-space factor~\cite{Kotila:2012zza,Stoica:2013lka}, $Z_i$ is the proton number of the initial nuclear state, and $E_{i(f)}$ is the energy of the initial (final) state. The matrix element ${\cal M}_{\rm GT}^{2\nu}$ is defined by the time-ordered product of two axial-vector currents  between the initial and final states, $i$ and $f$, as
\begin{eqnarray}
{\cal M}_{\rm GT}^{2\nu}=
6 \times \frac{1}{2}
\int dt \,{d^3x} \, {d^3y} 
\, 
\langle f |  T \left[ J^+_{3}(\vec{x},t) J^+_{3}(\vec{y},0)  \right] | i \rangle = 6\sum_{{\mathfrak n}}\frac{\langle f |  \tilde{J}_3^+ |{\mathfrak n}\rangle\langle {\mathfrak n} |  \tilde{J}_3^+ |i \rangle}{E_{\mathfrak n}-(E_{i}+E_{f})/2},
\label{eq:MGT}
\end{eqnarray}
where the spatial component of the $\Delta I_3=1$ zero-momentum axial-vector current in the $x_3$-direction is expressed as 
\begin{equation}
{J}_3^a(x)= \bar{q}(x) \frac{\gamma_3\gamma_5}{2} \tau^a q(x),
\qquad \tilde{J}_3^a   = \int d^3 x\, {J}_3^a(\vec{x},0), 
\end{equation}
where $\tau$ denotes a Pauli matrix in isospin space, and $\tau^+ = \frac{1}{\sqrt{2}} \left(\tau^1 + i\; \tau^2 \right)$.\footnote{Note that this normalization is different by a factor of $\sqrt{2}$ to that used above Eq.~\eqref{eq:ppMATdef} in Sec.~\ref{subsec:ppfusion}, in order to match the conventions of Refs.~\cite{Shanahan:2017bgi,Tiburzi:2017iux}.} In Eq.~(\ref{eq:MGT}), the  complete set of zero-momentum energy eigenstates is indexed by ${\mathfrak n}$, and the factor of $6$ in ${\cal M}_{\rm GT}^{2\nu}$ is a consequence of rotational symmetry and the normalization convention of the currents.

In Refs.~\cite{Shanahan:2017bgi,Tiburzi:2017iux}, the first LQCD calculation of ${\cal M}_{\rm GT}^{2\nu}$ was presented, focusing on the $nn\to ppe^-e^-\bar\nu_e\bar\nu_e$ transition. While this transition is not observed  in nature, the matrix element is well defined and occurs as an off-shell  subprocess in double-$\beta$ decays of larger nuclei. In this first calculation, many of the systematic uncertainties of LQCD methods were estimated rather than quantified: electromagnetism was neglected, a single lattice spacing and volume were used, and degenerate up, down, and strange quark masses corresponding to a larger-than-physical pion mass of $m_\pi = 806$ MeV were used for computational expediency. While  these systematics need further exploration, already this work brought to light an important qualitative conclusion regarding the importance of contributions to the double-$\beta$ decay of nuclei that do not enter single-$\beta$ decays, namely the isotensor axial polarizability, $\beta_A^{(2)}$, of the $\si$ two-nucleon system. This is an analog of the electric polarizabilities discussed in Sec.~\ref{sec:EM} and is defined through
\begin{equation}
\frac{1}{6}{\cal M}_{\rm GT}^{2\nu}(nn\to ppe^-e^-\bar\nu_e\bar\nu_e) = \beta_A^{(2)} - \frac{ |\langle pp|\tilde{J}_3^+|d\rangle|^2} {E_{pp}-E_d},
\label{eq:axial-polz}
\end{equation}
i.e., explicitly removing the Born contribution involving an intermediate deuteron from the Gamow-Teller amplitude. This isotensor axial polarizability is an intrinsically two-nucleon effect. Theoretical calculations of double-$\beta$ decay rates with fully quantified uncertainties will require constraint of the isotensor axial polarizabilities of nuclei. Refs.~\cite{Shanahan:2017bgi,Tiburzi:2017iux} also outlined how LQCD  can provide input to many-body methods by constraining second-order electroweak properties of light nuclear systems through determining the leading $\Delta I = 2$ LEC of pionless EFT.

In Refs.~\cite{Shanahan:2017bgi,Tiburzi:2017iux}, the matrix elements relevant to \tnubb\ were determined via the background-field technique discussed in Sec.~\ref{subsec:backgroundfield}. 
Because of the isotensor ($\Delta I = 2$) nature of the $nn \rightarrow pp$ transition, and because only a single insertion of the $\Delta I = 1$ axial-current operators is allowed on any single quark line, no disconnected  contractions of quark fields are required to evaluate the relevant axial-current matrix elements. The required matrix elements are therefore constructed from correlation functions formed from propagators computed in a background field corresponding to a single axial-current insertion, as described in Sec.~\ref{subsec:backgroundfield}.
For an axial current $J_3^a(x)$, expanding the background-field correlation functions from Eq.~\eqref{eq:bfcorr} to second order in the insertion of the current leads to a form analogous to Eq.~\eqref{eq:quad1}, with the inserted operator being ${J}_3^a(x)$.
The second-order term in the field strength, proportional to $\lambda_u^2$, can be extracted from analysis of calculations of the background-field correlation functions at multiple field strengths  $\lambda_u$, while an analogous procedure can be followed for the response to a $d$-flavored external field. The background-field correlation function defined in this way involves sums over the insertion times $t_{\{1,2\}}$ of the two axial currents, which is sufficient for a determination of the matrix element of the $nn\rightarrow pp$ transition. More refined methods with limited time windows of insertion will likely be necessary in extensions of this approach to calculations of the double-$\beta$ decay transitions of larger nuclei, and at  lighter values of the quark  masses where the spectra of excitations with the same quantum numbers in the initial, intermediate, and finals states becomes more dense.  

Using the isospin relation:
\begin{eqnarray}
\langle p p | J^+_{3}(x) J^+_{3}(y) | n n \rangle 
=
\langle n p | J_3^{(u)}(x) J_3^{(u)}(y) | n p \rangle 
-\frac{1}{2} \langle n n | J_3^{(u)}(x) J_3^{(u)}(y) + J_3^{(d)}(x)  J_3^{(d)}(y) | n n \rangle
\label{eq:recipe},
\quad
\end{eqnarray}
the $\Delta I_3 =2$ correlation function of relevance to the $nn \to pp$ transition can be extracted from  flavor-diagonal background fields in which the extended propagators defined in Sec.~\ref{subsec:backgroundfield} only include a single current insertion. Consequently, 
\begin{align}\label{eq:Ct}
C_{nn\to pp}(t)
=a^{11}\sum_{{\vec x},{\vec y},{\vec z}}\sum_{t_{1,2}=0}^t 
\langle 0| \chi_{pp}({\vec x},t) T\left[J_3^{+} ({\vec y},t_1)  
J_3^{+} ({\vec z},t_2) \right]   \chi^\dagger_{nn}(0) |0 \rangle.
\end{align}
can be extracted by combining the coefficients of the terms quadratic in the field strength in Eq.~\eqref{eq:quad1} and its $d$-flavored analog using the state combinations prescribed in Eq.~\eqref{eq:recipe}.
The use of the background field allows for straightforward evaluation of the large number contractions implicit in Eq.~\eqref{eq:Ct}.
Inserting three complete sets of states between the source and the currents, between the two currents, and between the sink and the currents, Eq.~\eqref{eq:Ct} can be written as 
\begin{align} 
C_{nn\to pp}(t) =& \,
2V
\sum_{\mathfrak{n},\mathfrak{m}, \mathfrak{l}'} 
\langle 0| \chi_{pp} | \mathfrak{n}\rangle \langle \mathfrak{m}| \chi_{nn}^\dagger | 0\rangle
e^{- E_\mathfrak{n} t} 
\frac{
	\langle \mathfrak{n}| \tilde{J}_3^{+} | \mathfrak{l}' \rangle 
	\langle \mathfrak{l}' | \tilde{J}_3^{+}  | \mathfrak{m} \rangle
}{E_{\mathfrak{l}'}- E_\mathfrak{m}} 
\nonumber \\&
\hspace*{3cm}\times \left(
\frac{
	e^{- \left(E_{\mathfrak{l}'} - E_\mathfrak{n}\right) t} -1
}
{E_{\mathfrak{l}'} - E_\mathfrak{n}}
+
\frac{
	e^{\left(E_\mathfrak{n}-E_\mathfrak{m}\right)t}-1
}
{E_\mathfrak{n}-E_\mathfrak{m}}\right), 
\label{eq:corr}
\end{align}
where the zero-momentum energy eigenstates with the quantum numbers of the $pp$, $nn$ and deuteron systems are denoted as $|\mathfrak{n} \rangle$, $|\mathfrak{m}\rangle$, and $|\mathfrak{l}' \rangle$, respectively.\footnote{Equations \eqref{eq:corr} and \eqref{eq:Ct} have a different normalization than that in Refs.~\cite{Shanahan:2017bgi,Tiburzi:2017iux}, for consistency with the choices in earlier sections.} $E_{\mathfrak{l}'}$ and $E_\mathfrak{n}$ are the energies of the $\mathfrak{l}'$th and $\mathfrak{n}$th  states in the spin-triplet and spin-singlet channels. 
Once isolated, the time dependence of this correlation function allows the short-distance isotensor axial polarizability, defined in Eq.~\eqref{eq:axial-polz}, to be extracted.
The long-distance contribution from the intermediate-state deuteron pole is determined by the square of the magnitude of the  single-current matrix element $\langle pp| \tilde{J}_3^{+}  |d\rangle$, and can be extracted most precisely from the linear response to the background field in Eq.~\eqref{eq:quad1} as discussed in the previous section. By forming appropriate ratios \cite{Shanahan:2017bgi,Tiburzi:2017iux}, the ``short-distance'' combination of  contributions from intermediate states other than the deuteron can be extracted, corresponding to the isotensor axial polarizability $\beta_A^{(2)}$. Together, these combine to give the full double Gamow-Teller transition matrix element.

The energy gaps at the large quark masses used in Refs.~\cite{Shanahan:2017bgi,Tiburzi:2017iux} were such that the above separation between the various terms could be cleanly performed. However, at smaller quark masses (including the physical values), difficulties will arise in this method. Notably, the initial and final states will no longer be  bound, complicating the relationship between the finite-volume bi-local matrix elements and the infinite-volume transition amplitudes, so  the formalism presented in Ref.~\cite{Davoudi:2020xdv} will be required.

The double Gamow-Teller transition matrix element for the $nn\rightarrow pp$ process can be used to constrain LECs in EFT descriptions of the same system. As discussed in Sec.~\ref{sec:eft}, with that achieved, the potentially key contribution from the isotensor axial polarizability can be included in phenomenological models and EFT-based many-body calculations of the decay rates of the larger nuclei that are used in experiment. For LQCD calculations undertaken with the large quark masses of Refs.~\cite{Shanahan:2017bgi,Tiburzi:2017iux}, and the low-energy kinematic of this process, it is natural to consider a pionless-EFT description~\cite{Beane:2000fi,Phillips:1999hh}. In Refs.~\cite{Shanahan:2017bgi,Tiburzi:2017iux}, the dibaryon formulation of the pionless EFT~\cite{Beane:2000fi,Phillips:1999hh} was used to match the LQCD results for the axial polarizability at $m_{\pi} = 806$ MeV to the LECs of the EFT characterizing the second-order response of the dibayon (representing the two-nucleon isosinglet or isotriplet systems) to an axial-vector background field. A similar framework for matching pionless EFT to LQCD correlation functions for the double-$\beta$ decay process was developed with nucleon degrees of freedom in Ref.~\cite{Davoudi:2020xdv}.   

At LO in the EFT, the interactions between the dibaryon field and the background field are momentum independent, and at first order in the background-field strength they are of the following form~\cite{Butler:1999sv, Butler:2000zp, Kong:2000px}:
\begin{align}
{\cal L}^{(1)}  = & -\frac{g_A}{2} N^\dagger \sigma_3 \left[ W_3^- \tau^++W^3_3 \tau^3+W_3^+ \tau^- \right] N
\nonumber\\
&- \frac{l_{1,A}} {2M_N\sqrt{r_s r_t}} \left[ W_3^- t_3^\dagger s^++W^3_3t_3^\dagger s^3+W_3^+ t_3^\dagger s^- +\text{h.c.} \right],
\label{eq:L-dibaryon-1}
\end{align}
where the superscript $(i)$ on the Lagrangian indicates the order of the terms in the background field strength.
Here, $W^a_i$ represents the background field, with  $a$ ($i$) denoting the isovector (vector) indices, and $W_\mu^{\pm} \equiv (W_\mu^1\pm i W_\mu^2)/\sqrt{2}$, $N$ is the nucleon field, $s^a$ is the $a$th isospin component of the isotriplet dibaryon field, and $t_j$ is the $j$th spin component of the isosinglet dibaryon field. $\sigma$ and $\tau$ refer to Pauli matrices in spin and isospin space, respectively. $l_{1,A}$ is the axial coupling of the dibaryon (characterizing the transition between the isotriplet and isosinglet channels).\footnote{This is the counterpart of $L_{1,A}$ in the nucleonic formulation that was introduced in Sec.~\ref{sec:weak}. For a relation between the two couplings, see e.g., Ref.~\cite{Briceno:2012yi}.} $M_N$ is the nucleon mass and $r_s$ ($r_t$) is the s-wave effective range in the isotriplet (isosinglet) two-nucleon channel. At second order in the background axial field and LO in the EFT momentum expansion, the only short-distance contribution to the $nn \to pp$ isotensor transition is 
\begin{align}
{\cal L}^{(2)}  = 
- \frac{h_{2,S}} {2M_Nr_s} (W_3^+)^2 s^{+\dagger} s^-,
\label{eq:L-dibaryon-2}
\end{align}
arising from coupling to an $I=2,~I_3=2$ combination of the background field.
Note that the coupling $h_{2,S}$ only contributes to, and can only be constrained from,  $\Delta I=2$ processes such as the  $nn \to pp$ transition amplitude.

\begin{figure}[!t]
	\includegraphics[width=0.95\columnwidth]{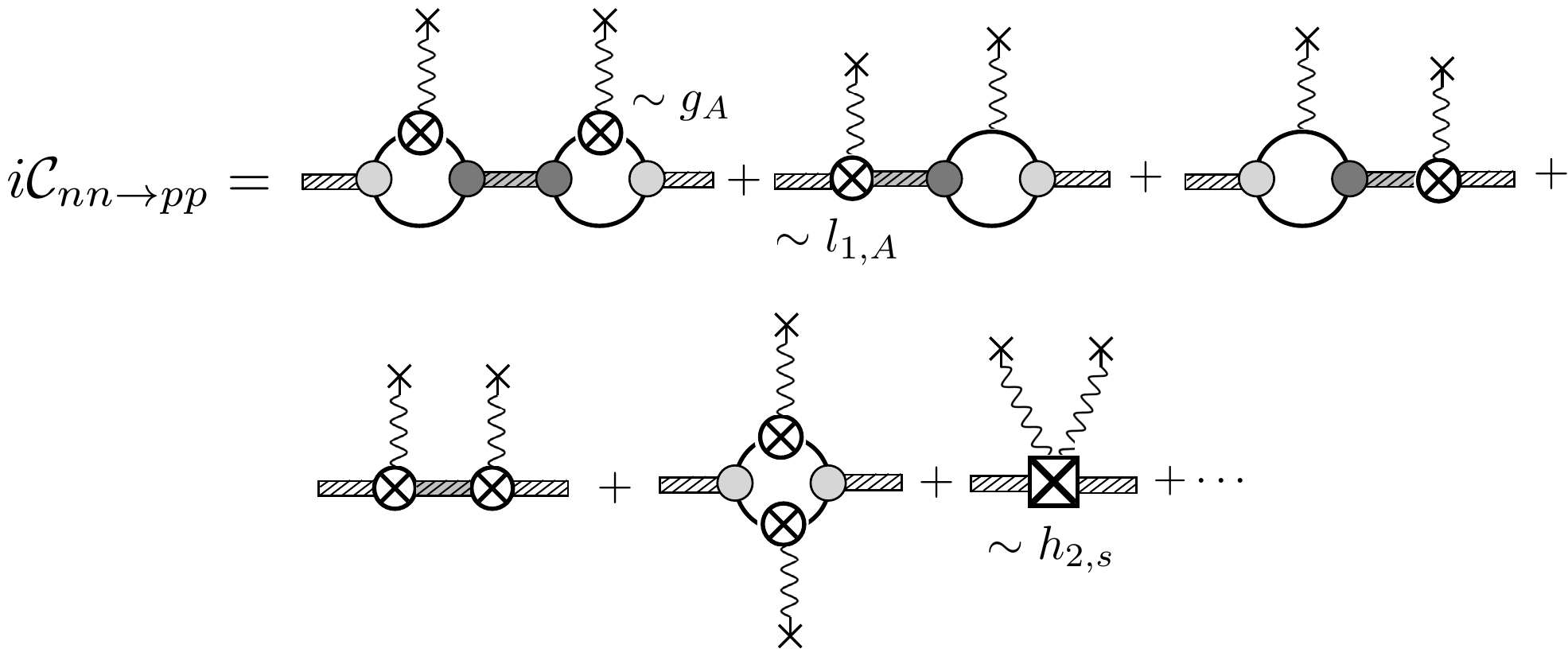}
	\caption{
		The correlation function corresponding to the $nn \to pp$ transition at second order in the axial background field in the dibaryon pionless EFT formalism. The small light (dark) gray circles denote the isotriplet (isosinglet) strong dibaryon coupling to two nucleons. 
		The dressed (by $s$-wave strong interactions) isotriplet (isosinglet) dibaryon propagator is represented by the thick dashed light (dark) gray lines, while the nucleon propagator is shown with thin black lines. 
		Finally, the doubly-weak dibaryon coupling to the background field ($\sim h_{2,S}$) is represented by the crossed square. 
		[Figure from Ref.~\cite{Tiburzi:2017iux}.]
	}
	\label{fig:cnnpp}
\end{figure}

The various interactions with the axial background field give rise to the $nn \to pp$ correlation function at second order in the background field. The contributions to this correlation function are  shown diagrammatically in Fig.~\ref{fig:cnnpp}. The finite-volume correlation function can be expressed using a similar expansion, but with a cubic spatial volume with PBCs assumed. The time-momentum representation of the correlation function when Wick-rotated to Euclidean space corresponds to the LQCD correlation function obtained at the same order in the background field strength. A substantial benefit of a LQCD study performed at a large pion mass is the bound nature of both the di-neutron(proton) and the deuteron, making the matching relation between finite and infinite volume exact up to exponential corrections in volume that are expected to be at the  percent level in the study of Refs.~\cite{Shanahan:2017bgi,Tiburzi:2017iux}. Furthermore, at the threshold kinematics that are considered, no intermediate two- or more-hadron states besides a bound deuteron contribute to the non-local amplitude, making the matching between Euclidean and Minkowski correlation functions straightforward, see Refs.~\cite{Tiburzi:2017iux,Briceno:2019opb,Davoudi:2020xdv}. Considering these features, taking the same ratio of the $nn \to pp$ transition correlation function to the zero-field two-point function as done for the LQCD calculation, and taking the second derivative with respect to the external-field strength, yields an  expression that can be matched precisely to the LQCD result. This leads to a matrix element that can be expressed as
\begin{align}
\mathcal{M}_{nn \to pp}
= 
-\frac{|g_A(1+S)+\mathbb{L}_{1,A}|^2}{\Delta}
\ +\
\frac{Mg_A^2}{4 \gamma_s^{2}}
-\mathbb{H}_{2,S}.
\label{eq:M-nnpp-EFTb}
\end{align}
The quantities $\mathbb{L}_{1,A}$ and $\mathbb{H}_{2,S}$ are directly related to the correlated two-nucleon single- and two-axial couplings $l_{1,A}$ and $h_{2,S}$, respectively (see Refs.~\cite{Shanahan:2017bgi,Tiburzi:2017iux} for further details), $\gamma_{s}$ is the $nn$ binding momentum, $\Delta$ is the energy difference between the dineutron and the deuteron as defined after Eq.~\eqref{eq:Jtil}, and $S$ is a known SU(4) Wigner symmetry-breaking factor.

The LECs in Eq.~(\ref{eq:M-nnpp-EFTb}) can be constrained using the LQCD calculations of Refs.~\cite{Shanahan:2017bgi,Tiburzi:2017iux}. The first term corresponds to the deuteron pole and provides $\sim 90\%$ of the full matrix element, with the remaining contributions from $\mathbb{H}_{2,S}$ and the similarly-sized $g_A^2$ term. Explicitly, the correlated two-nucleon doubly-weak axial coupling of the pionless EFT in the dibaryon approach is
$\mathbb{H}_{2,S}=4.7(1.3)(1.8)~{\rm fm}$
at $m_{\pi}= 806~{\rm MeV}$, where the two uncertainties are from statistical and systematic effects.
This result suggests that the contribution from the new short-distance coupling $\mathbb{H}_{2,S}$ may play an important role in analyses of double-$\beta$ decay processes of larger nuclei. The numerical value, however, remains unknown at the physical values of the quark masses and will be the subject of future LQCD studies.

\subsection{Neutrinoless double-$\beta$ decay}
\label{sec:znubb}

For light Majorana neutrinos, neutrinoless double-$\beta$ decay corresponds to weak interactions that are separated over length-scales that are significantly larger than the intrinsic QCD length scale, $\Lambda_{\rm QCD}^{-1}$. In this case, the non-locality of the weak interactions must be accounted for in LQCD calculations. 
A second possibility is that the $\Delta I=2$ LNV interactions arise from physics at energy scales that are considerably above the electroweak scale, $\Lambda_{\rm LNV} \gg  M_W$. At the hadronic scales relevant for LQCD calculations, this high-scale physics can be integrated out, resulting in the generation of local operators beyond those in the SM.
In this context, the SM forms the renormalizable sector of the so-called Standard Model Effective Field Theory (SMEFT) that additionally includes an infinite tower of higher-dimensional operators, ${\cal O}_n^{(d)}$, that are suppressed by powers of the scale, $\Lambda$, of BSM dynamics~\cite{Weinberg:1979sa,Wilczek:1979hc}: 
\begin{equation}
{\cal L}_{\rm SMEFT} = {\cal L}_{\rm SM}   +  \sum_{d \geq 5}  \sum_n\frac{C_n^{(d)}}{ \Lambda^{d-4} } \, {\cal O}_n^{(d)} \,,
\label{eq:SMEFT}
\end{equation}
where $d$ is the operator dimension and the $C_n^{(d)}$ are Wilson coefficients. The Weinberg operator that provides the simplest mechanism for generation of a neutrino Majorana mass term enters at $d=5$.
In various BSM scenarios, $\Lambda$ can be anywhere from the TeV scale to the GUT scale, $\Lambda_{\rm GUT}\sim 10^{16}$ GeV.
Recent reviews  \cite{Engel:2016xgb,Dolinski:2019nrj,Cirigliano:2020yhp} provide more detail on the phenomenological aspects of \znubb; here, the focus is on LQCD calculations relevant for both the light Majorana-neutrino exchange scenario and the local-operator scenario. 

The EFT description of the \znubb\ decay at the nucleon level originates in Ref. \cite{Prezeau:2003xn} in the context of short-distance LNV operators that contribute to a LNV two-nucleon potential within the Weinberg power counting of nuclear forces \cite{Weinberg:1990rz,Weinberg:1991um}. Although formally not renormalizable order-by-order \cite{Kaplan:1995yg}, such a treatment indicates that \znubb\ decay of a $\pi^+$ that is exchanged between the nucleons is the LO contribution to the \znubb\ two-nucleon potential, requiring constraints on the relevant pion matrix elements. At NLO, the LNV pion-nucleon coupling contributes, requiring the evaluation of the relevant matrix elements between the neutron and the proton-pion states. At NNLO, the contact two-nucleon LNV operator contributes, necessitating the evaluation of the matrix element in the two-nucleon states. Recently, it was shown~\cite{Cirigliano:2018hja} that Weinberg power counting for this process breaks down due to UV divergences in $s$-channel loops involving  neutrino exchange. Consequently, the contact two-nucleon LNV operator must enter at LO to absorb these divergences.\footnote{This is the same failure mode that precludes a chiral expansion for spin-singlet nuclear interactions using Weinberg's power counting \cite{Kaplan:1998tg,Kaplan:1998we}.} LQCD determinations of the three classes of matrix elements (pion matrix elements, mixed pion-nucleon matrix elements, and two-nucleon matrix elements) in both short-range and long-range \znubb\ scenarios will be required to confirm the suggested power counting of Ref.~\cite{Cirigliano:2018hja}, and to constrain the associated new short-distance LECs. At  present, calculations have been performed for purely pionic transitions by several groups \cite{Tuo:2019bue,Feng:2018pdq,Detmold:2020jqv,Nicholson:2018mwc}, with extensions to the pion-nucleon and $nn\to pp$ cases currently in progress.

\subsubsection{Long-distance matrix elements for \znubb} 

A light Majorana neutrino  can propagate over distances that are resolvable at the QCD scale, and the non-locality of the second-order weak process needs to be incorporated into the evaluation of the corresponding QCD matrix elements. This leads to more complicated calculations than those for the \tnubb\ process discussed in Sec.~\ref{sec:tnubb}, as the \znubb\ process includes integration over the momentum carried by the neutrino propagator.

The \znubb\ process between an initial state $i$ and final state $fe^-e^-$ occurs through two insertions of the $\Delta I=1$ weak Hamiltonian ${\cal H}_W$, that arises from integrating out SM physics above the mass of the bottom quark. This leads to the bi-local matrix element 
\begin{equation}
\int d^{4} x \, d^{4} y \left\langle f e e \right\vert T \left[ \mathcal{H}_{W}(x) \mathcal{H}_{W}(y) \right] \left\vert i \right\rangle = 4 m_{\beta \beta} G_{F}^{2} V_{ud}^{2} \int d^{4} x \, d^{4} y \, H_{\alpha \beta}(x,y) L_{\alpha \beta}(x,y),
\label{eqn:dbd_me}
\end{equation}
where $V_{ud}$ is a Cabbibo-Kobyashi-Masakowa matrix element, the leptonic tensor is given by
\begin{equation}
L_{\alpha \beta} \equiv \bar{e}_{L}(p_{1}) \gamma_{\alpha} S_{\nu}(x,y) \gamma_{\beta} e_{L}^{C}(p_{2}) e^{- i p_{1} \cdot x} e^{- i p_{2} \cdot y}
\label{eqn:leptonic_tensor}
\end{equation}
where $p_{1,2}$ are the electron momenta, $e_L$ is an electron spinor and $e_L^C$ is its charge conjugate, and the hadronic tensor is 
\begin{equation}
H_{\alpha \beta} \equiv \left\langle f \right\vert T \left[ J_{\alpha L}(x) J_{\beta L}(y) \right] \left\vert i \right\rangle,
\label{eqn:hadronic_tensor}
\end{equation}
with $J_{\mu L}(x)=\bar{q}_{u,L}(x) \gamma_{\mu} q_{d,L}(x)$ for left-handed quark fields. 
In Eq.~\eqref{eqn:leptonic_tensor},
\begin{eqnarray}
S_\nu(x,y)  = \int \frac{d^4q}{(2\pi)^4} \frac{ e^{iq\cdot (x-y)}}{q^2}\,,
\end{eqnarray}
where the small mass of the (SM) neutrino in this scenario is neglected in the denominator.
The convolution with the leptonic tensor and the integration over spacetime mean that evaluations of the hadronic tensor are required at all spacetime points. 
Since  LQCD calculations are performed in finite-volume Euclidean spacetime, extracting the infinite-volume Mikowski-space matrix elements requires a non-trivial matching, particularly when the initial, intermediate, and/or final states are multiple hadrons.

The \znubb\ matrix element that induces a transition between an initial $\pi^-$ state and a final $\pi^+e^-e^-$ state has been studied in Refs.~\cite{Detmold:2018zan,Detmold:2020jqv,Tuo:2019bue}. This transition is  unphysical due to the electron masses and the degeneracy of the $\pi^\pm$ states; however, it can contribute in \znubb\ decay of physical nuclei and can be studied at unphysical kinematics where all external particles are at zero momentum (note that the corresponding matrix element is equivalent to the kinematically-allowed charge-exchange zero-momentum scattering process $\pi^- e^+ \to \pi^+ e^-$). This is the simplest \znubb\ process to investigate in LQCD as, unlike the two-nucleon case ($nn\to pp ee$), there is no exponential StN problem at increasing Euclidean time, and it only involves single hadrons in the  initial and final states. In addition, there are chiral perturbation theory ($\chi$PT)  predictions for this low-energy process that depend on a single LEC at NLO, namely $g_{\nu}^{\pi \pi}$ \cite{Cirigliano:2017tvr, Tuo:2019bue}.

\begin{figure*}[!t]
	\subfigure[]{
		\raisebox{0mm}{\includegraphics[width=0.4\textwidth]{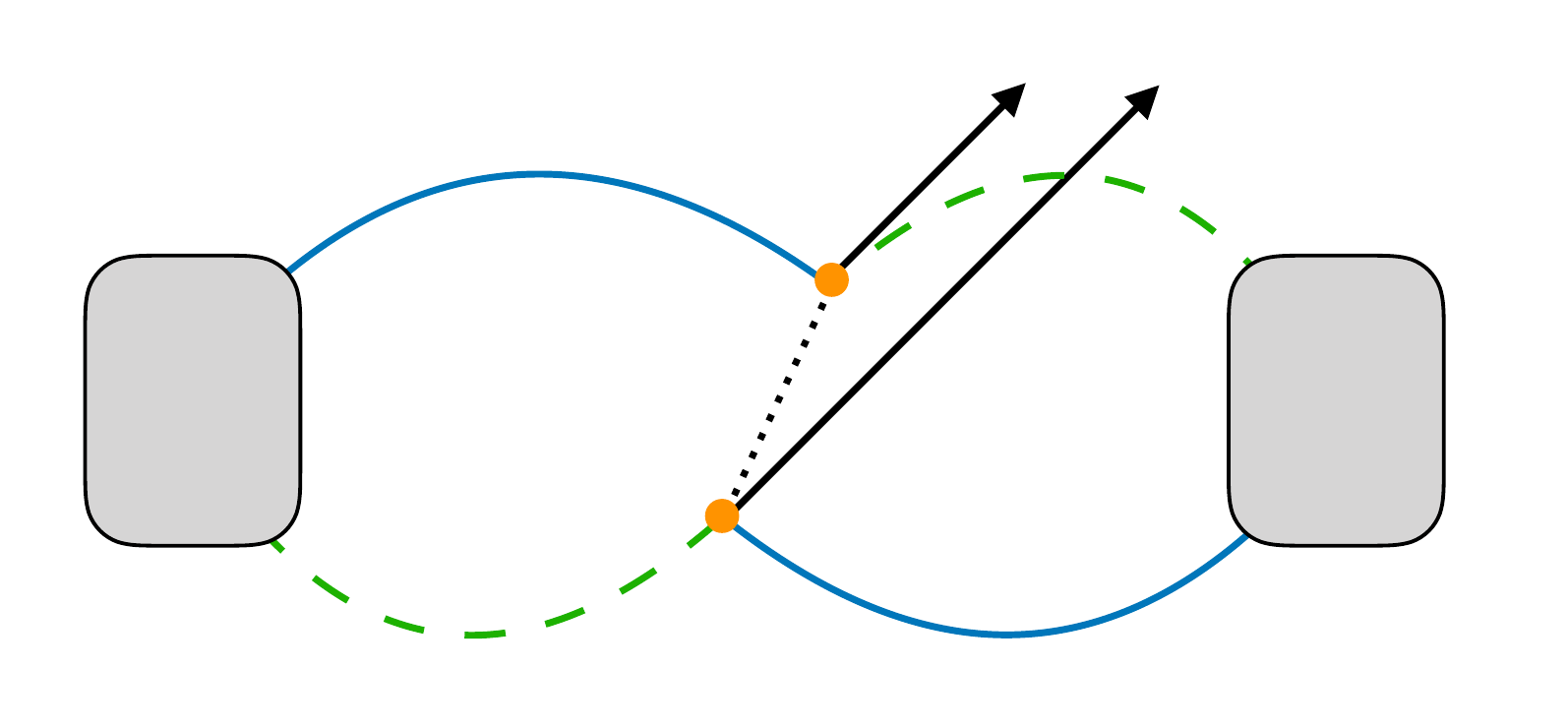}}
	}\qquad \qquad
	\subfigure[]{
		\centering
		\includegraphics[width=0.4\textwidth]{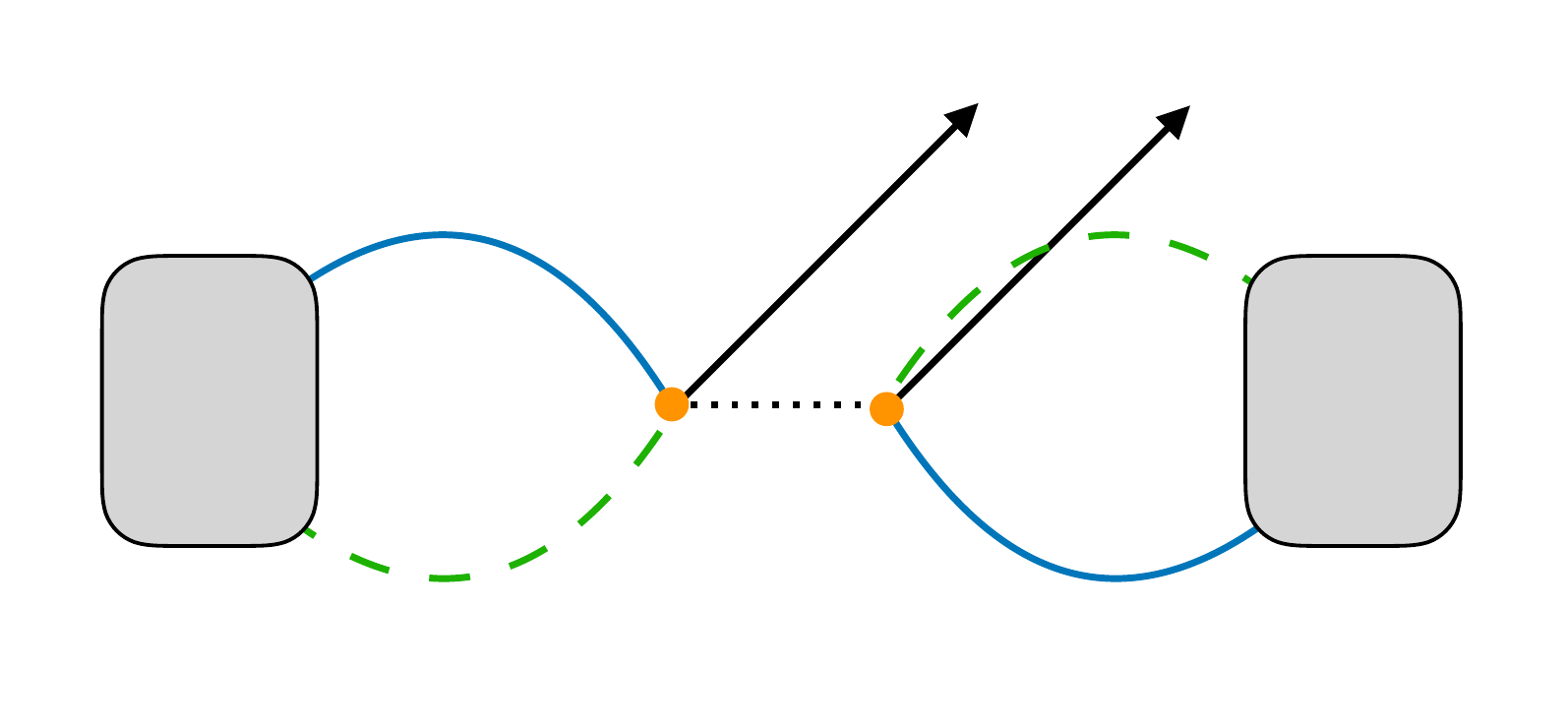}
	}   
	\centering
	\subfigure[]{
		\raisebox{0mm}{\includegraphics[width=0.38\textwidth]{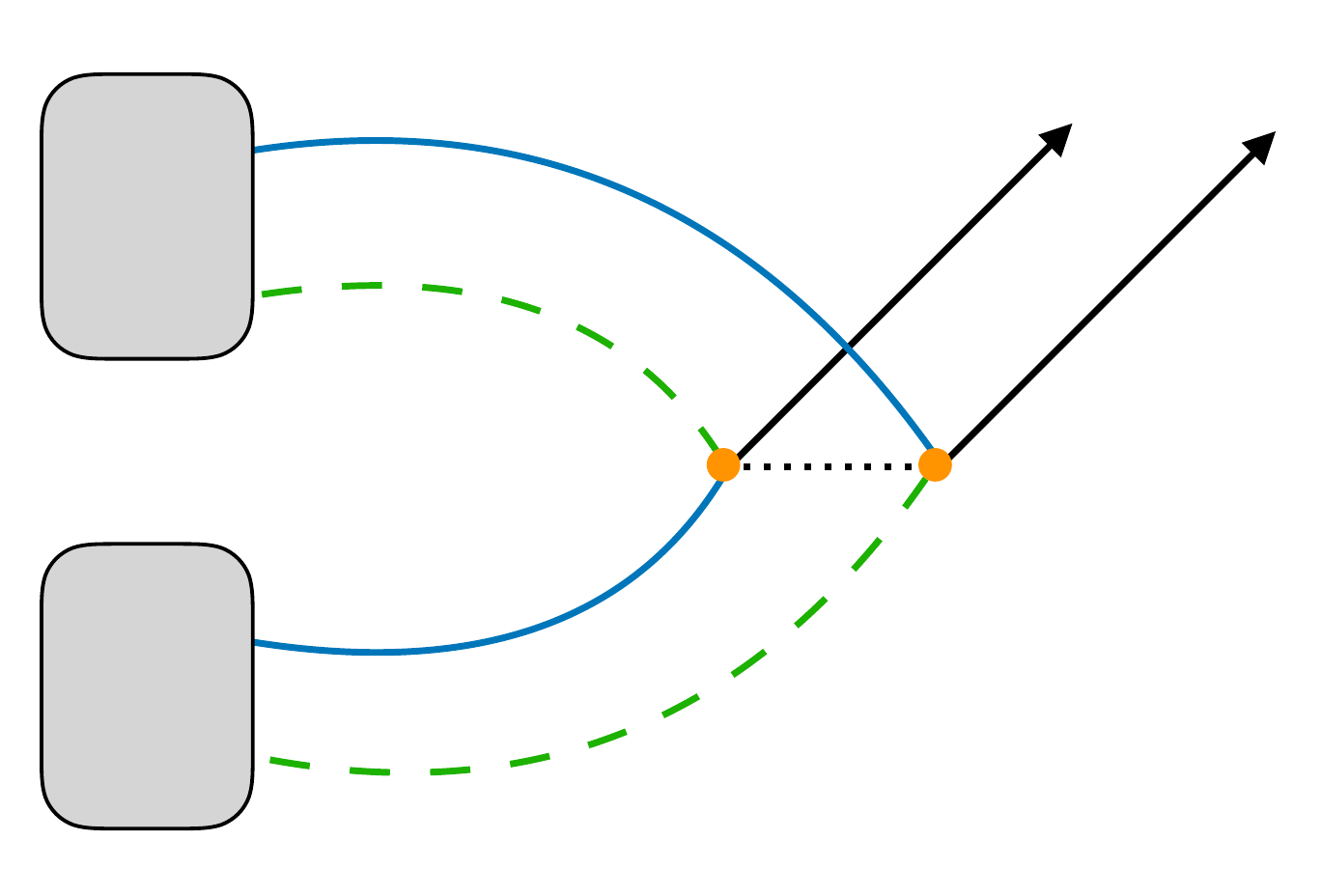}}
	}\qquad \qquad
	\subfigure[]{
		\centering
		\includegraphics[width=0.38\textwidth]{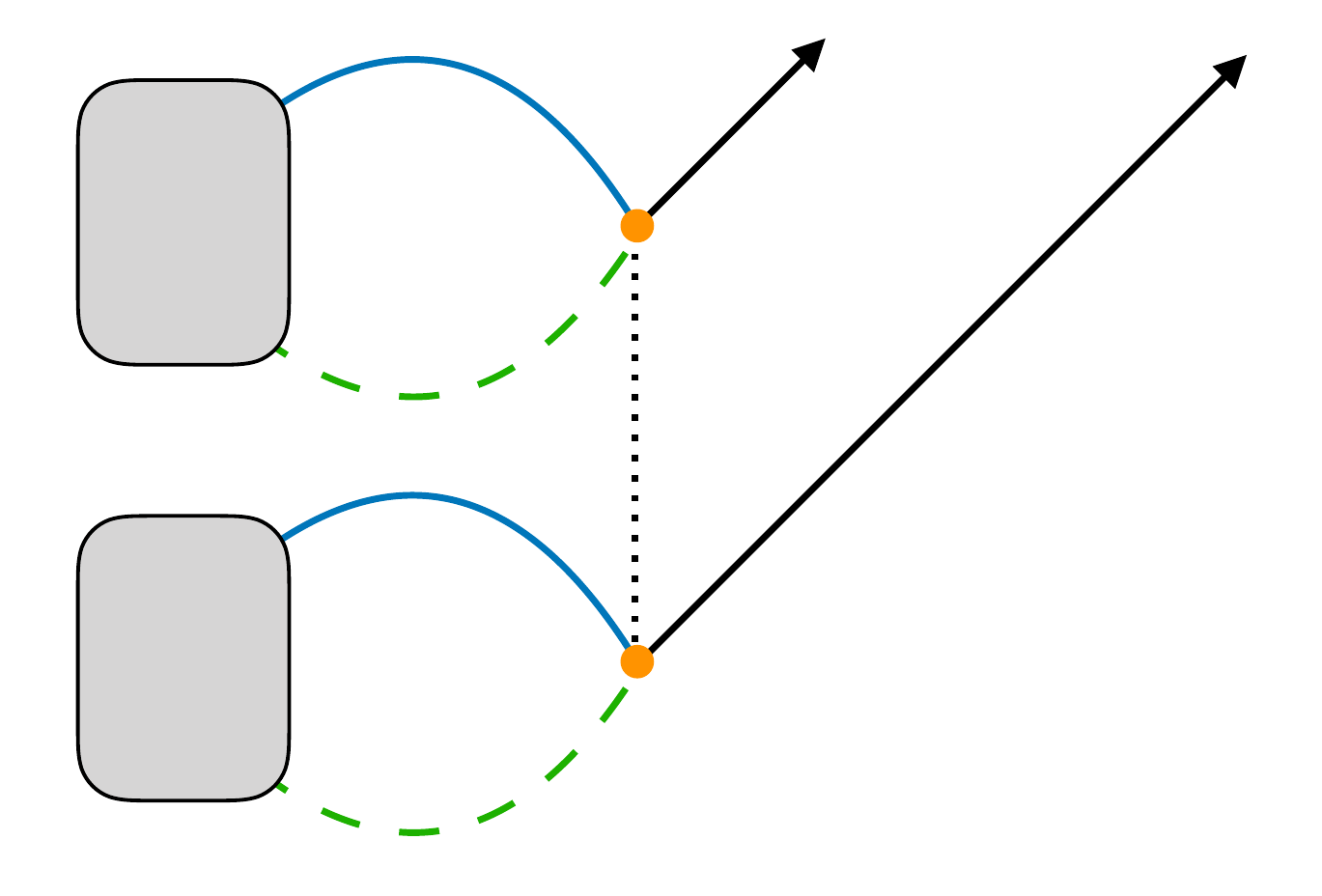}
	}   
	\caption{\label{fig:pi2piee}Contractions for the $\pi^- \to \pi^+ e^- e^-$ transition, (a) and (b), and for the $\pi^-\pi^- \to e^- e^-$ transition, (c) and (d). The solid blue and dashed green lines represent down and up quark propagators respectively, and the circles correspond to the $\Delta I=1$ weak vertices. The dotted and solid black lines represent the Majorana neutrino propagator and electron final states respectively. 
		[Figure modified from Ref.~\cite{Cirigliano:2020yhp}.]}
\end{figure*}

\begin{figure*}
	\centering
	\includegraphics[width=0.46\textwidth]{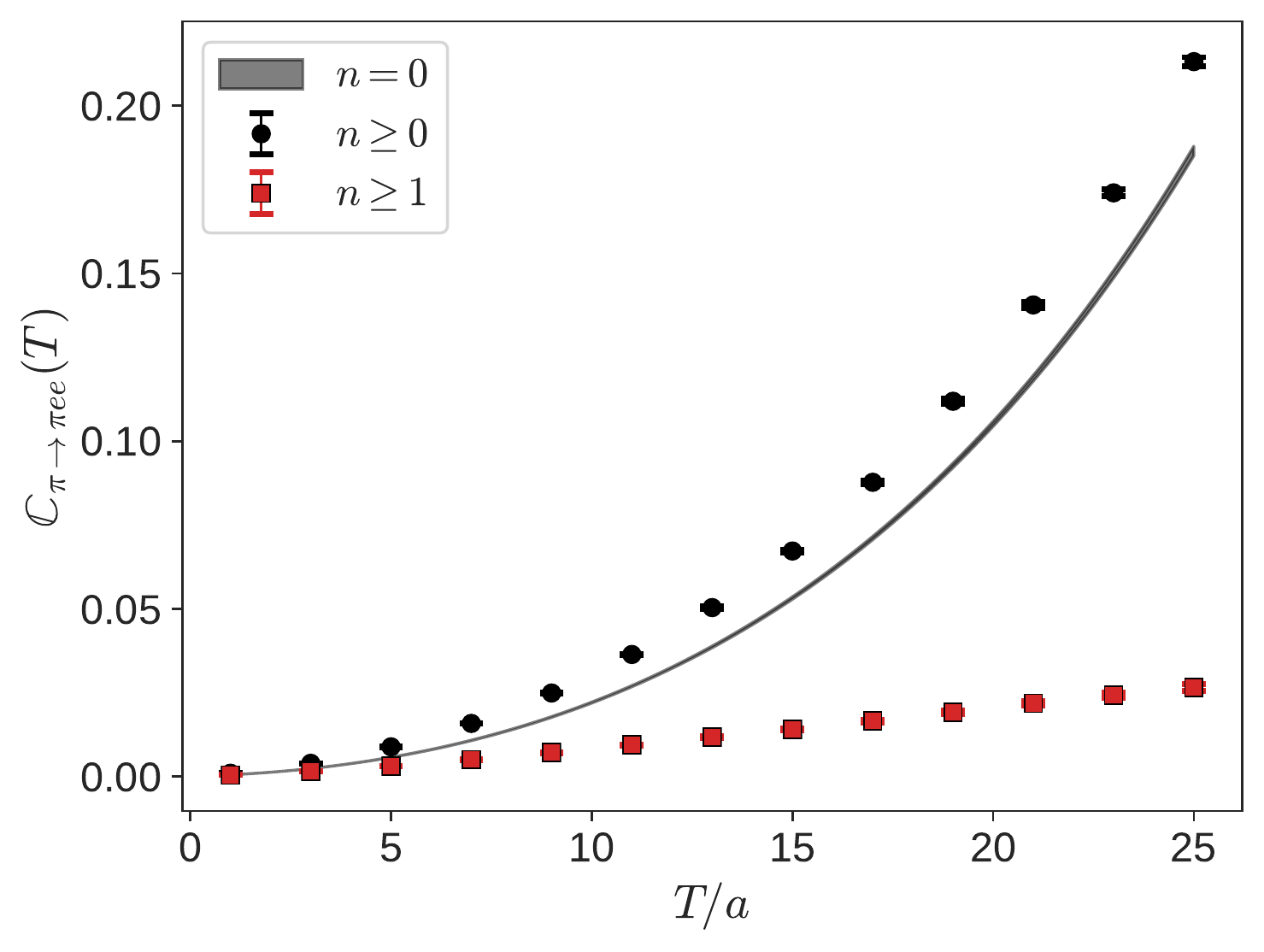}
	\qquad
	\includegraphics[width=0.47\textwidth]{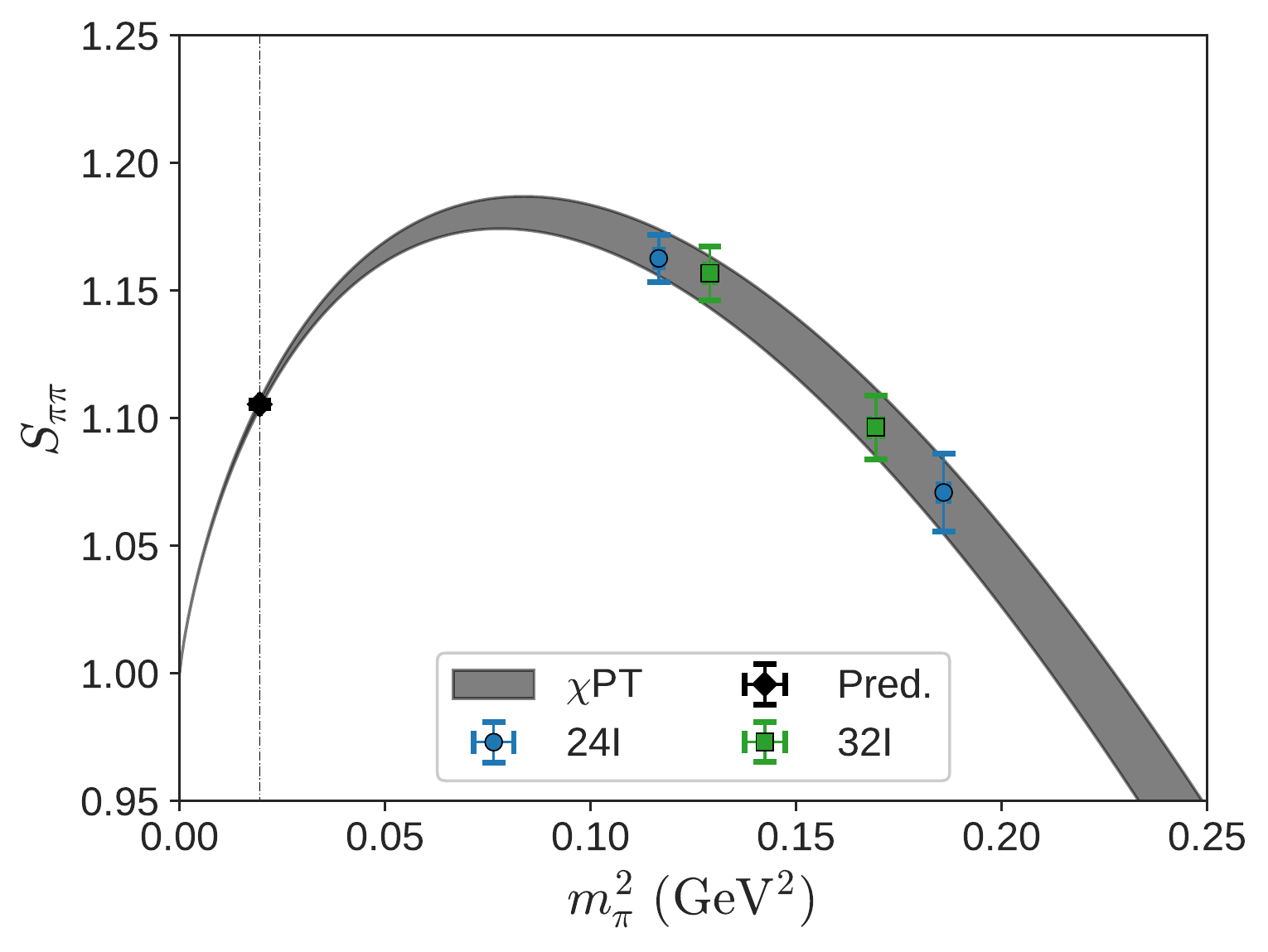}
	\caption{\label{fig:znubblongmurphy1} 
		The  integrated transition amplitude from Ref.~\cite{Detmold:2020jqv} for an example ensemble before and after subtracting the vacuum state contribution  in Eq.~(\ref{eq:largeT}) for fixed $t$ (left). Also shown is the pion mass dependence of the matrix element normalized by the matrix element in the chiral limit, $S_{\pi\pi}$, as constrained by  the chiral/continuum/finite volume extrapolation \cite{Detmold:2020jqv} (right). The dotted vertical line in the right panel indicates the physical pion mass. [Figures from Ref.~\cite{Detmold:2020jqv}.]}
\end{figure*}

Two independent studies of this process are presented in Refs.~\cite{Detmold:2018zan,Detmold:2020jqv,Tuo:2019bue} using domain-wall fermions. The calculations use techniques that build upon  studies  of rare kaon decays by the RBC collaboration \cite{Bai:2014cva,Bai:2018hqu} and the \tnubb\ process discussed above.  For  these particular initial and final states, the hadronic matrix elements of interest can be determined from the correlation function
\begin{equation}
\begin{split}
C_{\mu\nu}^{\pi\to\pi ee}(t_+,x,y,t_-)=\left\langle T\left[\chi_{\pi^+}(t_+)J_{\mu L}(x)J_{\nu L}(y)\chi^\dagger_{\pi^-}(t_-)\right]\right\rangle
\end{split}
\label{eq:pipi4pt}
\end{equation} 
where $\chi_{\pi^+}(t)=a^3\sum_{\vec{x}}\bar{q}_{u}(\vec{x},t)\gamma_5 q_d(\vec{x},t)$ and $\chi_{\pi^-}=\chi^\dagger_{\pi^+}$ are interpolating operators for zero-momentum pion states,  and terms with $\mu\leftrightarrow\nu$ and $x\leftrightarrow y$ are implied by the time-ordered product.
After integrating out the quarks, this correlation function produces different types of contractions as shown in  Fig.~\ref{fig:pi2piee}. 
By convolving with the lepton tensor, integrating over the  spatial positions of the currents and over some range of current insertion times, and inserting a complete set of states between the two currents in Eq.~(\ref{eq:pipi4pt}), it can be shown that
\begin{align}
{\mathbb C}_{\pi\to\pi ee}(t;T) &\equiv
a^8\sum_{{\vec x},{\vec y}} \sum_{t_x=0}^{T}\sum_{t_y=0}^{T} \frac{ L^{\mu\nu}(x,y)C_{\mu\nu}^{\pi \rightarrow \pi ee}(t_+,x,y,t_-) }{C_\pi(t)}  \nonumber \\
&\propto
\sum_{n}  \frac{\left\langle \pi e e \right\vert \mathcal{H}_{W} \left\vert n \right\rangle \left\langle n \right\vert \mathcal{H}_{W} \left\vert \pi \right\rangle}{E_n(E_{n} - m_{\pi})} \left[ T + \frac{e^{-(E_{n}-m_{\pi}) T} - 1}{E_{n} - m_{\pi}} \right]
\label{eq:largeT}
\end{align}
for pions at rest, where $T$ is the extent of the temporal integration window for the weak-current insertions and $t = \vert t_{+} - t_{-} \vert$ is the $\pi^{-} - \pi^{+}$ source-sink separation. To arrive at this expression,  the current insertions are assumed to be  sufficiently far from the pion source and sink 
($t_- \ll 0 \ll T \ll t_+$) 
in order that excited-state contributions before and after the integration window are  negligible. The infinite tower of states contributing to the sum are: $\{\left|e\bar\nu_e\right\rangle$, $\left|\pi e\bar\nu_e\right\rangle$, $\left|n=2 \right\rangle$, $\ldots\}$, with energies $E\sim m_e<m_\pi$, $E\sim m_\pi$ and $E > m_\pi$, respectively (the particle content of the $\left|n=2 \right\rangle$ and higher states indicated by the ellipsis are not specified). For the lowest-energy state, the terms in the square brackets in Eq.~(\ref{eq:largeT})  grow exponentially with $T$ and the matrix element is just the pion decay constant. For the second state, $\left|\pi e\bar\nu_e\right\rangle$, the term in the square brackets is approximately quadratic in $T$, while the remaining $n\ge 2$ terms should behave linearly at large $T$. By extracting these pieces individually, the matrix element governing \znubb, i.e.,
\begin{equation}
{\cal M}^{\pi\to\pi ee} = \sum_{n} \frac{\left\langle \pi e e \right\vert \mathcal{H}_{W} \left\vert n \right\rangle \left\langle n \right\vert \mathcal{H}_{W} \left\vert \pi \right\rangle}{E_n(E_{n} - m_{\pi})}\,,
\label{eqn:spectral_me}
\end{equation}
can be reconstructed, where ${\cal H}_W$ is the weak Hamiltonian density appearing in Eq.~\eqref{eqn:dbd_me}.
This is illustrated in Fig.~\ref{fig:znubblongmurphy1} in the approach of Ref.~\cite{Detmold:2020jqv} (Ref.~\cite{Tuo:2019bue} and Ref.~\cite{Detmold:2020jqv} use a similar overall approach but have technical differences, in particular in the way the neutrino propagator is implemented). Knowledge of the matrix element at various different quark masses, lattice spacings, and volumes is sufficient to determine the NLO $\chi$PT LEC
\begin{align}
g_{\nu}^{\pi \pi}(\mu=770\ {\rm MeV})&= - 11.96(31), && (\pi^-\pi^-\rightarrow e^-e^-) & \text{\cite{Feng:2018pdq} }
\nonumber \\
&=
-10.89(28)(33)_{L}(66)_{a}, && (\pi^-\rightarrow \pi^+e^-e^-)& \text{\cite{Tuo:2019bue}} \nonumber \\
&=-10.78(12)(51), &&(\pi^-\rightarrow \pi^+e^-e^-)& \text{\cite{Detmold:2020jqv}}
\label{eq:grhoresult}
\end{align}
where the first uncertainty in each case is statistical and the others are due to systematic effects, either in combination or broken into different contributions, as indicated by the subscripts $a$ and $L$ referring to lattice spacing and finite volume uncertainties, respectively.  The result of the preliminary study in Ref.~\cite{Feng:2018pdq} only includes a statistical uncertainty.
These values are in good agreement with each other and are more precise than, and in  reasonable agreement with, the large-$N_c$ estimate $ g_{\nu}^{\pi \pi}(\mu)|_{\mu=m_{\rho}}=-7.6 $~\cite{Cirigliano:2017tvr,Ananthanarayan:2004qk}. 

The same second-order $\Delta I=1$ weak interactions also generate the   $\pi^-\pi^-\to  e^- e^-$ transition (the crossed-channel analog of that discussed above). This
transition is kinematically allowed but is not accessible experimentally. As with $\pi^-\to\pi^+ e^- e^-$, however, it  provides a  useful theoretical arena in which to develop LQCD  techniques for second order processes. In Ref.~\cite{Feng:2018pdq},  the amplitude for this transition  in the light neutrino exchange scenario was investigated. Domain-wall fermion ensembles with quark masses corresponding to  pion masses $m_\pi=420$ and 140 MeV were used, and, while exploratory, this calculation demonstrated the feasibility of the methods used. 
As with  $\pi^-\to\pi^+ e^- e^-$, there are two types of contractions involved in constructing the LQCD correlation functions from which the transition amplitudes can be determined; these are also shown in Fig.~\ref{fig:pi2piee}.
An important complication in this calculation is that the initial state is a  two-particle state and the finite-volume state  must be converted to the desired infinite-volume state using the Lellouch-L\"uscher factor \cite{Lellouch:2000pv,Lin:2001ek}. This requires knowledge of the appropriate $I=2$ $\pi\pi$-scattering phase shifts, which can be extracted from spectroscopic calculations of $\pi^-\pi^-$ systems. As presented in Eq.~\eqref{eq:grhoresult} above, the results provide further constraints on $g_{\nu}^{\pi \pi}$ that  are compatible with those obtained from the $\pi^-\to \pi^+ e^- e^-$ transition amplitude.

\subsubsection{Short-distance matrix elements for \znubb}

As discussed  above, if  BSM physics contributes to \znubb\ at scales above the electroweak scale  (with or without light Majorana neutrinos), then  at lower scales the new physics can be integrated out and manifests  through  local composite operators built from SM fields.
A generic high-scale physics scenario will produce multiple different operators at low energies, 
of which the most phenomenologically relevant (having the lowest dimension) are the five four-quark scalar operators, along with four negative parity counterparts.\footnote{The four-quark vector operators are also relevant in constructing the two-nucleon \znubb\ potential as they couple the neutron state and the pion-proton state, inducing  LNV in the $nn\to pp$ transition at NLO in the power counting of Ref.~\cite{Prezeau:2003xn}.} Using the basis of Ref.~\cite{Nicholson:2018mwc}, the positive-parity operators can be expressed as
\begin{align}
\label{eq:Ops}
\mathcal{O}_{1+}^{++} &= \left(\bar{q}_L \tau^+ \gamma^{\mu}q_L\right)\left[\bar{q}_R \tau^+\gamma_{\mu} q_R \right] , \cr
\mathcal{O}_{2+}^{++} &= \left(\bar{q}_R \tau^+ q_L\right)\left[\bar{q}_R \tau^+ q_L \right] + \left(\bar{q}_L \tau^+ q_R\right)\left[\bar{q}_L \tau^+ q_R \right] , \cr
\mathcal{O}_{3+}^{++} &= \left(\bar{q}_L \tau^+ \gamma^{\mu}q_L\right)\left[\bar{q}_L \tau^+ \gamma_{\mu} q_L \right] + \left(\bar{q}_R \tau^+ \gamma^{\mu}q_R\right)\left[\bar{q}_R \tau^+ \gamma_{\mu} q_R \right] , \cr\mathcal{O}_{1+}^{'++} &= \left(\bar{q}_L \tau^+ \gamma^{\mu}q_L\right]\left[\bar{q}_R \tau^+\gamma_{\mu} q_R \right) , \cr
\mathcal{O}_{2+}^{'++} &= \left(\bar{q}_L \tau^+ q_L\right]\left[\bar{q}_L \tau^+  q_L \right) 
+ \left(\bar{q}_R \tau^+ q_R\right]\left[\bar{q}_R \tau^+ q_R \right),
\end{align}
where the  notation $()$ or $[]$ indicates which color indices are contracted together~\cite{Takahashi}. 
To determine the effects of these operators on \znubb\ rates in an EFT context \cite{Savage:1998yh,Prezeau:2003xn,Cirigliano:2019vdj}, the matrix elements $\langle \pi^+ |O^{++}_{i} | \pi^- \rangle$, $\langle\pi^+p|O_i^{++}|n\rangle$, and $\langle pp |O^{++}_{i} |nn \rangle$ are required. 
At present, only the pion matrix elements have been studied and these give access to a subset of the relevant EFT LECs. 

\begin{figure*}
	\centering
	\includegraphics[width=0.42\textwidth]{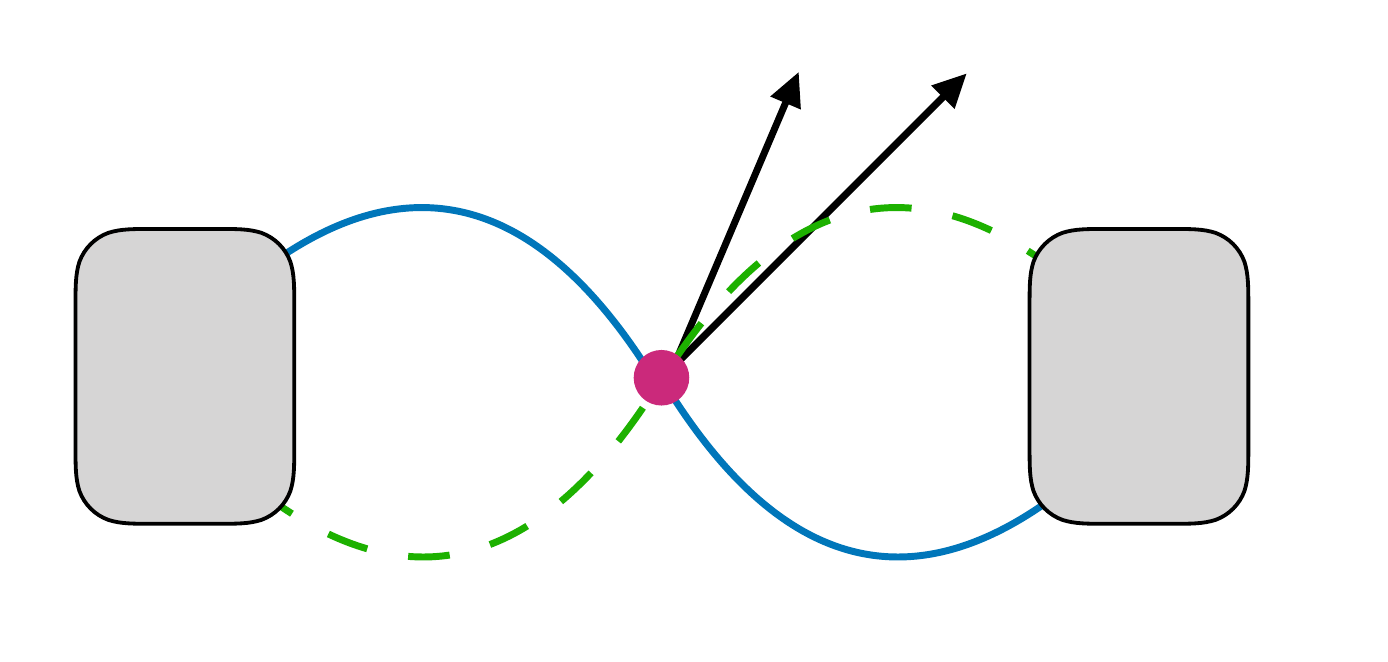}
	\caption{\label{fig:shortdist}Contractions for the $\pi^- \to \pi^+ e^- e^-$ transition induced by short-distance four-quark operators. The solid blue and dashed green lines represent down and up quark propagators, respectively, and the dark circle represents the $\Delta I=2$ operators introduced in Eq.~\eqref{eq:Ops}. The solid black lines represent the electron final states. [Figure from Ref.~\cite{Cirigliano:2020yhp}.]}
\end{figure*}

\begin{figure*}
	\centering
	\includegraphics[width=0.48\textwidth]{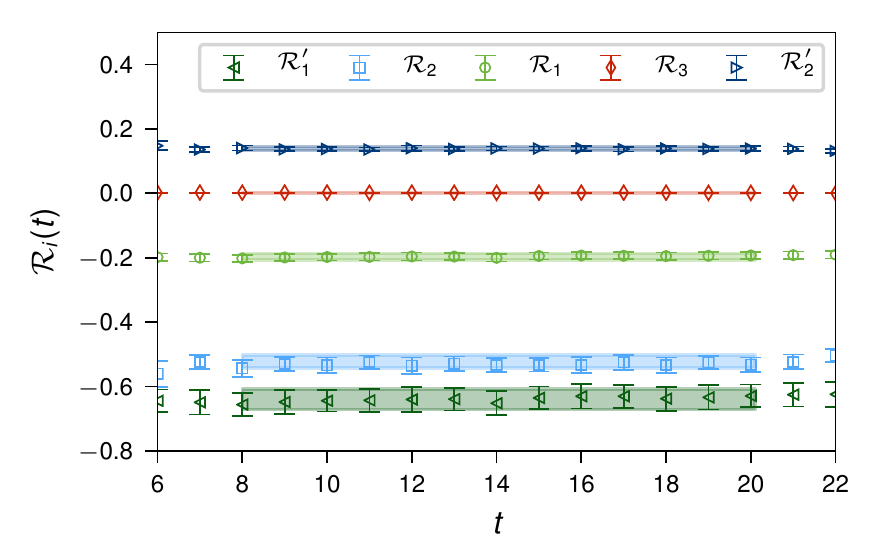}
	\includegraphics[width=0.48\textwidth]{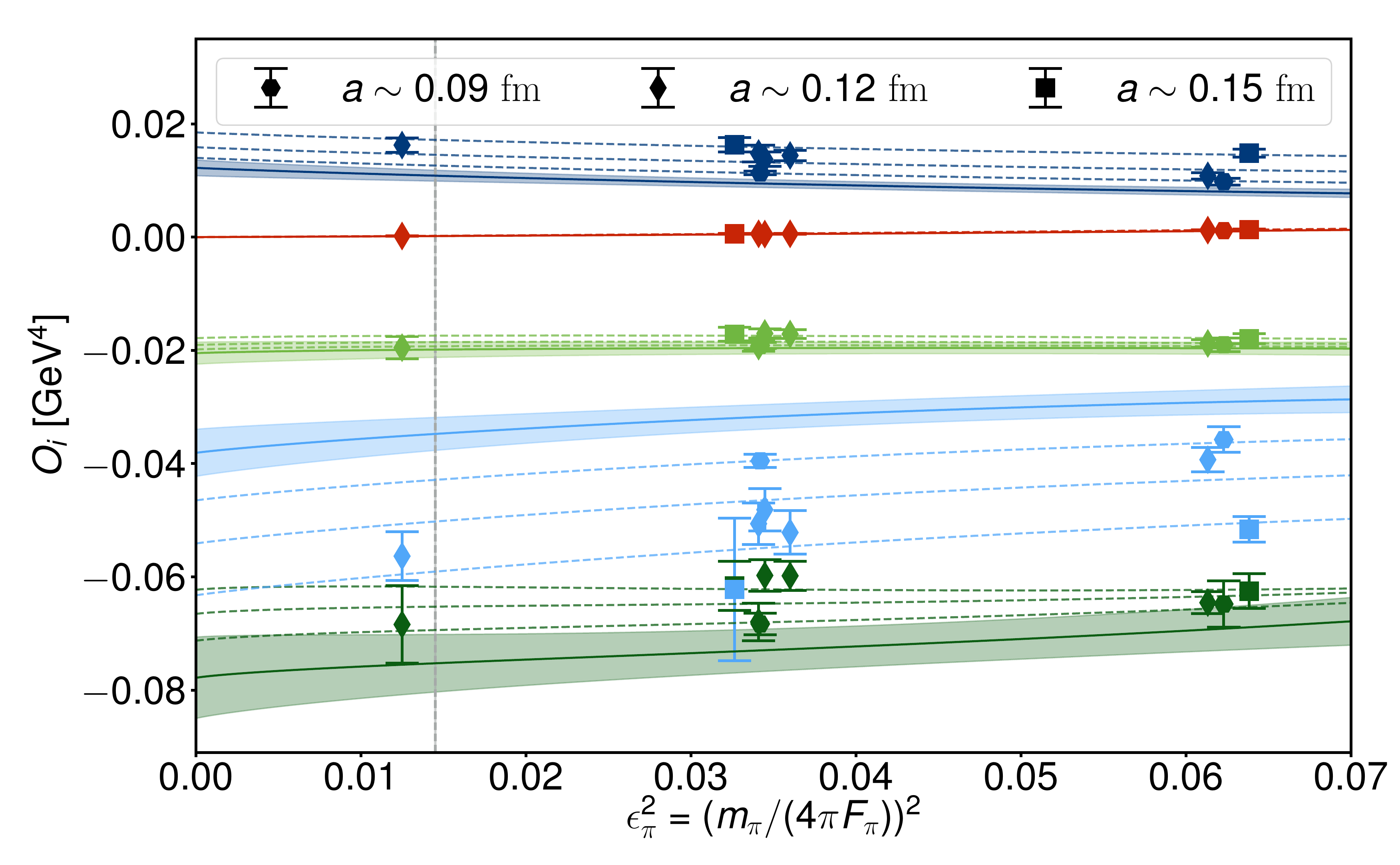}
	\caption{\label{fig:0nubbshortsigs}
		An example of the  ratio of correlation functions, $
		\mathcal{R}_i \equiv \mathcal{R}_{3/2,\mathcal{O}_i^{++}}^{(\pi)}$,
		for the five relevant \znubb\ operators on a near-physical pion-mass ensemble with $a \approx 0.12$ fm (left).  The horizontal bands are the ground-state contributions to $R_i$ extracted from single-state fits.  
		Interpolations/extrapolations of the pion matrix elements (right). The bands represent the 68\% confidence interval of the continuum, infinite-volume extrapolated values of the matrix elements. The vertical gray band indicates the physical pion mass. [Figures from Ref.~\cite{Nicholson:2018mwc}.]}
\end{figure*}

Ref.~\cite{Nicholson:2018mwc} presented a comprehensive calculation of these pion matrix elements, using multiple lattice spacings, lattice volumes and a range of quark masses including those very close to the physical values. 
To extract the relevant matrix elements, the four-quark operators are inserted between source and sink operators for the pion to build  three-point functions:
\begin{equation}\label{eq:C3pt_znubb}
C^{(\pi)}_{{\rm 3pt},\mathcal{O}_i^{++}}(t_i,t_f)=a^6\sum_{\vec{x},\vec{y}}\langle\chi_{\pi^{+}}(\vec{x},t_f) {\cal O}^{++}_i(\vec{0},0)\chi_{\pi^+}(\vec{y},t_i) \rangle\,,
\end{equation}
where $\chi_{\pi^+}(\vec{x},t_f)$ is an interpolating  operator with the quantum numbers of the $\pi^+$. The corresponding  contraction is shown in Fig.~\ref{fig:shortdist}. Note that here the operator is kept at a fixed spacetime point while the source and sink pion interpolating operators are inserted at arbitrary times, allowing for a complete exploration of the $t_i$ and $t_f$ dependence of this correlation function for the cost of a single quark inversion sourced at the operator.
Ratios 
\begin{eqnarray}
\mathcal{R}_{3/2,\mathcal{O}_i^{++}}^{(\pi)} = \frac{C^{(\pi)}_{{\rm 3pt},\mathcal{O}_i^{++}}(t_i,t_f)}{{C^{(\pi)}_\text{2pt}(t_i)C^{(\pi)}_\text{2pt}(t_f)}}\ \overset{t_i,t_f\to\infty}{\propto}\ \langle\pi^+|{O}^{++}_{i}|\pi^-\rangle,
\end{eqnarray}
which asymptote to the pion matrix elements of operators ${\cal O}^{++}_i$, can be formed by combining the above three-point function with the pion two-point correlation function, $C^{(\pi)}_\text{2pt}(t)$, Eq.~\eqref{eq:2pt0mom}.  In Fig.~\ref{fig:0nubbshortsigs}, an example of this ratio calculated in Ref.~\cite{Nicholson:2018mwc} is reproduced, showing the clear ground-state plateaus obtained with this method. 

After the matrix elements were extracted for different lattice spacings, masses, and volumes, extrapolations to the continuum, physical pion mass, and infinite-volume limits were performed.
These dependencies are determined by $\chi$PT \cite{Savage:1998yh,Cirigliano:2017ymo,Nicholson:2018mwc} generalized to incorporate finite lattice-spacing corrections~\cite{Sharpe:1998xm,Berkowitz:2017opd} and finite-volume effects~\cite{Gasser:1987zq}.
To extract the phenomenologically-relevant matrix elements, the bare quark operators must be renormalized and evolved to the appropriate scale for matching to particular BSM scenarios. As the scale is run between the electroweak and QCD scales, some of the operators defined in Eq.~\eqref{eq:Ops}  mix under renormalization, requiring  consideration of the full matrix of operators including the off-diagonal mixing. 
In Ref.~\cite{Nicholson:2018mwc}, the relevant matrix of renormalization constants was computed non-perturbatively following the Rome-Southampton method~\cite{Martinelli:1994ty}.
At the scale  $\mu=2$ GeV in the $\overline{\rm MS}$ scheme, that work found values of matrix elements that are in agreement with, and considerably more precise than, naive dimensional analysis estimates and also with indirect extractions from relations between $\pi^+$--$\pi^-$, $K$--$\bar{K}$ and $K \rightarrow \pi \pi$  matrix elements for which results exist in the literature~\cite{Savage:1998yh,Cirigliano:2017ymo}.

\subsection{Future impact}

For both \znubb\ and \tnubb, the existing LQCD studies described in the preceding subsections are stepping stones towards constraining phenomenologically-relevant matrix elements in nuclei. Short- and long-range matrix elements in pion states contribute to the LNV two-nucleon potential, but their effects might be of higher order \cite{Savage:1998yh,Prezeau:2003xn,Cirigliano:2019vdj}. For the neutrinoful $nn \rightarrow pp$ matrix elements, the constraints are as-yet at unphysically large quark masses, but new short-distance contributions that must be accounted for in future phenomenological studies have been identified.

Future studies using the approaches reviewed here have the potential to reduce the model dependence implicit in phenomenological calculations of double-$\beta$ decay rates. Such LQCD studies are required in order to connect calculations of double-$\beta$ decay matrix elements to the SM; there are unknown LO LECs in the nuclear EFT for both decay modes \cite{Shanahan:2017bgi,Cirigliano:2019vdj} that so far can only be constrained through LQCD calculations.
Nuclear EFTs used at present for double-$\beta$ decay analyses also suffer from ill-defined power countings, whose numerical validity requires testing.
Thus, even neglecting the systematic uncertainties inherent in the many-body methods built on top of nuclear EFTs, there are currently order-of-magnitude uncertainties in theory predictions of the double-$\beta$ decay matrix elements.
Moreover, LQCD can provide a set of benchmark quantities that phenomenological nuclear models can be compared with to constrain their input parameters or to assess their predictive power and their ability to quantify uncertainties. While experimental observation of \znubb\ would be a remarkable event, fully exploiting such an observation to reveal the underlying BSM physics mechanism will require precision calculations rooted in the SM.

Significant progress is  required to achieve complete LQCD calculations of double-$\beta$ decay matrix elements. As discussed above, performing the \tnubb\ LQCD calculation of Refs.~\cite{Shanahan:2017bgi,Tiburzi:2017iux}, or analogous $nn\to pp$ \znubb\ calculations, at physical quark masses presents technical challenges related to the calculation of non-local matrix elements in finite volumes \cite{Feng:2020nqj,Davoudi:2020xdv}.  Additionally, these studies should be extended to larger nuclear systems; calculations in unphysical $nnp\to ppp$ or $nnn\to npp$ systems as well as in nuclei such as $^4{\rm H} \to  {}^4{\rm Li}$ and $^6{\rm He} \to   {}^6{\rm Be}$ will  better constrain the LECs in the nuclear EFTs and phenomenological models capable of describing experimentally-relevant nuclei. Developments are needed in both  LQCD, nuclear EFTs,  and the many-body methods based on them, in order for this to become a realistic prospect. In particular, the issues of contraction complexity and noise discussed in Sec. \ref{sec:tools} must be ameliorated to enable LQCD calculations of these nuclear transitions, while the convergence properties and appropriate power countings of EFTs must be better understood.

\section{Nuclear matrix elements for beyond--Standard-Model physics}
\label{sec:BSM}
It is known that the Standard Model is not complete; definitive observational and experimental evidence confirms the existence of dark matter, the dominance of matter over antimatter in the visible universe, and a small but non-zero mass for neutrinos. In many phenomenologically-viable scenarios,  BSM physics originates at very high energy scales, and in that case, the SMEFT framework~\cite{Buchmuller:1985jz} (discussed in the previous section) can be built to encompass all BSM scenarios in which the SM emerges as the appropriate low-energy EFT, with BSM physics entering through towers of higher-dimensional operators with coefficients that depend inversely on the scale of new physics, see Eq.~\eqref{eq:SMEFT}. SM matrix elements of various operators in this framework then encode BSM interactions with the SM, and it is these matrix elements which must be constrained to interpret the results of experimental searches for signals of new physics. To enhance their sensitivity, experiments at the intensity frontier searching for signals of BSM physics often use targets constructed from nuclei of large atomic number. In these cases, the critical SM theory inputs that are needed to optimally exploit experimental results are therefore the nuclear, rather than nucleon, matrix elements of BSM operators.
QCD matrix elements for BSM-physics searches with nuclei thus fall into two classes: those for which the matrix elements of BSM operators in light nuclei are themselves of phenomenological interest, and those for which matrix elements of larger nuclei are crucially required, and where the impact of few-nucleon LQCD calculations is to provide key inputs and constraints for {\it ab initio} many-body approaches to calculating the matrix elements of large nuclei, as discussed in  Sec.~\ref{sec:forces}.

To date, only a small number of matrix elements relevant to BSM physics have been calculated in LQCD for light nuclei with $A\le 4$. In particular, the static responses of nuclei to scalar and tensor interactions have been calculated at quark masses corresponding to a pion mass of $m_\pi = 806$~MeV, with ongoing studies at lighter quark masses. While the systematic uncertainties in these studies are not yet fully controlled, the results are nevertheless of significant phenomenological interest. Knowledge of  scalar-current nuclear  matrix elements is key to interpreting the results of terrestrial dark-matter direct-detection experiments searching for weakly-interacting dark matter particles (WIMPs) with a spin-independent coupling to nuclei. The tensor current nuclear matrix elements determine the quark electric dipole moment (EDM) contributions to nuclear EDMs and are necessary to interpret corresponding searches for BSM CP violation. 
As these matrix elements are difficult to constrain using experiment, LQCD calculations provide key non-perturbative information that is not accessible by any other method.

\subsection{Scalar matrix elements}
\label{subsec:DM}

The form of potential non-gravitational interactions between dark matter and SM particles is unknown and depends on the BSM model considered~\cite{Alexander:2016aln}. Nevertheless, at low energy such interactions can often be parameterized using the SMEFT; in  most scenarios, operators mediating interactions between dark matter and quarks appear at operator dimension six and seven, and involve local quark bilinear operators~\cite{Bishara:2017pfq}. Constraining the nuclear matrix elements of these operators from LQCD provides a non-perturbative connection between the SMEFT description and nuclear models and EFTs, and eventually many-body methods~\cite{Fitzpatrick:2012ix,Cirigliano:2012pq,Menendez:2012tm,Hoferichter:2015ipa,Hoferichter:2016nvd,Korber:2017ery,Andreoli:2018etf,Krebs:2020plh}, since typically dark-matter direct-detection experiments involve large nuclei such as Xenon, Germanium, Iodine, and Argon~\cite{1805.12562,1708.06917,1404.3443,1707.01632,1708.08869,1204.3094,1611.01499,1510.07754,1702.07666,1509.01515,1701.08157,0804.2741,1002.1028,1805.10486,0804.2738,1710.06650,1802.06994,1802.06039,1603.02214,Bernabei:1998fta,Kozynets:2018dfo}. Searches for sub-GeV dark matter, where dark-matter interactions with nuclei are mediated by a new force carrier, have also been proposed based on small nuclei, for example using superfluid $^4$He~\cite{Hertel:2018aal,Hertel:2019thc,Battaglieri:2017aum}. 

Since the low-energy limit of a generic spin-independent interaction transforms as a scalar, a broad class of BSM scenarios can be constrained by determinations of nuclear matrix elements of scalar currents.
In this context, the leading  operators coupling spin-${1\over 2}$ dark matter $\chi$ with scalar quark bilinears can be expressed as
\begin{align} {\cal L}^\text{scalar}  = {G_F\over 2} \ \bar{\chi}\chi\ \left[
(a_S^{(u)}+a_S^{(d)})\, \bar{q}q  + (a_S^{(u)}-a_S^{(d)})\,
\bar{q}\tau^3 q  + 2 \ a_S^{(s)} \bar{q}_sq_s + \ldots\ \right],
\label{eq:scalarquarklevel}
\end{align}
where, as in previous sections, $q = (q_u,q_d)^T$, and $G_F$ is the Fermi constant included to normalize the  couplings of dark matter to the quarks. 
In many BSM scenarios, such as neutralino WIMPs in supersymmetric extensions of the SM~\cite{Goodman:1984dc,Jungman:1995df}, these are the most important operators. Scalar-isoscalar gluonic operators that mix with $\bar{q}q$ are also important in some theories, such as models of technibaryon dark matter~\cite{Nussinov:1992he,Chivukula:1992pn,Bagnasco:1993st}.
Nuclear matrix elements of the scalar currents $\tilde J_S^{(f)} = \int d^3 x\, \bar{q}_f(\vec{x},0) q_f(\vec{x},0)$ define the nuclear sigma terms:
\begin{equation}
\sigma_{Z,N}^{(f)}  \equiv g_S^{(f)}(Z,N) \,m_f \equiv
\langle Z,N| 
m_f \tilde J_S^{(f)}
| Z,N\rangle,
\label{eg:ZNsigdef}
\end{equation}
where $N$ and $Z$ are the neutron and proton numbers.
While  $ g_S^{(f)}(Z,N)$ is renormalization-scale dependent, $\sigma_{Z,N}^{(f)}$ is not.

In the impulse approximation, low-energy nuclear observables are dominated by the contributions from
individual nucleons, and the nucleon sigma terms dictate those of nuclei:
\begin{align}
\sigma^{(f)}_{Z,N}\big|_\text{impulse}  =  (N+Z)\sigma_N^{(f)},
\label{eq:naive}
\end{align}
where $\sigma_N^{(f)}$ defines the nucleon sigma term for a given quark flavor, $f$. A combination of the $u$- and $d$-quark nucleon sigma terms, $\sigma_{\pi N}=\frac{1}{2}(m_u+m_d)\langle N | \tilde J_S^{(u)}+\tilde J_S^{(d)} | N \rangle$, can be constrained by pion-nucleon scattering experiments, see Ref.~\cite{Hoferichter:2015tha} for a recent analysis. They can also be calculated from LQCD with competitive precision; in the last five years, computations with the physical values of the quark masses have been achieved with 10\%-15\% statistical and systematic uncertainties~\cite{Ren:2014vea,Yang:2015uis,Durr:2015dna,Bali:2016lvx,Abdel-Rehim:2016won,Shanahan:2016pla,Borsanyi:2020bpd}, although results for $\sigma_{\pi N}$ are in some tension with phenomenological analyses of experimental data~\cite{RuizdeElvira:2017stg}. In particular, the community consensus average of $N_f=2+1$ flavor LQCD calculations is $\sigma_{\pi N}=39.7(3.6)$~MeV (and $\sigma_{\pi N}=64.9(1.5)(13.2)$~MeV for $N_f=2+1+1$ flavor calculations)~\cite{Aoki:2019cca}, while the latest analyses of experimental data yield $\sigma_{\pi N}=58(5)$~MeV~\cite{RuizdeElvira:2017stg}.
Future more precise LQCD calculations will explore possible sources of this tension.
The strange-quark sigma term is best constrained from LQCD~\cite{Ren:2014vea,Yang:2015uis,Durr:2015dna,Bali:2016lvx,Abdel-Rehim:2016won,Borsanyi:2020bpd}, with an average of $N_f=2+1$ flavor calculations giving $\sigma_{s}=52.9(7.0)$~MeV (and a less precise result, $\sigma_{s}=41.0(8.8)~\mathrm{MeV}$, from $N_f=2+1+1$ flavor calculations)~\cite{Aoki:2019cca}, while heavy-quark contributions can be computed within the heavy-quark expansion (with LQCD providing a consistency check for the charm quark contribution~\cite{Borsanyi:2020bpd}). The impulse approximation, however, neglects multi-body contributions to the sigma terms that will be present at some level. In many aspects of nuclear structure, nuclear interactions modeled by meson-exchange currents in phenomenological models, and by higher-body operators and exchange currents in EFTs, are found to modify the impulse approximation at the few-percent level. For the sigma terms, however, it has been argued that such effects might be far more significant and, moreover, the $Z$ and $N$ dependence of the nuclear sigma terms could be significantly modified from an impulse approximation~\cite{Prezeau:2003sv,Hoferichter:2016nvd,Hoferichter:2019uwa}. In particular, although such a dependence is not expected in chiral EFT using Weinberg power counting where two-body scalar currents are subleading~\cite{Hoferichter:2015ipa}, inconsistencies in this power counting could remove this suppression in renormalizable EFTs, as seen for other operators~\cite{Valderrama:2014vra,Cirigliano:2018hja,deVries:2020loy}.
Clearly, the assumption of the impulse approximation must be tested, both to enable robust interpretation of the results of ongoing and planned direct dark-matter detection experiments, and in order to optimize the design of future experiments.

While at this stage the nuclear sigma terms of the large nuclei of experimental interest are not directly accessible from LQCD because of computational limitations, the static responses of light nuclei with $A\le 4$ to scalar currents can be computed. In particular, these quantities have been determined, via the Feynman-Hellmann theorem, from the quark-mass dependence of the masses of light nuclei~\cite{Beane:2013kca}, and more recently for $A\leq 3$ using a direct background-field method~\cite{Chang:2017eiq}. All studies to date have significant systematic uncertainties arising, for example, from quark masses that result in larger-than-physical values of the pion mass. Nevertheless, even the existing results are of phenomenological interest; as discussed above, constraining the size of deviations from the impulse approximation is particularly important to reliably convert limits on dark-matter--nucleus interaction cross-sections from experiments into a bound on the mass of dark matter particles. 

The first LQCD study of nuclear effects in the sigma terms was undertaken in Ref.~\cite{Beane:2013kca}, which focused on the light-quark contribution. Assuming isospin symmetry ($m_u=m_d=\bar{m}$), the combined $u$ and $d$ quark contribution to the nuclear sigma term can be expressed as 
\begin{align}
\sigma_{\pi (Z,N)} = \sigma^{(u+d)}_{Z,N}  = & \,\,
\bar{m}
\langle Z,N| 
J_S^{(u)}+J_S^{(d)}
| Z,N\rangle
= 
\bar{m}{d\over d\bar{m}} E_{Z,N}
\nonumber\\
= & 
\left[ 1 + {\cal O}\left(m_\pi^2\right) \right]
{m_\pi\over 2}{d\over d m_\pi}  E_{Z,N},
\label{eg:ZNsigdefisospin}
\end{align}
where $E_{Z,N}$ is the energy of the nuclear ground state, and the second line follows from the Gell-Mann--Oakes--Renner relation~\cite{GellMann:1968rz,Gasser:1983yg} (Ref.~\cite{Beane:2013kca} included a conservative uncertainty to account for deviations in that relation). In terms of the binding energy of the nucleus, $B_{Z,N}=A M_N -E_{Z,N} $, where $M_N$ is the isospin-averaged nucleon mass and $A=N+Z$, this can be re-expressed as
\begin{align}
\sigma^{(u+d)}_{Z,N}  = 
A \sigma^{(u+d)}_{N} + \sigma^{(u+d)}_{B_{Z,N}}
=  
A \sigma^{(u+d)}_{N} - {m_\pi\over 2}{d\over d m_\pi} B_{Z,N}.
\label{eg:ZNsigdefisospinbinding}
\end{align}
In this way, the light-quark nuclear sigma terms can be constrained using LQCD calculations of only the nucleon mass and the binding energies of light nuclei as a function of pion mass. Because of the high computational cost of of LQCD calculations of nuclei, the binding energies of light nuclei have, however, only been calculated at a few, widely-separated, larger-than-physical, values of the pion mass. Given this status, the finite differences used in Ref.~\cite{Beane:2013kca} to determine the slope of the binding energies with respect to the pion masses resulted in large uncertainties in the nuclear sigma terms. Nevertheless, the conclusion of that study was that nuclear effects in the light-quark sigma terms of nuclei with $A\le 4$ are at the few- to 10-percent level at $m_\pi=660$~MeV, and at the few-percent level at $m_\pi= 330$~MeV.

\begin{figure}[!h]
	\centering
	\includegraphics[width=0.75\columnwidth]{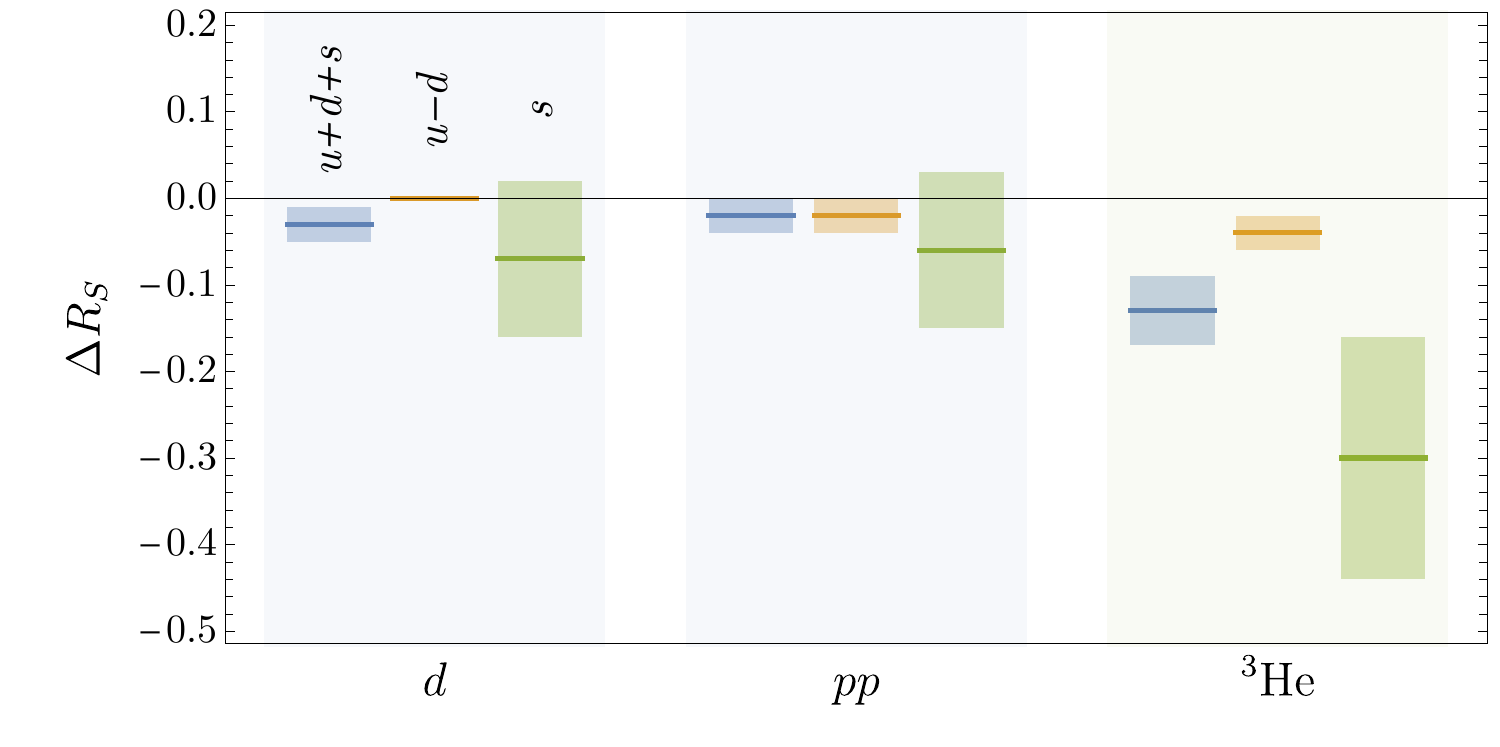}
	\includegraphics[width=0.75\columnwidth]{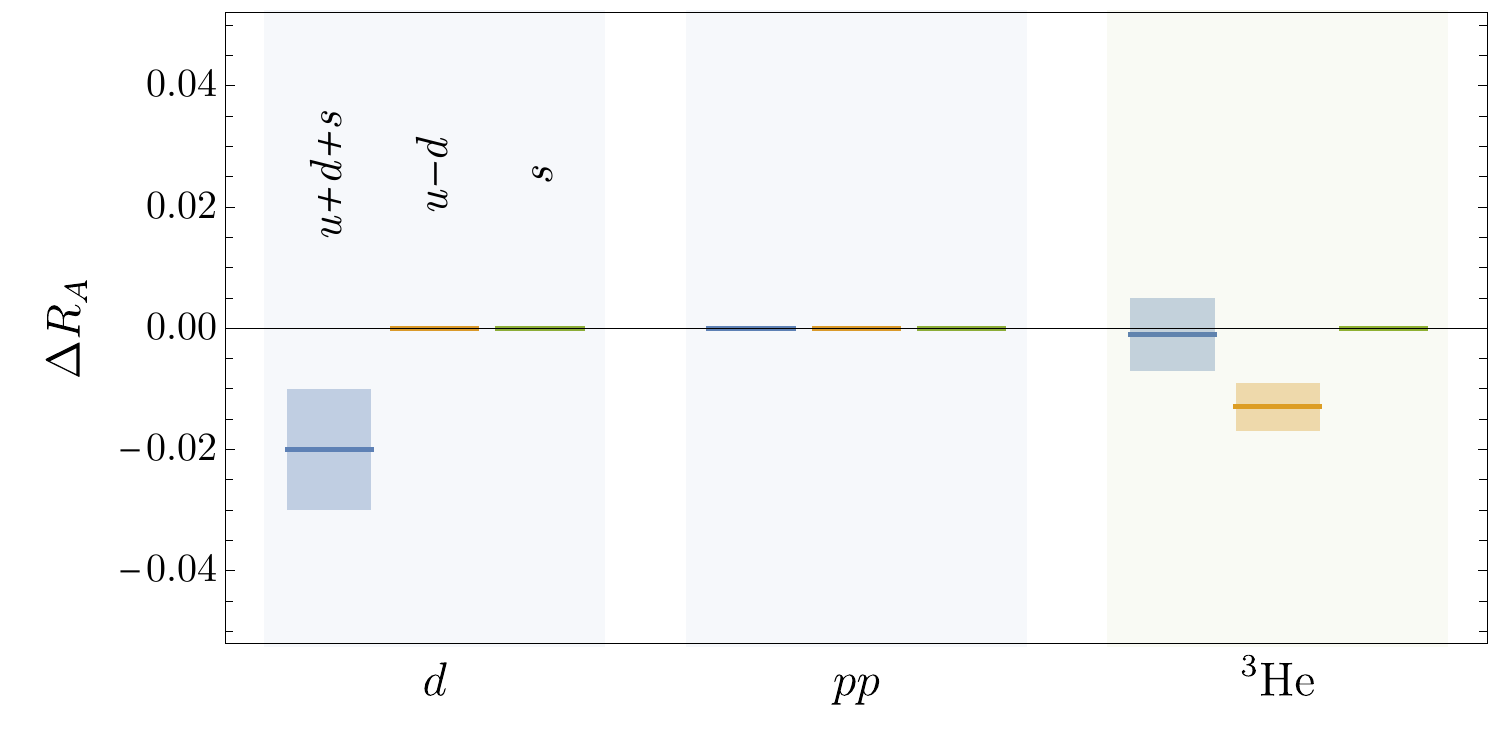}
	\includegraphics[width=0.75\columnwidth]{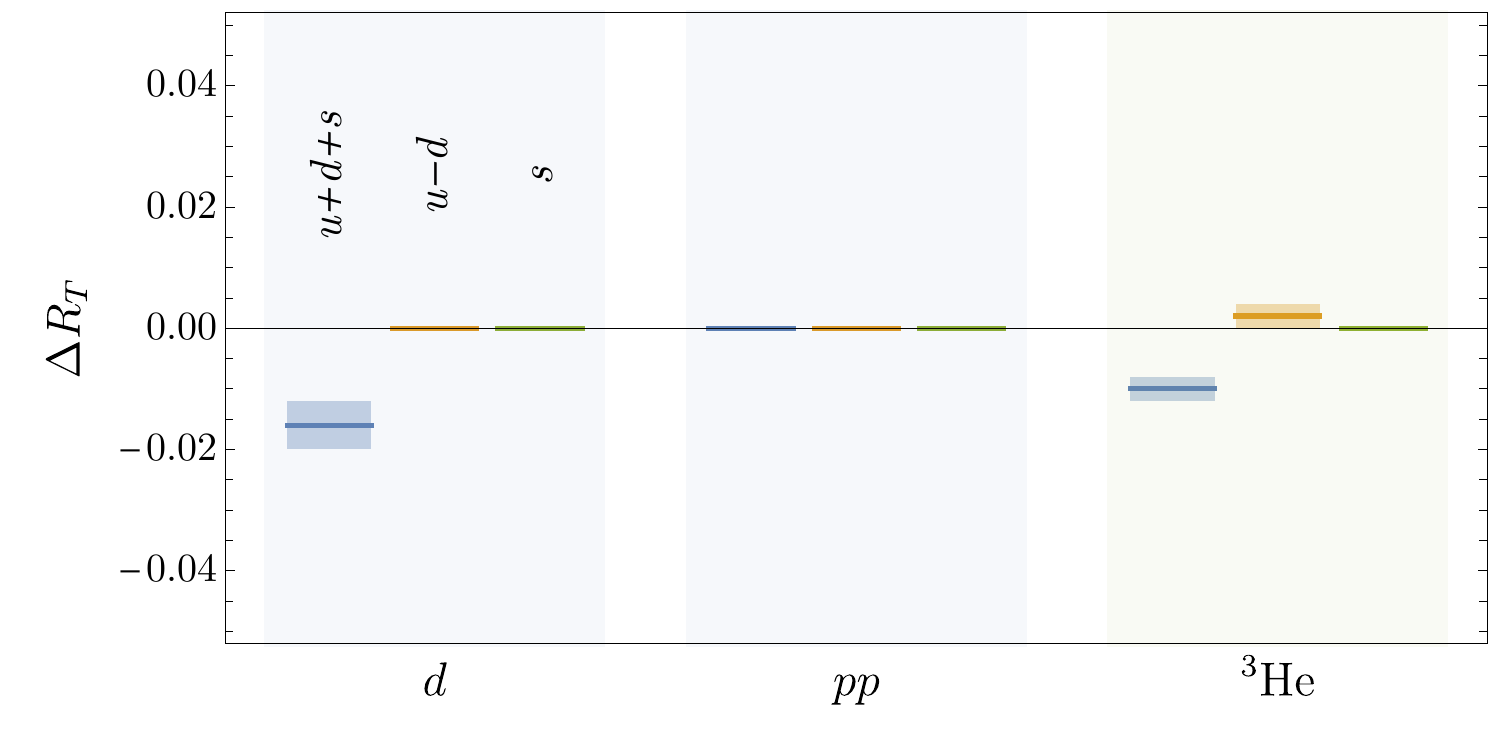}
	\caption{Differences $\Delta R_X$ between LQCD calculations of the ratios $R_X^{(f)}(A)=g_X^{(f)}(A)/g_X^{(f)}(p)$ for nuclear matrix elements $g_X^{(f)}(A)$ of nuclei $A$ with Dirac structure $X=\{S,A,T\}$ and flavor combination $f$, from their values calculated using nuclear ground states with non-interacting nucleons occupying only the lowest shell-model states. The strange-quark matrix elements are small and indistinguishable from zero for the axial and tensor matrix elements. Quantities that are identically zero are shown by lines at zero, and there is no associated uncertainty band. [Data from Ref.~\cite{Chang:2017eiq}.]
		\label{fig:SATMEs} }
\end{figure}

More recently, the complete flavor-decompositions of the nuclear sigma terms for $A\in\{2,3\}$ nuclei were calculated from LQCD for the first time using the background-field method described in Sec.~\ref{subsec:backgroundfield} to compute the matrix elements directly~\cite{Chang:2017eiq}, albeit also at larger-than-physical values of the quark masses corresponding to  $m_\pi= 806$~MeV, and in the limit of SU(3)$_f$-flavor symmetry. Moreover, in addition to the scalar matrix elements, axial and tensor Dirac structures were investigated. Explicitly, matrix elements of operators of the form $\bar{q} \Gamma_X\Lambda^{(j)} q\,$ were calculated, where $q=(q_u,q_d,q_s)^T$, $X\in\{S,A,T\}$ for Dirac structures $\Gamma_S=1$, $\Gamma_A=\gamma_5\gamma_\mu$, $\Gamma_T=i \sigma_{\mu\nu}$, and flavor structures  $\Lambda^{(3)}\equiv{\rm diag}(1,-1,0)$, $\Lambda^{(8)}\equiv{\rm diag}(1,1,-2)$, and the identity $\Lambda^{(0)}\equiv{\rm diag}(1,1,1)$.
In the notation of Ref.~\cite{Chang:2017eiq} (also as used above in Eq.~(\ref{eg:ZNsigdef})), these nuclear matrix elements are quantified by couplings $g_X^{(f)}$ times simple kinematic factors. Disconnected (equivalently for non-strange nuclei in the SU(3)$_f$ limit, the strange-quark) contributions, are defined by the difference $g_X^{(\rm disc)} = g_X^{(s)} = (g_X^{(0)} -g_X^{(8)})/3$. Of primary physical interest in the BSM-phenomenology context are the nuclear effects in these matrix elements; to isolate multi-nucleon contributions, the ratios of the nuclear matrix elements in a given nucleus $A$ to those in the proton can be calculated and compared with naive single-nucleon (NSN) estimates obtained using nuclear ground states with non-interacting nucleons occupying only the lowest shell-model states~\cite{Buck:1975ae,Krofcheck:1985fg,Chou:1993zz,Brown:1978zz,Wildenthal:1983zz}. Defining the matrix-element ratios as $R_X^{(f)}(A)=g_X^{(f)}(A)/g_X^{(f)}(p)$, the differences $\Delta R_X^{(f)}(A)=R_X^{(f)}(A)- R_X^{(f)}(A)_{\rm NSN}$ provide a measure of many-body nuclear effects in the matrix elements, since the NSN estimates $R_X^{(f)}(A)_{\rm NSN}$ correspond to the impulse approximation for each nuclear matrix element and are determined only by the baryon number, spin, and isospin quantum numbers of the nuclear state.
Figure~\ref{fig:SATMEs} summarizes the key results of Ref.~\cite{Chang:2017eiq}; multi-body effects in the nuclear matrix elements are at the few-percent level, except in the scalar matrix elements where deviations from the impulse approximation as large as $\sim10$\% are observed. 
While the nuclear effects in the axial matrix elements are found to be small in light nuclei, it is known that these effects scale with the size of the nucleus, becoming as large as 30\% in medium-mass nuclei~\cite{Wildenthal:1983zz,Buck:1975ae,MartinezPinedo:1996vz}. With the nuclear effects in the scalar matrix elements found to be at the 10\% level even in light nuclei, this comparison provides an indication of the significant level of uncertainty in the scalar nuclear matrix elements in isotopes of relevance to experiments.  
The strange-quark contributions to axial and tensor nuclear matrix elements are negligible, but are significant for the scalar matrix elements, which is consistent with studies of the same matrix elements in the proton~\cite{Green:2015wqa,Bhattacharya:2015wna,Alexandrou:2017qyt}. Moreover, for each Dirac structure, the nuclear modifications follow a scaling that is approximately dictated by the magnitude of the corresponding charge. 

In order to constrain the sigma terms of the large nuclei used in dark matter direct-detection experiments, these LQCD calculations of the relevant matrix elements in small nuclei must be matched to many-body methods based on EFTs or phenomenological models~\cite{Hoferichter:2016nvd,Hoferichter:2019uwa}, as discussed in Sec.~\ref{sec:eft} above. Computations of two-body systems can be used to determine the dominant multi-body operators, and the effects of these contributions can be verified in few-nucleon systems before the many-body techniques are used to extrapolate to the large nuclei relevant for experiment. Executing this program will require further systematic control of the LQCD calculations, and, in particular, studies with the physical values of the quark masses. Nevertheless, the existing calculations are already of some phenomenological value; the large nuclear effects which are observed in the scalar matrix elements urge caution in using the impulse approximation in the interpretation of direct searches for dark matter. If this feature persists in studies at the physical values of the quark masses and with with fully-controlled uncertainties, nuclear effects in these quantities should not be neglected.

\subsection{Tensor matrix elements}
\label{subsec:tensorMEs}

The observed matter-antimatter asymmetry in the universe provides a tantalizing hint of possible new physics; it is roughly nine orders of magnitude larger than that which could be created by purely SM interactions if the universe was matter-antimatter symmetric at the end of the inflationary epoch~\cite{Morrissey:2012db}. New interactions which violate change conjugation (C) and parity (P) symmetry, however, are naturally generated in many BSM scenarios and could explain this observation~\cite{Chupp:2017rkp,Engel:2013lsa,Kim:2008hd}. Permanent EDMs of fundamental and composite particles are CP violating as well as time-reversal symmetry (T) violating, and searches for such EDMs in systems ranging from free leptons to condensed-matter systems provide some of the most powerful probes of SM and BSM CP violation~\cite{Chupp:2017rkp,Engel:2013lsa}.
Since EDMs could have non-trivial isospin dependence, their measurement or constraint in different systems~\cite{Engel:2013lsa,Mereghetti:2015rra,Bsaisou:2014zwa,Guo:2012vf} are a key target of current and planned experiments~\cite{Semertzidis:2003iq,Semertzidis:2011qv,Pretz:2013us} aiming to place direct constraints on BSM physics. In particular, it was argued more than 30 years ago that T-violating nuclear forces could substantially enhance the EDMs~\cite{FLAMBAUM1986750} of nuclei. EFTs describing T violation in nuclei are under active investigation~\cite{Chupp:2017rkp,Yamanaka:2017mef,Bsaisou:2014oka,Maekawa:2011vs,deVries:2020loy}.

In SMEFT, T-violating quark-gluon interactions arise at dimension four, but the coupling $\bar{\theta}$ describing these interactions is constrained by neutron and nuclear EDM searches to be  $|\bar{\theta}|\lesssim 10^{-11}$~\cite{Kim:2008hd}. Assuming that $\bar{\theta}$ is suppressed due to Peccei-Quinn symmetry~\cite{Peccei:1977ur}, or another source of fine-tuning, the next sources of T violation in SMEFT arise from quark EDM operators at dimension five.
The quark EDM contributions to nuclear EDMs are encoded in tensor matrix elements through the dimension-five CP-odd operator $\bar{q}_f\sigma_{\mu\nu}\ \widetilde F^{\mu\nu}q_f$ (where $\widetilde F^{\mu\nu}=\frac{1}{2}\epsilon^{\mu\nu\rho\sigma}F_{\rho\sigma}$ is dual to the EM field-strength tensor $F^{\mu\nu}$). The tensor matrix elements of nuclei are thus needed to interpret proposed searches for EDMs in nuclear systems~\cite{Semertzidis:2003iq,Semertzidis:2011qv,Pretz:2013us,Engel:2013lsa,Mereghetti:2015rra,Wirzba:2016saz,Yamanaka:2016umw,Yamanaka:2017mef,Chupp:2017rkp}. 
As for the scalar matrix elements discussed in the previous subsection, the tensor charges of the nucleon and of light nuclei, i.e., the forward matrix elements of the tensor current discussed above, can be calculated from LQCD. For the up- and down-quark tensor charges of the nucleon, LQCD calculations have achieved a precision of 3--7\%~\cite{Aoki:2019cca}, while the strange-quark nucleon tensor charge is constrained to be much smaller than those of the light quarks~\cite{Gupta:2018lvp}. In Ref.~\cite{Chang:2017eiq}, nuclear effects in the tensor charges of light nuclei with $A< 4$ were resolved for the first time and found to be at the few-percent level in calculations with unphysically large values of the quark masses corresponding to $m_\pi= 806$~MeV. These nuclear effects were seen to be similar to the analogous nuclear effects in axial matrix elements, but far smaller than those in the scalar matrix elements. Figure~\ref{fig:SATMEs}, using data from Ref.~\cite{Chang:2017eiq}, summarizes these results. As well as their relevance to experimental searches for EDMs of light nuclei, the tensor charges also provide the hadronic input to dark-matter--nucleus scattering cross-sections in dark-matter models that generate tensor quark--dark-matter interactions~\cite{Bishara:2017pfq}. As discussed in Sec.~\ref{subsec:DM}, many direct searches for dark matter are undertaken using nuclei with $A\gg 4$; QCD-informed predictions of the tensor matrix elements of these large nuclei will require not only calculations of the nuclear tensor matrix elements with controlled systematic uncertainties, but the matching of the matrix elements for light nuclei to many-body methods based on EFTs and phenomenological models, as discussed above. If the small nuclear effects revealed in the tensor charges of light nuclei persist in controlled calculations at the physical quark masses, however, one might expect that impulse approximation will provide a better estimate for the cross-sections of scattering of models with tensor quark--dark-matter interactions than for spin-independent scattering, governed by the scalar matrix elements.

\subsection{Baryon-number violation}
\label{Sec:Bviolation}
The matter-antimatter asymmetry of the universe discussed above also motivates consideration of baryon-number-violating interactions, or more precisely interactions that violate the SM symmetry $B-L$, in the early universe \cite{Babu:2013jba}.
Interactions that violate $L$ and $B-L$ could give rise to a lepton asymmetry that is transferred to the baryon sector through electroweak sphalerons \cite{Fukugita:1986hr}. Low-energy signatures of such interactions are constrained by $0\nu\beta\beta$ searches as discussed in Sec.~\ref{sec:dbd}.
Alternatively, the baryon asymmetry could be generated through $B$ and $B-L$ violating interactions.
These interactions give rise to distinct experimental signatures including proton decay, neutron-antineutron ($n\bar{n}$) oscillations, and $B$-violating nuclear decays.
Proton decay and $|\Delta B|= 1$ nuclear decays receive dominant contributions in SMEFT from dimension-6 operators that preserve $B-L$ and, depending on the order of the electroweak phase transition, are therefore not directly relevant for baryogenesis, although they are still of interest for constraining Grand Unified Theories (GUTs) in particular.
LQCD calculations of $|\Delta B|=1$ matrix elements of the proton have a long history dating back to efforts to constrain the minimal SU(5) GUT in the 1980s~\cite{Hara:1986hk}.
Recently, direct calculations of the nucleon-to-meson transition amplitudes required to relate BSM physics parameters to the rates of decay processes, such as $p \rightarrow \pi^0 e^+$, have been performed using nearly-physical values of the light quark masses~\cite{Aoki:2017puj,Yoo:2018fyn}. Indirect calculations of proton-to-vacuum transition amplitudes related to the desired physical decay rates in chiral EFT have also been performed (see Ref.~\cite{Cirigliano:2019jig} for a review).
The same LQCD calculations can also be used to constrain searches for $B-L$ violating nucleon decays involving dimension-7 operators in SMEFT that give rise to $n\rightarrow \pi^+ e^-$ and other nucleon-decay processes~\cite{Heeck:2019kgr}.
To directly connect BSM-physics parameters to experimentally-observable decay rates in large-volume underground detectors, the matrix elements of $|\Delta B|=1$ operators should be calculated with nuclei in the initial and final states.
Such calculations have not yet been attempted in LQCD or nuclear EFT, and, for example, proton-decay constraints from Super Kamiokande~\cite{Nishino:2009aa} rely on nuclear models~\cite{Salcedo:1987md,Yamazaki:1999gz} in order to relate the proton and ${}^{16}\text{O}$ decay rates.

There has been recent progress in constraining $|\Delta B| = 2$ interactions in nuclei using a combination of LQCD and chiral EFT.
$n\bar{n}$ oscillations and $|\Delta B| = 2$ nuclear transitions are described in SMEFT by dimension-9 six-quark operators that violate $B-L$, and can arise as low-energy signatures of phenomenologically-viable baryogenesis models~\cite{Mohapatra2009,Babu:2013yww,Phillips2016}.
The nuclear matrix elements needed to constrain these models are defined by
\begin{equation}
\begin{split}
\mathcal{M}_I^{\Delta B = 2}(N,Z) = \langle N-2,Z| 
\cQ_I 
| N,Z\rangle , \qquad I\in\{1,\ldots 7\},
\end{split}\label{eq:DeltaB2}
\end{equation}
where at dimension 9 in SMEFT, the $\cQ_I$ include four SM gauge-singlet operators built from linear combinations of  $(q_u C P_{L,R} q_d)(q_u C P_{L,R} q_d)(q_d C P_{L,R} q_d)$ and  $(q_u C P_{L,R} q_u)(q_d C P_{L,R} q_d)(q_d C P_{L,R} q_d)$ with particular contractions of the color indices, where $C$ is the charge-conjugation matrix and $P_{L,R}$ project to left- and right-handed quark chiralities~\cite{Chang:1980ey,Kuo1980,Rao1984}.
The additional three independent six-quark operators arise at dimension 11 in SMEFT, accompanied by two powers of the Higgs field~\cite{Rinaldi:2019thf}, and are also of interest for BSM models of post-sphaleron baryogenesis~\cite{Babu:2006xc,Babu:2013yca}.
With $A=1$ and $Z=0$, Eq.~\eqref{eq:DeltaB2} corresponds to $n\bar{n}$ oscillations. The $n\bar{n}$ oscillation timescale is given by
\begin{equation}
\tau_{n\bar{n}} = \left| \sum_{I \in \{1,2,3,5\}} C_I \mathcal{M}_I^{\Delta B = 2}(1,0) \right|^{-1},
\end{equation}
where isospin symmetry has been used to reduce the number of independent matrix elements to five~\cite{Rinaldi:2019thf}, and the $C_I$ are  Wilson coefficients parameterizing the strength of $|\Delta B| = 2$ interactions at high scales.
The bound $\tau_{n\bar{n}} > 0.9 \times 10^8$~s obtained from cold-neutron-beam experiments~\cite{Baldo-Ceolin1994}, combined with LQCD determinations of $\mathcal{M}_I^{|\Delta B| = 2}(1,0)$, allow the $C_I$ for BSM theories of interest to be constrained.

Large-volume underground detectors provide much stronger constraints on $B$-violating nuclear decay half-lives of order $10^{31}$ years; however, nuclear lifetimes $\Gamma_{N,Z}^{-1}$ depend quadratically on the $n\bar{n}$ oscillation time in the impulse approximation as $\Gamma_{N,Z}^{-1} = R^{|\Delta B|=2}(N,Z) \tau_{n\bar{n}}^2$, where $R^{|\Delta B|=2}(N,Z)$ is a factor that must be calculated to relate constraints on $\Gamma_{N,Z}$ to constraints of the $C_I$.
Deuteron decay is the simplest $|\Delta B| = 2$ nuclear decay.
A search for deuteron decay at SNO provides a lower bound on the deuteron lifetime~\cite{Aharmim2017} by using an optical-potential model from Ref~\cite{Dover:1985hk} to determine $R^{|\Delta B|=2}(1,1)$.
In turn, this can be combined with LQCD calculations of $\mathcal{M}_I^{|\Delta B| = 2}(1,0)$~\cite{Rinaldi:2018osy,Rinaldi:2019thf} to constrain the fundamental parameters of BSM theories of $B$ violation.
More recently, chiral EFT with KSW power counting has been used in Ref.~\cite{Oosterhof:2019dlo} to verify that the impulse approximation result $\Gamma_d^{-1} = R^{|\Delta B|=2}(1,1) \tau_{n\bar{n}}^2$ is valid at LO, identify a $|\Delta B| = 2$ contact interaction arising at NLO, and calculate $R^{|\Delta B|=2}(1,1)$ in terms of the $\mathcal{M}_I^{|\Delta B| = 2}(1,0)$ matrix elements at NLO using a NDA  estimate of the unknown $|\Delta B| = 2$ LEC.
This chiral  EFT result for $R^{|\Delta B|=2}(1,1)$ is a factor of two larger than the earlier result of Ref.~\cite{Dover:1985hk}, and consequently turns the SNO constraints on $\Gamma_d$ into constraints on $\tau_{n\bar{n}}$ that are about a factor of two stronger than the cold-neutron-beam constraints.
However, chiral-EFT calculations using Weinberg power counting~\cite{Haidenbauer:2019fyd} show significant differences arising from non-perturbative one-pion-exchange effects in the deuteron initial state, and favor a value closer to the optical potential result.
Future LQCD calculations of the deuteron decay rate would provide a valuable test of the validity of various EFT power countings in $B$-violating amplitudes and allow LECs associated with $|\Delta B|=2$ contact interactions to be reliably determined.
Experimentally-relevant deuteron decay modes such as $d \rightarrow \pi^+ \pi^+ \pi^-$ include complicated FV effects associated with transition amplitudes for three-hadron states. Studies of three-hadron systems from LQCD may allow direct access to such matrix element in the  future, however more immediate progress in constraining $|\Delta B|=2$ chiral EFT interactions using LQCD may be possible by matching matrix-element results for processes such as $d \rightarrow \pi^+ \bar{\nu}$ that avoid these FV complications.
LQCD calculations of more inclusive processes such as the total deuteron decay rate could potentially avoid these challenges through the use of spectral reconstruction techniques~\cite{Bulava:2019kbi,Hansen:2017mnd} to extract total decay rates from Euclidean correlation functions, but these approaches face separate challenges related to inverting a Laplace transform.
Further studies are needed to explore the impact of LQCD constraints on EFTs and models of $B$ violation in nuclei on the interpretation of searches for $B$ violation at future detectors such as Hyper-Kamiokande and DUNE~\cite{Barrow:2019viz}.

\subsection{Future impact}

Interpreting the results of experimental searches for BSM physics in many scenarios requires matrix elements which encode BSM couplings to the SM. For intensity-frontier experiments using nuclear targets, it is the corresponding nuclear matrix elements of BSM operators that are needed. 
As discussed in Secs.~\ref{subsec:DM} and \ref{subsec:tensorMEs}, calculations of the scalar and tensor matrix elements of light nuclei are primarily important as constraints on nuclear many-body approaches. While studies with the physical quark masses are necessary to undertake such a matching program, the recent LQCD calculations of the scalar and tensor matrix elements in light nuclei with $A\leq 3$, and unphysically-large values of the quark masses, have already provided phenomenologically-relevant information about dark-matter--nucleus scattering cross-sections in scenarios where dark matter couples spin-independently to nuclei, and about the quark contributions to nuclear EDMs. In particular, large nuclear effects in the scalar matrix elements in light nuclei urge caution in using an impulse approximation to estimate scalar matrix elements of the large nuclei used in dark-matter direct-detection experiments. Calculations of these matrix elements with fully-controlled systematic uncertainties can be anticipated within the next decade. Complete systematic control will require not only studies with quark masses tuned to match the physical hadron masses, but also investigation of lattice-spacing and volume dependence, and studies of operator mixing with gluon operators under renormalization (in the case of isoscalar matrix elements). Ultimately, controlled LQCD determinations of these matrix elements will reduce the theory uncertainty in the response of nuclei to probes relevant to BSM-physics scenarios and allow a rigorous uncertainty quantification in the interpretation of BSM-physics searches.

There is also potential for calculations of scalar matrix elements in light nuclei that are within reach of near-future LQCD studies to provide direct input to experimental searches. For example, BSM physics that produces additional interactions between atomic electrons and the nucleus would lead to small shifts in atomic energy levels that can be tested through optical measurements of frequency shifts between pairs of isotopes of hydrogen and helium atoms, light ions including lithium and nitrogen, as well as heavy atoms and ions~\cite{Berengut:2017zuo,Delaunay:2017dku}. Constraining the most relevant contributions requires SM knowledge of scalar-current matrix elements in these light nuclei, as well as the differences between charge radii of the isotope pairs. 

In addition to the scalar and tensor matrix elements, other key nuclear matrix elements will be calculable using LQCD on the same timescale, including nuclear matrix elements of operators such as $\bar{q}_f \gamma_{\{\mu} D_{\nu\}} q_f$ (where the braces indicate symmetrization and trace-subtraction) which will constrain models of velocity-dependent dark matter,  and nuclear matrix elements of dimension-9 operators relevant for $B-L$ violating decays~\cite{Oosterhof:2019dlo,Abe:2011ky}. As discussed in Sec.~\ref{Sec:Bviolation}, the most stringent constraints on $B-L$ violation through the $n\bar{n}$ oscillation process are obtained by experiments searching for its occurrence inside nuclei such as the deuteron and oxygen~\cite{Babu:2013yww}. Accurate interpretation of observations of $B$-violating decays requires SM knowledge of the nuclear matrix elements of the six-quark operators responsible for the decay. LQCD calculations of these matrix elements in light nuclei will constrain nuclear EFTs and phenomenological models, and will thereby provide valuable input to these experimental programs in the coming years.

Finally, another possible role for LQCD studies in informing searches for BSM physics is in the context of the upcoming muon-to-electron-conversion experiment (mu2e) at Fermilab \cite{Bartoszek:2014mya}, which will search for lepton-flavor violation through the conversion of a muon to an electron in the field of an aluminum nucleus. While neutrino oscillations allow for this process in the SM, their contribution is many orders of magnitude smaller than the sensitivity of the experiment \cite{Cheng:1980tp}. In many BSM models, a larger lepton-flavor-violation signal is expected and a significant role may be played by four-fermion operators of the form $(\bar{e}\Gamma\mu)(\bar{q}\Gamma'q)$ \cite{Kuno:1999jp}; nuclear matrix elements of the hadronic part of these four-fermion operators is thus a key target of nuclear-physics studies. This is again an example where the primary role of future LQCD studies will be to constrain nuclear many-body approaches which can reach larger nuclei than will be feasible to study in LQCD directly, and success will require continued investment and advances both in LQCD and in nuclear many-body methods, bridged by phenomenological models or nuclear EFTs.

\section{Outlook}
\label{sec:future}
\noindent
Building on more than forty years of intense development, a new era in first-principles studies of the SM of particle physics has emerged.
Fully-controlled LQCD calculations, including QCD and QED effects, have become tractable for aspects of single-hadron structure, and encouraging first calculations of nuclear structure, reactions, and matrix elements involving light nuclei have been undertaken. The field is entering a period in which first-principles calculations of low-energy aspects of nuclear structure and matrix elements, with complete quantification of uncertainties, are becoming practical.  This new capability is anticipated to have far-reaching impacts on diverse experimental programs in the form of SM information about nuclear effects that can be used as input into experimental measurements of more complicated processes, and as SM benchmarks in searches for new physics. With sufficiently precise calculations, the significant impact that LQCD has had in the realm of quark-flavor physics, and the resulting tight coupling of fundamental theory with experiment, will be extended into the nuclear realm. 

Theoretical control of electroweak and BSM nuclear matrix elements in particular will impact experiments designed to refine our understanding, and predictive capabilities, of the evolution of the universe and to hunt for evidence of BSM physics. This review has highlighted the last decade of progress on this front, as well as key results that can be anticipated in the future. In the context of the electromagnetic properties of nuclei, controlled first-principles calculations will improve constraints on nuclear polarizabilities and the electromagnetic radii of nuclei.  Experiments to constrain neutron properties are typically undertaken using the deuteron, and in this context the control of the nuclear effects in light nuclei afforded by LQCD will be particularly impactful.
Speculatively, future lattice QCD+QED calculations, in which the strength of electromagnetic interaction is an input parameter,
may also improve our understanding of nuclear astrophysics and the BBN chain reaction via theory constraints on reaction cross-sections of light nuclei.
Calculations of the nuclear matrix elements of electroweak currents will improve our understanding of light nuclei and hypernuclei, and impact astrophysical investigations into the nature of dense matter. Simultaneously, calculations of the axial matrix elements and form factors of light nuclei are projected to provide important constraints on neutrino-nucleus scattering cross-sections. In particular, improved theory constraints are required to maximize the potential of long-baseline neutrino  experiments,  such as  DUNE, whose mission is to determine the neutrino mass hierarchy and oscillation parameters, and search for baryon-number violation and other BSM processes. By constraining nuclear many-body methods required for accurate neutrino-energy reconstruction, LQCD will have an important role to play in this arena.
Theoretical understanding of weak nuclear matrix elements is also important to the interpretation of atomic and nuclear parity-violation experiments and constraining parity-violating pion-nucleon and nucleon-nucleon interactions. With first-principles calculations of 
important subprocesses in $2\nu\beta\beta$ and  $0\nu\beta\beta$
decay, and of BSM matrix elements in the context of dark-matter--nucleus interactions, baryon-number violating decays, and permanent electric dipole moments, LQCD will ultimately provide input to key experimental design decisions, such as the choices of target materials, to maximize the potential of BSM physics discovery. 

Beyond electroweak and BSM nuclear MEs, modifications to the 
partonic structure of nuclei will be informed by LQCD calculations that are presently in  the early stages of development, quantitatively elucidating the QCD origin of important aspects of nuclear structure such as the EMC effect.
For example, at the EIC facility, planned for construction at Brookhaven National Laboratory over the next decade, LQCD calculations will provide SM benchmarks for measurements of the quark and gluon structure of nuclei, including moments of parton distribution functions and their flavor dependence. 
With theoretical control of nuclear effects from LQCD, measurements of nuclear PDFs are expected to improve the flavor separation of proton PDFs and thereby reduce the effect of what is currently a leading uncertainty in searches for new physics in proton-proton collisions at the LHC and in deep-inelastic scattering experiments.

Critical to achieving the promise of first-principles nuclear physics is the continued development and support of the HPC hardware and software needed to undertake these demanding computations. The availability of exascale computing resources to basic nuclear-physics research is crucial to this mission. Optimal use of these resources requires continued development of software and algorithms for LQCD, and further investigation of novel tools such as machine learning, and new hardware such as field-programmable gate arrays and various forms of quantum computing. 
Moreover, since it is anticipated that the controlled first-principles calculations of nuclear matrix elements that will be achieved in the next decade will remain limited to studies of light nuclei with $A\lesssim 6$, impact on experiments involving heavier nuclei will likely continue to require matching to nuclear EFTs and phenomenological models.
To capitalize on the coming era of controlled LQCD for nuclear physics will thus require continued effort to develop robust methods for propagating uncertainties from LQCD, with the corresponding systematic extrapolations in lattice spacing and volume, through matching to many-body methods, to the effects of the theory uncertainties on experiment. In many cases, these pipelines do not yet exist, and assumptions from nuclear models are embedded into analysis frameworks.
Achieving the full promise of LQCD calculations of nuclear structure and interactions will thus require coupled efforts in computing and algorithms, theory and phenomenology, and lattice field theory itself.

In summary, the status and prospects of using LQCD to calculate the response of nuclei to electroweak and BSM interactions have been reviewed.  Such calculations have a central role in the search for new physics and in precision tests of the SM.
First LQCD calculations in light nuclei have now emerged and,
with expected increases in HPC resources, the next decade will see precision calculations performed in light nuclei with complete uncertainty quantification at the physical quark masses.

\section*{Acknowledgements}

We thank the many members of the international nuclear, high-energy, and computational physics communities who have undertaken or supported, directly or indirectly, the work reviewed here. We also thank our various collaborators, both inside and outside the NPLQCD Collaboration, who have contributed to our understanding of the subject of this review over many years. We express our particular gratitude to Silas Beane, Marc Illa, David B Kaplan, and Brian Tiburzi. 

ZD is supported by Alfred P. Sloan fellowship, by the U.S. Department of Energy (DOE) Office of Science Early Career Award, under award no. DESC0020271, and by Maryland Center for Fundamental Physics at the University of Maryland, College Park. WD and PES acknowledge support from the U.S.~DOE grant DE-SC0011090.
WD is also supported within the framework of the TMD Topical Collaboration of the U.S.~DOE Office of Nuclear Physics, and  by the SciDAC4 award DE-SC0018121.
PES is additionally supported by the National Science Foundation under CAREER Award 1841699 and under EAGER grant 2035015, by a NEC research award, and by the Carl G and Shirley Sontheimer Research Fund. 
KO was supported in part by U.S.~DOE grant \mbox{
	DE-FG02-04ER41302} and in part by the Jefferson
Science Associates, LLC under U.S.~DOE Contract DE-AC05-06OR23177.
AP acknowledges support from the Spanish Ministerio de Econom\'{\i}a y Competitividad (MINECO) under the project MDM-2014-0369 of ICCUB (Unidad de Excelencia “Mar\'{\i}a de Maeztu”) and from additional European FEDER funds under the contract FIS2017-87534-P.
MJS is supported by the Institute for Nuclear Theory under U.S.~DOE grant DE-FG02-00ER41132. 
This manuscript has been authored by Fermi Research Alliance, LLC under Contract No. DE-AC02-07CH11359 with the U.S. Department of Energy, Office of Science, Office of High Energy Physics.

\pagebreak

\bibliography{review}
\end{document}